\documentclass[12pt,a4paper,headings=twolinechapter]{book}
\usepackage[utf8]{inputenc}
\usepackage[english]{babel}
\usepackage{amsmath}
\usepackage{amssymb}
\usepackage{amsfonts}
\usepackage{amsthm}
\usepackage{slashed}
\usepackage{MnSymbol}
\makeatletter
\g@addto@macro\bfseries{\boldmath}
\makeatother
\newtheorem{thm}{Theorem}[section]

\theoremstyle{plain}
\newcommand{\thistheoremname}{}
\newtheorem{genericthm}[thm]{\thistheoremname}

\newtheorem{definition}{Definition}
\usepackage{graphicx}
\usepackage{subcaption}
\graphicspath{{images/}} 
\newcommand\x{.75}

\newcommand\vertexsize{.07}
\newcommand\scaleCC{1.3}

\newcommand{\IllustrationIntro}{\begin{tikzpicture}[baseline={([yshift=-1ex]current bounding box.center)}]
\draw[wilson] (0,2) -- (0,-2);
\draw[fill=white] (0,1) circle (.2);
\draw[fill=white] (0,-1) circle (.2);
\draw[fill=white] (2,0) circle (.2);
\draw[scalar,-Latex] (0,2.2) -- (0,3);
\node[] at (.35,2.5) {$\tau$};
\node[] at (2.6,-.05) {$\Op_1$};
\node[] at (-.65,1) {$\Oh_2$};
\node[] at (-.65,-1) {$\Oh_3$};
\end{tikzpicture}}

\newcommand{\OPE}{\begin{tikzpicture}[baseline={([yshift=-1ex]current bounding box.center)}]
\draw[wilson] (-1.5,0) -- (1.5,0);
\draw[wilson,SCRed] (-.75,0) -- (.75,0);
\draw[fill=white] (-.75,0) circle (.1);
\draw[fill=white] (.75,0) circle (.1);
\draw[fill=white] (0,1.25) circle (.1);
\node[] at (0,1.75) {$\Op_{\Delta_1}$};
\node[] at (-.75+.1,-.5) {$\Oh_{\Dh_2}$};
\node[] at (.75+.1,-.5) {$\Oh_{\Dh_3}$};
\node[scale=1.5] at (3.4,0) {$= \sum\limits_{\Dh} \lambdah_{\Dh_2 \Dh_3 \Dh}$};
\draw[wilson] (-.75+6,0) -- (.75+6,0);
\draw[fill=white] (0+6,0) circle (.1);
\draw[fill=white] (0+6,1.25) circle (.1);
\node[] at (0+6,1.75) {$\Op_{\Delta_1}$};
\node[] at (0+.1+6,-.5) {$\Oh_{\Dh}$};
\end{tikzpicture}}

\newcommand{\WittenDiagrams}{\scalebox{2.5}{\begin{tikzpicture}[baseline={([yshift=-.55ex]current bounding box.center)}]
 \pgfmathsetmacro{\x}{.8*sqrt(2)/2}
 \pgfmathsetmacro{\y}{.8*sqrt(2)/2}
 \pgfmathsetmacro{\z}{.8*.5}
 \draw [wilson, SCRed] (-\x,\y) to[out=-60,in=60] (-\x,-\y);
 \draw [thick] (0,0) circle [radius=.8];
 \draw[fill=white] (.8, 0) circle (.075);
 \draw[fill=white] (-\x,\y) circle (.075);
 \draw[fill=white] (-\x,-\y) circle (.075);
 \node[scale=.5] at (.2,0) {AdS$_5$};
 \node[scale=.5,SCRed] at (-.15,-.35) {AdS$_2$};
 \node[scale=.5] at (1,.45) {$\pd$AdS$_5$};
 \node[scale=.5,SCRed] at (-1.6,-\y) {$\pd$AdS$_2$};
 \draw[-Latex,SCRed] (-1.25,-\y)--(-.75,-\y);
 \node[scale=.5] at (1.1,0) {$\Op_{\Delta_1}$};
 \node[scale=.5,SCRed] at (-\x,\y+.3) {$\Oh_{\Dh_2}$};
 \node[scale=.5,SCRed] at (-\x,-\y-.25) {$\Oh_{\Dh_3}$};
\end{tikzpicture}}}

\newcommand{\WittenDiagramOne}{\begin{tikzpicture}[baseline={([yshift=-.55ex]current bounding box.center)}]
 \pgfmathsetmacro{\x}{.8*sqrt(2)/2}
 \pgfmathsetmacro{\y}{.8*sqrt(2)/2}
 \pgfmathsetmacro{\z}{.8*.5}
 \draw [wilson, SCRed] (-\x,\y) to[out=-60,in=60] (-\x,-\y);
 \draw [thick] (0,0) circle [radius=.8];
 \draw [scalar] (-\z, 0) -- (.8, 0);
 \draw [scalar] (-\x,\y) -- (-\x,-\y);
 \draw[fill=WCOrange,WCOrange] (-.4,0) circle (\vertexsize);
 \draw[fill=white] (.8, 0) circle (.075);
 \draw[fill=white] (-\x,\y) circle (.075);
 \draw[fill=white] (-\x,-\y) circle (.075);
\end{tikzpicture}}
\newcommand{\WittenDiagramTwo}{\begin{tikzpicture}[baseline={([yshift=-.55ex]current bounding box.center)}]
 \pgfmathsetmacro{\x}{.8*sqrt(2)/2}
 \pgfmathsetmacro{\y}{.8*sqrt(2)/2}
 \pgfmathsetmacro{\z}{.8*.5}
 \draw [wilson, SCRed] (-\x,\y) to[out=-60,in=60] (-\x,-\y);
 \draw [thick] (0,0) circle [radius=.8];
 \draw [scalar] (-\z, 0) -- (.8, 0);
 \draw [scalar] (-.4,0) .. controls (-.3,.4) and (-.3,.2) .. (-\x,\y);
 \draw [scalar] (-\x,\y) -- (-\x,-\y);
 \draw[fill=WCOrange,WCOrange] (-.4,0) circle (\vertexsize);
 \draw[fill=white] (.8, 0) circle (.075);
 \draw[fill=white] (-\x,\y) circle (.075);
 \draw[fill=white] (-\x,-\y) circle (.075);
\end{tikzpicture}}
\newcommand{\WittenDiagramThree}{\begin{tikzpicture}[baseline={([yshift=-.55ex]current bounding box.center)}]
 \pgfmathsetmacro{\x}{.8*sqrt(2)/2}
 \pgfmathsetmacro{\y}{.8*sqrt(2)/2}
 \pgfmathsetmacro{\z}{.8*.5}
 \draw [wilson, SCRed] (-\x,\y) to[out=-60,in=60] (-\x,-\y);
 \draw [thick] (0,0) circle [radius=.8];
 \draw [scalar] (-\x, \y) -- (.8, 0);
 \draw [scalar] (-\x, -\y) -- (.8, 0);
 \draw[fill=white] (.8, 0) circle (.075);
 \draw[fill=white] (-\x,\y) circle (.075);
 \draw[fill=white] (-\x,-\y) circle (.075);
\end{tikzpicture}}
\newcommand{\WittenDiagramFour}{\begin{tikzpicture}[baseline={([yshift=-.55ex]current bounding box.center)}]
 \pgfmathsetmacro{\x}{.8*sqrt(2)/2}
 \pgfmathsetmacro{\y}{.8*sqrt(2)/2}
 \pgfmathsetmacro{\z}{.8*.5}
 \draw [wilson, SCRed] (-\x,\y) to[out=-60,in=60] (-\x,-\y);
 \draw [thick] (0,0) circle [radius=.8];
 \draw [scalar] (-\x, \y) -- (0, 0);
 \draw [scalar] (-\x, -\y) -- (0, 0);
 \draw [scalar] (0, 0) -- (.8, 0);
 \draw[fill=WCOrange,WCOrange] (0,0) circle (\vertexsize);
 \draw[fill=white] (.8, 0) circle (.075);
 \draw[fill=white] (-\x,\y) circle (.075);
 \draw[fill=white] (-\x,-\y) circle (.075);
\end{tikzpicture}}

 \newcommand{\BulkDefectDefectLO}{\begin{tikzpicture}[baseline={([yshift=-2ex]current bounding box.center)}]
\draw[wilson] (0,0) -- (1.1,0);
\draw[scalar] (.42,0) -- (.42,1);
\draw[scalar] (.68,0) -- (.68,1);
\draw[fill=white] (.55,1) circle (.18);
\draw[fill=white] (.42,.88) circle (.075);
\draw[fill=white] (.68,.88) circle (.075);
\draw[fill=white] (.42,0) circle (.075);
\draw[fill=white] (.68,0) circle (.075);
\end{tikzpicture}}

\newcommand{\BulkDefectDefectNLOSEOne}{\begin{tikzpicture}[baseline={([yshift=-1ex]current bounding box.center)}]
\draw[wilson] (0,0) -- (1.1,0);
\draw[scalar] (.2,0) -- (.42,.88);
\draw[scalar] (.9,0) -- (.68,.88);
\draw[fill=white] (.55,1) circle (.18);
\draw[fill=white] (.42,.88) circle (.075);
\draw[fill=white] (.68,.88) circle (.075);
\draw[fill=white] (.2,0) circle (.075);
\draw[fill=white] (.9,0) circle (.075);
\draw[fill=SEColor,SEColor] (.315,.44) circle (.15);
\end{tikzpicture}}
\newcommand{\BulkDefectDefectNLOSETwo}{\begin{tikzpicture}[baseline={([yshift=-1ex]current bounding box.center)}]
\draw[wilson] (0,0) -- (1.1,0);
\draw[scalar] (.2,0) -- (.42,.88);
\draw[scalar] (.9,0) -- (.68,.88);
\draw[fill=white] (.55,1) circle (.18);
\draw[fill=white] (.42,.88) circle (.075);
\draw[fill=white] (.68,.88) circle (.075);
\draw[fill=white] (.2,0) circle (.075);
\draw[fill=white] (.9,0) circle (.075);
\draw[fill=SEColor,SEColor] (.785,.44) circle (.15);
\end{tikzpicture}}
\newcommand{\BulkDefectDefectNLOX}{\begin{tikzpicture}[baseline={([yshift=-1ex]current bounding box.center)}]
\draw[wilson] (0,0) -- (1.1,0);
\draw[scalar] (.35,0) -- (.68,.88);
\draw[scalar] (.75,0) -- (.42,.88);
\draw[fill=white] (.55,1) circle (.18);
\draw[fill=white] (.42,.88) circle (.075);
\draw[fill=white] (.68,.88) circle (.075);
\draw[fill=white] (.35,0) circle (.075);
\draw[fill=white] (.75,0) circle (.075);
\draw[fill=VertexColor,VertexColor] (.55,.53) circle (\vertexsize);
\end{tikzpicture}}
\newcommand{\BulkDefectDefectNLOH}{\begin{tikzpicture}[baseline={([yshift=-1ex]current bounding box.center)}]
\draw[wilson] (0,0) -- (1.1,0);
\draw[scalar] (.2,0) -- (.42,.88);
\draw[scalar] (.9,0) -- (.68,.88);
\draw[gluon] (.31,.44) -- (.79,.44);
\draw[fill=white] (.55,1) circle (.18);
\draw[fill=white] (.42,.88) circle (.075);
\draw[fill=white] (.68,.88) circle (.075);
\draw[fill=white] (.2,0) circle (.075);
\draw[fill=white] (.9,0) circle (.075);
\draw[fill=VertexColor,VertexColor] (.31,.44) circle (\vertexsize);
\draw[fill=VertexColor,VertexColor] (.79,.44) circle (\vertexsize);
\end{tikzpicture}}
\newcommand{\BulkDefectDefectNLOYOne}{\begin{tikzpicture}[baseline={([yshift=-1ex]current bounding box.center)}]
\draw[wilson] (0,0) -- (1.1,0);
\draw[scalar] (.35,0) -- (.42,.88);
\draw[scalar] (.75,0) -- (.68,.88);
\draw[gluon] (.385,.44) -- (0,.44);
\draw[fill=white] (.55,1) circle (.18);
\draw[fill=white] (.42,.88) circle (.075);
\draw[fill=white] (.68,.88) circle (.075);
\draw[fill=white] (.35,0) circle (.075);
\draw[fill=white] (.75,0) circle (.075);
\draw[fill=VertexColor,VertexColor] (.385,.44) circle (\vertexsize);
\draw[fill=VertexColor,VertexColor] (.15,0) circle (\vertexsize);
\draw[fill=VertexColor,VertexColor] (.95,0) circle (\vertexsize);
\draw[fill=VertexColor,VertexColor] (.55,0) circle (\vertexsize);
\end{tikzpicture}}
\newcommand{\BulkDefectDefectNLOYTwo}{\begin{tikzpicture}[baseline={([yshift=-1ex]current bounding box.center)}]
\draw[wilson] (0,0) -- (1.1,0);
\draw[scalar] (.35,0) -- (.42,.88);
\draw[scalar] (.75,0) -- (.68,.88);
\draw[gluon] (.715,.44) -- (1.1,.44);
\draw[fill=white] (.55,1) circle (.18);
\draw[fill=white] (.42,.88) circle (.075);
\draw[fill=white] (.68,.88) circle (.075);
\draw[fill=white] (.35,0) circle (.075);
\draw[fill=white] (.75,0) circle (.075);
\draw[fill=VertexColor,VertexColor] (.715,.44) circle (\vertexsize);
\draw[fill=VertexColor,VertexColor] (.15,0) circle (\vertexsize);
\draw[fill=VertexColor,VertexColor] (.95,0) circle (\vertexsize);
\draw[fill=VertexColor,VertexColor] (.55,0) circle (\vertexsize);
\end{tikzpicture}}
\newcommand{\BulkDefectDefectNLOFoneOne}{\begin{tikzpicture}[baseline={([yshift=-2ex]current bounding box.center)}]
\draw[wilson] (-.5,0) -- (1.6,0);
\draw[scalar] (-.12,0) -- (.42,.88);
\draw[scalar] (.95,.44) -- (.68,.88);
\draw[fill=white] (.55,1) circle (.18);
\draw[fill=white] (.42,.88) circle (.075);
\draw[fill=white] (.68,.88) circle (.075);
\draw[fill=white] (.35,0) circle (.075);
\draw[fill=white] (.75,0) circle (.075);
\draw[fill=VertexColor,VertexColor] (-.12,0) circle (\vertexsize);
\draw[fill=VertexColor,VertexColor] (.125,0) circle (\vertexsize);
\draw[fill=VertexColor,VertexColor] (1.2,0) circle (\vertexsize);
\path[clip] (-.4,0) rectangle (1.9,1.2);
\draw[scalar] (.55,0) circle (.2);
\draw[fill=white] (.55,1) circle (.18);
\draw[fill=white] (.42,.88) circle (.075);
\draw[fill=white] (.68,.88) circle (.075);
\draw[fill=white] (.35,0) circle (.075);
\draw[fill=white] (.75,0) circle (.075);
\draw[fill=VertexColor,VertexColor] (-.12,0) circle (\vertexsize);
\draw[fill=VertexColor,VertexColor] (.125,0) circle (\vertexsize);
\draw[fill=VertexColor,VertexColor] (1.2,0) circle (\vertexsize);
\end{tikzpicture}}

\newcommand{\BulkDefectDefectNLOFoneTwo}{\begin{tikzpicture}[baseline={([yshift=-.2ex]current bounding box.center)}]
\draw[wilson] (-.25,0) -- (1.35,0);
\draw[scalar] (.42,0) -- (.42,.88);
\draw[scalar] (.68,.44) -- (.68,.88);
\draw[fill=white] (.55,1) circle (.18);
\draw[fill=white] (.42,.88) circle (.075);
\draw[fill=white] (.68,.88) circle (.075);
\draw[fill=white] (0,0) circle (.075);
\draw[fill=white] (1.1,0) circle (.075);
\draw[fill=VertexColor,VertexColor] (.42,0) circle (\vertexsize);
\draw[fill=VertexColor,VertexColor] (.72,0) circle (\vertexsize);
\path[clip] (-.4,0) rectangle (1.9,-.41);
\draw[scalar] (.55,.2) circle (.6);
\draw[wilson] (-.1,0) -- (1.2,0);
\draw[scalar] (.42,0) -- (.42,.88);
\draw[scalar] (.68,.44) -- (.68,.88);
\draw[fill=white] (.55,1) circle (.18);
\draw[fill=white] (.42,.88) circle (.075);
\draw[fill=white] (.68,.88) circle (.075);
\draw[fill=white] (0,0) circle (.075);
\draw[fill=white] (1.1,0) circle (.075);
\draw[fill=VertexColor,VertexColor] (.42,0) circle (\vertexsize);
\draw[fill=VertexColor,VertexColor] (.72,0) circle (\vertexsize);
\end{tikzpicture}}

\newcommand{\BulkDefectDefectNLOFoneThree}{\begin{tikzpicture}[baseline={([yshift=-1ex]current bounding box.center)}]
\draw[wilson] (-.5,0) -- (1.6,0);
\draw[scalar] (.15,.44) -- (.42,.88);
\draw[scalar] (1.22,0) -- (.68,.88);
\draw[fill=white] (.55,1) circle (.18);
\draw[fill=white] (.42,.88) circle (.075);
\draw[fill=white] (.68,.88) circle (.075);
\draw[fill=white] (.35,0) circle (.075);
\draw[fill=white] (.75,0) circle (.075);
\draw[fill=VertexColor,VertexColor] (1.22,0) circle (\vertexsize);
\draw[fill=VertexColor,VertexColor] (1.45,0) circle (\vertexsize);
\path[clip] (-.4,0) rectangle (1.9,1.2);
\draw[scalar] (.55,0) circle (.2);
\draw[fill=white] (.55,1) circle (.18);
\draw[fill=white] (.42,.88) circle (.075);
\draw[fill=white] (.68,.88) circle (.075);
\draw[fill=white] (.35,0) circle (.075);
\draw[fill=white] (.75,0) circle (.075);
\draw[fill=VertexColor,VertexColor] (1.22,0) circle (\vertexsize);
\draw[fill=VertexColor,VertexColor] (1.45,0) circle (\vertexsize);
\end{tikzpicture}}

\newcommand{\BulkDefectDefectGeneralLeadingOrder}{\begin{tikzpicture}[baseline={([yshift=-1ex]current bounding box.center)}]
\draw[wilson] (-1.2,0) -- (3.2,0);
\draw[multiscalar] (.9,1.5) to[out=50+180,in=75] (0,0);
\draw[multiscalar] (1.1,1.5) to[out=-50,in=15+90] (2,0);
\draw[multiscalar] (0,0) to[out=-75,in=-15-90] (2,0);
\draw[scalar] (-.4,0) to[out=75,in=180] (1,1.55);
\draw[scalar] (-1,0) to[out=90,in=180-25] (1,1.6);
\draw[fill=VertexColor,VertexColor] (-.4,0) circle (\vertexsize);
\draw[fill=VertexColor,VertexColor] (-1,0) circle (\vertexsize);
\node[scale=.7] at (-.5,.65) {\footnotesize $\ldots$};
\draw[scalar] (.5,0) to[out=90,in=-90] (.95,1.5);
\draw[scalar] (1.5,0) to[out=90,in=-90] (1.05,1.5);
\draw[fill=VertexColor,VertexColor] (.5,0) circle (\vertexsize);
\draw[fill=VertexColor,VertexColor] (1.5,0) circle (\vertexsize);
\node[scale=.7] at (1,.5) {\footnotesize $\ldots$};
\draw[scalar] (2.4,0) to[out=15+90,in=0] (1,1.55);
\draw[scalar] (3,0) to[out=90,in=25] (1,1.6);
\draw[fill=VertexColor,VertexColor] (2.4,0) circle (\vertexsize);
\draw[fill=VertexColor,VertexColor] (3,0) circle (\vertexsize);
\node[scale=.7] at (2.5,.65) {\footnotesize $\ldots$};
\node[scale=.7] at (.225,1) {\footnotesize $a_1$};
\node[scale=.7] at (2-.225,1) {\footnotesize $a_2$};
\node[scale=.7] at (1,-.35) {\footnotesize $a_3$};
\node[scale=.7] at (2.7,1.2) {\footnotesize $a_4$};
\draw[fill=white] (1,1.5) circle (.18);
\draw[fill=white] (0,0) circle (.18);
\draw[fill=white] (2,0) circle (.18);
\end{tikzpicture}}

\newcommand{\DiagramIntro}{\begin{tikzpicture}[baseline={([yshift=0ex]current bounding box.center)}]
\draw[-Latex] (-.65,0)--(-.65,-2+.35);
\draw[-Latex] (.65,0)--(.65,-2+.35);
\draw[-Latex] (-.65,-2)--(-.65,-4+.35);
\draw[-Latex] (.65,-2)--(.65,-4+.35);
\draw[-Latex] (0,-4)--(0,-6+.35);
\draw[-Latex] (-5.25+1,0)--(-1,0);
\draw[-Latex] (-5.25+1,-4)--(-1,-4);
\draw[-Latex] (-5.25+1,-6)--(-1,-6);
\draw[-Latex] (5.25,-4)--(5.25,-2+.75-.35);
\draw[] (-5.25,-2+.75) -- (0-1.65-.25,-2+.75);
\draw[-Latex] (0-1.65-.25,-2+.75)--(0-1.65-.25,-2+.35);
\draw[] (-5.25,-2-.75) -- (0+1.65,-2-.75);
\draw[-Latex] (0+1.65,-2-.75)--(0+1.65,-2-.35);
\draw[] (5.25,-2+.75) -- (0-1.65+.25,-2+.75);
\draw[-Latex] (0-1.65+.25,-2+.75)--(0-1.65+.25,-2+.35);
\draw[-Latex] (5.25+1,0)--(1,0);
\draw[-Latex] (5.25,-4)--(1,-4);
\draw[-Latex] (5.25-1,-4+.05)--(2,-2.35);
\draw[-Latex] (5.25-1,-4-.05)--(1,-6);
\draw[rounded corners,fill=VeryLightBlueGrey] (0-1, 0-.35) rectangle (0+1,0+.35) {};
\node[] at (0,0) {$\vev{1111}$};
\draw[rounded corners,fill=VeryLightBlueGrey] (0-1.5-1.65, -2-.35) rectangle (0+1.5-1.65,-2+.35) {};
\node[] at (0-1.65,-2) {$\vev{1122}$};
\draw[rounded corners,fill=VeryLightBlueGrey] (0-1.5+1.65, -2-.35) rectangle (0+1.5+1.65,-2+.35) {};
\node[] at (0+1.65,-2) {$\vev{1212}$};
\draw[rounded corners,fill=VeryLightBlueGrey] (0-1, -4-.35) rectangle (0+1,-4+.35) {};
\node[] at (0,-4) {$\vev{11112}$};
\draw[rounded corners,fill=VeryLightBlueGrey] (0-1, -6-.35) rectangle (0+1,-6+.35) {};
\node[] at (0,-6) {$\vev{111111}$};
\draw[rounded corners,fill=VeryLightOrange] (-5.25-1, 0-.35) rectangle (-5.25+1,0+.35) {};
\node[] at (-5.25,0) {$\Fds_{1111}$};
\draw[rounded corners,fill=VeryLightOrange] (-5.25-1,-2-.35+.75) rectangle (-5.25+1,-2+.35+.75) {};
\node[] at (-5.25,-2+.75) {$\Fds_{1122}$};
\draw[rounded corners,fill=VeryLightOrange] (-5.25-1,-2-.35-.75) rectangle (-5.25+1,-2+.35-.75) {};
\node[] at (-5.25,-2-.75) {$\Fds_{1212}$};
\draw[rounded corners,fill=VeryLightOrange] (-5.25-1, -4-.35) rectangle (-5.25+1,-4+.35) {};
\node[] at (-5.25,-4) {$\Fds_{11112}$};
\draw[rounded corners,fill=VeryLightOrange] (-5.25-1, -6-.35) rectangle (-5.25+1,-6+.35) {};
\node[] at (-5.25,-6) {$\Fds_{111111}$};
\draw[rounded corners,fill=VeryLightRed] (5.25-1.5,-.35) rectangle (5.25+1.5,.35) {};
\node[] at (5.25,0) {1 $R$-channel};
\draw[rounded corners,fill=VeryLightRed] (5.25-1.5,-2+.75-.35) rectangle (5.25+1.5,-2+.75+.35) {};
\node[] at (5.25,-2+.75) {1 $R$-channel};
\draw[rounded corners,fill=VeryLightRed] (5.25-1.8, -4-.35) rectangle (5.25+1.8,-4+.35) {};
\node[] at (5.25,-4) {Train track integral};
\end{tikzpicture}}
\newcommand{\MethodFivePoint}{\begin{tikzpicture}[baseline={([yshift=0ex]current bounding box.center)}]
\draw[-Latex] (0,0)--(0,-1.5+.35);
\draw[-Latex] (0,-1.5)--(0,-3+.35);
\draw[-Latex] (0,-3)--(0,-4.5+.35);
\draw[-Latex] (0,-4.5)--(0,-6+.35);
\draw[-Latex] (0,-6)--(0,-7.5+.35);
\draw[-Latex] (-5.25+1,-1.5)--(-1.5,-1.5);
\draw[-Latex] (-5.25+1,-3)--(-1.5,-3);
\draw[-Latex] (5.25-1,-4.5)--(1.5,-4.5);
\draw[-Latex] (5.25-1,-6)--(1.5,-6);
\draw[-Latex] (-5.25-1,-7.5)--(-1.5,-7.5);
\draw[rounded corners,fill=VeryLightBlueGrey] (0-1.5, 0-.35) rectangle (0+1.5,0+.35) {};
\node[] at (0,0) {$6$ $R$-channels};
\draw[rounded corners,fill=VeryLightBlueGrey] (0-1.5, -1.5-.35) rectangle (0+1.5,-1.5+.35) {};
\node[] at (0,-1.5) {$2$ $R$-channels};
\draw[rounded corners,fill=VeryLightBlueGrey] (0-1.5, -3-.35) rectangle (0+1.5,-3+.35) {};
\node[] at (0,-3) {$1$ $R$-channel};
\draw[rounded corners,fill=VeryLightBlueGrey] (0-1.5, -4.5-.35) rectangle (0+1.5,-4.5+.35) {};
\node[] at (0,-4.5) {$231$ coefficients};
\draw[rounded corners,fill=VeryLightBlueGrey] (0-1.5, -6-.35) rectangle (0+1.5,-6+.35) {};
\node[] at (0,-6) {$20$ coefficients};
\draw[rounded corners,fill=VeryLightBlueGrey] (0-1.5, -7.5-.35) rectangle (0+1.5,-7.5+.35) {};
\node[] at (0,-7.5) {fully fixed};
\draw[rounded corners,fill=VeryLightOrange] (-5.25-1.5, -1.5-.35) rectangle (-5.25+1.5,-1.5+.35) {};
\node[] at (-5.25,-1.5) {SCWI $+ \Fds_{11112}$};
\draw[rounded corners,fill=VeryLightOrange] (-5.25-1.5, -3-.35) rectangle (-5.25+1.5,-3+.35) {};
\node[] at (-5.25,-3) {Crossing};
\draw[rounded corners,fill=VeryLightRed] (5.25-1.9, -4.5-.35) rectangle (5.25+1.9,-4.5+.35) {};
\node[] at (5.25,-4.5) {Ansatz};
\draw[rounded corners,fill=VeryLightRed] (5.25-1.9, -6-.35) rectangle (5.25+1.9,-6+.35) {};
\node[] at (5.25,-6) {Train track integral};
\draw[rounded corners,fill=VeryLightOrange] (-5.25-2.5, -7.5-.35) rectangle (-5.25+2.5,-7.5+.35) {};
\node[] at (-5.25,-7.5) {Pinching to $\vev{1212}$, $\vev{1122}$};
\end{tikzpicture}}
\newcommand{\MethodSixPoint}{\begin{tikzpicture}[baseline={([yshift=0ex]current bounding box.center)}]
\draw[-Latex] (0,0)--(0,-1.5+.35);
\draw[-Latex] (0,-1.5)--(0,-3+.35);
\draw[-Latex] (0,-3)--(0,-4.5+.35);
\draw[-Latex] (0,-4.5)--(0,-6+.35);
\draw[-Latex] (0,-6)--(0,-7.5+.35);
\draw[-Latex] (-5.25+1,-1.5)--(-1.5,-1.5);
\draw[-Latex] (-5.25+1,-3)--(-1.5,-3);
\draw[-Latex] (5.25-1,-4.5)--(1.5,-4.5);
\draw[-Latex] (5.25-1,-6)--(1.5,-6);
\draw[-Latex] (-5.25-1,-7.5)--(-1.5,-7.5);
\draw[rounded corners,fill=VeryLightBlueGrey] (0-1.5, 0-.35) rectangle (0+1.5,0+.35) {};
\node[] at (0,0) {$15$ $R$-channels};
\draw[rounded corners,fill=VeryLightBlueGrey] (0-1.5, -1.5-.35) rectangle (0+1.5,-1.5+.35) {};
\node[] at (0,-1.5) {$4$ $R$-channels};
\draw[rounded corners,fill=VeryLightBlueGrey] (0-1.5, -3-.35) rectangle (0+1.5,-3+.35) {};
\node[] at (0,-3) {$1$ $R$-channel};
\draw[rounded corners,fill=VeryLightBlueGrey] (0-1.5, -4.5-.35) rectangle (0+1.5,-4.5+.35) {};
\node[] at (0,-4.5) {$1406$ coefficients};
\draw[rounded corners,fill=VeryLightBlueGrey] (0-1.5, -6-.35) rectangle (0+1.5,-6+.35) {};
\node[] at (0,-6) {$21$ coefficients};
\draw[rounded corners,fill=VeryLightBlueGrey] (0-1.5, -7.5-.35) rectangle (0+1.5,-7.5+.35) {};
\node[] at (0,-7.5) {fully fixed};
\draw[rounded corners,fill=VeryLightOrange] (-5.25-1.5, -1.5-.35) rectangle (-5.25+1.5,-1.5+.35) {};
\node[] at (-5.25,-1.5) {SCWI $+ \Fds_{111111}$};
\draw[rounded corners,fill=VeryLightOrange] (-5.25-1.5, -3-.35) rectangle (-5.25+1.5,-3+.35) {};
\node[] at (-5.25,-3) {Crossing};
\draw[rounded corners,fill=VeryLightRed] (5.25-1.9, -4.5-.35) rectangle (5.25+1.9,-4.5+.35) {};
\node[] at (5.25,-4.5) {Ansatz};
\draw[rounded corners,fill=VeryLightRed] (5.25-1.9, -6-.35) rectangle (5.25+1.9,-6+.35) {};
\node[] at (5.25,-6) {Train track integral};
\draw[rounded corners,fill=VeryLightOrange] (-5.25-1.9, -7.5-.35) rectangle (-5.25+1.9,-7.5+.35) {};
\node[] at (-5.25,-7.5) {Pinching to $\vev{11112}$};
\end{tikzpicture}}

\newcommand{\ScalarPropagator}{\begin{tikzpicture}[baseline={([yshift=-.85ex]current bounding box.center)}]
\draw[scalar] (0,.5) -- (1,.5);
\draw[fill=white] (0,.5) circle (.075);
\draw[fill=white] (1,.5) circle (.075);
\node[] at (0,.1) {\footnotesize \makebox[\widthof{$\smash{\mu,a}$}][c]{$\smash{I,a}$}};
\node[] at (0,.9) {\footnotesize $1$};
\node[] at (1,.1) {\footnotesize \makebox[\widthof{$\smash{\mu,a}$}][c]{$\smash{J,b}$}};
\node[] at (1,.9) {\footnotesize $2$};
\end{tikzpicture}}
\newcommand{\GluonPropagator}{\begin{tikzpicture}[baseline={([yshift=-.85ex]current bounding box.center)}]
\draw[gluon] (0,.5) -- (1,.5);
\draw[fill=white] (0,.5) circle (.075);
\draw[fill=white] (1,.5) circle (.075);
\node[] at (0,.1) {\footnotesize \makebox[\widthof{$\mu,a$}][c]{$\smash{\mu,a}$}};
\node[] at (0,.9) {\footnotesize $1$};
\node[] at (1,.1) {\footnotesize \makebox[\widthof{$\nu,b$}][c]{$\smash{\nu,b}$}};
\node[] at (1,.9) {\footnotesize $2$};
\end{tikzpicture}}
\newcommand{\FermionPropagator}{\begin{tikzpicture}[baseline={([yshift=-.85ex]current bounding box.center)}]
\draw[scalar] (0,.5) -- (1,.5);
\draw[fill=white] (0,.5) circle (.075);
\draw[fill=white] (1,.5) circle (.075);
\draw[-Latex] (.62,.5)--(.63,.5);
\node[] at (0,.1) {\footnotesize \makebox[\widthof{$\smash{\mu,a}$}][c]{$\smash{a}$}};
\node[] at (0,.9) {\footnotesize $1$};
\node[] at (1,.1) {\footnotesize \makebox[\widthof{$\smash{\nu,b}$}][c]{$\smash{b}$}};
\node[] at (1,.9) {\footnotesize $2$};
\end{tikzpicture}}
\newcommand{\GhostPropagator}{\begin{tikzpicture}[baseline={([yshift=-.85ex]current bounding box.center)}]
\draw[ghost] (0,.5) -- (1,.5);
\draw[fill=white] (0,.5) circle (.075);
\draw[fill=white] (1,.5) circle (.075);
\node[] at (0,.1) {\footnotesize \makebox[\widthof{$\smash{\mu,a}$}][c]{$\smash{a}$}};
\node[] at (0,.9) {\footnotesize $1$};
\node[] at (1,.1) {\footnotesize \makebox[\widthof{$\smash{\nu,b}$}][c]{$\smash{b}$}};
\node[] at (1,.9) {\footnotesize $2$};
\end{tikzpicture}}

\newcommand{\VertexScalarScalarGluon}{\begin{tikzpicture}[baseline={([yshift=-.35ex]current bounding box.center)}]
\draw[gluon] (0,.5) -- (.75,.5);
\draw[scalar] (.75,.5) -- (1.25,1);
\draw[scalar] (.75,.5) -- (1.25,0);
\draw[fill=VertexColor,VertexColor] (.75,.5) circle (\vertexsize);
\draw[fill=white] (0,.5) circle (.075);
\draw[fill=white] (1.25,1) circle (.075);
\draw[fill=white] (1.25,0) circle (.075);
\node[anchor=west] at (1.3,1) {\footnotesize $I,a$};
\node[] at (1.25,1.3) {\footnotesize $1$};
\node[anchor=west] at (1.3,0) {\footnotesize $J,b$};
\node[] at (1.25,-.3) {\footnotesize $2$};
\node[] at (0,.1) {\footnotesize $\mu,c$};
\node[] at (0,.9) {\footnotesize $3$};
\end{tikzpicture}}
\newcommand{\VertexFermionFermionScalar}{\begin{tikzpicture}[baseline={([yshift=-.45ex]current bounding box.center)}]
\draw[scalar] (0,.5) -- (.75,.5);
\draw[scalar] (.75,.5) -- (1.25,1);
\draw[-Latex] (1.05,.8)--(.95,.7);
\draw[scalar] (.75,.5) -- (1.25,0);
\draw[-Latex] (1,.25)--(1.1,.15);
\draw[fill=VertexColor,VertexColor] (.75,.5) circle (\vertexsize);
\draw[fill=white] (0,.5) circle (.075);
\draw[fill=white] (1.25,1) circle (.075);
\draw[fill=white] (1.25,0) circle (.075);
\end{tikzpicture}}
\newcommand{\VertexFermionFermionGluon}{\begin{tikzpicture}[baseline={([yshift=-.35ex]current bounding box.center)}]
\draw[gluon] (0,.5) -- (.75,.5);
\draw[scalar] (.75,.5) -- (1.25,1);
\draw[-Latex] (1.05,.8)--(.95,.7);
\draw[scalar] (.75,.5) -- (1.25,0);
\draw[-Latex] (1,.25)--(1.1,.15);
\draw[fill=VertexColor,VertexColor] (.75,.5) circle (\vertexsize);
\draw[fill=white] (0,.5) circle (.075);
\draw[fill=white] (1.25,1) circle (.075);
\draw[fill=white] (1.25,0) circle (.075);
\end{tikzpicture}}
\newcommand{\VertexGhostGhostGluon}{\begin{tikzpicture}[baseline={([yshift=-.35ex]current bounding box.center)}]
\draw[gluon] (0,.5) -- (.75,.5);
\draw[ghost] (.75,.5) -- (1.25,1);
\draw[ghost] (.75,.5) -- (1.25,0);
\draw[fill=VertexColor,VertexColor] (.75,.5) circle (\vertexsize);
\draw[fill=white] (0,.5) circle (.075);
\draw[fill=white] (1.25,1) circle (.075);
\draw[fill=white] (1.25,0) circle (.075);
\end{tikzpicture}}
\newcommand{\VertexGluonGluonGluon}{\begin{tikzpicture}[baseline={([yshift=-.35ex]current bounding box.center)}]
\draw[gluon] (0,.5) -- (.75,.5);
\draw[gluon] (.75,.5) -- (1.25,1);
\draw[gluon] (.75,.5) -- (1.25,0);
\draw[fill=VertexColor,VertexColor] (.75,.5) circle (\vertexsize);
\draw[fill=white] (0,.5) circle (.075);
\draw[fill=white] (1.25,1) circle (.075);
\draw[fill=white] (1.25,0) circle (.075);
\end{tikzpicture}}
\newcommand{\VertexFourScalars}{\begin{tikzpicture}[baseline={([yshift=-.35ex]current bounding box.center)}]
\draw[scalar] (0,0) -- (1.25,1);
\draw[scalar] (0,1) -- (1.25,0);
\draw[fill=VertexColor,VertexColor] (.625,.5) circle (\vertexsize);
\draw[fill=white] (0,0) circle (.075);
\draw[fill=white] (0,1) circle (.075);
\draw[fill=white] (1.25,1) circle (.075);
\draw[fill=white] (1.25,0) circle (.075);
\node[anchor=west] at (1.3,1) {\footnotesize $I,a$};
\node[] at (1.25,1.3) {\footnotesize $1$};
\node[anchor=west] at (1.3,0) {\footnotesize $J,b$};
\node[] at (1.25,-.3) {\footnotesize $2$};
\node[anchor=east] at (-0.05,1) {\footnotesize $K,c$};
\node[] at (0,1.3) {\footnotesize $3$};
\node[anchor=east] at (-0.05,0) {\footnotesize $L,d$};
\node[] at (0,-.3) {\footnotesize $4$};
\end{tikzpicture}}

\newcommand{\VertexScalarScalarGluonGluon}{\begin{tikzpicture}[baseline={([yshift=-.35ex]current bounding box.center)}]
\draw[scalar] (.625,.5) -- (1.25,1);
\draw[scalar] (.625,.5) -- (1.25,0);
\draw[gluon] (0,0) -- (.625,.5);
\draw[gluon] (0,1) -- (.625,.5);
\draw[fill=VertexColor,VertexColor] (.625,.5) circle (\vertexsize);
\draw[fill=white] (0,0) circle (.075);
\draw[fill=white] (0,1) circle (.075);
\draw[fill=white] (1.25,1) circle (.075);
\draw[fill=white] (1.25,0) circle (.075);
\end{tikzpicture}}
\newcommand{\VertexGluonGluonGluonGluon}{\begin{tikzpicture}[baseline={([yshift=-.35ex]current bounding box.center)}]
\draw[gluon] (.625,.5) -- (1.25,1);
\draw[gluon] (.625,.5) -- (1.25,0);
\draw[gluon] (0,0) -- (.625,.5);
\draw[gluon] (0,1) -- (.625,.5);
\draw[fill=VertexColor,VertexColor] (.625,.5) circle (\vertexsize);
\draw[fill=white] (0,0) circle (.075);
\draw[fill=white] (0,1) circle (.075);
\draw[fill=white] (1.25,1) circle (.075);
\draw[fill=white] (1.25,0) circle (.075);
\end{tikzpicture}}

\newcommand{\SelfEnergyNoText}{\begin{tikzpicture}[baseline={([yshift=-.35ex]current bounding box.center)}]
\draw[scalar] (0,.5) -- (1,.5);
\draw[fill=white] (0,.5) circle (.075);
\draw[fill=white] (1,.5) circle (.075);
\draw[fill=SEColor,SEColor] (.5,.5) circle (.15);
\end{tikzpicture}}
\newcommand{\SelfEnergyDiagramOne}{\begin{tikzpicture}[baseline={([yshift=-.4ex]current bounding box.center)}]
\draw[scalar] (0,.5) -- (2,.5);
\draw[gluon] (.5,.5) arc (180:-20:.5);
\draw[fill=white] (0,.5) circle (.075);
\draw[fill=white] (2,.5) circle (.075);
\draw[fill=VertexColor,VertexColor] (.5,.5) circle (\vertexsize);
\draw[fill=VertexColor,VertexColor] (1.5,.5) circle (\vertexsize);
\end{tikzpicture}}
\newcommand{\SelfEnergyDiagramTwo}{\begin{tikzpicture}[baseline={([yshift=-.5ex]current bounding box.center)}]
\draw[scalar] (0,.5) -- (.5,.5);
\draw[scalar] (1.5,.5) -- (2,.5);
\draw[scalar,postaction={decorate}] (.5,.5) arc (180:0:.5);
\draw[-Latex] (1.1,1)--(1.11,1);
\draw[scalar,postaction={decorate}] (.5,.5) arc (-180:0:.5);
\draw[-Latex] (.9,0)--(.89,0);
\draw[fill=white] (0,.5) circle (.075);
\draw[fill=white] (2,.5) circle (.075);
\draw[fill=VertexColor,VertexColor] (.5,.5) circle (\vertexsize);
\draw[fill=VertexColor,VertexColor] (1.5,.5) circle (\vertexsize);
\end{tikzpicture}}
\newcommand{\SelfEnergyDiagramThree}{\begin{tikzpicture}[baseline={([yshift=-.75ex]current bounding box.center)}]
\draw[scalar] (0,.5) -- (2,.5);
\draw[scalar] (1,1) circle (.5);
\draw[fill=white] (0,.5) circle (.075);
\draw[fill=white] (2,.5) circle (.075);
\draw[fill=VertexColor,VertexColor] (1,.5) circle (\vertexsize);
\end{tikzpicture}}
\newcommand{\SelfEnergyDiagramFour}{\begin{tikzpicture}[baseline={([yshift=-1.05ex]current bounding box.center)}]
\draw[scalar] (0,.5) -- (2,.5);
\draw[gluon] (1,1) circle (.5);
\draw[fill=white] (0,.5) circle (.075);
\draw[fill=white] (2,.5) circle (.075);
\draw[fill=VertexColor,VertexColor] (1,.5) circle (\vertexsize);
\end{tikzpicture}}

\newcommand{\DefectVertexOnePointScalar}{\begin{tikzpicture}[baseline={([yshift=-3ex]current bounding box.center)}]
\draw[wilson] (0,0) -- (1.5,0);
\draw[scalar] (.75,0) -- (.75,1);
\draw[fill=white] (.75,1) circle (.075);
\draw[fill=white] (.3,0) circle (.075);
\draw[fill=white] (1.2,0) circle (.075);
\draw[fill=VertexColor,VertexColor] (.75,0) circle (\vertexsize);
\node[anchor=east] at (.7,1) {\footnotesize $I$};
\node[] at (1.05,1) {\footnotesize $1$};
\node[] at (.3,-.3) {\footnotesize $2$};
\node[] at (1.2,-.3) {\footnotesize $3$};
\end{tikzpicture}}
\newcommand{\DefectVertexOnePointGluon}{\begin{tikzpicture}[baseline={([yshift=-1ex]current bounding box.center)}]
\draw[wilson] (0,0) -- (1.5,0);
\draw[gluon] (.75,0) -- (.75,1);
\draw[fill=white] (.75,1) circle (.075);
\draw[fill=white] (.3,0) circle (.075);
\draw[fill=white] (1.2,0) circle (.075);
\draw[fill=VertexColor,VertexColor] (.75,0) circle (\vertexsize);
\node[anchor=east] at (.7,1) {\footnotesize $\mu$};
\node[] at (1.05,1) {\footnotesize $1$};
\node[] at (.3,-.3) {\footnotesize $2$};
\node[] at (1.2,-.3) {\footnotesize $3$};
\end{tikzpicture}}

\newcommand{\TrainTrack}{\begin{tikzpicture}[baseline={([yshift=-.5ex]current bounding box.center)}]
\draw[scalar] (0,0) -- (2.25,0);
\draw[scalar] (.75,-.75) -- (.75,.75);
\draw[scalar] (1.5,-.75) -- (1.5,.75);
\draw[fill=white] (0,0) circle (.075);
\draw[fill=white] (2.25,0) circle (.075);
\draw[fill=white] (.75,-.75) circle (.075);
\draw[fill=white] (.75,.75) circle (.075);
\draw[fill=white] (1.5,-.75) circle (.075);
\draw[fill=white] (1.5,.75) circle (.075);
\draw[fill=VertexColor,VertexColor] (.75,0) circle (\vertexsize);
\draw[fill=VertexColor,VertexColor] (1.5,0) circle (\vertexsize);
\node[] at (.75,1.1) {\footnotesize $1$};
\node[] at (-.25,0) {\footnotesize $2$};
\node[] at (.75,-1.1) {\footnotesize $3$};
\node[] at (1.5,-1.1) {\footnotesize $4$};
\node[] at (2.5,0) {\footnotesize $5$};
\node[] at (1.5,1.1) {\footnotesize $6$};
\end{tikzpicture}}
\newcommand{\Hintegral}{\begin{tikzpicture}[baseline={([yshift=-.5ex]current bounding box.center)}]
\draw[scalar] (.75,0) -- (1.5,0);
\draw[scalar] (.75,-.75) -- (.75,.75);
\draw[scalar] (1.5,-.75) -- (1.5,.75);
\draw[fill=white] (.75,-.75) circle (.075);
\draw[fill=white] (.75,.75) circle (.075);
\draw[fill=white] (1.5,-.75) circle (.075);
\draw[fill=white] (1.5,.75) circle (.075);
\draw[fill=VertexColor,VertexColor] (.75,0) circle (\vertexsize);
\draw[fill=VertexColor,VertexColor] (1.5,0) circle (\vertexsize);
\node[] at (.75,1.1) {\footnotesize $1$};
\node[] at (.75,-1.1) {\footnotesize $2$};
\node[] at (1.5,-1.1) {\footnotesize $3$};
\node[] at (1.5,1.1) {\footnotesize $4$};
\end{tikzpicture}}

\newcommand{\NLOX}{\begin{tikzpicture}[baseline={([yshift=-2ex]current bounding box.center)}]
\draw[wilson] (0,0) -- (2,0);
\draw[fill=white] (.25,0) circle (.075);
\draw[fill=white] (.75,0) circle (.075);
\draw[fill=white] (1.25,0) circle (.075);
\draw[fill=white] (1.75,0) circle (.075);
\draw[fill=VertexColor,VertexColor] (1,.43) circle (\vertexsize);
\path[clip] (0,0) rectangle (2,1);
\draw[scalar] (.75,0) circle (.5);
\draw[scalar] (1.25,0) circle (.5);
\draw[wilson] (0,0) -- (2,0);
\draw[fill=white] (.25,0) circle (.075);
\draw[fill=white] (.75,0) circle (.075);
\draw[fill=white] (1.25,0) circle (.075);
\draw[fill=white] (1.75,0) circle (.075);
\draw[fill=VertexColor,VertexColor] (1,.43) circle (\vertexsize);
\end{tikzpicture}}
\newcommand{\NNLOXXOne}{\begin{tikzpicture}[baseline={([yshift=-2ex]current bounding box.center)}]
\draw[wilson] (0,0) -- (2,0);
\draw[fill=white] (.25,0) circle (.075);
\draw[fill=white] (.75,0) circle (.075);
\draw[fill=white] (1.25,0) circle (.075);
\draw[fill=white] (1.75,0) circle (.075);
\draw[fill=VertexColor,VertexColor] (1,.43) circle (\vertexsize);
\path[clip] (0,0) rectangle (2,1);
\draw[scalar] (.75,0) circle (.5);
\draw[scalar] (1.25,0) circle (.5);
\draw[wilson] (0,0) -- (2,0);
\draw[scalar] (1.25,0) .. controls (1.25-.65,.3) and (1.5,.4) .. (1.75,0);
\draw[fill=white] (.25,0) circle (.075);
\draw[fill=white] (.75,0) circle (.075);
\draw[fill=white] (1.25,0) circle (.075);
\draw[fill=white] (1.75,0) circle (.075);
\draw[fill=VertexColor,VertexColor] (1,.43) circle (\vertexsize);
\draw[fill=VertexColor,VertexColor] (1.25-0.075,.25) circle (\vertexsize);
\end{tikzpicture}}
\newcommand{\NNLOXXTwo}{\begin{tikzpicture}[baseline={([yshift=-2ex]current bounding box.center)}]
\draw[wilson] (0,0) -- (2,0);
\draw[fill=white] (.25,0) circle (.075);
\draw[fill=white] (.75,0) circle (.075);
\draw[fill=white] (1.25,0) circle (.075);
\draw[fill=white] (1.75,0) circle (.075);
\draw[fill=VertexColor,VertexColor] (1,.43) circle (\vertexsize);
\path[clip] (0,0) rectangle (2,1);
\draw[scalar] (.75,0) circle (.5);
\draw[scalar] (1.25,0) circle (.5);
\draw[wilson] (0,0) -- (2,0);
\draw[scalar] (1.75,0) .. controls (1.75+.65,.3) and (1.5,.4) .. (1.25,0);
\draw[fill=white] (.25,0) circle (.075);
\draw[fill=white] (.75,0) circle (.075);
\draw[fill=white] (1.25,0) circle (.075);
\draw[fill=white] (1.75,0) circle (.075);
\draw[fill=VertexColor,VertexColor] (1,.43) circle (\vertexsize);
\draw[fill=VertexColor,VertexColor] (1.75-0.075,.25) circle (\vertexsize);
\end{tikzpicture}}
\newcommand{\NNLOXHOne}{\begin{tikzpicture}[baseline={([yshift=-2ex]current bounding box.center)}]
\draw[wilson] (0,0) -- (2,0);
\draw[scalar] (1.75,0) arc (20:160:.265);
\draw[gluon] (1.125,.35) .. controls (1.25,.35) and (1.35,.3) .. (1.5,.15);
\draw[fill=white] (.25,0) circle (.075);
\draw[fill=white] (.75,0) circle (.075);
\draw[fill=white] (1.25,0) circle (.075);
\draw[fill=white] (1.75,0) circle (.075);
\draw[fill=VertexColor,VertexColor] (1,.43) circle (\vertexsize);
\path[clip] (0,0) rectangle (2,1);
\draw[scalar] (.75,0) circle (.5);
\draw[scalar] (1.25,0) circle (.5);
\draw[wilson] (0,0) -- (2,0);
\draw[fill=white] (.25,0) circle (.075);
\draw[fill=white] (.75,0) circle (.075);
\draw[fill=white] (1.25,0) circle (.075);
\draw[fill=white] (1.75,0) circle (.075);
\draw[fill=VertexColor,VertexColor] (1,.43) circle (\vertexsize);
\draw[fill=VertexColor,VertexColor] (1.175,.275) circle (\vertexsize);
\draw[fill=VertexColor,VertexColor] (1.5,.15) circle (\vertexsize);
\end{tikzpicture}}
\newcommand{\NNLOXHTwo}{\begin{tikzpicture}[baseline={([yshift=-2ex]current bounding box.center)}]
\draw[wilson] (0,0) -- (2,0);
\draw[scalar] (1.75,0) arc (20:160:.265);
\draw[gluon] (1.25,.505) -- (1.5,.15);
\draw[fill=white] (.25,0) circle (.075);
\draw[fill=white] (.75,0) circle (.075);
\draw[fill=white] (1.25,0) circle (.075);
\draw[fill=white] (1.75,0) circle (.075);
\draw[fill=VertexColor,VertexColor] (1,.43) circle (\vertexsize);
\path[clip] (0,0) rectangle (2,1);
\draw[scalar] (.75,0) circle (.5);
\draw[scalar] (1.25,0) circle (.5);
\draw[wilson] (0,0) -- (2,0);
\draw[fill=white] (.25,0) circle (.075);
\draw[fill=white] (.75,0) circle (.075);
\draw[fill=white] (1.25,0) circle (.075);
\draw[fill=white] (1.75,0) circle (.075);
\draw[fill=VertexColor,VertexColor] (1,.43) circle (\vertexsize);
\draw[fill=VertexColor,VertexColor] (1.225,.5) circle (\vertexsize);
\draw[fill=VertexColor,VertexColor] (1.5,.15) circle (\vertexsize);
\end{tikzpicture}}
\newcommand{\NNLOSubtractFactorOne}{\begin{tikzpicture}[baseline={([yshift=.3ex]current bounding box.center)}]
\draw[wilson] (1.1,0) -- (1.9,0);
\draw[scalar] (1.75,0) arc (20:160:.265);
\draw[fill=white] (1.25,0) circle (.075);
\draw[fill=white] (1.75,0) circle (.075);
\end{tikzpicture}}
\newcommand{\NNLOSubtractFactorTwo}{\begin{tikzpicture}[baseline={([yshift=-.6ex]current bounding box.center)}]
\draw[wilson] (.8,0) -- (2.2,0);
\draw[gluon] (1.5,.175) -- (1.5,.5);
\draw[scalar] (1.75,0) arc (20:160:.265);
\draw[fill=white] (1.25,0) circle (.075);
\draw[fill=white] (1.75,0) circle (.075);
\draw[fill=VertexColor,VertexColor] (1,0) circle (\vertexsize);
\draw[fill=VertexColor,VertexColor] (2,0) circle (\vertexsize);
\draw[fill=VertexColor,VertexColor] (1.5,.175) circle (\vertexsize);
\end{tikzpicture}}
\newcommand{\NNLOXYSubtractOne}{\begin{tikzpicture}[baseline={([yshift=-2ex]current bounding box.center)}]
\draw[wilson] (0,0) -- (2,0);
\draw[gluon] (1.125,.35) .. controls (1.25,.35) and (1.35,.3) .. (1.5,0);
\draw[fill=white] (.25,0) circle (.075);
\draw[fill=white] (.75,0) circle (.075);
\draw[fill=white] (1.25,0) circle (.075);
\draw[fill=white] (1.75,0) circle (.075);
\draw[fill=VertexColor,VertexColor] (1,.43) circle (\vertexsize);
\draw[fill=VertexColor,VertexColor] (1.5,0) circle (\vertexsize);
\path[clip] (0,0) rectangle (2,1);
\draw[scalar] (.75,0) circle (.5);
\draw[scalar] (1.25,0) circle (.5);
\draw[wilson] (0,0) -- (2,0);
\draw[fill=white] (.25,0) circle (.075);
\draw[fill=white] (.75,0) circle (.075);
\draw[fill=white] (1.25,0) circle (.075);
\draw[fill=white] (1.75,0) circle (.075);
\draw[fill=VertexColor,VertexColor] (1,.43) circle (\vertexsize);
\draw[fill=VertexColor,VertexColor] (1.175,.275) circle (\vertexsize);
\draw[fill=VertexColor,VertexColor] (1.5,0) circle (\vertexsize);
\end{tikzpicture}}
\newcommand{\NNLOXYSubtractTwo}{\begin{tikzpicture}[baseline={([yshift=-2ex]current bounding box.center)}]
\draw[wilson] (0,0) -- (2,0);
\draw[gluon] (1.25,.505) -- (1.5,0);
\draw[fill=white] (.25,0) circle (.075);
\draw[fill=white] (.75,0) circle (.075);
\draw[fill=white] (1.25,0) circle (.075);
\draw[fill=white] (1.75,0) circle (.075);
\draw[fill=VertexColor,VertexColor] (1,.43) circle (\vertexsize);
\draw[fill=VertexColor,VertexColor] (1.5,0) circle (\vertexsize);
\path[clip] (0,0) rectangle (2,1);
\draw[scalar] (.75,0) circle (.5);
\draw[scalar] (1.25,0) circle (.5);
\draw[wilson] (0,0) -- (2,0);
\draw[fill=white] (.25,0) circle (.075);
\draw[fill=white] (.75,0) circle (.075);
\draw[fill=white] (1.25,0) circle (.075);
\draw[fill=white] (1.75,0) circle (.075);
\draw[fill=VertexColor,VertexColor] (1,.43) circle (\vertexsize);\draw[fill=VertexColor,VertexColor] (1.225,.5) circle (\vertexsize);
\draw[fill=VertexColor,VertexColor] (1.5,0) circle (\vertexsize);
\end{tikzpicture}}
\newcommand{\NNLOFourPtTrainTrack}{\begin{tikzpicture}[baseline={([yshift=-2ex]current bounding box.center)}]
\draw[wilson] (0,0) -- (2,0);
\draw[scalar] (1.75,0) arc (35:150:.6);
\draw[fill=white] (.25,0) circle (.075);
\draw[fill=white] (.75,0) circle (.075);
\draw[fill=white] (1.25,0) circle (.075);
\draw[fill=white] (1.75,0) circle (.075);
\draw[fill=VertexColor,VertexColor] (1,.43) circle (\vertexsize);
\path[clip] (0,0) rectangle (2,1);
\draw[scalar] (.75,0) circle (.5);
\draw[scalar] (1.25,0) circle (.5);
\draw[wilson] (0,0) -- (2,0);
\draw[fill=white] (.25,0) circle (.075);
\draw[fill=white] (.75,0) circle (.075);
\draw[fill=white] (1.25,0) circle (.075);
\draw[fill=white] (1.75,0) circle (.075);
\draw[fill=VertexColor,VertexColor] (1,.43) circle (\vertexsize);
\draw[fill=VertexColor,VertexColor] (1.25-0.075,.25) circle (\vertexsize);
\end{tikzpicture}}

\newcommand{\NNLOFivePt}{\begin{tikzpicture}[baseline={([yshift=-2ex]current bounding box.center)}]
\draw[wilson] (0,0) -- (2.5,0);
\draw[scalar] (1.75,0) arc (10:170:.775);
\draw[scalar] (2.25,0) arc (10:170:.775);
\draw[scalar] (2.25,0) arc (10:170:.5);
\draw[fill=white] (.25,0) circle (.075);
\draw[fill=white] (.75,0) circle (.075);
\draw[fill=white] (1.25,0) circle (.075);
\draw[fill=white] (1.75,0) circle (.075);
\draw[fill=white] (2.25,0) circle (.075);
\draw[fill=VertexColor,VertexColor] (1.25,.58) circle (\vertexsize);
\draw[fill=VertexColor,VertexColor] (1.57,.375) circle (\vertexsize);
\end{tikzpicture}}

\newcommand{\NNLOSixPt}{\begin{tikzpicture}[baseline={([yshift=-2ex]current bounding box.center)}]
\draw[wilson] (0,0) -- (3,0);
\draw[scalar] (2.25,0) arc (10:170:1.015);
\draw[scalar] (2.75,0) arc (10:170:.775);
\draw[scalar] (1.75,0) arc (10:170:.5);
\draw[fill=white] (.25,0) circle (.075);
\draw[fill=white] (.75,0) circle (.075);
\draw[fill=white] (1.25,0) circle (.075);
\draw[fill=white] (1.75,0) circle (.075);
\draw[fill=white] (2.25,0) circle (.075);
\draw[fill=white] (2.75,0) circle (.075);
\draw[fill=VertexColor,VertexColor] (1.42,.37) circle (\vertexsize);
\draw[fill=VertexColor,VertexColor] (1.86,.625) circle (\vertexsize);
\end{tikzpicture}}

\newcommand{\NNNLOEightPt}{\begin{tikzpicture}[baseline={([yshift=-3ex]current bounding box.center)}]
\draw[wilson] (0,0) -- (4,0);
\draw[scalar] (2.25,0) arc (10:170:1.015);
\draw[scalar] (2.75,0) arc (10:170:.5);
\draw[scalar] (3.25,0) arc (10:170:1.015);
\draw[scalar] (3.75,0) arc (10:170:1.525);
\draw[fill=white] (.25,0) circle (.075);
\draw[fill=white] (.75,0) circle (.075);
\draw[fill=white] (1.25,0) circle (.075);
\draw[fill=white] (1.75,0) circle (.075);
\draw[fill=white] (2.25,0) circle (.075);
\draw[fill=white] (2.75,0) circle (.075);
\draw[fill=white] (3.25,0) circle (.075);
\draw[fill=white] (3.75,0) circle (.075);
\draw[fill=VertexColor,VertexColor] (2.1,.375) circle (\vertexsize);
\draw[fill=VertexColor,VertexColor] (1.75,.7) circle (\vertexsize);
\draw[fill=VertexColor,VertexColor] (1.2,.83) circle (\vertexsize);
\end{tikzpicture}}

\newcommand{\CombChannel}{\scalebox{\scaleCC}{\begin{tikzpicture}[baseline={([yshift=-0ex]current bounding box.center)}]
\draw[scalar] (-4.5,0) -- (0.5,0);
\draw[scalar] (-3.75,0) -- (-3.75,0.7);
\draw[scalar] (-2.75,0) -- (-2.75,0.7);
\draw[scalar] (-1.25,0) -- (-1.25,0.7);
\draw[scalar] (-0.25,0) -- (-0.25,0.7);
\node[] at (-3.2,.2) {\tiny $\Dh_1$};
\node[] at (-.75,.2) {\tiny $\Dh_{n-3}$};
\node[] at (-2,.5) {$\ldots$};
\draw[MediumBlueGrey,fill=MediumBlueGrey] (-3.75,0) circle (.05);
\draw[MediumBlueGrey,fill=MediumBlueGrey] (-2.75,0) circle (.05);
\draw[MediumBlueGrey,fill=MediumBlueGrey] (-1.25,0) circle (.05);
\draw[MediumBlueGrey,fill=MediumBlueGrey] (-0.25,0) circle (.05);
\end{tikzpicture}}}
\newcommand{\SnowflakeChannel}{\scalebox{\scaleCC}{\begin{tikzpicture}[baseline={([yshift=-0ex]current bounding box.center)}]
\draw[scalar] (0,0) -- (0,.7);
\draw[scalar] (0,.7) -- (.7,1.15);
\draw[scalar] (0,.7) -- (-.7,1.15);
\draw[scalar] (0,0) -- (-.7,-.45);
\draw[scalar] (-.7,-.45) -- (-1.4,0);
\draw[scalar] (-.7,-.45) -- (-.7,-1.15);
\draw[scalar] (0,0) -- (.7,-.45);
\draw[scalar] (.7,-.45) -- (1.4,0);
\draw[scalar] (.7,-.45) -- (.7,-1.15);
\node[] at (-.5,0) {\tiny $\Dh_1$};
\node[] at (.2,-.4) {\tiny $\Dh_2$};
\node[] at (.25,.35) {\tiny $\Dh_3$};
\draw[MediumBlueGrey,fill=MediumBlueGrey] (0,0) circle (.05);
\draw[MediumBlueGrey,fill=MediumBlueGrey] (0,.7) circle (.05);
\draw[MediumBlueGrey,fill=MediumBlueGrey] (-.7,-.45) circle (.05);
\draw[MediumBlueGrey,fill=MediumBlueGrey] (.7,-.45) circle (.05);
\end{tikzpicture}}}

\newcommand{\DefectSSSSTwoLoopsSelfEnergyOne}{\begin{tikzpicture}[baseline={([yshift=-2ex]current bounding box.center)}]
\draw[wilson] (0,0) -- (2,0);
\draw[fill=white] (.25,0) circle (.075);
\draw[fill=white] (.75,0) circle (.075);
\draw[fill=white] (1.25,0) circle (.075);
\draw[fill=white] (1.75,0) circle (.075);
\draw[fill=VertexColor,VertexColor] (1,.43) circle (\vertexsize);
\path[clip] (0,0) rectangle (2,1);
\draw[scalar] (.75,0) circle (.5);
\draw[scalar] (1.25,0) circle (.5);
\draw[wilson] (0,0) -- (2,0);
\draw[fill=white] (.25,0) circle (.075);
\draw[fill=white] (.75,0) circle (.075);
\draw[fill=white] (1.25,0) circle (.075);
\draw[fill=white] (1.75,0) circle (.075);
\draw[fill=VertexColor,VertexColor] (1,.43) circle (\vertexsize);
\draw[fill=SEColor,SEColor] (.5,.4) circle (.1);
\end{tikzpicture}}
\newcommand{\DefectSSSSTwoLoopsSelfEnergyTwo}{\begin{tikzpicture}[baseline={([yshift=-2ex]current bounding box.center)}]
\draw[wilson] (0,0) -- (2,0);
\draw[fill=white] (.25,0) circle (.075);
\draw[fill=white] (.75,0) circle (.075);
\draw[fill=white] (1.25,0) circle (.075);
\draw[fill=white] (1.75,0) circle (.075);
\draw[fill=VertexColor,VertexColor] (1,.43) circle (\vertexsize);
\path[clip] (0,0) rectangle (2,1);
\draw[scalar] (.75,0) circle (.5);
\draw[scalar] (1.25,0) circle (.5);
\draw[wilson] (0,0) -- (2,0);
\draw[fill=white] (.25,0) circle (.075);
\draw[fill=white] (.75,0) circle (.075);
\draw[fill=white] (1.25,0) circle (.075);
\draw[fill=white] (1.75,0) circle (.075);
\draw[fill=VertexColor,VertexColor] (1,.43) circle (\vertexsize);
\draw[fill=SEColor,SEColor] (.82,.24) circle (.1);
\end{tikzpicture}}
\newcommand{\DefectSSSSTwoLoopsSelfEnergyThree}{\begin{tikzpicture}[baseline={([yshift=-2ex]current bounding box.center)}]
\draw[wilson] (0,0) -- (2,0);
\draw[fill=white] (.25,0) circle (.075);
\draw[fill=white] (.75,0) circle (.075);
\draw[fill=white] (1.25,0) circle (.075);
\draw[fill=white] (1.75,0) circle (.075);
\draw[fill=VertexColor,VertexColor] (1,.43) circle (\vertexsize);
\path[clip] (0,0) rectangle (2,1);
\draw[scalar] (.75,0) circle (.5);
\draw[scalar] (1.25,0) circle (.5);
\draw[wilson] (0,0) -- (2,0);
\draw[fill=white] (.25,0) circle (.075);
\draw[fill=white] (.75,0) circle (.075);
\draw[fill=white] (1.25,0) circle (.075);
\draw[fill=white] (1.75,0) circle (.075);
\draw[fill=VertexColor,VertexColor] (1,.43) circle (\vertexsize);
\draw[fill=SEColor,SEColor] (1.18,.24) circle (.1);
\end{tikzpicture}}
\newcommand{\DefectSSSSTwoLoopsSelfEnergyFour}{\begin{tikzpicture}[baseline={([yshift=-2ex]current bounding box.center)}]
\draw[wilson] (0,0) -- (2,0);
\draw[fill=white] (.25,0) circle (.075);
\draw[fill=white] (.75,0) circle (.075);
\draw[fill=white] (1.25,0) circle (.075);
\draw[fill=white] (1.75,0) circle (.075);
\draw[fill=VertexColor,VertexColor] (1,.43) circle (\vertexsize);
\path[clip] (0,0) rectangle (2,1);
\draw[scalar] (.75,0) circle (.5);
\draw[scalar] (1.25,0) circle (.5);
\draw[wilson] (0,0) -- (2,0);
\draw[fill=white] (.25,0) circle (.075);
\draw[fill=white] (.75,0) circle (.075);
\draw[fill=white] (1.25,0) circle (.075);
\draw[fill=white] (1.75,0) circle (.075);
\draw[fill=VertexColor,VertexColor] (1,.43) circle (\vertexsize);
\draw[fill=SEColor,SEColor] (1.5,.4) circle (.1);
\end{tikzpicture}}

\newcommand{\DefectSSSSTwoLoopsXXOne}{\begin{tikzpicture}[baseline={([yshift=-2ex]current bounding box.center)}]
\draw[scalar] (.5,.365) -- (1.5,.365);
\draw[wilson] (0,0) -- (2,0);
\draw[fill=white] (.25,0) circle (.075);
\draw[fill=white] (.75,0) circle (.075);
\draw[fill=white] (1.25,0) circle (.075);
\draw[fill=white] (1.75,0) circle (.075);
\draw[fill=VertexColor,VertexColor] (.5,.365) circle (\vertexsize);
\draw[fill=VertexColor,VertexColor] (1.5,.365) circle (\vertexsize);
\path[clip] (0,0) rectangle (2,1);
\draw[scalar] (.5,.085) circle (.28);
\draw[scalar] (1.5,.085) circle (.28);
\draw[wilson] (0,0) -- (2,0);
\draw[fill=white] (.25,0) circle (.075);
\draw[fill=white] (.75,0) circle (.075);
\draw[fill=white] (1.25,0) circle (.075);
\draw[fill=white] (1.75,0) circle (.075);
\draw[fill=VertexColor,VertexColor] (.5,.365) circle (\vertexsize);
\draw[fill=VertexColor,VertexColor] (1.5,.365) circle (\vertexsize);
\path[clip] (0,.365) rectangle (2,1);
\draw[scalar] (1,0) circle (.62);
\draw[fill=VertexColor,VertexColor] (.5,.365) circle (\vertexsize);
\draw[fill=VertexColor,VertexColor] (1.5,.365) circle (\vertexsize);
\end{tikzpicture}}
\newcommand{\DefectSSSSTwoLoopsXXTwo}{\begin{tikzpicture}[baseline={([yshift=-2ex]current bounding box.center)}]
\draw[wilson] (0,0) -- (2,0);
\draw[fill=white] (.25,0) circle (.075);
\draw[fill=white] (.75,0) circle (.075);
\draw[fill=white] (1.25,0) circle (.075);
\draw[fill=white] (1.75,0) circle (.075);
\draw[fill=VertexColor,VertexColor] (1,.6) circle (\vertexsize);
\draw[fill=VertexColor,VertexColor] (1,.245) circle (\vertexsize);
\path[clip] (0,0) rectangle (2,1);
\draw[scalar] (1,-.2) circle (.8);
\draw[scalar] (1,-.025) circle (.27);
\draw[wilson] (0,0) -- (2,0);
\draw[fill=white] (.25,0) circle (.075);
\draw[fill=white] (.75,0) circle (.075);
\draw[fill=white] (1.25,0) circle (.075);
\draw[fill=white] (1.75,0) circle (.075);
\draw[fill=VertexColor,VertexColor] (1,.6) circle (\vertexsize);
\draw[fill=VertexColor,VertexColor] (1,.245) circle (\vertexsize);
\path[clip] (.8,.245) rectangle (1.2,.6);
\draw[scalar] (.8,.4225) circle (.27);
\draw[scalar] (1.2,.4225) circle (.27);
\draw[fill=VertexColor,VertexColor] (1,.6) circle (\vertexsize);
\draw[fill=VertexColor,VertexColor] (1,.245) circle (\vertexsize);
\end{tikzpicture}}

\newcommand{\DefectSSSSTwoLoopsXHOne}{\begin{tikzpicture}[baseline={([yshift=-2ex]current bounding box.center)}]
\draw[gluontwo] (.4,.36) -- (.81,.225);
\draw[wilson] (0,0) -- (2,0);
\draw[fill=white] (.25,0) circle (.075);
\draw[fill=white] (.75,0) circle (.075);
\draw[fill=white] (1.25,0) circle (.075);
\draw[fill=white] (1.75,0) circle (.075);
\draw[fill=VertexColor,VertexColor] (1,.43) circle (\vertexsize);
\draw[fill=VertexColor,VertexColor] (.4,.36) circle (\vertexsize);
\draw[fill=VertexColor,VertexColor] (.81,.225) circle (\vertexsize);
\path[clip] (0,0) rectangle (2,1);
\draw[scalar] (.75,0) circle (.5);
\draw[scalar] (1.25,0) circle (.5);
\draw[wilson] (0,0) -- (2,0);
\draw[fill=white] (.25,0) circle (.075);
\draw[fill=white] (.75,0) circle (.075);
\draw[fill=white] (1.25,0) circle (.075);
\draw[fill=white] (1.75,0) circle (.075);
\draw[fill=VertexColor,VertexColor] (1,.43) circle (\vertexsize);
\draw[fill=VertexColor,VertexColor] (.4,.36) circle (\vertexsize);
\draw[fill=VertexColor,VertexColor] (.81,.225) circle (\vertexsize);
\end{tikzpicture}}
\newcommand{\DefectSSSSTwoLoopsXHTwo}{\begin{tikzpicture}[baseline={([yshift=-2ex]current bounding box.center)}]
\draw[gluontwo] (.81,.225) -- (1.19,.225);
\draw[wilson] (0,0) -- (2,0);
\draw[fill=white] (.25,0) circle (.075);
\draw[fill=white] (.75,0) circle (.075);
\draw[fill=white] (1.25,0) circle (.075);
\draw[fill=white] (1.75,0) circle (.075);
\draw[fill=VertexColor,VertexColor] (1,.43) circle (\vertexsize);
\draw[fill=VertexColor,VertexColor] (.81,.225) circle (\vertexsize);
\draw[fill=VertexColor,VertexColor] (1.19,.225) circle (\vertexsize);
\path[clip] (0,0) rectangle (2,1);
\draw[scalar] (.75,0) circle (.5);
\draw[scalar] (1.25,0) circle (.5);
\draw[wilson] (0,0) -- (2,0);
\draw[fill=white] (.25,0) circle (.075);
\draw[fill=white] (.75,0) circle (.075);
\draw[fill=white] (1.25,0) circle (.075);
\draw[fill=white] (1.75,0) circle (.075);
\draw[fill=VertexColor,VertexColor] (1,.43) circle (\vertexsize);
\draw[fill=VertexColor,VertexColor] (.81,.225) circle (\vertexsize);
\draw[fill=VertexColor,VertexColor] (1.19,.225) circle (\vertexsize);
\end{tikzpicture}}
\newcommand{\DefectSSSSTwoLoopsXHThree}{\begin{tikzpicture}[baseline={([yshift=-2ex]current bounding box.center)}]
\draw[gluontwo] (1.6,.36) -- (1.19,.225);
\draw[wilson] (0,0) -- (2,0);
\draw[fill=white] (.25,0) circle (.075);
\draw[fill=white] (.75,0) circle (.075);
\draw[fill=white] (1.25,0) circle (.075);
\draw[fill=white] (1.75,0) circle (.075);
\draw[fill=VertexColor,VertexColor] (1,.43) circle (\vertexsize);
\draw[fill=VertexColor,VertexColor] (1.6,.36) circle (\vertexsize);
\draw[fill=VertexColor,VertexColor] (1.19,.225) circle (\vertexsize);
\path[clip] (0,0) rectangle (2,1);
\draw[scalar] (.75,0) circle (.5);
\draw[scalar] (1.25,0) circle (.5);
\draw[wilson] (0,0) -- (2,0);
\draw[fill=white] (.25,0) circle (.075);
\draw[fill=white] (.75,0) circle (.075);
\draw[fill=white] (1.25,0) circle (.075);
\draw[fill=white] (1.75,0) circle (.075);
\draw[fill=VertexColor,VertexColor] (1,.43) circle (\vertexsize);
\draw[fill=VertexColor,VertexColor] (1.6,.36) circle (\vertexsize);
\draw[fill=VertexColor,VertexColor] (1.19,.225) circle (\vertexsize);
\end{tikzpicture}}
\newcommand{\DefectSSSSTwoLoopsXHFour}{\begin{tikzpicture}[baseline={([yshift=-2ex]current bounding box.center)}]
\draw[wilson] (0,0) -- (2,0);
\draw[fill=white] (.25,0) circle (.075);
\draw[fill=white] (.75,0) circle (.075);
\draw[fill=white] (1.25,0) circle (.075);
\draw[fill=white] (1.75,0) circle (.075);
\draw[fill=VertexColor,VertexColor] (1,.43) circle (\vertexsize);
\draw[fill=VertexColor,VertexColor] (.4,.36) circle (\vertexsize);
\draw[fill=VertexColor,VertexColor] (1.6,.36) circle (\vertexsize);
\path[clip] (0,0) rectangle (2,1);
\draw[scalar] (.75,0) circle (.5);
\draw[scalar] (1.25,0) circle (.5);
\draw[wilson] (0,0) -- (2,0);
\draw[fill=white] (.25,0) circle (.075);
\draw[fill=white] (.75,0) circle (.075);
\draw[fill=white] (1.25,0) circle (.075);
\draw[fill=white] (1.75,0) circle (.075);
\draw[fill=VertexColor,VertexColor] (.4,.36) circle (\vertexsize);
\draw[fill=VertexColor,VertexColor] (1.6,.36) circle (\vertexsize);
\path[clip] (0,.36) rectangle (2,1);
\draw[gluontwo] (1,.15) circle (.65);
\draw[fill=VertexColor,VertexColor] (1,.43) circle (\vertexsize);
\draw[fill=VertexColor,VertexColor] (.4,.36) circle (\vertexsize);
\draw[fill=VertexColor,VertexColor] (1.6,.36) circle (\vertexsize);
\end{tikzpicture}}

\newcommand{\DefectSSSSTwoLoopsSpider}{\begin{tikzpicture}[baseline={([yshift=-2ex]current bounding box.center)}]
\draw[wilson] (0,0) -- (2,0);
\draw[fill=white] (.25,0) circle (.075);
\draw[fill=white] (.75,0) circle (.075);
\draw[fill=white] (1.25,0) circle (.075);
\draw[fill=white] (1.75,0) circle (.075);
\draw[fill=VertexColor,VertexColor] (1,.43) circle (\vertexsize);
\path[clip] (0,0) rectangle (2,1);
\draw[scalar] (.75,.1) circle (.52);
\draw[scalar] (1.25,.1) circle (.52);
\draw[wilson] (0,0) -- (2,0);
\draw[fill=white] (.25,0) circle (.075);
\draw[fill=white] (.75,0) circle (.075);
\draw[fill=white] (1.25,0) circle (.075);
\draw[fill=white] (1.75,0) circle (.075);
\draw[scalar, fill=white] (1,.45) circle (.275);
\draw[-Latex] (1,.725)--(1.1,.725);
\draw[-Latex] (1,.18)--(.9,.18);
\draw[-Latex] (.725,.43)--(.725,.53);
\draw[-Latex] (1.275,.44)--(1.275,.34);
\draw[fill=VertexColor,VertexColor] (.775,.615) circle (\vertexsize);
\draw[fill=VertexColor,VertexColor] (.785,.25) circle (\vertexsize);
\draw[fill=VertexColor,VertexColor] (1.225,.615) circle (\vertexsize);
\draw[fill=VertexColor,VertexColor] (1.215,.25) circle (\vertexsize);
\end{tikzpicture}}

\newcommand{\DefectSSSSTwoLoopsXYOne}{\begin{tikzpicture}[baseline={([yshift=-2ex]current bounding box.center)}]
\draw[wilson] (-.2,0) -- (2.2,0);
\draw[gluon] (.4,.36) -- (.1,.66);
\draw[fill=white] (.25,0) circle (.075);
\draw[fill=white] (.75,0) circle (.075);
\draw[fill=white] (1.25,0) circle (.075);
\draw[fill=white] (1.75,0) circle (.075);
\draw[fill=VertexColor,VertexColor] (1,.43) circle (\vertexsize);
\draw[fill=VertexColor,VertexColor] (.4,.36) circle (\vertexsize);
\draw[fill=VertexColor,VertexColor] (0,0) circle (\vertexsize);
\draw[fill=VertexColor,VertexColor] (.5,0) circle (\vertexsize);
\draw[fill=VertexColor,VertexColor] (2,0) circle (\vertexsize);
\path[clip] (0,0) rectangle (2,1);
\draw[scalar] (.75,0) circle (.5);
\draw[scalar] (1.25,0) circle (.5);
\draw[wilson] (0,0) -- (2,0);
\draw[fill=white] (.25,0) circle (.075);
\draw[fill=white] (.75,0) circle (.075);
\draw[fill=white] (1.25,0) circle (.075);
\draw[fill=white] (1.75,0) circle (.075);
\draw[fill=VertexColor,VertexColor] (1,.43) circle (\vertexsize);
\draw[fill=VertexColor,VertexColor] (.4,.36) circle (\vertexsize);
\draw[fill=VertexColor,VertexColor] (0,0) circle (\vertexsize);
\draw[fill=VertexColor,VertexColor] (.5,0) circle (\vertexsize);
\draw[fill=VertexColor,VertexColor] (2,0) circle (\vertexsize);
\end{tikzpicture}}

\newcommand{\DefectSSSSTwoLoopsXYTwo}{\begin{tikzpicture}[baseline={([yshift=-2ex]current bounding box.center)}]
\draw[wilson] (0,0) -- (2,0);
\draw[gluon] (.81,.225) -- (.46,.225);
\draw[fill=white] (.25,0) circle (.075);
\draw[fill=white] (.75,0) circle (.075);
\draw[fill=white] (1.25,0) circle (.075);
\draw[fill=white] (1.75,0) circle (.075);
\draw[fill=VertexColor,VertexColor] (1,.43) circle (\vertexsize);
\draw[fill=VertexColor,VertexColor] (.81,.225) circle (\vertexsize);
\draw[fill=VertexColor,VertexColor] (.5,0) circle (\vertexsize);
\draw[fill=VertexColor,VertexColor] (1,0) circle (\vertexsize);
\path[clip] (0,0) rectangle (2,1);
\draw[scalar] (.75,0) circle (.5);
\draw[scalar] (1.25,0) circle (.5);
\draw[wilson] (0,0) -- (2,0);
\draw[fill=white] (.25,0) circle (.075);
\draw[fill=white] (.75,0) circle (.075);
\draw[fill=white] (1.25,0) circle (.075);
\draw[fill=white] (1.75,0) circle (.075);
\draw[fill=VertexColor,VertexColor] (1,.43) circle (\vertexsize);
\draw[fill=VertexColor,VertexColor] (.81,.225) circle (\vertexsize);
\draw[fill=VertexColor,VertexColor] (.5,0) circle (\vertexsize);
\draw[fill=VertexColor,VertexColor] (1,0) circle (\vertexsize);
\end{tikzpicture}}

\newcommand{\DefectSSSSTwoLoopsXYThree}{\begin{tikzpicture}[baseline={([yshift=-2ex]current bounding box.center)}]
\draw[wilson] (0,0) -- (2,0);
\draw[gluon] (1.19,.225) -- (1.54,.225);
\draw[fill=white] (.25,0) circle (.075);
\draw[fill=white] (.75,0) circle (.075);
\draw[fill=white] (1.25,0) circle (.075);
\draw[fill=white] (1.75,0) circle (.075);
\draw[fill=VertexColor,VertexColor] (1,.43) circle (\vertexsize);
\draw[fill=VertexColor,VertexColor] (1.19,.225) circle (\vertexsize);
\draw[fill=VertexColor,VertexColor] (1.5,0) circle (\vertexsize);
\draw[fill=VertexColor,VertexColor] (1,0) circle (\vertexsize);
\path[clip] (0,0) rectangle (2,1);
\draw[scalar] (.75,0) circle (.5);
\draw[scalar] (1.25,0) circle (.5);
\draw[wilson] (0,0) -- (2,0);
\draw[fill=white] (.25,0) circle (.075);
\draw[fill=white] (.75,0) circle (.075);
\draw[fill=white] (1.25,0) circle (.075);
\draw[fill=white] (1.75,0) circle (.075);
\draw[fill=VertexColor,VertexColor] (1,.43) circle (\vertexsize);
\draw[fill=VertexColor,VertexColor] (1.19,.225) circle (\vertexsize);
\draw[fill=VertexColor,VertexColor] (1.5,0) circle (\vertexsize);
\draw[fill=VertexColor,VertexColor] (1,0) circle (\vertexsize);
\end{tikzpicture}}

\newcommand{\DefectSSSSTwoLoopsXYFour}{\begin{tikzpicture}[baseline={([yshift=-2ex]current bounding box.center)}]
\draw[wilson] (-.2,0) -- (2.2,0);
\draw[gluon] (1.6,.36) -- (1.9,.66);
\draw[fill=white] (.25,0) circle (.075);
\draw[fill=white] (.75,0) circle (.075);
\draw[fill=white] (1.25,0) circle (.075);
\draw[fill=white] (1.75,0) circle (.075);
\draw[fill=VertexColor,VertexColor] (1,.43) circle (\vertexsize);
\draw[fill=VertexColor,VertexColor] (1.6,.36) circle (\vertexsize);
\draw[fill=VertexColor,VertexColor] (0,0) circle (\vertexsize);
\draw[fill=VertexColor,VertexColor] (1.5,0) circle (\vertexsize);
\draw[fill=VertexColor,VertexColor] (2,0) circle (\vertexsize);
\path[clip] (0,0) rectangle (2,1);
\draw[scalar] (.75,0) circle (.5);
\draw[scalar] (1.25,0) circle (.5);
\draw[wilson] (0,0) -- (2,0);
\draw[fill=white] (.25,0) circle (.075);
\draw[fill=white] (.75,0) circle (.075);
\draw[fill=white] (1.25,0) circle (.075);
\draw[fill=white] (1.75,0) circle (.075);
\draw[fill=VertexColor,VertexColor] (1,.43) circle (\vertexsize);
\draw[fill=VertexColor,VertexColor] (1.6,.36) circle (\vertexsize);
\draw[fill=VertexColor,VertexColor] (0,0) circle (\vertexsize);
\draw[fill=VertexColor,VertexColor] (1.5,0) circle (\vertexsize);
\draw[fill=VertexColor,VertexColor] (2,0) circle (\vertexsize);
\end{tikzpicture}}

 \newcommand{\ONDiagramNLO}{\begin{tikzpicture}[baseline={([yshift=-2ex]current bounding box.center)}]
\draw[wilson] (0,0) -- (1.1,0);
\draw[scalar] (.3,0) -- (.55,.5);
\draw[scalar] (.8,0) -- (.55,.5);
\draw[scalar] (.3,0) -- (.55,.5);
\draw[scalar] (.95,0.3) -- (.55,.5);
\draw[scalar] (.55,.5) -- (.55,1);
\draw[fill=white] (.55,1) circle (.1);
\draw[fill=white] (.55,.85) circle (.075);
\draw[fill=white] (.3,0) circle (.075);
\draw[fill=white] (.8,0) circle (.075);
\draw[fill=VertexColor,VertexColor] (.55,.48) circle (\vertexsize);
\draw[fill=VertexColor,VertexColor] (1,0) circle (\vertexsize);
\draw[fill=VertexColor,VertexColor] (.1,0) circle (\vertexsize);
\draw[fill=VertexColor,VertexColor] (.55,0) circle (\vertexsize);
\end{tikzpicture}}

\newcommand{\XIntegral}{\begin{tikzpicture}[baseline={([yshift=-.35ex]current bounding box.center)}]
\draw[scalar] (0,0) -- (1.25,1);
\draw[scalar] (0,1) -- (1.25,0);
\draw[fill=VertexColor,VertexColor] (.625,.5) circle (\vertexsize);
\draw[fill=white] (0,0) circle (.075);
\draw[fill=white] (0,1) circle (.075);
\draw[fill=white] (1.25,1) circle (.075);
\draw[fill=white] (1.25,0) circle (.075);
\end{tikzpicture}}
\newcommand{\YIntegral}{\begin{tikzpicture}[baseline={([yshift=-.45ex]current bounding box.center)}]
\draw[scalar] (0,.5) -- (.75,.5);
\draw[scalar] (.75,.5) -- (1.25,1);
\draw[scalar] (.75,.5) -- (1.25,0);
\draw[fill=VertexColor,VertexColor] (.75,.5) circle (\vertexsize);
\draw[fill=white] (0,.5) circle (.075);
\draw[fill=white] (1.25,1) circle (.075);
\draw[fill=white] (1.25,0) circle (.075);
\end{tikzpicture}}

\newcommand{\IntegralYOneTwoTwo}{\scalebox{\x}{\begin{tikzpicture}[baseline={([yshift=-.55ex]current bounding box.center)}]
\draw[scalar] (0,.5) -- (1,.5);
\draw[scalar] (1.5,.5) circle (.5);
\draw[fill=white] (0,.5) circle (.075);
\draw[fill=white] (2,.5) circle (.075);
\draw[fill=VertexColor,VertexColor] (1,.5) circle (\vertexsize);
\end{tikzpicture}}}
\newcommand{\IntegralXOneTwoThreeThree}{\scalebox{\x}{\begin{tikzpicture}[baseline={([yshift=-.55ex]current bounding box.center)}]
\draw[scalar] (.5,1) -- (1,.5);
\draw[scalar] (.5,0) -- (1,.5);
\draw[scalar] (1.5,.5) circle (.5);
\draw[fill=white] (.5,1) circle (.075);
\draw[fill=white] (.5,0) circle (.075);
\draw[fill=white] (2,.5) circle (.075);
\draw[fill=VertexColor,VertexColor] (1,.5) circle (\vertexsize);
\end{tikzpicture}}}
\newcommand{\IntegralXOneOneTwoTwo}{\scalebox{\x}{\begin{tikzpicture}[baseline={([yshift=-.55ex]current bounding box.center)}]
\draw[scalar] (.5,.5) circle (.5);
\draw[scalar] (1.5,.5) circle (.5);
\draw[fill=white] (0,.5) circle (.075);
\draw[fill=white] (2,.5) circle (.075);
\draw[fill=VertexColor,VertexColor] (1,.5) circle (\vertexsize);
\end{tikzpicture}}}
\newcommand{\IntegralFOneThreeTwoThree}{\scalebox{\x}{\begin{tikzpicture}[baseline={([yshift=-.55ex]current bounding box.center)}]
\draw[scalar] (0,0) -- (.625,0);
\draw[scalar] (0,1) -- (.625,1);
\draw[scalar] (.625,0) arc (270:450:.5);
\draw[scalar] (.625,0) -- (.625,1);
\draw[fill=VertexColor,VertexColor] (.625,0) circle (\vertexsize);
\draw[fill=VertexColor,VertexColor] (.625,1) circle (\vertexsize);
\draw[fill=white] (0,0) circle (.075);
\draw[fill=white] (0,1) circle (.075);
\draw[fill=white] (1.125,.5) circle (.075);
\end{tikzpicture}}}

\newcommand{\KiteIntegral}{\begin{tikzpicture}[baseline={([yshift=-.35ex]current bounding box.center)}]
\draw[scalar] (0,0) -- (2,0);
\draw[scalar] (.5,0) -- (1,1);
\draw[scalar] (.5,0) -- (1,-1);
\draw[scalar] (1.5,0) -- (1,1);
\draw[scalar] (1.5,0) -- (1,-1);
\draw[ghost] (1,-1) -- (1,1);
\draw[fill=VertexColor,VertexColor] (.5,0) circle (\vertexsize);
\draw[fill=VertexColor,VertexColor] (1.5,0) circle (\vertexsize);
\draw[fill=white] (0,0) circle (.075);
\draw[fill=white] (2,0) circle (.075);
\draw[fill=white] (1,1) circle (.075);
\draw[fill=white] (1,-1) circle (.075);
\node[] at (-.3,0) {{\footnotesize $2$}};
\node[] at (2.3,0) {{\footnotesize $4$}};
\node[] at (1,1.3) {{\footnotesize $1$}};
\node[] at (1,-1.3) {{\footnotesize $3$}};
\end{tikzpicture}}

\usepackage{tikz}
\usetikzlibrary{
  positioning,
  arrows,
  calc,
  arrows.meta,
  decorations.pathmorphing,
  decorations.markings,
  shapes.geometric,
  patterns
}
\tikzset{
  snake it/.style={
    decorate, 
    decoration=snake,
    segment length=3
  }
}
\usepackage[compat=1.1.0]{tikz-feynman} 
\usepackage[sorting=none,backend=biber,style=numeric-comp,refsection=chapter]{biblatex}
\usepackage[
  a4paper,
  twoside,
  inner=25mm,
  outer=20mm,
  top=25mm,
  bottom=25mm
]{geometry}

\usepackage{caption}
\usepackage{titlesec}
\usepackage{titletoc}
\titlecontents*{subsubsection}[7.1em]
{\small}
{\thecontentslabel. }
{}
{, \thecontentspage}
[.---][.]
\usepackage{tabularx}
\usepackage{booktabs}
\usepackage{caption}
\usepackage{afterpage}
\title{
\textbf{\Large{Analytical Methods in Quantum Field Theories}}\\
\textbf{\large{From Loop Integrals to Defect Correlators}}\\[.618cm]
  \textbf{\Huge{DISSERTATION}\\ 
  \Large{zur Erlangung des akademischen Grades} \\
  \Huge{DOCTOR RERUM NATURALIUM}\\
  \Large{(Dr. rer. nat.)}}\\[1cm]
  {\large im Fach: Physik\\
   \large Spezialisierung: Theoretische Physik}\\[0.618cm]
  {\large eingereicht an der \\ Mathematisch-Naturwissenschaftliche Fakult\"at \\ der Humboldt‐Universit\"at zu Berlin}
}
\author{\large{\textbf{von}}\\\huge{\textbf{Daniele Artico}}\\[0.618cm]
		\small{Pr\"asidentin der Humboldt-Universit\"at zu Berlin}\\\large{Prof. Dr. Julia von Blumenthal}\\
		\small{Dekan der Mathematisch-Naturwissenschaftliche Fakult\"at}\\\large{Prof. Dr. Emil J. W. List-Kratochvil}\\[.618cm]
		\small{Gutachter*innen}\\\large{{Prof. Dr. Jan C. Plefka}}\\\large{{Prof. Dr. Charlotte F. Kristjansen}}\\\large{{Prof. Dr. Lorenzo Magnea}}}
\date{\small{Verteidigung:}\\\large{{29.09.2025}}}

\definecolor{DarkBlueGrey}{RGB}{76,94,107}
\definecolor{MediumBlueGrey}{RGB}{110,135,153}
\definecolor{LightBlueGrey}{RGB}{134,163,184}
\definecolor{VeryLightBlueGrey}{RGB}{242,249,255}
\definecolor{WCOrange}{RGB}{242,146,29}
\definecolor{VeryLightOrange}{RGB}{255,245,233}
\definecolor{SCRed}{RGB}{179,48,48}
\definecolor{VeryLightRed}{RGB}{255,239,239}
\definecolor{VertexColor}{RGB}{242,146,29}
\definecolor{GluonColor}{RGB}{255,172,172}
\definecolor{SEColor}{RGB}{134,163,184}

\definecolor{BGBox}{RGB}{255,254,230}
\definecolor{PlaneColor}{RGB}{230,230,230}
\definecolor{BlobColor}{RGB}{190,180,230}

\def\shuffle{\sqcup\mathchoice{\mkern-7mu}{\mkern-7mu}{\mkern-3.2mu}{\mkern-3.8mu}\sqcup}

\newcommand{\pd}{\partial}
\newcommand{\spd}{\slashed{\partial}}
\newcommand{\sx}{\slashed{x}}

\newcommand{\Li}{{\normalfont\text{Li}}}

\newcommand{\vev}[1]{\langle\, #1 \, \rangle}

\newcommand{\ub}{\bar{u}}

\def\veps{\varepsilon}

\newcommand{\Wl}{\mathcal{W}_\ell}

\newcommand{\Op}{\mathcal{O}}
\newcommand{\Oh}{\hat{\mathcal{O}}}
\newcommand{\Dh}{{\hat{\Delta}}}
\newcommand{\uh}{{\hat{u}}}
\newcommand{\nh}{{\hat{n}}}

\newcommand{\lambdah}{\hat{\lambda}}

\def\Am{{\mathcal{A}}}
\def\Bm{{\mathcal{B}}}
\def\Cm{{\mathcal{C}}}

\def\Gm{{\mathcal{G}}}

\def\Im{{\mathcal{I}}}

\def\Km{{\mathcal{K}}}
\def\Lm{{\mathcal{L}}}

\def\Nm{{\mathcal{N}}}
\def\Om{{\mathcal{O}}}
\def\Pm{{\mathcal{P}}}

\def\Sm{{\mathcal{S}}}

\def\Wm{{\mathcal{W}}}

\def\Fds{{\mathbb{F}}}

\def\Ids{{\mathbb{I}}}

\def\Tds{{\mathbb{T}}}

\def\veps{\varepsilon}

\newcommand\psib{{\bar{\psi}}}
\newcommand\cb{{\bar{c}}}

\newcommand\zb{{\bar{z}}}

\newif\ifstartcompletesineup
\newif\ifendcompletesineup
\pgfkeys{
    /pgf/decoration/.cd,
    start up/.is if=startcompletesineup,
    start up=true,
    start up/.default=true,
    start down/.style={/pgf/decoration/start up=false},
    end up/.is if=endcompletesineup,
    end up=true,
    end up/.default=true,
    end down/.style={/pgf/decoration/end up=false}
}
\pgfdeclaredecoration{complete sines}{initial}
{
    \state{initial}[
        width=+0pt,
        next state=upsine,
        persistent precomputation={
            \ifstartcompletesineup
                \pgfkeys{/pgf/decoration automaton/next state=upsine}
                \ifendcompletesineup
                    \pgfmathsetmacro\matchinglength{
                        0.5*\pgfdecoratedinputsegmentlength / (ceil(0.5* \pgfdecoratedinputsegmentlength / \pgfdecorationsegmentlength) )
                    }
                \else
                    \pgfmathsetmacro\matchinglength{
                        0.5 * \pgfdecoratedinputsegmentlength / (ceil(0.5 * \pgfdecoratedinputsegmentlength / \pgfdecorationsegmentlength ) - 0.499)
                    }
                \fi
            \else
                \pgfkeys{/pgf/decoration automaton/next state=downsine}
                \ifendcompletesineup
                    \pgfmathsetmacro\matchinglength{
                        0.5* \pgfdecoratedinputsegmentlength / (ceil(0.5 * \pgfdecoratedinputsegmentlength / \pgfdecorationsegmentlength ) - 0.4999)
                    }
                \else
                    \pgfmathsetmacro\matchinglength{
                        0.5 * \pgfdecoratedinputsegmentlength / (ceil(0.5 * \pgfdecoratedinputsegmentlength / \pgfdecorationsegmentlength ) )
                    }
                \fi
            \fi
            \setlength{\pgfdecorationsegmentlength}{\matchinglength pt}
        }] {}
    \state{downsine}[width=\pgfdecorationsegmentlength,next state=upsine]{
        \pgfpathsine{\pgfpoint{0.5\pgfdecorationsegmentlength}{0.5\pgfdecorationsegmentamplitude}}
        \pgfpathcosine{\pgfpoint{0.5\pgfdecorationsegmentlength}{-0.5\pgfdecorationsegmentamplitude}}
    }
    \state{upsine}[width=\pgfdecorationsegmentlength,next state=downsine]{
        \pgfpathsine{\pgfpoint{0.5\pgfdecorationsegmentlength}{-0.5\pgfdecorationsegmentamplitude}}
        \pgfpathcosine{\pgfpoint{0.5\pgfdecorationsegmentlength}{0.5\pgfdecorationsegmentamplitude}}
}
    \state{final}{}
}

\tikzset{
corner/.style={line width=1pt,dashed,draw=black,dash pattern=on 6pt off 4pt},
scalar/.style={line width=1pt,draw=black},
gluon/.style={line width=1pt,decorate, draw=GluonColor,
    decoration={complete sines,aspect=0,amplitude=1.25mm,segment length=1.5mm,start up,end up}},
gluontwo/.style={line width=1pt,decorate, draw=GluonColor,
    decoration={complete sines,aspect=0,amplitude=.7mm,segment length=1mm,start up,end up}},
ghost/.style={line width=1pt,loosely dotted,draw=black},
wilson/.style={line width=2pt,draw=black},
multiscalar/.style={line width=3pt,draw=black},
 }

\NewDocumentCommand\semiloop{O{black}mmmO{}O{above}}
{%
\draw[#1] let \p1 = ($(#3)-(#2)$) in (#3) arc (#4:({#4+180}):({0.5*veclen(\x1,\y1)})node[midway, #6] {#5};)
}

\newcommand{\beq}{\begin{equation}}
\newcommand{\eeq}{\end{equation}}
\newcommand{\nn}{\nonumber}
\newcommand{\tr}{\text{tr}}

\DeclareUnicodeCharacter{0301}{\'{e}}
\DeclareUnicodeCharacter{2212}{\textendash}
 \DeclareUnicodeCharacter{202F}{\,}
 
\usepackage{calc}
\usepackage[pdftitle={Analytical Methods in Quantum Field Theories: From Loop Integrals to Defect Correlators},
  pdfauthor={Daniele Artico},
  pdfsubject={Dissertation},
  pdfkeywords={Quantum Field Theory, Loop Integrals, Defect Correlators, Defect CFT, Conformal Bootstrap, Dissertation, Physics, Quantenfeldtheorie, Defekt-CFT, Physik}
  pdfcreator={LaTeX},
  pdfproducer={pdfLaTeX},
  pdflang={en-US},
  citecolor=blue,
  linkcolor=blue,
  colorlinks=true,
  linktocpage=true,
  bookmarksopen=true,
  bookmarksopenlevel=1,
  pdfpagemode=UseOutlines,
  pdfstartview=Fit]{hyperref}
\usepackage{bookmark}
\usepackage{cleveref}
\crefformat{equation}{#2equation~#1#3}
\Crefformat{equation}{#2Equation~#1#3}

\begin{document}
\pagenumbering{gobble} 
\vspace*{\fill}
\begin{center}
\includegraphics[scale=1]{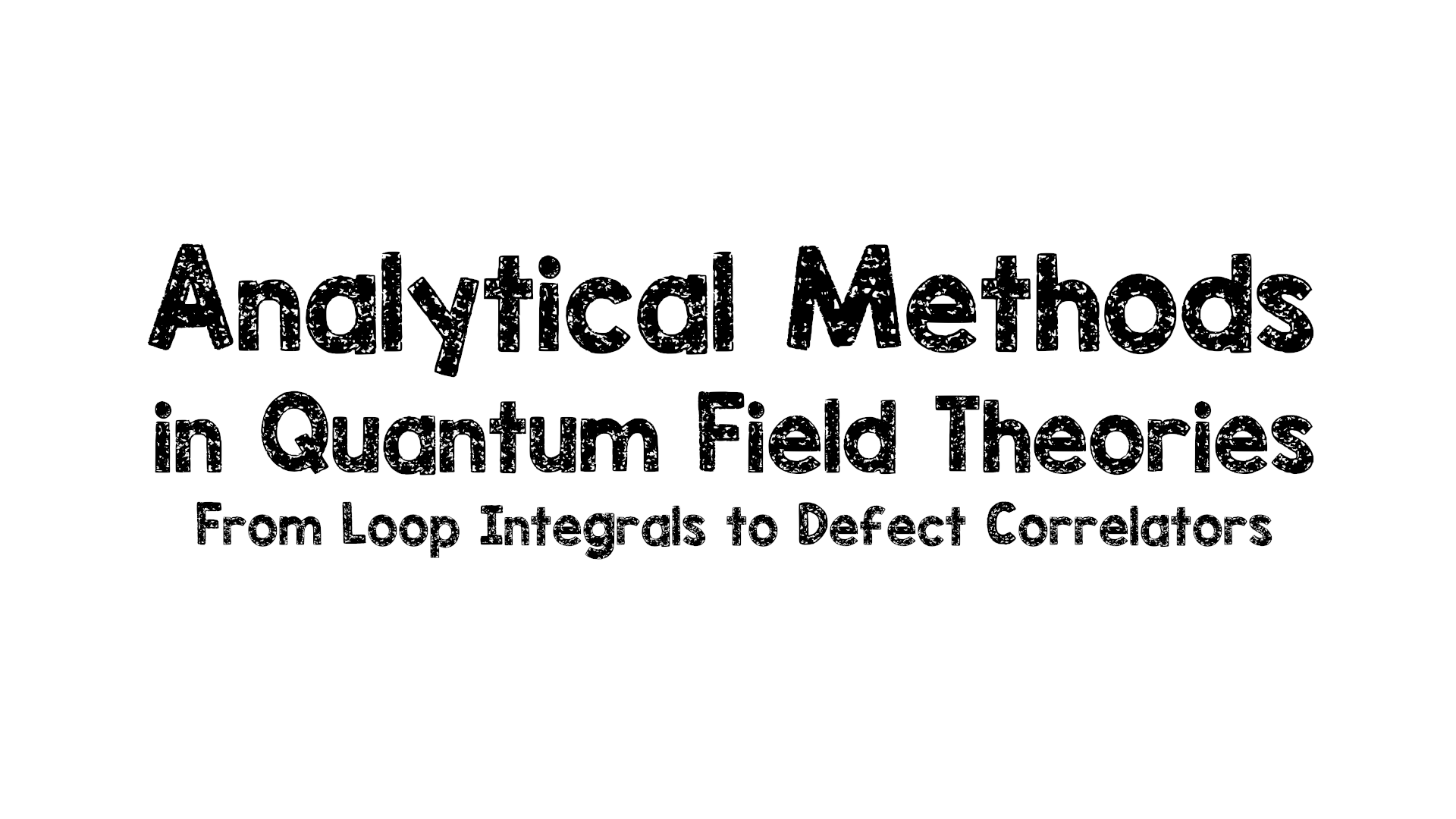}
\end{center}
\vspace*{\fill}

\maketitle

\noindent
\pagebreak
\begin{minipage}[t]{.49\textwidth}
\end{minipage}
\hfill
\begin{minipage}[t]{.49\textwidth}
\vspace{5cm}
\textit{This thesis is dedicated to Refaat Alareer, Sufian Tayeh and all the scholars and students killed in the genocide in Gaza.}
\end{minipage}
\newpage 

\ 

\newpage

\begin{center}
\textbf{Declaration of authorship}
\end{center}
\small\textit{
I declare that I have completed the thesis independently using only the aids and tools specified. I have not applied for a doctor’s degree in the doctoral subject elsewhere and do not hold a corresponding doctor’s degree. I have taken due note of the Faculty of Mathematics and Natural Sciences PhD Regulations, published in the Official Gazette of Humboldt-Universität zu Berlin no. 42/2018 on 11/07/2018.}\\[0.5cm]
This thesis is an original monograph based on the research works
\begin{itemize}
\item[J1] \textbf{D. Artico}, J. Barrat, and G. Peveri, “Perturbative bootstrap of the Wilson-line defect 	CFT: Multipoint correlators,” \textit{JHEP} 02 (2025) 190, arXiv: 2410.08271 [hep-th].
	Published: Feb 27, 2025
\item[J2] \textbf{D. Artico}, J. Barrat, and Y. Xu, “Perturbative bootstrap of the Wilson-line defect 	CFT:Bulk-defect-defect correlators,” \textit{JHEP} 03 (2025) 191, arXiv: 2410.08273 [hep-th].
	Published: Mar 26, 2025
\item[J3] \textbf{D. Artico} and L. Magnea, “Integration-by-parts identities and diﬀerential equations 	for parametrised Feynman integrals,” \textit{JHEP} 03 (2024) 096, arXiv: 2310.03939 	[hep-ph].
	Published: Mar 15, 2024
\item[P1] \textbf{D. Artico} and L. Magnea, “IBPs and diﬀerential equations in parameter space,” vol. 	RADCOR2023,2024, p. 058., 
	arXiv: 2311.02457 [hep-ph].
	Published: Mar 15, 2024
\end{itemize}
I, the author, am aware of the University’s regulations concerning plagiarism, including those regulations concerning disciplinary actions that may result from plagiarism. Any use of the works of any other author, in any form, is properly acknowledged at their point of use. The following research works have also been published during my time as a PhD candidate at Humboldt-Universit\"at zu Berlin but are not part of the scientific findings discussed in this thesis:
\begin{itemize}
\item[J4] \textbf{D. Artico}, S. Durham, L. Horn, F. Mezzenzana, M. Morrison, and A. Norberg, 	“«Beyond being analysts of doom»: Scientists on the frontlines of climate action,”
	Frontiers in Sustainability, vol. 4, 2023, issn:2673-4524. 
	Published: Jun 01, 2023
\item[P2] \textbf{D. Artico}, “Loop Tree Duality with generalized propagator powers: numerical UV 	subtraction for two-loop Feynman Integrals,” vol. LL2024, 2024, p. 025., 
	arXiv: 2409.01313 [hep-ph].
	Published: Sep 17, 2024
\end{itemize}

The journal paper J3 ad the conference proceeding P1 are the ones the first chapter of my thesis is based on. My co-author and I share equal contribution to the scientific findings. The introduction and conclusions of paper J3 and the proceeding P2 were drafted by Lorenzo Magnea and edited by me. The central body of paper J3 were 	drafted by me and edited by Lorenzo Magnea. The journal papers J1 and J2 contain the core findings reported in the second and third chapters of my thesis. All authors equally contributed to the scientific findings in the papers. The paper J1 was drafted by Julien Barrat, edited by me and reviewed by Giulia 	Peveri. The paper J2 was drafted by me, edited by Julien Barrat and reviewed by 	Yingxuan Xu. The journal paper J4 received equal scientific and writing contributions by all authors, as stated in the paper. The conference proceeding P2 is the result of my independent work of research and writing. I am grateful to Dirk Kreimer for useful comments on the ideas.
\begin{flushright}
\end{flushright}
\normalsize
\pagebreak
\begin{center}
\textbf{Abstract}
\end{center}
\small
Correlation functions hold a prominent position among the mathematical tools used to test our models of nature. In particle physics, scattering amplitudes are closely connected to the probability for a given particle interaction to happen; in black hole physics, they can contribute to the modeling of gravitational waves; in cosmology they can help investigate the properties of the inflationary universe and in conformal field theory correlation functions of local operators contain information on measurable quantities such as scaling dimensions. A wide variety of computational techniques corresponds to this multitude of applications, whose range of validity spans from weak coupling (e.g., Feynman diagrams) to strong coupling (e.g., bootstrap techniques) to any value of the coupling (e.g., integrability-based approaches). This thesis expands the available techniques at weak coupling by investigating the linear space of Feynman integrals and the role that (super)symmetry plays in reducing the number of integrals necessary to calculate correlators in the presence of a one-dimensional extended operator -- the line defect. In the first part, linear relations among Feynman parametrized integrals are derived from their properties as projective forms; these relations are then tested on one- and multi-loop examples, and their connection to the algebra of polynomial ideals is uncovered. In the second part, made of two chapters, the defect CFT formed by the $\Nm = 4$ super Yang-Mills theory in the presence of a Maldacena-Wilson line is studied through bulk-defect-defect and multipoint correlation functions up to next-to-next-to-leading order in the perturbative expansion at weak coupling. The investigations into this defect CFT lead to the identification of classes of integrals containing all the perturbative information necessary to compute the correlators, which turn out to be either rational functions or Goncharov polylogarithms.\\

Korrelationsfunktionen sind das zentrale mathematische Instrument, um Theorien, die die Natur beschreiben, zu testen. In der Teilchenphysik stehen Streuamplituden in engem Zusammenhang mit der Wahrscheinlichkeit, dass eine bestimmte Teilchenwechselwirkung stattfindet. In der Gravitationsphysik können sie zur Modellierung von Gravitationswellen beitragen. Kosmologische Korrelatoren können helfen, die Eigenschaften des inflationären Universums zu untersuchen und in konformen Feldtheorien (CFT) enthalten Korrelationsfunktionen von lokalen Operatoren Informationen über messbare Größen wie die Skalendimensionen. Diese Vielzahl von Anwendungen hat die Entwicklung einer breiten Palette von Berechnungstechniken motiviert, deren Gültigkeitsbereich von schwacher Kopplung (z. B. Feynman-Diagramme) über starke Kopplung (z. B. Bootstrap-Techniken) bis hin zu beliebigen Werten der Kopplung (z. B. auf Integrabilität basierende Ansätze) reicht. Diese Arbeit erweitert die verfügbaren Methoden bei schwacher Kopplung: (Super-) Symmetriebeschränkungen, die zur Berechnung von Korrelatoren in Gegenwart eines eindimensionalen erweiterten Operators - des Liniendefekts - erforderlich sind, können den liniaren Raum und die Anzahl der Feynman-Integrale reduzieren. Im ersten Teil werden lineare Beziehungen zwischen parametrisierten Feynman-Integralen von ihren Eigenschaften als projektive Formen abgeleitet. Diese werden dann an Ein- und Mehrschleifenbeispielen getestet, und ihre Verbindung zu nichtlinearer Algebra von Polynomidealen wird herausgearbeitet. Der zweite Teil, der aus zwei Kapiteln besteht, handelt von der Defekt-CFT, die von der $\Nm = 4$ Super-Yang-Mills-Theorie in Anwesenheit einer Maldacena-Wilson-Linie gebildet wird. Sie wird anhand von Bulk-Defekt-Defekt- und Mehrpunkt-Korrelationsfunktionen bis zur zweiten Ordnung in der Störungstheorie bei schwacher Kopplung untersucht. Die gewonnenen Erkenntnisse der Defekt-CFT erlauben es, Integralklassen zu identifizieren, die alle notwendigen Störungsinformationen für die Berechnung der Korrelatoren enthalten und sich entweder als rationale Funktionen oder Goncharov-Polylogarithmen ausdrücken lassen.
\normalsize
\pagebreak

\begin{center}
\textbf{Acknowledgments}
\end{center}
\small
It is difficult to condense in a few paragraphs the amount of gratefulness I feel for the large number of people who contributed to my scientific growth in the last five years, and even impossible to name all the extraordinary individuals who shaped my personal growth. If you are reading these acknowledgments, you may likely be among them. I am grateful for everything we shared. What follows is a list of people who have supported my work in these years, to whom I address my professional and personal gratitude.\\

My doctoral studies were advised and mentored by an unusual number of scientific staff, whom I mention in chronological order. To \textsc{Markus Schultze}, \textsc{Peter Uwer} and the Phenomenology of Particle Physics group of Humboldt-Universit\"at zu Berlin goes my gratitude for welcoming me in their group, for sharing a lot of their knowledge about particle physics, for supporting my independent research even when it detached from the direction of the group, and for teaching me the value of autonomy in carrying on a research project. I sincerely thank \textsc{Dirk Kreimer} for our extended discussions on Feynman integrals divergences and for his precious career advice. I express my deep gratitude to \textsc{Jan Plefka} for accepting me as his PhD student when I decided to focus my research on defect CFT and for the strong support I received in my search for a postdoc position. I am profoundly grateful to \textsc{Lorenzo Magnea} for a collaboration that started as a thesis supervision during my years at Universit\`a degli Studi di Torino and evolved into a scientific cooperation where I am treated as a peer. I learned a lot from Lorenzo both from a scientific and personal point of view and I am looking forward to our next research enterprise.\\

This thesis would not have been possible without the guidance and support of \textsc{Julien Barrat}. We shared much more than the authorship of our papers: the student seminar organization, the creation of an outreach YouTube channel, a quantum devil performance, deep conversations about physics and life, a number of lunches at the falafel stand, and, last but not least, joyous walks with his family. There is a word to summarize all of this: we shared a friendship that I am honored to have.\\

In my years at Humboldt-Universit\"at I was lucky to be part of the doctoral program RTG2575 -- Rethinking Quantum Field Theory, funded by the Deutsche Forschungsgemeinschaft with project number 41753389. Within the senior staff of the RTG, \textsc{Agostino Patella} holds a prominent place in my acknowledgments: I thank him for having listened to my concerns in the moments of most need, for always providing critical insights and kindly voicing disagreements, and for having been available to envision together the role of mentor in the second round of the RTG. A special thanks goes to \textsc{Valentina Forini} for the insightful and enjoyable conversations way beyond physics, for the recommendation letters sent around, and for being available as the president of the defense committee. All the students of the RTG have had an impact on my scientific path and consequently on this thesis, through the seminars, the informal discussions, and the lunch breaks. I thank in particular \textsc{Jasper Rosmalee Nepveu} and \textsc{Yingxuan Xu} for the wonderful years together in the phenomenology group, \textsc{Giulia Peveri} for being an excellent collaborator, discussion companion, and recently chess opponent, and all the members of the YouTube project \textsc{Non Standard Models} for having joined forces in trying to make our field more accessible and inclusive. Among the students I met at Humboldt-Univerist\"at, a special thanks goes to \textsc{Giuseppe Casale}, \textsc{Tomas Codina, Alessandro Cotellucci}, \textsc{Maddalena Ferragatta}, \textsc{Moritz Kade}, \textsc{Maria Kallimani}, \textsc{Camilla Lavino}, and \textsc{Allison Pinto}.\\

I want to express my gratitude to \textsc{Charlotte Kristjansen} for accepting to be one of the referees of this thesis, and to her and \textsc{Adam Chalabi} for the great hospitality at the Niels Bohr Institute in Copenhagen. I also thank \textsc{Gabriel Bliard} and \textsc{Philine Van Vliet} who welcomed me to ENS in  Paris, \textsc{Andrea Cavagli\`a} who welcomed me back in Torino and who was always available for insightful discussions; and \textsc{Carlo Meneghelli} and my future group in Parma for the enthusiasm and interest shown during my visit there.\\

\textsc{Lou Chevrollier} and \textsc{Claudio Iuliano} have contributed to the revision of some crucial parts of my research works, and much more. {Lou} has represented to me a model of freedom, integrity and courage since the day we met, and has been a fundamental reference point for some of the most difficult decisions of these years. If teaching is provoking the right questions, {Lou} has undoubtedly been a teacher to me. I am grateful to {Claudio} for having represented an example of resilience in an academic world that is always more toxic and competitive and that puts a lot of weight on students and precarious researchers. Sometimes, the joy of research is just in discovering together how to integrate products of Bessel functions. A sincere thanks goes to \textsc{Valerie Gu\'ed\'e} for helping me structure the intense work of this last year while keeping an eye on my well-being, and for reminding me to choose my fights. Our conversations have made this final manuscript real.\\

My experience as a worker in science would have been missing a fundamental part if I did not join the group \textsc{Scientist Rebellion}. There are people in this world who have decided to walk the talk and to act according to the emergency represented by the fast-progressing climate collapse. People who are risking their job and their freedom to make it clear that we need a radical change, because there are people already now suffering the much worse consequences of our model of over-consumption. I am grateful to all these people for making me reflect every day on my role as a scientist and, most importantly, as a human being. I cannot nominate all of these extraordinary people, so a very non-exhaustive list will have to suffice. My gratitude goes to \textsc{Rose Abramoff, Fabian Dablander, Elodie Duyck, Nana-Maria Grüning, Julia Halder, Laura Horn, Lorenzo Masini, Marta Matos, Lorenzo Maria Perrone, Fernando Racimo, Nate Rugh, Teresa Santos,} and \textsc{Laura Stalenhoef}.\\

The final mention of deep gratitude goes to my parents \textsc{Antonello Artico} and \textsc{Anna Maria Lovreglio}, to my sister \textsc{Giorgia Artico}, and to my friends \textsc{Sara Cocito, Ilaria Costa, Giacomo Milano, Antonia Lopiccolo, Andrea Oliva, Luca Tallone, Mattia Tallone} and \textsc{Lorenza Valsania} for the constant unconditional support. They have contributed to making me the person I am in many more ways than I can mention. \\

The resistance of the philosopher mentioned in the Introduction \textsc{Albert Lautman} against any form of reductionism in science had to become soon in his life the resistance against an oppressive regime. Member of the French Liberation units and head of a network of escape routes between Toulouse and Spain, he was arrested and executed on August 1st, 1944. His fight against the Nazi conquest echoed his resistence against the `rejection of the spirit of research in favour of violent attitudes that reason has nothing to do with'. It is from this deep love for freedom and research that I want to take inspiration.   \\
\vspace*{\fill}
\pagebreak

\pagenumbering{arabic}
\setcounter{page}{1}
\tableofcontents

\chapter*{Introduction}
\markboth{Introduction}{}
\addcontentsline{toc}{chapter}{Introduction}
The research question that unites all the chapters of this thesis is, deep down, a question that recurs throughout the history of physics -- in the broad sense of the discipline trying to explain a vast number of natural phenomena with mathematical models. This question can be expressed as follows: given a physical model, with its rules and range of applicability, what are the mathematical functions that will appear in the description of phenomena according to this model? Of course, no scientific work of finite length can ever answer this question in an exhaustive way, and the concrete scientific findings presented in each chapter will answer much more concise questions. This interest for the classes of functions that are necessary to describe observables stands at the foundation of this thesis, especially for the observables of theories presenting symmetries that we can classify. A brilliant exposition of this deep relationship present in several branches of mathematics is contained in the work of the French philosopher Albert Lautman, which we briefly quote for the sharpness of his description.\\

The years 1935-1937 were marked by the rise and dominance of the neopositivism of the Vienna Circle and analytic philosophy in the realm of philosophy of mathematics. Rather than developing a deep analysis of the exciting developments in mathematics at the time -- the works of Noether, Galois, Riemann, and Poincaré all appeared in the previous century -- the majority of philosophers in the field were interested in reducing mathematics to the mere theory of sets and logic. It is against this current that the young Albert Lautman raised his arguments during his intervention at the \textit{Congrès Descartes} of 1937 in Paris. When describing the reality inherent to mathematical theories \cite{Lautman:1937gg}, Lautman states that
\begin{quote}
\textit{In all these purely mathematical examples, we always see a way of structuring a basic sector that can be interpreted in terms of existence for certain new entities, functions, transformations, and numbers that the structure of the sector thus seems to preform.}
\end{quote}
The quoted movement inside mathematical theories, the relationship between the structure of a sector and the existence of new entities, refers to the remarkable ability of a mathematical theory to create new objects starting from axioms and abstract ideas. The examples Lautman brings in his text testify to his deep knowledge of the mathematics of his time. The relationship between the genus of a surface and the number of independent finite integrals defined on such surface, between the number of classes of finite groups and its irreducible representations, and finally between the features of a physical model and the functions characterizing phenomena are all examples of ways in which the defining elements of a field find a correspondence in outcomes and observables. This study of the functions that need to be used to describe physical phenomena is indeed a constant in the history of science. All physical theories have tried to model reality via principles from which functions can be derived and then compared with measurements: from the circular motions of planets described by Greek, Hellenistic, Persian, and Arab astronomers to Kepler's law, from the law describing the period of the pendulum to plane waves, from Coulomb and Ampere laws to Maxwell equations. Quantum field theory, the most complete description of subatomic particles we have so far, makes no distinction: via the analytic and numeric calculation of scattering amplitudes and correlation functions we develop and test our model of microscopic reality, and the understanding of the functions appearing in this model is one of the interesting aspect of our research. During the years, the space of functions grew richer with the connections to Goncharov polylogarithms \cite{Duhr:2011zq}, elliptic polylogarithms \cite{Broedel:2019hyg}, and functions defined on Calabi-Yau varieties to recently describe scattering among black holes \cite{Driesse:2024feo}. In this thesis, we consider two topics within this broad perspective of understanding the space of functions describing different quantum field theories: the study of the vector space of Feynman integrals, and the analytic bootstrap of defect conformal field theories. Both topics are extensive research fields in their own right, and their relevance cannot be overstated. In the next paragraph, we report some of the motivations making these two fields particularly interesting in present-day research.\\		
\paragraph{The Standard Model and the precision program at particle detectors} \phantom{a}\\
Feynman integrals are a key tool in our current theoretical predictions of scattering amplitudes between elementary particles. The Standard Model of particle physics (SM) currently represents the most complete and advanced model of the subatomic world, with major successes such as the theoretical prediction of heavy quarks, weak-force bosons, and the Higgs particle before their experimental discoveries \cite{Glashow:1961tr, Weinberg:1967tq, Salam:1968rm, Higgs:1964pj, Englert:1964et}. It represents the union of several theories, including the Glashow–Salam–Weinberg model of electroweak unification \cite{Glashow:1961tr, Weinberg:1967tq, Salam:1968rm}, the Glashow–Iliopoulos–Maiani (GIM) mechanism for flavour-changing neutral currents \cite{Glashow:1970gm}, and quantum chromodynamics (QCD), which describes the strong interaction \cite{Fritzsch:1973pi, Gross:1973id, Politzer:1973fx}. The particle content of the SM consists of 17 particles classified according to the representations of the Lorentz group: 12 fermions—6 quarks and 6 leptons—4 kinds of gauge bosons, and 1 scalar particle. While providing a rigorous and convincing explanation for the behavior of subatomic particles, the model still presents some challenges and depends on 19 parameters which all have to be determined experimentally: 9 fermion masses—the neutrino mass being one of the open problems, as it is set to zero in the SM—3 mixing angles, one CP-violating phase, 3 gauge couplings, one QCD vacuum angle, the Higgs mass, and the Higgs vacuum expectation value. In Table \ref{tab:SM}, we list all the experimentally fixed parameters with their current bet fit. Beyond the mentioned problem of neutrino masses \cite{Super-Kamiokande:1998kpq, SNO:2001kpb}, the challenges of the SM include the absence of an explanation for gravity and dark matter, as well as the hierarchy problem. Following the recent exposition in \cite{Peskin:2025lsg}, we identify the hierarchy problem as the absence of a compelling explanation for electroweak symmetry breaking (EWSB), rather than a mere problem of scales. The Higgs potential in the SM takes the form
\begin{equation*}
V(\phi) \sim \mu^2\phi^2+\lambda \phi^4
\end{equation*}
where the parameters are inserted by hand with no deeper explanation. Quantum corrections require precise cancellations to keep the scale of EWSB far from the Planck scale \cite{tHooft:1979rat, Veltman:1980mj}, but perhaps it is more insightful to consider how previous scale hierarchy problems -- such as that in QCD -- were resolved through the discovery of underlying dynamical mechanisms of scale generation \cite{Dimopoulos:1981zb, Weinberg:1975gm}. All of these problems motivate extensions of the SM, either via UV completions or effective field theories, which model unknown interactions via known fields and higher-dimension operators \cite{Buchmuller:1985jz, Grzadkowski:2010es}. Both approaches benefit from high-precision theoretical predictions, such as scattering calculations at second (or higher) order in the weak coupling perturbative expansion. \\
\begin{table}[h!]
\centering
\renewcommand{\arraystretch}{1.3}
\begin{tabular}{|l|c|c|l|}
\hline
\textbf{Parameter} & \textbf{Symbol} & \textbf{Value} & \textbf{Units / Scheme} \\
\hline
Electron mass & $m_e$ & $0.510998950 \pm 0.000000015$ & MeV \\
Muon mass & $m_\mu$ & $105.6583755 \pm 0.0000023$ & MeV \\
Tau mass & $m_\tau$ & $1776.86 \pm 0.12$ & MeV \\
Up quark mass & $m_u$ & $1.9 \pm 0.2$ & MeV @ 2 GeV \\
Down quark mass & $m_d$ & $4.4 \pm 0.2$ & MeV @ 2 GeV \\
Strange quark mass & $m_s$ & $87 \pm 6$ & MeV @ 2 GeV \\
Charm quark mass & $m_c$ & $1320 \pm 20$ & MeV \\
Bottom quark mass & $m_b$ & $4240 \pm 20$ & MeV \\
Top quark mass & $m_t$ & $173500 \pm 800$ & MeV (on-shell) \\
\hline
CKM angle $\theta_{12}$ & $\theta_{12}$ & $13.1^\circ \pm 0.1^\circ$ & – \\
CKM angle $\theta_{23}$ & $\theta_{23}$ & $2.38^\circ \pm 0.06^\circ$ & – \\
CKM angle $\theta_{13}$ & $\theta_{13}$ & $0.201^\circ \pm 0.011^\circ$ & – \\
CKM CP phase & $\delta$ & $0.995 \pm 0.030$ & radians \\
\hline
U(1) gauge coupling & $g_1$ & $0.357$ & @ $\mu = m_Z$ \\
SU(2) gauge coupling & $g_2$ & $0.652$ & @ $\mu = m_Z$ \\
SU(3) gauge coupling & $g_3$ & $1.221$ & @ $\mu = m_Z$ \\
QCD $\theta$ angle & $\theta_{\mathrm{QCD}}$ & $\lesssim 10^{-10}$ & – \\
Higgs VEV & $v$ & $246.22 \pm 0.12$ & GeV \\
Higgs mass & $m_H$ & $125.25 \pm 0.17$ & GeV \\
\hline
\end{tabular}
\caption{The 19 free parameters of the Standard Model and their current best-fit values. Values are compiled from PDG 2024 \cite{PhysRevD.110.030001} and current literature.}
\label{tab:SM}
\end{table}
Particle physicists began drafting a wishlist of theoretical predictions of scattering amplitudes between 2004 and 2005 at Fermilab and at the Les Houches Workshop \textit{Physics at TeV Colliders 2005}, Standard
Model and Higgs working group. The original list included processes at next-to-leading order involving six external particles as the cutting edge. The so-called NLO revolution led to the automation of one-loop calculations and therefore to completing the wishlist early into the 2010s, making space for the new ambitious challenge of NNLO calculations involving multiloop integrals \cite{Heinrich:2020ybq}. The outstanding theoretical achievements involving amplitude and unitarity based techniques, symbol bootstrap and master integral decomposition are still not enough to guarantee effortless evaluation of scattering process up to two orders in the perturbative expansion; therefore, any progress in the understanding of the mathematics of loop integrals and in their fast evaluation represents a potential new step into the next generation of theoretical predictions. To understand the current level of pressure on the existing tools, we list here some of the most interesting processes studying the Higgs sector (whose relevance we can infer from the previous paragraph) with their level of theoretical precision at the time of publishing of Ref. \cite{Heinrich:2020ybq}: Higgs production in gluon fusion (with full top-quark mass dependence up to NLO \cite{Anastasiou:2009kn}, up to $N^3LO$ in the heavy top limit \cite{Chen:2021isd}), Higgs production in bottom quark fusion (up to $N^3LO$ in the five-flavour scheme \cite{Duhr:2019kwi}, up to NLO in the four-flavour scheme \cite{Dittmaier:2003ej}) and the Higgs production associated with a top-quark pair (available up to NLO for both strong and electroweak corrections \cite{Broggio:2019ewu}, with NNLO corrections available under some approximations \cite{Catani:2022mfv}) particularly relevant for the study of the EWSB, as multiple new physics models assume a special role for the heaviest quark. Further connections between the study of loop integrals and physical quantities of interest include the recent literature on gravitational waves \cite{Mogull:2020sak,Jakobsen:2021zvh,Jakobsen:2021smu,Driesse:2024xad,Driesse:2024feo,Hoogeveen:2025tew,Mogull:2025cfn} and cosmological correlators \cite{Green:2023ids,Arkani-Hamed:2023kig,Arkani-Hamed:2023bsv,DuasoPueyo:2023kyh,Salcedo:2024smn} for the study of inflation models. We will introduce some key concepts for both topics later in the thesis. For a more comprehensive review of the topics, the interested reader is referred to the mentioned references and their bibliographies. \\

\begin{figure}
\begin{center}
\includegraphics[scale=.5]{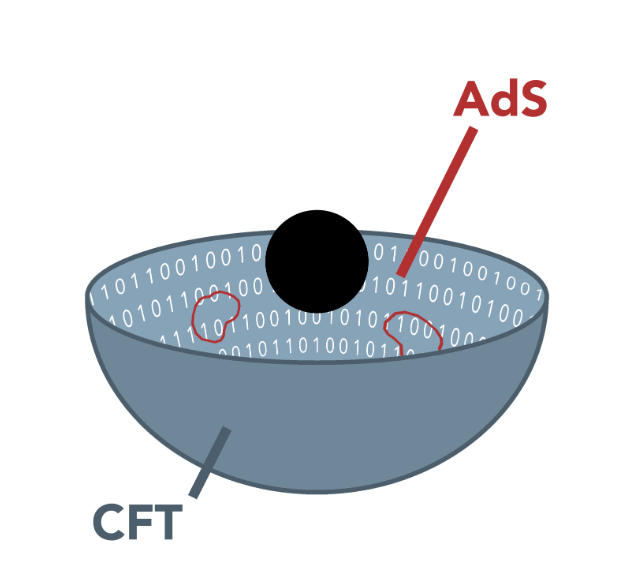}
\end{center}
\caption{A visual representation of the holographic principle. We thank Julien Barrat for the permission to use his drawing. The original can be found in Ref. \cite{Barrat:2024nod}.}
\label{fig:Holo}
\end{figure}

\paragraph{The holographic principle and the AdS/CFT correspondence}\phantom{a}\\
We can introduce the other field of interest in this thesis -- conformal field theory, in particular in the presence of a defect -- by referring to one of the challenges in the SM we mentioned above: the lack of a quantum theory for gravity. To introduce the connection between models of gravity and field theory, consider the famous paper by S. Hawking \cite{Hawking:1975vcx} proving how quantum effects can cause black holes to create and emit particles as if they were bodies with a finite non-zero temperature. Black holes are some of the most fascinating phenomena of nature, with an event horizon separating the spacetime into causally disconnected regions. Around black holes, gravity cannot be neglected with respect to other forces\footnote{More precisely, the surface gravity is very large near the event horizon.} and therefore there are quantum phenomena that require an explanation from the still missing quantum theory of gravity (e.g. the famous information paradox, see Ref. \cite{Hawking:2015qqa}). Despite still missing for black holes, the measurement of the radiation predicted in Ref. \cite{Hawking:1975vcx} for sonic analogues of black holes \cite{Steinhauer:2015saa} still represents an extraordinary achievement of theoretical physics in predicting phenomena before their experimental detection. When interpreted as thermodynamic objects, the entropy of black holes is proportional to the area, and not the volume \cite{Bekenstein:1973ur,Hawking:1975vcx}: in other words, the description of a volume of space can be thought of as encoded on a lower-dimensional boundary to the region. This statement as enounced represents an instance of the holographic principle \cite{tHooft:1993dmi, Susskind:1994vu}, schematically depicted in Figure \ref{fig:Holo}. More generally, the holographic principle postulates that a higher-dimensional theory with gravity can be fully described by a lower-dimensional theory without gravity. A particularly precise realization of this idea is given by the AdS/CFT correspondence \cite{Maldacena:1997re}, which relates a conformal field theory in flat space-time to a gravitational theory in a higher-dimensional Anti-de Sitter space. \\

We are going to introduce conformal symmetry and some features of conformal field theory in Chapter \ref{ch:BDD}, as well as providing some further motivation for our interest in CFT in the next paragraph on renormalization group flow and critical phenomena. In this introduction, however, we simply mention that a conformal field theory is a theory that is invariant under transformations preserving the angles, e.g. a global re-scaling of space-time. The AdS/CFT correspondence introduced in \cite{Maldacena:1997re} traces a correspondence between a gravitational theory in $d+1$ dimensions and a conformal field theory defined on the boundary of the AdS space. The most common example involves the maximally supersymmetric four-dimensional theory $\Nm = 4$ super Yang-Mills with gauge group $SU(N)$ -- a supersymmetric cousin of the strong force sector of the SM -- and introduces an important characteristic for the duality: the weak-strong coupling correspondence. The strongly coupled quantum field theory, a sector which is hardly accessible from standard perturbative calculations, is unlocked by its correspondence to a small perturbation of the free theory of gravity in the AdS space, which is possible to treat with the usual framework of perturbative expansion. This has led to progress both in particle physics \cite{Erdmenger:2007bn,Erdmenger:2007cm} and in condensed matter physics \cite{Evans:2001ab,Hartnoll:2008kx}.

\paragraph{Phase transitions and renormalization group flow}\phantom{a}\\
The relevance of conformal field theories is certainly not limited to their presence in the AdS/CFT correspondence. On the contrary, this duality represents possibly the most recent instance of an ubiquitous presence in physical models. The possibly strongest motivation of the interest for conformally symmetric theories comes from their connection to statistical physics and second-order phase transition.  To briefly characterize a second-order phase transition, consider the intuitive definition of correlation length of a system: the length characterizing until how far the fluctuations of the elements of the model in a point influence another point. We can imagine that, under ordinary conditions, thermal fluctuations of any quantity will have a local effect, thus giving origin to a finite correlation length; this is the case for fluids under ordinary conditions and also of gas-liquid mixes under so called first-order phase transitions, where the two phases mix but are still distinct. Under special conditions, for example under a specific temperature and pressure for fluids, the correlation length becomes infinite and the fluctuations of a quantity will have long-range effects as well.
\begin{figure}[!h]
\begin{center}
\includegraphics[scale=0.75]{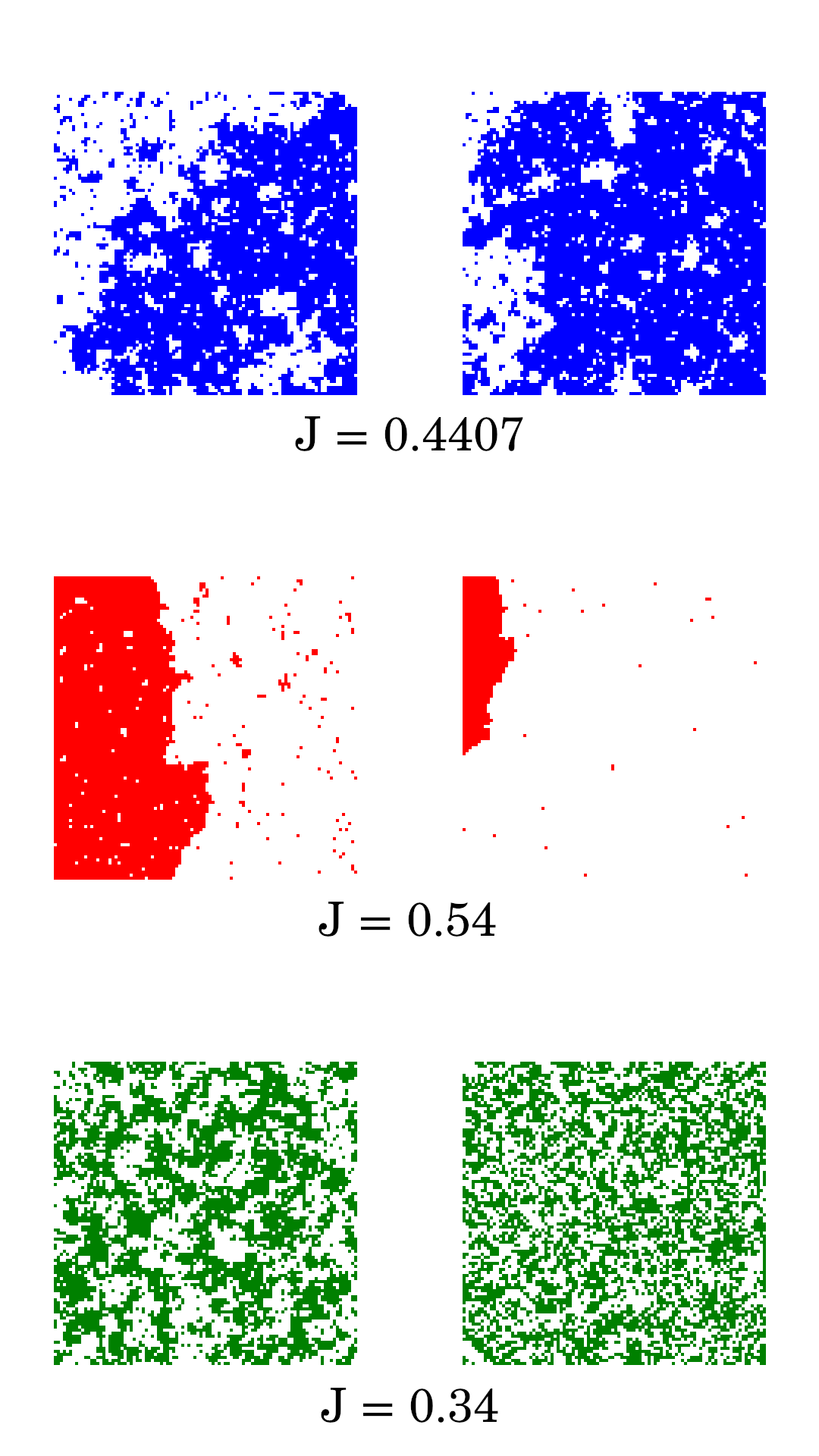}
\end{center}
\caption{$3\times3$ block-spin transformations for the $2d$ Ising model at the critical point (blue), above it (red), and below it (green). The images were generated by our version of the code by Douglas Stanford at \textit{http://large.stanford.edu/courses/2008/ph372/stanford1/c/}. On the left we reproduce the section of the original lattice; on the right, the block-spin transformed one. The value of the reduced coupling is reported below each image.}
\label{fig:Ising}
\end{figure}
In order to see this visually, consider the plots of the two-dimensional Ising model represented in Figure \ref{fig:Ising}. They are generated, respectively, at the critical point (blue color), above the critical point (red color), and below the critical point (green color) on a toroidal lattice with 2000x2000 sites. On the left, we represent a region of size 100x100 after a Monte Carlo simulation has been run for 5000 cycles. On the right, we represent the result of a block-spin transformation -- an averaging procedure between 9 neighboring sites conceptually analogous to zooming out from the area. What we observe is that at the critical points, the two images look fairly similar, with clusters of equal valued spins of different sizes that do not appear or disappear when applying a block spin transformation: this means that the points all influence each other and that the correlation length of the model diverges. Both above and beyond the critical point, we see the model either \textit{flowing} towards a so-called ferromagnetic state (large clusters of equal spins, the red image) or towards the so-called paramagnetic phase (disordered state, in green). The key questions to answer now to understand the role of conformal field theory in statistical physics and critical phenomena are two: how did we start discussing fluid phase transitions and end up with names that remind magnetization phenomena? And is there a reason why we chose the verb \textit{to flow}?  \\

The answer to both questions resides in the notion of the \textit{renormalization group flow}. Consider again Fig. \ref{fig:Ising}: if we keep the process of zooming out, ignoring more and more the short-range physics and focusing on large scales, what we technically do is study a theory that has a similar Hamiltonian, but a different coupling. Intuitively, when we integrate out some of the degrees of freedom of the two-dimensional Ising model above the most likely configurations are the ones where neighboring points have similar spins. The interaction between two blocks will then roughly have a new coupling strength that is multiplied by 3 (the number of pairs we integrate out) and therefore, if we start from a large enough coupling, the repetition of the block-spin transformation will transform the theory into a zero temperature one, where neighboring spins are strongly correlated. On the other hand, for high temperatures (small coupling) the system has to be in a paramagnetic state, which is therefore another attraction point of the block-spin transformation. The process of studying a process at a different scale where some small-scale degrees of freedom are removed is part of the framework of ideas known as renormalization group flow, and its immediate consequences are simple to sketch. This represents the answer to the question on the terminology of \textit{flow} we presented above: under scale transformations theories change and flow into different ones. In details, if we consider the linearized transformation of the new couplings $K'_a$ next to an attraction (also called a critical) point $K^*_a$, we can write
\begin{equation*}
K'_a-K^*_a = \sum_{b} T_{ab} (K_b-K^*_b)
\end{equation*}
and the eigenvalues of the matrix $T_{ab}$ will determine whether the coupling flows towards the critical point (irrelevant variables) or away from it (relevant variables). This idea offers an answer to the first question listed above: we could use magnetization-related terminology after speaking about fluid phase transitions because the renormalization group flow connects physical theories with different microscopic dynamics to classes of equivalence of the large-scale behavior, called \textit{universality classes}. The behavior of fluid systems and ferromagnetic materials next to a second phase transition is the same; more precisely, the scaling behavior of physical observables such as free energy, susceptibility, and correlation functions is the same. The nature of the fixed point of the renormalization group flow can be inferred from the perspective we adopted (which is the one in \cite{Cardy_1996}) of the renormalization flow as a result of the analysis of a theory at different scales. The fixed point of the flow will be theories that are invariant under scale transformations, \textit{i.e.} conformal field theories that then assume a role of signposts in the space of quantum field theories. The connection between the renormalization group flow and CFT led to landmark results such as the ones in Refs. \cite{Zamolodchikov:1986gt,Cardy:1988cwa}.\\

\paragraph{A tail for the spherical cow}\phantom{a}\\
It is difficult to trace back the origin of the long-standing \textit{spherical cow} joke among theoretical physicists, hinting at the tendency of our models of nature to reduce everything to the essential features. Seen from afar, every shape, object, human, and non-human animal will resemble a point or a sphere. It is a metaphor that hints at the renormalization flow mechanism, where the smaller characteristics that distinguish one model from another are neglected at sufficiently large scales \cite{Kadanoff:1966wm}. As simple as it may seem, this idea led to crucial developments in physics, such as the understanding that we can investigate the physics of the primordial universe by looking at energy density correlators at large scales, blind to the local inhomogeneities represented by planets, stars, and galaxies. Once again, this preamble emphasizes the importance of understanding the endpoints of the renormalization group flows in classifying theories based on their universal behaviors. According to the previous section, the study of conformal field theories is enough for this scope; however, the invariance under rescaling carries a consequence that justifies the introduction of the last element of the theories studied in this thesis: defects. \\
\begin{figure}[!h]
\begin{center}
\includegraphics[scale=0.5]{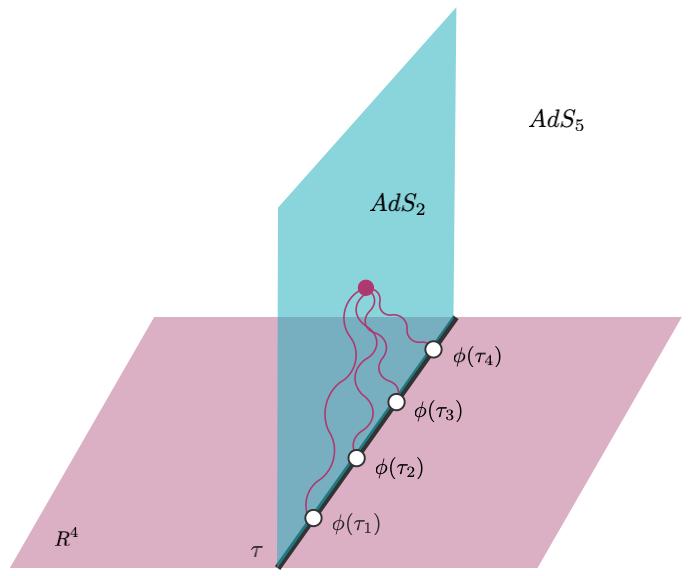}
\end{center}
\caption{Four-point function of local operators inserted on the Wilson line from a
Witten diagram on the AdS2 worldsheet. Credits to Giulia Peveri from the image taken with permission from \cite{Peveri:2023qip}.}
\label{fig:AdSdefect}
\end{figure}

When considering a non-conformal theory with a finite correlation length, we can justify an infinite volume assumption as the exponential decay of correlators justifies the notion of points being `far enough' from any boundary. In other words, the influence of the non-local extended operator represented by a boundary can be neglected for correlators in a non-conformal theory, assuming the fields we consider are not near this boundary. This is not the case anymore for conformal field theory, where correlators have a polynomial -- not exponential -- decay and where the presence of extended operators will influence correlation functions \cite{McAvity:1995zd,Andrei:2018die}. The framework developed to study the effect of non-local operators on a CFT is called defect conformal field theory and focuses on a particular kind of extended operators: the ones breaking the conformal symmetry in a controlled way \cite{Billo:2013jda,Liendo:2012hy,Billo:2016cpy}. In the introduction to Chapter \ref{ch:BDD} we will introduce more extensively this framework and the particular case we will analyze; nevertheless, it is worth introducing already here what we mean by breaking the symmetry in a controlled way. Let us consider a $4d$ CFT and a straight line defect: there will still be a symmetry preserved by the rotations around the defect
\begin{equation*}
\text{4d CFT} \rightarrow SO(3)\,.
\end{equation*} 
Assuming we can define a QFT on the defect, the operators here will carry a quantum number related to the $SO(3)$ representation they belong to. However, we can still do better: we can actually define a line defect equipped with a $1d$ CFT, a \textit{conformal defect}. This will correspond to the breaking
\begin{equation*}
\text{4d CFT} \rightarrow \text{1d CFT}\otimes SO(3)\,
\end{equation*} 
and therefore operators defined on the defect will still be part of a CFT and will be studied with all the available techniques developed for CFTs. This also means that the motivation for being interested in CFT is still valid for the defect, and a question we can ask in particular is whether this one-dimensional CFT corresponds to some two-dimensional theory of gravity, according to the holographic principle stated above. The answer is that indeed, the line defect that we will consider (a supersymmetric defect, as it will be clarified in Chapter \ref{ch:BDD}) is the boundary to an $AdS_2$ space and corresponds to a $D3$-brane stretching in the extra dimension of the $AdS_5$. One of the many consequences of this is the possibility of studying correlators of operators living on the defect at strong-coupling via perturbative calculations in the dual theory as depicted in Fig. \ref{fig:AdSdefect},  which can also be combined with the \textit{bootstrap approach} investigating the implications of the residual symmetry on the system \cite{Ferrero:2021bsb,Ferrero:2023znz,Ferrero:2023gnu}. \\

As for conformal field theories, the study of line defects also has implications and connections to statistical field theory and condensed matter models, where a line defect extended in the time direction can model a defective atom in a crystal lattice. An interesting example, which we mention here without further elaboration as it will not play a role in the thesis, is the Yukawa model that is conjectured to describe graphene sheets at the critical point \cite{Herbut:2006cs,Vozmediano:2010zz}. For a description of correlator in the defect Yukawa CFT we refer to \cite{Giombi:2022vnz,Barrat:2024nod}.
\section*{The structure of the thesis}

This thesis presents the results of four years of research guided by the broad perspective of understanding the space of functions that describe different quantum field theories. We have mentioned that this overarching interest puts together research works otherwise dealing with different subjects and techniques: the functional space of Feynman integrals and defect conformal field theory. The thesis consists of three chapters, each referring mostly to one of the scientific papers published together with other scientists who equally contributed to the presented research. Having considered that the main subject of each chapter can be studied on its own, regardless of the overall perspective reported in this preface, we organize the references according to their appearance in each chapter. This is the reason why the bibliography at the end of this thesis matches this organization. Each chapter also contains an introduction presenting the research problem studied and its specific context, as well as a conclusion summarizing the main results and the research directions that are interested to follow. A particular emphasis is put on describing how each chapter fits into the broader landscape of research on such a topic, to highlight the results achieved, the elements of novelty, or the alternative perspectives brought. In the following paragraph, we spend some words on each chapter to summarize its content and its fitness to the study of the space of functions describing correlators in quantum field theory.\\

The first chapter is dedicated to the study of linear relations among the functions belonging to the class of Feynman parametric integrals, and it expands the results obtained in the paper \textit{Integration-by-parts identities and differential equations for parametrised Feynman integrals} \cite{Artico:2023jrc}. First introduced in Ref.~\cite{Feynman:1948ur}, Feynman integrals are the most widespread tool to perform calculations in particle physics and beyond~\cite{Heinrich:2020ybq}. Their integral representation, the space of functions they belong to, and the geometry associated with a given integral -- in the simplest case, a complex curve of genus zero, in more complicated cases, of higher genus -- has been a topic of study since their introduction and has given access to a vast amount of knowledge on this subject \cite{Weinzierl:2022eaz}. The core of this chapter is the investigation of linear relations among the functions describing Feynman integrals, relations that have been known since the late 1960s~\cite{Regge:1968rhi}; their use has particularly spread since the calculation of $\beta$-functions at 4-loop \cite{Chetyrkin:1981qh}. Working in the integral representation known as Feynman-Schwinger parametrization, it is shown in this first chapter how to establish a correspondence between linear relations among integrals and the study of a certain polynomial ideal, with the aim of optimizing in the future the construction of the Gelfand-Kapranov-Zelevinsky differential equation systems these functions obey~\cite{GKZ1}. \\

The second and third chapters focus on a particular set-up where studying the symmetries and the kind of functions that can appear allows the computation of different correlators up to next-to-next-to-leading order in the weak-coupling perturbative expansion. The set-up we consider is the $\Nm = 4$ supersymmetric Yang-Mills theory with gauge group $SU(N)$ for large values of $N$, a theory which presents the largest amount of supersymmetry allowed in a renormalizable four-dimensional theory \cite{muller2010introduction}; furthermore it presents a conformal symmetry that is preserved at the quantum level and for large $N$ it is an integrable quantum field theory. To this framework, we add the extended operator called Maldacena-Wilson line \cite{Maldacena:1998im} breaking half of the supersymmetric charges and reducing the four-dimensional conformal symmetry into a one-dimensional one (times the three-dimensional rotations around the line). The second chapter of this thesis in particular is based on the results presented in \textit{Perturbative bootstrap of the Wilson-line defect CFT: Bulk-defect-defect correlators} \cite{Artico:2024wnt}. It focuses on the simplest non-kinematically fixed correlation functions, the ones involving one operator of the bulk theory and two superconformal primary excitations of the defect. To compute these correlators, we use all the residual symmetries to set up a bootstrap problem at weak coupling that highly constrains the functional form that such quantities can assume. If expressed as functions of the right combination of space-time coordinates, these correlators are described only by rational functions up to next-to-leading order; the expected logarithmic behavior characteristic of the operator product expansion at NLO is absent due to non-trivial cancellations that simplify the functional space and that provide a guide for the future research on this subject.\\

Building on the techniques presented in the second chapter, in the third chapter we present and elaborate on the topic of correlation functions among multiple scalar defect excitations of scaling dimension protected from quantum corrections by the residual supersymmetry. The results for such correlators up to next-to-next-to-leading order at weak coupling were first obtained in the paper \textit{Perturbative bootstrap of the Wilson-line defect CFT: Multipoint correlators} \cite{Artico:2024wut}. While now presented in an ideal progressive sequence, it is important to mention that Refs. \cite{Artico:2024wnt} and \cite{Artico:2024wut} were developed at the same time, and therefore some intuitions that are part of Chapter 2 are related to the work presented in Chapter 3 and vice versa. In this thesis, we tried to reformulate the arguments leading to the results in a way that would highlight the progression of ideas and mathematical difficulty; reading the two chapters keeping in mind their parallel development will however be beneficial to the reader. The study of multipoint correlators of the supersymmetric Maldacena-Wilson line using supersymmetry constraints has been the focus of multiple techniques: among the approaches at strong coupling, we list Refs. \cite{Liendo:2016ymz,Liendo:2018ukf,Ferrero:2021bsb,Ferrero:2023znz,Ferrero:2023gnu}; at weak coupling we refer to \cite{Barrat:2021tpn,Barrat:2022eim}; and for all values of the coupling constant remarkable results were obtained in \cite{Cavaglia:2021bnz,Cavaglia:2022qpg,Cavaglia:2022yvv,Cavaglia:2023mmu}. The list is far from comprehensive, as a more exhaustive account of the contributions would require a section on its own, as we try to do at the beginning of Chapter 3. When coming to the overarching interest of this thesis regarding the space of functions describing correlators in a QFT, the main result of the efforts presented in Chapter 3 is the dependence of correlators up to six points on a special kind of integrals called train-track integrals \cite{Bourjaily:2017bsb,Bourjaily:2018ycu,Ananthanarayan:2020ncn,Loebbert:2020glj,Kristensson:2021ani,Morales:2022csr,McLeod:2023qdf}. Such integrals are known to be impossible to express in terms of Goncharov polylogarithms when the external points are in general position; however when the points are aligned such feature has been proven to disappear up to six-external points \cite{Rodrigues:2024znq}, thus making it possible to write the polylogarithmic ansatz necessary to bootstrap multipoint correlators up to NNLO at weak coupling. To the best of our knowledge, no perturbative result has ever found evidence of the presence of a different class of functions for multipoint correlators; however, proof that this feature will not break for a higher perturbative order or number of external legs is still missing. It is relevant to mention that non-polylogarithmic train-track integrals also appear in pure $\Nm = 4$ sYM for the scattering of 10 massless particles, providing the first appearance of a higher genus function in a theory that for a long time was considered to be completely described by Goncharov polylogarithms.\\ 

Having described the motivation of our interest in Feynman integrals and defect CFT and the structure of the present thesis, we now proceed with the main chapters.

\chapter{Integration by parts identities in Feynman parameter space}
\markboth{Integration by parts identities in Feynman parameter space}{}
\label{chapter:IBP}
\section{Introduction and preliminaries}
\label{sec:PreliminariesFeyn}
\subsection{The role Feynman integrals and the structure of the chapter}
\label{subsec:DiffEq}
This chapter is based on Refs. \cite{Artico:2023jrc} and \cite{Artico:2023bzt} written by the author of this thesis together with Lorenzo Magnea. These studies approach the problem of reducing a set of Feynman integrals to a basis of master integrals using a representation for the integrals unusual in these kinds of problems, the Feynman parametric representation. Before diving into the core subject of the chapter and defining the conventions we choose regarding the Feynman parametrisation, we dedicate this introduction to reviewing the importance of Feynman integrals in present-day theoretical physics, focusing on the nowadays most important techniques to solve them analytically. In this section, we also present some of the motivations that led us to consider the study of the reduction problem for Feynman parametric integrals, providing some comparisons with the most common techniques. Such comparisons, together with our result, will open some perspective into a research direction aiming at optimizing parametric integral reduction problems with the help of algebraic techniques.\\


For decades, the calculation of high-order Feynman integrals has been the cornerstone of the precision
physics program at present and future particle accelerators~\cite{Heinrich:2020ybq}. In the last 20 years, however, their range of application has considerably expanded, with application to cosmology \cite{Chen:2010xka,Wang:2013zva} and gravitational wave physics where the emergence of Calabi–Yau manifolds in black holes scattering was recently predicted \cite{Driesse:2024feo}\footnote{See references therein for an overview of the rich literature that has emerged on the subject of gravitational waves.}. In the second and third chapters of this thesis, Feynman integrals will appear in the calculation of correlation functions of local operators in a defect conformal field theory -- further examples of this application can be found for example in \cite{Barrat:2024nod,Peveri:2023qip}. While the color algebra in gauge theory can be promptly simplified via algorithms -- based, for example, on FORM \cite{Vermaseren:2000nd} -- Feynman integrals often present a tensor structure that is generally addressed via tensor decomposition techniques, such as the Passarino-Veltman decomposition \cite{Passarino:1978jh}, employing physical projector operators \cite{Peraro:2019cjj}, building an orthogonal dual basis \cite{Anastasiou:2023koq}, and the orbit partition approach implemented in \cite{Goode:2024cfy}.\\

Once all reduction procedures are completed, we are left with a set of scalar integrals to compute. The systematic development of modern methods to compute scalar Feynman integrals, going beyond a direct evaluation of their parametric expression, began with the identification and explicit construction of Integration-by-Parts (IBP) identities in dimensional regularisation, in Refs. \cite{Tkachov:1981wb,Chetyrkin:1981qh}, and reached a further degree of sophistication with the development of the method of differential equations~\cite{Kotikov:1990kg,Remiddi:1997ny,Gehrmann:1999as}\footnote{A pedagogical review of the differential equation method is presented in \cite{Henn:2014qga}. The lecture notes on scattering amplitudes in quantum field theory \cite{Badger:2023eqz} present the topic in a comprehensive overview of contemporary techniques in the research on scattering amplitudes.}. These two sets of ideas can be combined into powerful algorithms~\cite{Laporta:2000dsw}, and the procedure further streamlined and optimized by the identification of the linear functional spaces where (classes of) Feynman integrals live~\cite{Duhr:2011zq,Duhr:2019wtr,Abreu:2022mfk}, and by taking maximal advantage of dimensional regularisation~\cite{Henn:2013pwa}. The combined use of these tools has dramatically extended the range of processes for which high-order calculations are available and has broadened our understanding of the mathematics of Feynman integrals, as reviewed for example in~\cite{Heinrich:2020ybq,Weinzierl:2022eaz}. The current state-of-the-art algorithms \cite{Wu:2025aeg,Lange:2025fba} have been used in the remarkable efforts to compute the scattering angle and impulse of classical black hole scattering at fifth post-Minkowskian order \cite{Driesse:2024xad} and the complete function space for planar two-loop six-particle scattering amplitudes \cite{Henn:2025xrc}.\\

It is an interesting historical fact that the idea of studying and eventually computing Feynman integrals by means of IBPs and differential equations pre-dates all the 
developments just discussed, and was originally proposed not in the momentum 
representation, but in Feynman-parameter space. Studies of Feynman diagrams flourished in the $S$-matrix era, as illustrated in the classic textbook~\cite{Eden:1966dnq}, and recently a renewed interest in the monodromy properties of Feynman integrals is surging \cite{Kreimer:2021jni,Bourjaily:2020wvq}.
In particular, the projective nature of
Feynman parameter integrands, and the importance of the monodromy properties 
of Feynman integrals under analytic continuation around their singularities, were
soon uncovered, and attracted the attention of mathematicians~\cite{Pham,Lascoux}
and physicists~\cite{Regge:1968rhi}. In this context, Tullio Regge and collaborators
published a series of papers~\cite{Ponzano:1969tk,Ponzano:1970ch,Regge:1972ns}  
studying the monodromy ring of interesting classes of Feynman graphs: first
the ones we would at present describe as multi-loop sunrise graphs in 
Ref.~\cite{Ponzano:1969tk}, then generic one-particle irreducible $n$-point 
one-loop graphs in Ref.~\cite{Ponzano:1970ch}, and finally the natural 
combination of these two classes, in which each propagator of the one-loop 
$n$-point diagram is replaced by a $k$-loop sunrise~\cite{Regge:1972ns}. 
All of these papers employ the parameter representation as a starting point, 
and make heavy use of the projective nature of the integrand.
At the time, these studies by Regge and collaborators did not immediately 
yield computational methods, but it is interesting to notice that, at least at the 
level of conjectures, several deep insights that have emerged in greater detail
in recent years were already present in the old literature. For example Regge,
in Ref.~\cite{Regge:1968rhi}, argues, on the basis of homology arguments, that 
all Feynman integrals must belong to a suitably generalized class of hypergeometric 
functions, an insight that was sharpened much more recently with the introduction 
of the Lee-Pomeransky representation~\cite{Lee:2013hzt} of Feynman integrals 
and the application of the GKZ theory of hypergeometric functions~\cite{GKZ1,GKZ2,
delaCruz:2019skx,Klausen:2019hrg,Feng:2019bdx,Klausen:2021yrt,Chestnov:2022alh,Ananthanarayan:2022ntm,
Klausen:2023gui}. Regge further argues that such functions obey sets of (possibly) 
high-order differential equations, which he describes as `a slight generalization of the 
well-known Picard-Fuchs equations', also a recurrent theme in present-day research~\cite{Lairez:2022zkj,Muller-Stach:2012tgj,Mishnyakov:2024rmb}.\\

While general algorithms were not developed at the time, two of Regge's collaborators,
Barucchi and Ponzano, were able to construct a concrete application of the general 
formalism for one-loop diagrams~\cite{Barucchi:1973zm,Barucchi:1974bf}. In those
papers, they show that for one-loop diagrams it is always possible to organize the 
relevant Feynman integrals into sets and find 
a system of linear homogeneous differential equations in the Mandelstam invariants
that closes on these sets, with the maximum required size of the system being 
$2^{n}-1$ for graphs with $n$ propagators\footnote{This counting has been reproduced
with modern (and more general) methods in \cite{Bitoun:2017nre,Bitoun:2018afx,
Mizera:2021icv}.}. These systems of differential equations were of interest 
to Barucchi and Ponzano because they effectively determine the singularity structure of 
the solutions, and thus the monodromy ring, in agreement with the general results of
Regge's earlier work. From a modern viewpoint, it is perhaps just as interesting
to use the system directly for the evaluation of the integrals, as done with
the usual momentum-space approach: this is the direction that we pursued in  \cite{Artico:2023jrc,Artico:2023bzt}, and in the current chapter. Starting from the ideas of Refs.~\cite{Regge:1968rhi,Ponzano:1969tk,
Ponzano:1970ch,Regge:1972ns} and the concrete results of Barucchi and 
Ponzano~\cite{Barucchi:1973zm,Barucchi:1974bf}, we describe a projective framework 
to derive IBP identities and systems of linear differential equations for Feynman integrals. 
To do so, we need to generalize the Barucchi-Ponzano results in several directions. 
First of all, those results predate the widespread use of dimensional regularisation \cite{Leibbrandt:1975dj}, and 
do not in principle apply directly to infrared-divergent integrals. Fortunately, the projective 
framework naturally involves the (integer) powers of the propagators appearing in the 
diagram. These can be continued to complex values, providing a regularisation that is 
readily mapped to dimensional regularisation\footnote{Regge and collaborators also use 
this regularisation, having in mind mostly ultraviolet divergences, since the framework 
at the time was constructed for generic massive particles. They refer to the complex
values of the powers of the propagators as `Speer parameters', whereas we would 
now refer to this procedure as analytic regularisation.}.The projective framework applies directly to IR divergent integrals, as the examples we provide in the chapter show. The procedure to derive IBP identities in projective space 
generalizes naturally to higher loops.\\

Clearly, at two loops and beyond it would be of 
paramount interest to have a generalization of the Barucchi-Ponzano theorem, guaranteeing 
the closure of a system of linear differential equations, providing an upper limit for its 
size, and giving a constructive procedure to build the system. This could require a much 
deeper understanding of the monodromy ring of higher-loop integrals, which certainly points to promising avenues for future research. We can nonetheless apply the parameter-space IBP technique to multiloop examples, and directly derive sets of differential equations on a case-by-case basis. Indeed, we show that the method 
can be successfully applied to multi-loop integrals to build differential equations systems, and we provide examples, including 
the two-loop equal-mass sunrise for which we recover the appropriate elliptic differential 
equation. Our application of the projective framework highlights 
the importance of boundary terms in IBP identities: contrary to the momentum-space 
approach in dimensional regularisation, boundary terms do not in general vanish in 
the projective framework: on the contrary, they play an important role in linking 
complicated integrals to simpler ones, as we will see in concrete examples. Beyond the presence of boundary terms, a major difference between the parameter-space and momentum-space method lies in the absence of auxiliary propagators in parametric multi-loop integrals, a well-known necessity in momentum space where irreducible numerators are a byproduct of tensor decomposition. This absence is compensated by the presence of parametric integrals corresponding to Feynman integrals across different space-time dimensions: it will become clear from the examples that the space-time dimension special meaning in momentum space is lost in parameter space, where it is simply re-absorbed in the definitions of the exponents of the so-called Symanzik polynomials. A promising research question is represented by the study of the impact of these changes on the efficiency of reduction algorithms; in this direction, it is worth anticipating that the structure of Feynman parametrized integrals offers connections to the ideal algebra of polynomials that can be relevant to optimize the reduction to master integrals, as we will discuss in the last part of the chapter.\\

We note that the work presented in this chapter is part of a recent revival of interest 
in the mathematical structure of Feynman integrals in parameter space, and presents
interesting potential connections to several current research topics in this context. 
In particular, a study of IBP relations from the viewpoint of D-modules, starting from 
the Feynman parameter representation was carried on in \cite{Bitoun:2017nre,
Bitoun:2018afx}; other relevant connections include the applications of intersection 
theory~\cite{Mastrolia:2018uzb,Frellesvig:2019kgj,Frellesvig:2019uqt,Frellesvig:2020qot,
Chestnov:2022xsy}, the use of syzygy relations in reduction algorithms~\cite{Agarwal:2020dye,
Wang:2023nvh,Wu:2025aeg}, the study of generalised hypergeometric systems~\cite{Munch:2022ouq}, 
and the reduction of tensor integrals in parameter space~\cite{Chen:2019mqc,Chen:2019fzm,
Chen:2020wsh}. More generally, for the first time in several decades, we are witnessing
a rapid growth of our understanding of the mathematical properties of Feynman integrals, 
in particular with regards to analyticity and monodromy (see, for example,~\cite{Bourjaily:2020wvq,
Mizera:2021icv,Hannesdottir:2022bmo,Hannesdottir:2022xki,Britto:2023rig}, and the lectures 
in Ref.~\cite{Mizera:2023tfe}), with potential applications to questions of phenomenological 
interest, such as the study of infrared singularities~\cite{Arkani-Hamed:2022cqe} and
the development of efficient methods of numerical integration~\cite{Borinsky:2020rqs,
Borinsky:2023jdv}.\\

The structure of this chapter is the following. In Section \ref{subsec:Notat} we introduce our notation for the parameter representation of Feynman integrals and Symanzik polynomials, briefly 
reviewing well-known material for the sake of completeness. In section \ref{subsec:ProjectiveForms} we introduce projective forms, and in Section \ref{subsec:GeneralFormula} we use their differentiation and integration to lay the groundwork for the construction of IBP identities for generic projective integrals. In sections \ref{subsec:One-loopEx} and \ref{subsec:Multi-loopEx} we validate our results by discussing several concrete examples at one-loop and multi-loop levels. A proposal for the optimization of our parameter-based reduction algorithm based on polynomial ideals is presented in Section \ref{sec:Ideals} and applied to the two- and three-loop so-called banana integral in Section \ref{sub:BananaIdeal}. Our results and some future research directions are summarized in Section \ref{sec:Conclusions1}.
\subsection{Notations for parametrised Feynman integrals}
\label{subsec:Notat}

In this section, we describe the notation for Feynman integrals in momentum and parameter space that we will use throughout this chapter. Using this notation, we define the two Symanzik polynomials, summarise some basic properties of Feynman parametrized integrals, and provide some explicit examples for further clarification. In all the chapter, we adopt the notation used in \cite{Artico:2023jrc,Artico:2023bzt} and in \cite{Bogner:2007mn,Bogner:2010kv,Weinzierl:2022eaz}.
\\

Consider a connected Feynman graph $G$, with $l$ loops, $n$ internal lines\footnote{Internal lines connect two internal vertices of a graph, \textit{i.e.} vertices serving as endpoints for a number of edges larger than 1.} labeled by the index $i$ and carrying momenta $q_i$ and masses $m_i$ ($i = 1, \ldots, n$), and $e$ external lines labeled by the index $j$ and carrying momenta $p_j$ ($j = 1, \ldots, e$). At this stage we do not need to impose restrictions on external masses, so $p_j^2$ is unconstrained. On the other hand, momentum is conserved at all vertices of $G$: for each vertex, this imposes one linear constraint among the momenta assigned to each line. A momentum assignation respecting such constraints consists of parametrising the graph momenta assigning $l$ 
independent loop momenta $k_r$ ($r = 1, \ldots, l$) to suitable internal edges. 
The line momenta are then given by

\beq 
  q_i \, = \, \sum_{r = 1}^l \alpha_{ir} k_r + \sum_{j = 1}^e \beta_{ij} p_j \, ,
\label{eq:Incidence}
\eeq

where the elements of the {\it incidence matrices}, $\alpha_{ir}$ and $\beta_{ij}$, take values
in the set $\{-1,0,1\}$. The solution is not unique, as any set of internal $l$ lines with the rank of the incidence matrix  $\alpha_{ir}$ equal to $l$ can be chosen as a new set of loop momenta  $k'_r$. Once assigned to each internal line an integer number $\nu_i$, one can associate the graph $G$, the set of numbers $\nu_i$ and the real number $d$ altogether to the (scalar) Feynman integral

\beq
  I_G \left( \nu_i, d \right) \, = \, (\mu^2)^{\nu - l d/2} \int \prod_{r = 1}^{l} \frac{d^d k_r}{{\rm i} 
  \pi^{d/2}} \, \prod_{i = 1}^n \frac{1}{\left( - q_i^2 + m_i^2 \right)^{\nu_i}} \, , 
\label{eq:FeyInt}
\eeq

where we defined $\nu \equiv \sum_{i = 1}^n \nu_i$, and the integration must be performed 
by circling the poles in the complex plane of the loop energy variables according to Feynman's 
prescription for causality. The variable $\mu$ has the dimension of a squared mass so that the prefactor in \eqref{eq:FeyInt} makes the integral dimensionless\footnote{In a slight abuse of notation, the dimension $d$ and the differential operator $d$ are here denoted by the same symbol. We considered the chance of the meaning of $d$ not to be clear as very low, taking into account that the space-time dimension will not play a significant role in this chapter.}. The invariance under the change of loop vector basis of the momentum assignation can here be understood as the invariance under the change of integration variables of the integral \eqref{eq:FeyInt}. The integration over the loop momenta in \eqref{eq:FeyInt} can be performed in full generality using the Feynman parameter technique \cite{Feynman:1949}, using the identity

\beq
\label{eq:FPT}
  \prod_{i = 1}^n \frac{1}{\left (- q_i^2 + m_i^2 \right)^{\nu_i} } \, = \, 
  \frac{\Gamma(\nu)}{\prod_{j = 1}^n \Gamma(\nu_j)} \int_{z_j \geq 0} d^n z \,
  \delta \left(1 - \sum_{j = 1}^n z_j \right) \, \frac{ \prod_{j = 1}^n z_j^{\nu_j -1}}{\left( 
  \sum_{j = 1}^n z_j \left( -q_j^2 + m_j^2 \right) \right)^{\! \nu}} \, .
\eeq

By virtue of \eqref{eq:Incidence}, the sum in the denominator of the \textit{rhs} of the integrand in \eqref{eq:FPT} is a quadratic form in the loop momenta $k_r$, and can be written as 

\beq
\label{eq:Mdef}
  \sum_{j = 1}^n z_j \left( - q_j^2 + m_j^2 \right) \, = \, - \sum_{r,s = 1}^l  M_{rs} \, k_r  \cdot k_s + 
  2 \sum_{r = 1}^l k_r \cdot Q_r + J \, ,
\eeq

where $M$ is an $l \times l$ matrix with dimensionless entries which are linear in the Feynman 
parameters $z_i$, $Q$ is an $l$-component vector whose entries are linear combinations of 
the external momenta $p_j$, and $J$ is a linear combination of the Mandelstam invariants
$p_i \cdot p_j$ and the squared masses $m_j^2$. To perform the $d-$dimensional integration over the loop momenta we use the translation invariance of \textit{dimensional regularization}: we keep $d \in \mathbb{R}$ (or $d \in \mathbb{C}$) and we assume that the value of dimensional regularized integrals is invariant under the translation of the loop momenta. We can therefore complete the square in \eqref{eq:Mdef} and perform the integration obtaining that

\beq
\label{eq:FP}
  I_G \left( \nu_i, d \right) \, = \, \frac{\Gamma(\nu - l d/2)}{\prod_{j = 1}^n \Gamma(\nu_j)}
  \int_{z_j \geq 0} d^n z \, \delta \left(1 - \sum_{j = 1}^n z_j \right) 
  \left( \prod_{j = 1}^n z_j^{\nu_j -1}\right) \,
  \frac{\mathcal{U}^{\, \nu - (l+1) d/2}}{\mathcal{F}^{\, \nu - l d/2}} \, ,
\eeq

where the functions

\beq
\label{eq:SymPol}
  \mathcal{U} \, = \, \mathcal{U} (z_i) \, = \, \det M \, , \qquad \quad 
  \mathcal{F} \, = \, \mathcal{F} \left( z_i, \frac{p_i \cdot p_j}{\mu^2}, \frac{m_i^2}{\mu^2}
  \right) \, = \, \det M \, \left( J + Q M^{-1} Q \right)/\mu^2 \, ,
\eeq

are called graph polynomials or Symanzik polynomials. Refs. \cite{Bogner:2007mn,
Bogner:2010kv,Weinzierl:2022eaz} discuss in detail the properties of graph polynomials: here we only note that both polynomials are homogeneous in the set of Feynman parameters,
$z_i$, with ${\cal U}$ being of degree $l$ and ${\cal F}$ of degree $l+1$. These homogeneity properties set the stage for employing the tools of projective geometry, as discussed below.\\

\begin{figure}[!h]
\vspace{-2mm}
\centering
\includegraphics[scale=.7]{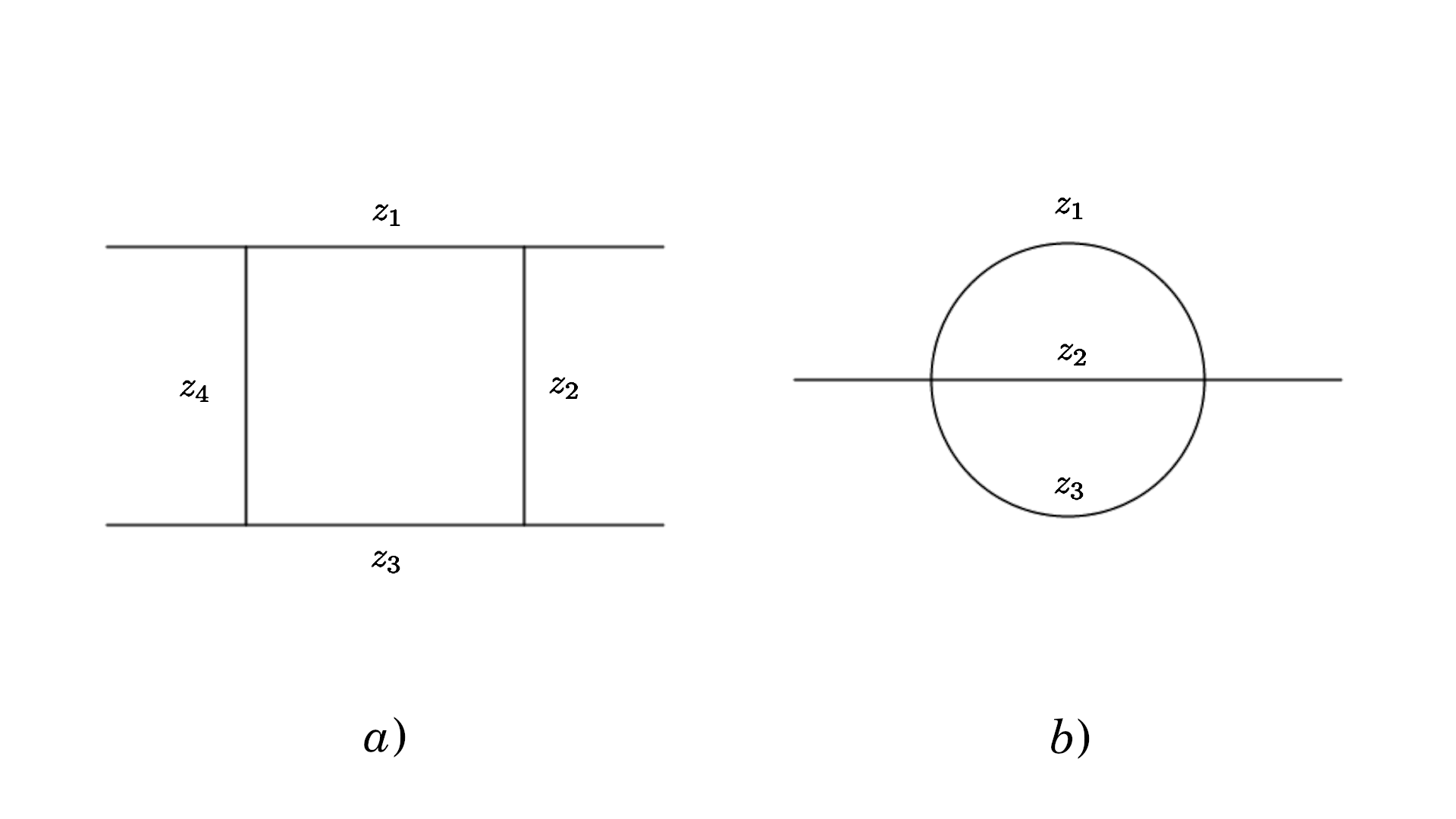}
\caption{a) One-loop box diagram b) Two-loop sunrise diagram}\label{fig:ExampleDiag}

\end{figure}

The computation of the Symanzik polynomials requires simple matrix operations that have been automatized in computer programs such as \textsc{PySecDec} \cite{Carter:2010hi,Borowka:2017idc} based on sector decomposition \cite{Heinrich:2008si}. Remarkably, Symanzyk polynomials can be constructed directly from the connectivity properties of the underlying Feynman graph. To do so, let us denote by $\mathcal{I}_G$
the set of the internal lines of $G$, each endowed with a Feynman parameter $z_i$. We now proceed to define the \textit{co-trees} and \textit{cuts} of a graph in a similar fashion to Refs. \cite{Artico:2023bzt,Bogner:2007mn}.
\begin{definition}
  A co-tree $\mathcal{T}_G \subset \mathcal{I}_G$ is a set of internal lines of $G$ 
such that that the lines in its complement $\overline{\mathcal{T}}_G \subset \mathcal{I}_G$ 
form a spanning tree, {\it i.e.} a graph with no closed loops which contains all the vertices 
of $G$.
\end{definition}

Note that, in the case of an $l$-loop graph, one needs to omit precisely $l$ lines to obtain a spanning tree: the number of elements of each co-tree is, therefore, $l$.
\begin{definition}
  A cut $\mathcal{C}_G \subset \mathcal{I}_G$ is a set of internal lines of $G$ with the
property that, upon omitting the lines of ${\cal C}_G$ from $G$, the graph becomes a
disjoint union of two connected subgraphs.
\end{definition}

Note that, in the case of an $l$-loop graph, one needs to omit precisely $l+1$ lines to obtain a disjoint union of two connected subgraphs: the number of elements of each cut is, therefore, $l+1$.\\

From the set of all co-trees, we compute the first Symanzik polynomial as 

\beq
  \label{eq:U}
  \mathcal{U} \, = \, \sum_{\mathcal{T}_G} \prod_{i \in \mathcal{T}_G} z_i \, .
\eeq

From the cardinality of the set representing each co-tree, we can conclude that the first Symanzik polynomial is homogeneous of degree $l$. It is also linear in each parameter $z_i$.\\

To define the second Symanzik polynomial from graph properties we associate to each cut the invariant mass $s \left({\cal C}_G \right)$, obtained by squaring the sum of the momenta flowing in (or out) one of the two remaining subgraphs -- by momentum conservation, it does not matter which subgraph we choose. The second polynomial $\mathcal{F}$ is then defined by 

\beq
\label{eq:F}
  \mathcal{F} \, = \, \sum_{\mathcal{C}_G} \frac{\hat{s} \left( \mathcal{C}_G \right)}{\mu^2} \, 
  \prod_{i \in \mathcal{C}_G} z_i \, - \, \mathcal{U} \sum_{i \in \mathcal{I}_G} 
  \frac{m_i^2}{\mu^2} \, z_i \, ,
\eeq

and given the cardinality of co-tree and cut sets ${\cal F}$ is homogeneous of degree $l+1$ in the Feynman parameters. To clarify the definitions of co-trees, cuts, and Symanzik polynomials two simple examples follow. We consider the two diagrams in Figure \ref{fig:ExampleDiag}.\\

For one-loop diagrams such as the box in Figure \ref{fig:ExampleDiag} a), each internal leg represents a co-tree. The first Symanzik polynomial is then given by the sum over all the Feynman parameters and for the considered diagram becomes
\beq
\label{eq:Ubox}
  \mathcal{U} \, = \, z_1 + z_2 + z_3 + z_4 \, .
\eeq

The polynomial $\mathcal{U}$ only relies on the connectivity properties of the diagram and not on any kinematic data, which instead must be specified to define the polynomial $\mathcal{F}$. If we consider a box diagram where all external legs are massless and all internal legs have equal mass (the kind of diagram given by a massive quark loop in gluon scattering after tensor reduction, for example), the only cuts with non-zero invariant mass are given by cutting $z_1$ and $z_3$ -- invariant mass $s$ -- or $z_2$ and $z_4$ -- invariant mass $t$. This leads to

\beq
\label{eq:FboxM}
  \mathcal{F} \, = \, \frac{s}{\mu^2} \, z_1 z_3 + \frac{t}{\mu^2} \, z_2 z_4 - 
  \frac{m^2}{\mu^2} \, \left( z_1 + z_2 + z_3 + z_4 \right)^2 \, .
\eeq

For the two-loop two-point diagram in Figure \ref{fig:ExampleDiag} b), each internal leg represents a spanning tree of the diagram. As a consequence, the first Symanzik polynomial is the sum of the monomials we obtain by removing one Feynman parameter by the product of all the parameters

\beq
\label{eq:U2}
  \mathcal{U} \, = \, z_1 z_2 + z_2 z_3 + z_1 z_3 \, ,
\eeq

while the polynomial $\mathcal{F}$ can be constructed by the only cut of the diagram, consisting of the set of all internal edges. Considering for the sake of simplicity all internal edges have equal mass  we get:

\beq
\label{eq:F2}
  \mathcal{F} \, = \, \frac{p^2}{\mu^2} \, z_1 z_2 z_3 - \frac{m^2}{\mu^2} \left( z_1 z_2 + 
  z_2 z_3 + z_1 z_3 \right) \left( z_1 + z_2 + z_3 \right) \, .
\eeq

Both examples show how $\mathcal{U}$ and $\mathcal{F}$ are polynomials of respective degree $l$ and $l+1$, as was already mentioned. This property makes the integrand in \eqref{eq:FP} invariant under uniform dilatation of the variables: indeed, one easily verifies that a change of variables of the form $z_i \to \lambda z_i$,
with $\lambda > 0$, leaves the integrand invariant -- except for the argument of the $\delta$ 
function. This invariance offers the possibility of studying Feynman parametrized integrals using tools from projective geometry: in the next sections this statement will be substantiated with the introduction of projective forms and the Cheng-Wu theorem.

\subsection{Projective forms and their differentials}
\label{subsec:ProjectiveForms}
In this section, we introduce the mathematical language we will use to define linear relations between Feynman parametrized integrals. To do so, we present a brief introduction to projective forms and their integration and differentiation. This section is somewhat formal, so it is useful to keep in mind from 
the beginning the correspondence between projective forms and parameter integrands, which we will highlight with explicit examples. The notations and definitions in this section are based on \cite{Lascoux:1968bor,Regge:1968rhi}: from a mathematical standpoint, this approach to the introduction of projective forms is to some extent historical, but we find it useful, in that all calculations are very explicit.
\\ 

Let us begin by considering the Grassman algebra of exterior forms in the differentials $dz_i$, 
where $i \in D \equiv \left \{ 1, 2, ..., N \right\}$, for a positive integer $N$. Let $A$ 
be a subset of $D$, of cardinality $|A| = a$, and let $\omega_A$ be its ordered volume form

\beq
\label{eq:wA}
  \omega_A \, = \, d z_{i_1} \wedge ... \wedge dz_{i_a} \, ,
\eeq

with $i_j \in A $, and $i_1 < i_2 < ... < i_a$: for example, if $A = \left\{ 1,2,3\right\}$, 
then $\omega_A = dz_{1} \wedge dz_2 \wedge dz_{3}$. Consider now the $a-1$ form  $\eta_A$ defined as

\beq
\label{eq:eta}
  \eta_A \, = \, \sum_{i \in A} \epsilon_{i, A - i} \, z_i  \, \omega_{A - i}
\eeq

where $A - i$ denotes the set $A$ with $i$ omitted, and we defined the signature factor 
$\epsilon_{k, B}$, for any $B \subseteq D$, and for any $k \notin B$, by means of

\beq
\label{eq:epsilon}
  \epsilon_{k,B} \, = \, (-1)^{|B_k|} \, ,  \qquad   B_k \, = \, \left\{ i \in B, i < k \right\} \, ,
\eeq 

while $\epsilon_{k,B} = 0$ if $k \in B$. Using the properties of the boundary operator $d$,
one easily verifies that the differential of $\eta_A$ is proportional to $\omega_A$. Indeed the differential operator yields

\beq
\label{eq:deta}
  d \eta_A \, = \, a \, \omega_A \, .
\eeq

thus showing that the form  $\eta_A$ can be consider the integral form of $\omega_A$. As an example, consider again $A = \left\{ 1, 2, 3 \right\}$: the form $\eta_A$ is 
then given by

\beq
\label{eq:eta example}
  \eta_{ \{1,2,3\} } \, = \, z_1 \, dz_2 \wedge dz_3 - z_2 \, dz_1 \wedge dz_3 + z_3 \, dz_1 
  \wedge dz_2 \, ,
\eeq

and its differential is equal to $3 \, dz_1\wedge dz_2 \wedge dz_3$. We define next 
{\it affine} $q$-forms, defined by

\beq
\label{eq:Affine}
  \psi_q \, = \, \sum_{|A| = q} R_A (z_i) \, \omega_A \, ,
\eeq
where $R_A$ is a homogeneous rational function of the of the variables $z_i$ with degree 
$- |A|  = - q$. Note that this function may depend on external parameters as well; in the case of Feynman diagrams such parameters will represent kinematic invariants of the diagram under consideration.  Note also that, to make room for dimensional 
regularisation of Feynman integrals, we will slightly generalize this definition to include polynomial factors raised to non-integer powers in both the numerator and the denominator of the functions $R_A$, while preserving the homogeneity requirement. The name affine form is a reference to their invariance under dilatations 
of all variables. Equation \eqref{eq:Affine} is readily seen to imply that also the $(q+1)$-form $d \psi_q$ 
is affine. An example of an affine form with $q = 2$ is given by the form

\begin{eqnarray}
\label{eq:Affine example}
 \psi_2 \left( \nu_i, \lambda, r \right)  =   d z_2 \wedge d z_3  \frac{z_1^{\nu_1 } \, z_2^{\nu_2 - 1} \, 
  z_3^{\nu_3 - 1} \, (z_1 z_2 + z_2 z_3 + z_3 z_1)^\lambda}{\left( r \, z_1 z_2 z_3 - 
  (z_1 z_2 + z_2 z_3 + z_3 z_1) (z_1 + z_2 + z_3) \right))^{\frac{2 \lambda + \nu}{3}}} ,
\end{eqnarray}

where, as before, $\nu = \nu_1 + \nu_2 + \nu_3$. The parameter $\lambda$ can acquire a linear dependence on the dimensional 
regularisation parameter $\epsilon$ in the case of Feynman integrals, as discussed above.
With this extension, the integrand will no longer be strictly speaking a rational function, but all the 
definitions and the relevant properties remain valid. A fundamental definition for the rest of the chapter is represented by the definition of \textit{projective forms}.

\begin{definition}
An affine form is defined to be {\it projective} if it can be identically re-written as a linear
combination of forms proportional to the volume $\eta_A$, defined in \eqref{eq:eta}.
\end{definition}

Explicitly, this means that
\beq
\label{eq:ProjFor}
  \psi_q \, = \, \sum_{|B| = q+1} T_B (z_i) \, \eta_B \, . 
\eeq

where the homogeneous functions $T_B (z_i)$ are obtained by suitably combining 
the functions $R_A (z_i)$ in \eqref{eq:Affine} with appropriate factors of $z_i$ arising from 
the definition of $\eta_A$ in \eqref{eq:eta}. An example of a projective form appearing in the study of Feynman integrals is the integrand of the one-loop massless box integral, which reads

\beq
\label{eq:BoxExample}
  \psi_3 \left( \lambda, r \right) \, = \,
  \frac{(z_1 + z_2 + z_3 + z_4)^{\lambda}}{\left(r \, z_1 z_3 + z_2 z_4 
  \right)^{2 + \lambda/2}} \, \,
  \eta_{ \left\{ 1, 2, 3, 4 \right\}} \, ,
\eeq

where for concrete applications one has $\lambda = 2 \epsilon$ and $r = t/s$ the ratio of the canonical Mandelstam variables. \\

To define linear relations among Feynman integrals in parameter space, we will apply the differential operator $d$ to a Feynman integral written as a projective form. This generates the desired linear relations provided that we prove that the differential of a projective form is a linear combination of other projective forms, which we can then identify as Feynman parametrized integrals \cite{Regge:1968rhi}. In the remaining part of this section, we prove this result.

\begin{thm}
\label{th:dp}
The differential of a projective form is itself projective.
\end{thm}
\begin{proof}

First, consider an operator $p$ trasforming an affine $q$-form into a projective (and therefore 
also affine) $(q - 1)$-form, according to 
\beq
\label{eq:Defp}
  p \, : \,  \sum_{|A| = q} R_A (z_i) \, \omega_A \quad \rightarrow \quad
  \sum_{|A| = q} R_A (z_i) \, \eta_A \, .
\eeq
The operator $p$ is nilpotent, \textit{i.e.} $p^2 = 0$. This can be shown for a single term in \eqref{eq:Defp}, $R_A (z_i) \, \omega_A$, and the generalization 
is then straightforward. In fact
\begin{eqnarray}
\label{eq:p^2}
  p^2 \Big( R_A (z_i) \, \omega_A \Big) & = & p \left( R_A (z_i) \sum_{i \in A} z_i \, 
  \epsilon_{i, A - i} \, \omega_{A - i} \right) \nonumber \\
  & = & R_A (z_i) \sum_{i > j, \left\{ i,j \right\} \in A} \! z_i z_j 
  \left( \epsilon_{i, A - i} \, \epsilon_{j, A - i - j} + 
  \epsilon_{j, A - j} \, \epsilon_{i, A - j - i} \right) \, = \, 0 \, .
\end{eqnarray}
The cancellation in the last step of \eqref{eq:p^2} follows from the definition of $\epsilon_{j, B}$ \eqref{eq:epsilon} and can be clarified by considering an explicit example starting from a $3$-form:
\begin{eqnarray}
\label{eq:p^2ex}
  p^2 \Big( R_A (z_i) \, d z_1 \wedge d z_2 \wedge d z_3  \Big) & = & 
  R_A (z_i)  \, p \Big( z_1 \, d z_2 \wedge d z_3 - z_2 \, d z_1 \wedge d z_3 
  + z_3 \, d z_1 \wedge d z_2 \Big) \quad \\
  && \hspace{-3cm} = \, R_A (z_i) \, \big( z_1 z_2 d z_3 - z_1 z_3 d z_2 - z_2 z_1 d z_3 + 
  z_2 z_3 d z_1 + z_3 z_1 d z_2 - z_3 z_2 d z_1 \big) \, = \, 0 \, .
\nonumber
\end{eqnarray}
The cancellation works since there are always 
two terms in the sum that are proportional to $z_i z_j$, and they contribute with opposite 
signs. Considering for example $i < j$, when the factor of $z_j$ is generated by the first
action of $p$, the sign of the term is given by the position of the indices $i$ and $j$
in the ordered list of the elements of $A$. On the other hand, when the factor $z_i$ 
is generated by the first action of $p$, what is relevant is the position of $j$ in $A-i$.\\

The crucial part of the proof of Theorem \ref{th:dp} is that the nilpotent operator $p$ can be composed with exterior differentiation to map affine $q$-forms into affine $q$-forms, and in particular the following formal identity holds 
\beq
\label{eq:dppd}
  d \circ p + p \circ d \, = \, 0 \, ,
\eeq
when acting on any affine $q$-form $\psi_q$. Equation \eqref{eq:dppd} determines that the differential $d$ of a projective $(q-1)$-form $p(\psi_q)$ can be obtained by applying the operator $p$ to the $(q+1)$-form $-d(\psi_q)$, therefore proving the differential of a projective form is again projective. The application to an affine form  $R_A \omega_A$ of each of the summand in the LHS of \eqref{eq:dppd} brings

\beq
\label{eq:pd}
  \big( p \circ d \big) \psi_q \, = \, \sum_{i \notin A} \, \frac{\partial R_A}{\partial z_i} \,
  (-1)^{|A_i|} \, \eta_{A \cup i} \, = \, \sum_{i \notin A} \sum_{j \in A \cup i} 
  \frac{\partial R_A}{\partial z_i} \, z_j \, (-1)^{|A_i|} \, (-1)^{|(A\cup i)_j|} \, 
  \omega_{A \cup i - j} \, , \quad
\eeq

for the second piece and 
\beq
\label{eq:dp}
  \big( d \circ p \big) \psi_q \, = \, \sum_{j \in A} \,\, \sum_{i \notin A \, \vee \, i = j } 
  \left( \frac{\partial R_A}{\partial z_i} \, z_j + R_A (z_\ell) \, \delta_{i,j} \right) 
  (-1)^{|A_j|} \, d z_i \wedge \omega_{A - j} \, 
\eeq

for the first one. Combining them to show the cancellation requires manipulating the indices of the sums and combining terms into two terms that manifestly cancel
\begin{eqnarray}
\label{eq:dppd2}
  \Big( d \circ p + p \circ d \Big) \, \psi_q & = & \sum_{j \in A, \, i \notin A} 
  \frac{\partial R_A}{\partial z_i} \, z_j \left( (-1)^{|A_i|} \, (-1)^{| (A \cup i)_j |} + 
  (-1)^{| A_j |} \, (-1)^{| (A - j)_i |} \right) \omega_{A \cup i - j} + \nonumber \\ 
  && + \sum_{j \in A, \, i = j} R_A \, \delta_{i,j} \, \omega_A \,
  + \, \sum_{i \in D} \frac{\partial R_A}{\partial z_i} \, z_i \, \omega_A \, = \, 0 \, ,
\end{eqnarray}

where each line cancels separately. We named the set of all variables as $D$ in the indices of the sums. Given that such cancellation may be difficult to visualize, we provide an example to show it explicitly. Consider the form 
\beq
\label{eq:exampleth}
  \psi_2 \, = \, \frac{z_1 + z_3}{(z_1 + z_2)^3} \, d z_1 \wedge d z_2 \, ,
\eeq

which implies
	
\beq
  d \psi_2 \, = \, \frac{1}{(z_1+z_2)^3} \, d z_1 \wedge d z_2 \wedge d z_3 
  \quad \longrightarrow \quad \big(p \circ d \big) \psi_2 \, = \, \frac{1}{(z_1+z_2)^3} 
  \, \eta_{\left\{ 1,2,3 \right\}} \, .
\eeq

On the other hand, one easily verifies that

\begin{eqnarray}
  \big(d \circ p \big) \psi_2 & = &
  d \left( \frac{z_1 (z_1 + z_3)}{(z_1 + z_2)^3} \, d z_2 - 
  \frac{z_2 (z_1 + z_3)}{(z_1 + z_2)^3} \, d z_1 \right) \nn \\
  & = & - \frac{z_3}{(z_1 + z_2)^3} \, d z_1 \wedge d z_2 - \frac{z_1}{(z_1 + z_2)^3} \, 
  d z_2 \wedge d z_3 + \frac{z_2}{(z_1 + z_2)^3	} \,  d z_1 \wedge d z_3 \nn \\ 
  & = & - \frac{1}{(z_1 + z_2)^3} \, \eta_{\left\{ 1,2,3 \right\}} \, ,
\end{eqnarray}

as desired. Equation \eqref{eq:dppd}, just established, is sufficient to conclude the proof of the theorem.
\end{proof}

In the context of Feynman parametric integration, the theorem is significant
for the following reason: given that Feynman integrals in the parameter representation
are integrals of projective forms on a simplex (as discussed below), applying the 
boundary operator $d$ on the integrand generates relations among forms with the 
same properties, \textit{i.e.} other Feynman integrands, or generalizations thereof.
These relations take the form of linear difference equations, which in turn can be 
used to build closed systems of differential equations to ultimately compute the
integrals, just as normally done in the momentum-space representation \cite{Henn:2014qga}.\\

To conclude this section, we highlight how from Equation \eqref{eq:p^2} one may be tempted to look for a co-homology structure with $p$ as boundary operator, defining $p$-closed and $p$-exact forms. Such defined co-homology would be a trivial one given that all $p$-closed forms are also $p$-exact. An example can clarify this statement. Consider 
a generic affine two-form

\beq
\label{eq:example for co-homology}
  \psi_2 \, = \, R_{12} \, dz_1 \wedge dz_2 + R_{13} \,dz_1 \wedge dz_3 + 
  R_{23} \, dz_2 \wedge dz_3 \, ,
\eeq

where $R_{ij}$ are homogeneous rational functions of degree $-2$ in the 
variables $z_1$, $z_2$ and $z_3$. By imposing that $p(\psi_2) = 0$ it is 
immediate to obtain 

\beq
\label{example for co-homology 2}
  \psi_2 \, = \, \frac{R_{12}}{z_3} z_3 dz_1\wedge dz_2 - \frac{R_{12}}{z_3} z_2 dz_1 
  \wedge dz_3 + \frac{R_{12}}{z_3} z_1 dz_2 \wedge dz_3 \, ,
\eeq

which is a projective form. The generalization to affine $n$-forms is straightforward, 
as the condition generates a system of linear equations that is enough to fix all the 
rational functions appearing in the form, but one (the overall coefficient function 
multiplying the projective volume form).

\subsection{Feynman integrals as projective forms and Cheng-Wu theorem}
\label{subsec:ChengWu}
In the last section of this introductory part, we focus on presenting the connection between Feynman integrals and projective forms. This serves as the foundation for the next part of the chapter, where we use the Theorem \ref{th:dp} to write a set of generic relations among parametric integrands that include and generalize those appearing in Feynman integrals. To do this, we first present how Feynman integrands belong to the class of projective forms and we present the proof of the so-called Cheng-Wu theorem \cite{Cheng:1987ga} regarding their integration\footnote{The fact that the Cheng-Wu theorem follows from 
the projective nature of parameter integrands was shown in Ref.~\cite{Panzer:2015ida},
and is discussed for example in Ref.~\cite{Weinzierl:2022eaz}.}.\\

To begin with, consider the projective form 
\beq
\label{eq:ProjectiveForm}
  \alpha_{n-1} \, = \, \eta_{n-1} \,\, \frac{Q \big( \left\{ z_i \right\} 
  \big)}{D^P \big( \left\{ z_i \right\} \big)} \, , 
\eeq
where $\eta_{n-1}$ is the complete volume form \eqref{eq:eta} of the set of $n$ variables $z_i$; $Q(\left\{ z_i \right\})$ is a polynomial in the vabiables $z_i$ of degree $(l + 1) P - n$; and $D(\left\{ z_i \right\})$ a polynomial of degree $(l + 1)$. We can establish a correspondence between this form and the integrand of Feynman parametrized integrals by recognizing that the integrand of \eqref{eq:FP} is a specific instance of \eqref{eq:ProjectiveForm}, with the polynomial $D$ given by the second Symanzik polynomial of the graph ${\cal F}$, with $P = \nu - l d/2$ and with the numerator $Q(\left\{ z_i \right\})$ equal to the numerator in \eqref{eq:FP}.\\

To find a correspondence to Feynman integrals, it is necessary to consider the integral of the form \eqref{eq:ProjectiveForm}. The integral in \eqref{eq:FP} is integrated over the $(n-1)$-dimensional simplex defined by the argument of the Dirac delta function. The following theorem clarifies that the integral of the form \eqref{eq:ProjectiveForm} is always the same when the integration happens over a domain that has the same image of the $(n-1)$-dimensional simplex in the projective space $\mathbb{PC}^{n - 1}$  -- thus also possibly clarifying the reason we called such forms \textit{projective}.

\begin{thm}[Cheng-Wu theorem]
\label{th:ChengWu}
Given two integration domains, $O, O' \in \mathbb{C}^n$,
if their image in $\mathbb{PC}^{n - 1}$ is the same simplex, 
then $\int_O \alpha_{n - 1} \, = \, \int_{O'} \alpha_{n-1}$.
\end{thm}
\begin{proof}
This follows immediately from the fact that
\begin{itemize}
\item[i)] $\alpha_{n-1}$ is a closed form, as immediately clear from theorem \ref{th:dp};
\item[ii)] $\eta_{n-1}$ is null on each surface defined by $z_i = 0$
\end{itemize}
Indeed, if we denote by $\Delta$ the subset of $\mathbb{C}^n$ given by the surface 
connecting points in the boundaries of $O$ and $O'$ that have a common image in 
the projective space, then $\int_{O + \Delta - O'} \alpha_{n - 1} = 0$ because of statement 
i), while $\int_{\Delta} \alpha_{n-1} = 0$ because of statement ii).
\end{proof}

\begin{figure}[!h]
\begin{center}
\includegraphics[scale=0.4]{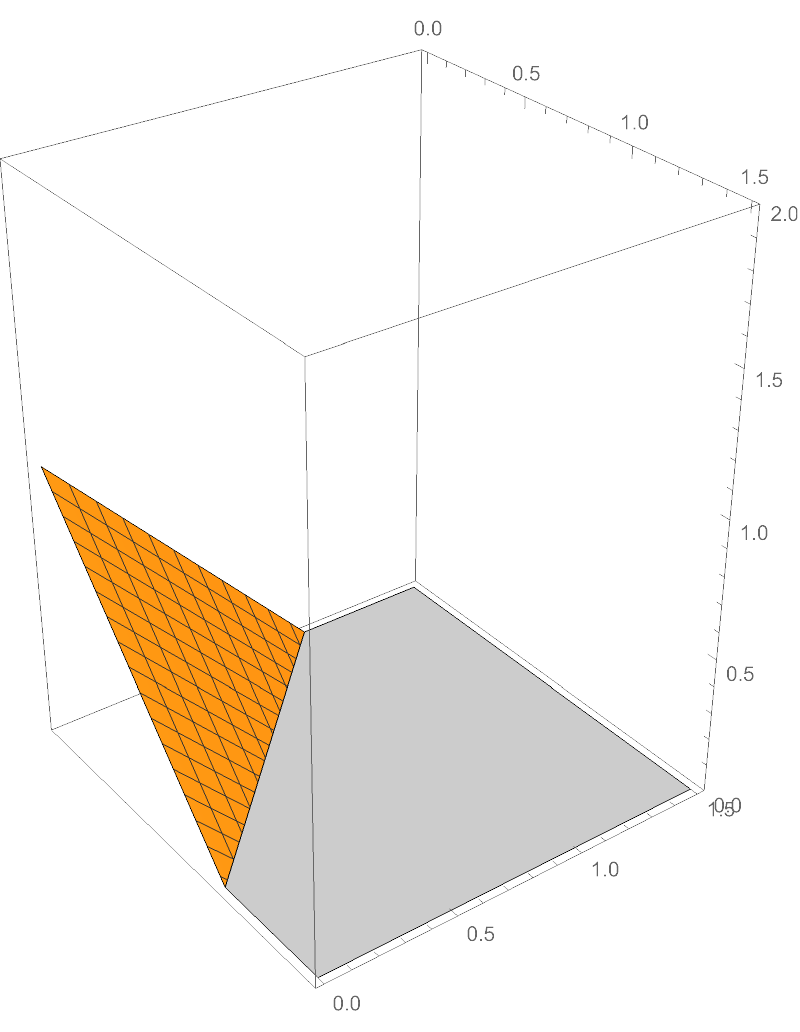}
\includegraphics[scale=0.4]{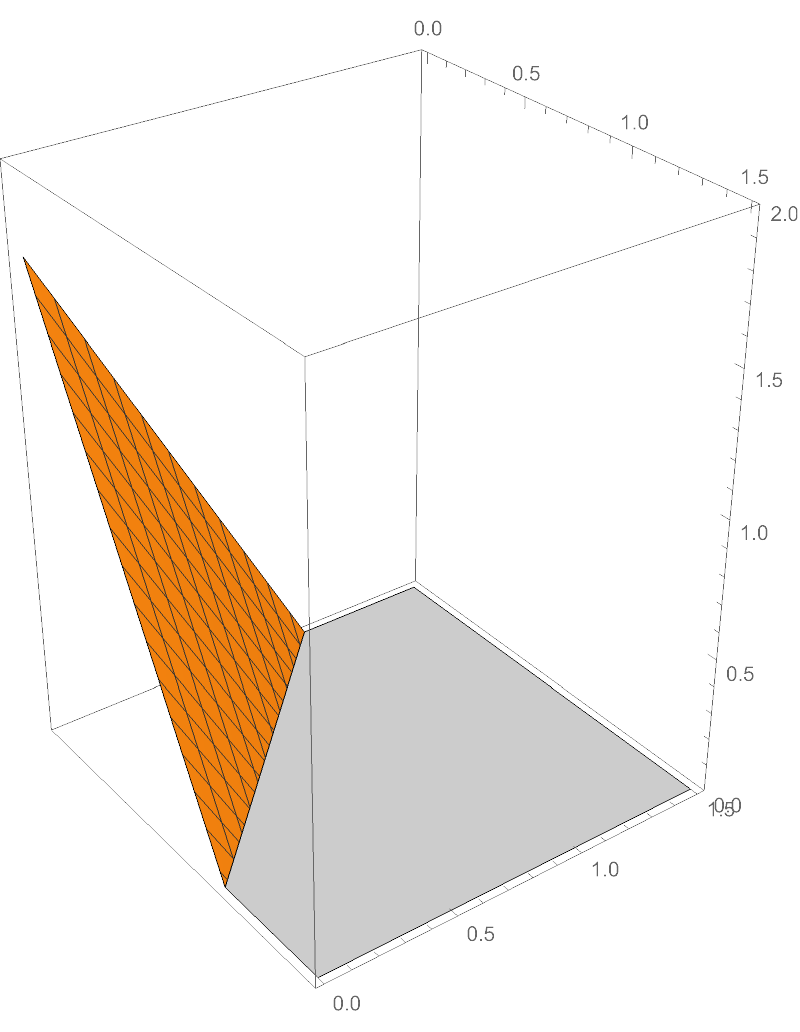}
\end{center}
\caption{Example of two domains in $\mathbb{R}^3$ that are equivalent under 
projective transformations.}\label{fig:simplex}

\end{figure}

The Cheng-Wu theorem, in its 
original form, states that in the argument of the delta function in the Feynman-parametrised 
expression, \eqref{eq:FP}, it is possible to restrict the sum to an arbitrary subset of Feynman 
parameters $z_i$. Consider the integration of \eqref{eq:ProjectiveForm} on the projection 
of the $n$-dimensional simplex, $S_{n-1} \equiv \left\{ z_i  \, | \, \sum_{i=1}^n z_i = 1 \right\}$. 
This choice of integration domain is arbitrary (within the set of projectively equivalent domains), 
as was proven above. This means that, for example, the set defined by $\{z_i \, | \, 
\sum_{i = 1}^n z_i = 1\}$ and the set $\{z_i \, | \, z_n + t  \sum_{i = 1}^{n-1} z_i = t\}$ 
are equivalent to any positive value of $t$. In the limit $t \to \infty$ the integration 
domain becomes independent of $z_n$, and becomes a semi-infinite $(n-1)$-dimensional
surface based on the simplex $\{z_i \, | \, \sum_{i = 1}^{n-1} z_i = 1\}$. Figure \ref{fig:simplex} provides 
an example in three dimensions of the two mentioned surfaces. When using the conventional choice of $S_{n-1}$ as integration domain, 
\beq
\label{eq:sumdz}
  d z_n \, = \, - \sum_{i = 1}^{n-1} d z_i \, ,
\eeq
we get an integral of the form
\beq
\label{eq:projective F}
  \int_{S_{n - 1}} \, \eta_{n - 1} \, \frac{Q(z)}{D^P(z)} \, = \, \int_{z_i \geq 0}
  d z_1 \ldots d z_n \, \delta \! \left(1 - \sum_{i=1}^n z_i \right) \, \frac{Q(z)}{D^P(z)} \, ,
\eeq

where we use the shorthand notation $z$ for the set $\{z_i\}$. Any consistent choice
of the polynomials $Q(z)$ and $D(z)$, yielding a projective form, provides a natural
generalisation of the Feynman integral \eqref{eq:FP}. 


\section{Integration by parts for parametrised integrals}
\label{sec:IBP}
\subsection{General formula for any number of loops}
\label{subsec:GeneralFormula}
After establishing a correspondence between generalized Feynman parametric integrals and projective forms in Section \ref{subsec:ChengWu}, we can proceed to deriving linear relations among such forms that are analogous to integration by parts identities in momentum space. The key concept, as we anticipated, is the Theorem \ref{th:dp} stating that the boundary of a projective
$q$-form is a projective $q+1$-form. We can now use this fact to construct integration-by-parts identities in Feynman parameter space in full generality. To this end, consider the projective 
$(n-2)$-form 
\beq
\label{eq:w}
  \omega_{n-2} \, \equiv \, \sum_{i  = 1}^{n} (-1)^i \, \eta_{\{z\} - z_i} \,
  \frac{H_i (z)}{(P - 1) \, \big( D(z) \big)^{P - 1}} \, ,
\eeq
with any suitable choice of the polynomials $H_i (z)$ and $D(z)$. We use the notation $\eta_{\{z\} - z_i}$ to refer to the projective volume $(n-2)$-form made from the set of all Feynman parameters except for $z_i$.  Dimensionally regularized UV and IR divergent Feynman integrals 
can be treated without difficulties: these divergences are regulated by the parameter 
$\epsilon$, which features in the exponents of the parameters; on the other side, technically, we need to restrict the choice of $D(z)$ so that the singularities of $\omega_{n-2}$ lie in a general 
position with respect to the simplex integration domain $S_{n-1}$, and in particular they 
do not touch the sub-simplexes forming the boundaries of $S_{n-1}$. This case was labeled as case (A) in \cite{Regge:1968rhi} and in general represents a restriction on the external kinematics.\\

Acting with the boundary operator $d$ on the form $\omega_{n - 2}$ and using \eqref{eq:dppd} to write the result in terms of projective forms we get
\beq
\label{eq:dw}
  d \omega_{n-2} \, = \, \frac{1}{(P-1) \, \big( D(z) \big)^{P - 1}} \, \, \eta_{\{z\}} \, 
  \sum_{i = 1}^n \frac{\partial H_i(z)}{\partial z_i} \, - \, \frac{\eta_{\{z\}}}{\big( D(z) \big)^P} \, 
  \sum_{i = 1}^n  H_i \, \frac{\partial D(z)}{\partial z_i} \, .
\eeq
This is the all-loop master equation sought. By integrating over a projective equivalent of the $(n-1)$-dimensional simplex the integrand Equation \eqref{eq:dw} becomes a linear relation among Feynman integrals; in particular, we highlight that linear combinations of integrals having denominator exponent $P$ can be reduced to integrals having a smaller value of denominator exponent $(P-1)$ modulo some boundary terms, that can be integrated using Stokes theorem.  It is important to stress that the possible presence of non-vanishing boundary terms represents a substantial difference with respect to the conventional IBP identities in momentum space: in Feynman parametrisation, boundary terms do not in general integrate to zero. In terms of Feynman diagrams, the integration over sub-simplexes represents the shrinking of a line of the diagram to a point, introducing sub-topologies in the algorithm. Equation \eqref{eq:dw} will be the basis for all the applications in the following sections.\\

With the goal in mind of using the integrated version of \eqref{eq:dw} in master integral reduction algorithms, we can highlight how the \textit{seeding} phase of momentum space-based algorithms here consists of suitable choices of the polynomials $H_i(z)$
\beq
\label{eq:formQ}
  Q \big( z \big) \, = \, \sum_{i = 1}^n  H_i(z) \, 
  \frac{\partial D(z)}{\partial z_i} \, .
\eeq
For example, by letting $H_i(z)$ be non-zero only for a particular value of $i$, one can get as many different equations as the number of parameters. As we will discuss further in the chapter, any attempt at reducing the number of necessary equations reduces to clever choices of the polynomials $H_i$. In order to explore applications of Equation \eqref{eq:dw} we are temporarily ignoring this direction to focus on the naive generation of IBP relations using the mentioned choice 
\beq
\label{eq:formH}
  H_i = H(z) \delta_{i,j}
\eeq
for particular values of $j$. We begin our examples by specialising to one-loop graphs.

\subsection{One-loop examples}
\label{subsec:One-loopEx}
In the following section, we will present two simple one-loop examples to show how IBP identities in parameter space concretely look like, and how subtopologies naturally appear as boundary terms. These examples are the one-loop massless box integral and the one-loop massless pentagon integral. Before proceeding with these concrete examples, we introduce the notation we use for one-loop Feynman parametrised integrals. Indeed for one-loop integrals, the Symanzik polynomials can  be written in a more explicit form, allowing for Equation \eqref{eq:dw} to be written in a form suitable for fast generation of IBP relations.\\

Consider \eqref{eq:FP} for a one-loop diagram with $n$ internal propagators. In this 
case we can write
\beq
\label{eq:1LFP}
  I_G (\nu_i, d) \, = \, \frac{\Gamma(\nu - d/2)}{\prod_{j = 1}^n \Gamma(\nu_j)}
  \int_{z_j \geq 0} \! d^n z \, \delta \bigg(1 - z_{n+1} \bigg) \, 
  \frac{\prod_{j = 1}^{n+1} z_j^{\nu_j -1}}{\bigg(
  \sum_{i=1}^{n+1} \sum_{j=1}^{i-1} s_{ij} z_i z_j \bigg)^{\nu - d/2}} \, , \quad
\eeq
where we introduced the matrix $s_{ij}$ $(i,j = 1, \ldots, n+1)$, defined by 
\beq
\label{eq:sij}
  s_{ij} \, = \, \frac{(q_j - q_i)^2}{\mu^2} \quad (i,j = 1, \ldots, n) \, , \qquad \quad
  s_{i, n+1} \, = \, s_{n+1, i} \, \equiv \, - \frac{m_i^2}{\mu^2} \, ,
\eeq
as well as the auxiliary quantities
\beq
\label{eq:zn+1}  z_{n+1} \, \equiv \, \sum_{i =1}^n z_i \, , \qquad \quad \nu_{n+1} \, \equiv \, \nu - d + 1 \, .
\eeq
We highlight that the auxiliary quantities just introduced are the first Symanzik polynomial and its exponent for one-loop integrals, thus simplifying the form of the integrand. The $s_{ij}$ represent the invariant squared masses of the combinations of external momenta flowing in or out of the diagram between line $i$ and line $j$, with internal masses included as invariant quantities associated with $z_{n+1}$.\\

As we mentioned in \ref{subsec:GeneralFormula}, for the moment we consider separately in \eqref{eq:dw} the $n$ forms obtained by omitting from the projective volume the variable $z_i$, ($i = 1, \ldots, n$), and choose for $H_i$ simply the numerator of 
\eqref{eq:1LFP}. This amounts to setting
\beq
\label{eq:chooseHi}
  H_i \, = \, \delta_{ih} \left( \prod_{j = 1}^n z_j^{\nu_j -1} \right) 
  \left(\sum_{k =1}^n z_k \right)^{\nu - d} 
  \, = \, \, \delta_{ih} \, \prod_{j = 1}^{n+1} z_j^{\nu_j -1} \, ,
 \eeq
for some $h \in \left\{ 1,...,n\right\}$. For the time being, we set $\nu_h > 1$; we can however anticipate that soon we will include the case $\nu_h = 1$ by considering also boundary terms. With the current constrain on $\nu_h$, Equation \eqref{eq:dw} becomes
\begin{eqnarray}
\label{eq:1LIBP} 
  && d \left( (-1)^h  \, \eta_{\{z\} - z_h} \frac{ \prod_{j = 1}^{n+1} z_j^{\nu_j -1}}{\big( 
  \nu - (d + 1)/2 \big) \left( \sum_{i = 1}^{n+1} 
  \sum_{j=1}^{i-1} s_{ij} z_i z_j \right)^{\nu - (d+1)/2 }} \right) \, = 
  \nonumber \\
  && = \, \frac{\eta_{\{z\}}}{ \big( \nu - (d+1)/2 \big) \left( \sum_{i = 1}^{n+1} 
  \sum_{j=1}^{i-1} s_{ij} z_i z_j  \right)^{\nu - (d+1)/2 }} \,
  \left[ (\nu_h -1) \frac{H_i}{z_h} + (\nu - d) \, \frac{H_i}{z_{n+1}} \right] + 
  \nonumber \\
  &&  - \, \frac{ \eta_{\{z\}} }{\left( \sum_{i = 1}^{n+1} 
  \sum_{j=1}^{i-1} s_{ij} z_i z_j \right)^{\nu - (d-1)/2 }} \, H_i \, \left( \sum_{k = 1}^{n+1} 
  \big( s_{k h} + s_{k, n+1} \big) z_k \right) \,.
\end{eqnarray}
where $s_{n+1,n+1} = 0$. Equation \eqref{eq:1LIBP} is still too complicated to immediately identify the structure of the difference equation. In order to clarify the structure of this expression and make it more similar to its momentum space counterpart, we introduce an index notation, following Ref.~\cite{Barucchi:1973zm}. We first define the one-loop integrand as
\beq
\label{eq:notforf1}
  f \big( \left\{ \nu_1, \ldots, \nu_{n+1} \right\} \big) \, \equiv \, 
  f \big( \left\{ {\cal R} \right\} \big) \, = \, \eta_{\{z\}} \, \frac{
  \prod_{j = 1}^{n+1} z_j^{\nu_j -1}}{\left( 
  \sum_{i = 1}^{n+1} \sum_{j=1}^{i-1} s_{ij} z_i z_j \right)^{\nu - d/2 }} \, .
\eeq
Then we write
\beq
\label{eq:notforf2}
  f \big( \left\{ {\cal I} \right\}_{-1}, \left\{ {\cal J} \right\}_{0}, \left\{ {\cal K} \right\}_{1})
\eeq
to denote the same function as in \eqref{eq:notforf1}, where however the indices $\nu_i \in \left\{ {\cal I}, {\cal J}, {\cal K} \right\}$ have been respectively decreased by one, left untouched, and increased by one. Note that we consider sets such that $ \left\{ {\cal I} \right\} \cup \left\{ {\cal J} \right\} \cup \left\{ {\cal K} \right\} = \mathcal{R}$. Given this condition, one of the three sets is not necessary and we will therefore adopt the convention that the set $\cal J$ can be omitted. Furthermore, the raising and lowering operations are defined in order to preserve the character of $f$ as a projective form, so the exponent of the denominator is re-determined after raising and lowering the indices.\\

Under these assumptions for the convention, Equation \eqref{eq:1LIBP} becomes
\begin{eqnarray}
\label{eq:1LIBP2}
  d \omega_{n-2}  + \sum_{k = 1}^{n+1} (s_{k h} + s_{k, n+1}) \,
  f \big( \left\{ k \right\}_{1} \big) & = & 
  \frac{\nu_h - 1}{\nu - (d+1)/2} \, f \big( \left\{ h \right\}_{-1} \big) \\
  & + &  \frac{\nu - d}{\nu - (d+1)/2} \, f \big( \left\{ n+1 \right\}_{-1} \big) \, . \nonumber
\end{eqnarray}
This is the desired integration by parts identity at one loop, which at this stage is kept at the integrand level to emphasise the fact that the boundary integral is not a priori vanishing. Notice that at the one-loop level, one also has the constraints
\beq
\label{eq:Uconsistency}
  \sum_{i = 1}^n f \big( \left\{ i \right\}_{1} \big) \, = \,  
  f \big(  \left\{ n+1 \right\}_{1} \big) \, 
\eeq
and
\beq
\label{eq:Fconsistency}
\sum_{i = 1}^{n+1} \sum_{j=1}^{i-1} s_{ij} f \big( \left\{ i,j \right\}_{1} \big) \, = \, 
f \big( \left\{ {\cal R}_0 \right\} \big) \ ,
\eeq
immediately following from the definitions of $f$ and the one-loop Symanzik polynomials: Equation \eqref{eq:Uconsistency} is the consistency condition related to the definition of the first Symanzik polynomial while Equation \eqref{eq:Fconsistency} relies on the definition of the second Symanzik polynomial.\\

Ref.~\cite{Barucchi:1973zm} shows that, when the boundary term is zero, \eqref{eq:1LIBP2}, \eqref{eq:Uconsistency} and \eqref{eq:Fconsistency} allow for the systematic construction of a closed system of first-order differential equations. The procedure they use is constructive and algorithmic, but one notices empirically that the number of integrals in the system is often higher than the number of actually independent master integrals, in concrete cases with specific mass assignments. In practical examples, the integrals used in the proof can still be used as a starting point, trying then to minimize the over-completeness of the resulting basis. We can show how the reduction works for the example of the one-loop massless box integral.

\subsubsection{One-loop massless box}
\label{ssubsec:OlMaB}

Let us consider the integral in \eqref{eq:1LFP} for the one-loop massless box integral, 
where $n = 4$, and for simplicity we set the renormalisation scale as $\mu^2 = 
(p_1+p_4)^2 \equiv s$, while $(p_1+p_2)^2 \equiv t$ (all momenta are considered to be incoming). 
In particular, we focus on the simple case where all $\nu_i = 1$, and we define
\beq
\label{eq:massless box}
  I_{\textrm{box}} \, = \, \Gamma (2 + \epsilon) \, \int_{S_{n-1}} \eta_{\{ z \}} \,
  \frac{ (z_1 + z_2 + z_3 + z_4)^{2 \epsilon}}{\left( r  z_1 z_3 + z_2 z_4 
  \right)^{2 + \epsilon}} \, \equiv \, \Gamma( 2 + \epsilon) \, I(1,1,1,1; 2 \epsilon) \, ,
\eeq
where we defined $r = t/s$, and the notation for four-point 
integrals is from now on $I(\nu_1, \nu_2, \nu_3, \nu_4; \nu_5)$. This 
notation is set up so that the arguments of the function correspond directly to 
the exponents of the propagators in the corresponding Feynman diagrams. 
Notice that in this framework, as it is well known, the dimension of spacetime 
becomes simply a parameter related to the exponents of the first Symanzik 
polynomial, and dimensional shift identities are naturally encoded in the parameter-based IBP equation, \eqref{eq:1LIBP2}. The matrix $s_{ij}$ for $I_{\rm box}$ reads 
\begin{equation}
s_{ij} = \left( \begin{array}
{c c c c c}

0 & 0 & r & 0 & 0 \\ 
 
0 & 0 & 0 & 1 & 0 \\

r & 0 & 0 & 0 & 0 \\

0 & 1 & 0 & 0 & 0 \\

0 & 0 & 0 & 0 & 0 \\

\end{array} \right)
\end{equation}
For this simple example, we aim at constructing a differential equation involving integral \eqref{eq:massless box} and other two master integrals using the parameter based IBP equations \eqref{eq:1LIBP}; we will assume the number of elements of the basis to be known, as for this simple example it is easy to proceed iteratively to reduce any integral to a linear combinations of 3 master integrals.
The construction of the overcomplete basis in Ref.~\cite{Barucchi:1973zm} shows that the integrals that appear in the final differential equations system are $I(1,1,1,1;2\epsilon)$ and 
the ones obtained from it by raising by one the powers $\nu_i$ for an even 
number of parameters.  Based on the result of Ref.~\cite{Barucchi:1973zm}, 
we take a basis of master integrals to be given by
\beq
\label{eq:basisbox}
  \Big\{ I(1, 1, 1, 1; 2 \epsilon), I(2, 1, 2, 1; 2 \epsilon),
  I(2, 2, 2, 2; 2 \epsilon) \Big\} \,.
\eeq
To verify that this is indeed a basis, and that we can close the 
system, we consider first the derivative of $I(1,1,1,1; 2 \epsilon)$ with respect 
to $r$,
\beq
\label{eq:dzI1}
  \partial_r I (1, 1, 1, 1; 2 \epsilon) \, = \, - (2 + \epsilon) \, I(2, 1, 2, 1; 2 \epsilon) \, ,
\eeq
which indeed contains only integrals belonging to the desired set. On the other hand
\beq
\label{eq:dzI2}
  \partial_r  I (2, 1, 2, 1; 2 \epsilon) \, = \, - ( 3 + \epsilon) \, I (3, 1, 3, 1; 2 \epsilon) \, .
\eeq
In order to proceed, it is necessary to express the integral $I(3, 1, 3, 1; 2 \epsilon)$ 
in terms of integrals belonging to the chosen set. \eqref{eq:1LIBP2} for $\nu_1 = 3$, 
$\nu_{3} = 2$, $\nu_2 = \nu_4 = 1$ and $h = 1$ becomes
\beq
\label{eq:2010}
  r I(3, 1, 3, 1; 2 \epsilon) + \int d \omega_{n-2} \, = \, 
  \frac{2}{3 + \epsilon} I(2, 1, 2, 1; 2 \epsilon) + \frac{2 \epsilon}{3 + \epsilon} 
  I(3, 1, 2, 1; - 1 + 2 \epsilon) \, .                                                                                                                                                                                                                                                                                                                                                                                
\eeq
The boundary term is 
\beq
\label{eq:vanbou}
  \frac{z_1^2 z_3 
  \left(z_1 + z_2 + z_3 + z_4 \right)^{2 \epsilon}}{(3 + \epsilon)
  (r z_1 z_3 + z_2 z_4)^{3 + \epsilon}}
  \big( z_2 d z_3 \wedge dz_4 - z_3 d z_2 \wedge d z_4 + z_4 
  d z_2 \wedge d z_3 \big) \Bigg|_{\partial S_{n-1}}
  \, = \, 0 \, ,
\eeq
which follows from the fact that the projective form $\eta^{234}$ vanishes on all boundary 
sub-simplexes, except the one defined by $z_1 = 0$, where however the integrand is zero. This observation justifies the constraint $\nu_h > 1$ which we anticipated being the condition to avoid boundary terms. This property holds for all the identities used in this section. Consider now \eqref{eq:1LIBP2} for $\nu_1 = \nu_2 = \nu_3 = 2$, $\nu_4 = 1$ 
and $h = 2$, as well as the sum rule in \eqref{eq:Uconsistency} for $I(2, 1, 2, 1; - 1 + 2 \epsilon)$. 
One finds
\begin{eqnarray}
\label{eq:1110}
  I(2, 2, 2, 2; 2 \epsilon) & = & \frac{1}{3 + \epsilon} I(2, 1, 2, 1; 2 \epsilon) + 
  \frac{2 \epsilon}{3 + \epsilon} I(2, 2, 2, 1; - 1 + 2 \epsilon) \, , \nonumber \\
  I(2, 1, 2, 1; 2 \epsilon) & = & 2 I(3, 1, 2, 1; - 1 + 2 \epsilon) + 2 I(2, 2, 2, 1; - 1 + 2 \epsilon) \, ,
\end{eqnarray}
where the symmetry of the integrand under the exchange of $(z_1, z_3)$ with $(z_2, z_4)$ 
has already been taken into account. This system and \eqref{eq:2010} allow to find a solution 
for $I(3, 1, 3, 1; 2 \epsilon)$, given by
\beq
\label{eq:I2020}
  I (3, 1, 3, 1; 2 \epsilon) \, = \, \frac{1}{r} \Big[ I(2, 1, 2, 1; 2 \epsilon) -	 
  I(2, 2, 2, 2; 2 \epsilon) \Big] \, ,
\eeq
involving only integrals we selected as elements of the basis \eqref{eq:basisbox}.\\

The last derivative to be computed in terms of the chosen set of basis integrals
is $\partial_r I(2, 2, 2, 2; 2 \epsilon)$, which is proportional to $I(3, 2, 3, 2; 2 \epsilon)$. 
Using the same procedure adopted so far, it is possible to get the linear system of equations
\beq
\footnotesize
\begin{cases}
\label{boxsystem2}
\frac{2}{(3+\epsilon)} I(2,1,2,1,2 \epsilon )+
\frac{2\epsilon}{(3+\epsilon)} I(3,1,2,1,-1+2\epsilon)
-I(3,1,3,1,2\epsilon)r = 0   \\
\frac{1}{(3+\epsilon)} I(1,2,1,2,2\epsilon)+
\frac{2\epsilon}{(3+\epsilon)}I(2,2,1,2,-1+2\epsilon)-I(2,2,2,2,2\epsilon)r = 0  \\
\frac{1}{(3+\epsilon)} I(2,1,2,1,2\epsilon)+
\frac{2\epsilon}{(3+\epsilon)} I(2,2,2,1,-1+2\epsilon)-I(2,2,2,2,2\epsilon) = 0  \\
\frac{2}{(4+\epsilon)} I(2,2,2,2,2\epsilon)+
\frac{2\epsilon}{(4+\epsilon)} I(3,2,2,2,-1+2\epsilon)
-I(3,2,3,2,2\epsilon)r = 0  \\
\frac{1}{(3+\epsilon)} I(2,2,1,2,-1+2\epsilon)
-\frac{1}{(3+\epsilon)} I(2,2,2,1,-1+2\epsilon) 
+I(2,3,2,2,-1+2\epsilon)-I(3,2,2,2,-1+2\epsilon)r = 0  \\
\frac{2}{(2+\epsilon)} I(1,1,1,1,2\epsilon)+
\frac{2\epsilon}{(2+\epsilon)} I(1,2,1,1,-1+2\epsilon)+ 
\frac{2\epsilon}{(2+\epsilon)} I(2,1,1,1,-1+2\epsilon)-I(1,2,1,2,2\epsilon)-
I(2,1,2,1,2\epsilon)r = 0  \\
-I(1,1,1,1,2\epsilon)+2I(1,2,1,1,-1+2\epsilon)+2I(2,1,1,1,-1+2\epsilon) = 0  \\
-I(2,1,2,1,2\epsilon)+2I(2,2,2,1,-1+2\epsilon)+2I(3,1,2,1,-1+2\epsilon) = 0 \\
-I(2,2,2,2,2\epsilon)+2I(2,3,2,2,-1+2\epsilon)+2I(3,2,2,2,-1+2\epsilon) = 0 
\end{cases}     
\eeq
whose solution for the desired integral is
\beq
\label{eq:I2121}
  I(3, 2, 3, 2; 2 \epsilon) \, = \frac{I(2, 1, 2, 1; 2 \epsilon) - I(1, 2, 1, 2; 2 \epsilon) + 
  (3 + \epsilon) (1 + \epsilon + 3 r) I(2, 2, 2, 2; 2 \epsilon)}{(3 + \epsilon)
  (4 + \epsilon) \, r (1 + r)}\,.
\eeq
The auxiliary integral $I(1, 2, 1, 2; 2 \epsilon) $ is easily removed by use of Equation \eqref{eq:Fconsistency} for the one-loop box integral
\beq
\label{eq:newrel}
  I(1,2,1,2,2\epsilon) \, = \, I(1,1,1,1,2\epsilon) - r I(2,1,2,1,2\epsilon) \, .
\eeq 
The system of differential equations for our basis set of integrals is now complete, 
and it reads
\beq
\label{eq:system}
  \partial_r \textbf{b} \, \equiv \, \partial_r \left(
  \begin{array}{c}
  I(1, 1, 1, 1; 2 \epsilon) \\
  I(2, 1, 2, 1; 2 \epsilon) \\
  I(2, 2, 2, 2; 2 \epsilon)
  \end{array}
  \right) =  \left(
  \begin{array}{c c c c}
  0 & - (2 + \epsilon)  & 0 \\
  0 & - \frac{3 + \epsilon}{r}  &  \frac{3  + \epsilon}{r} \\
  0 & 0  & - (3 + \epsilon) \\
  \frac{1}{(3 + \epsilon) } \left( \frac{1}{r}-\frac{1}{r+1} \right) & -\frac{1}{(3 + \epsilon) r } & -\frac{\epsilon +1}{r}+\frac{\epsilon -2}{r+1}
  \end{array}
  \right)
  \textbf{b} \, .
\eeq
Despite this system still not being in the canonical form \cite{Henn:2013pwa}, for obtaining which several techniques are available \cite{Lee:2014ioa,Prausa:2017ltv,Gituliar:2017vzm,Lee:2020zfb, Meyer:2017joq,Meyer:2018feh}, we can already read from the form we chose the letters of the symbol alphabeth of the solution, $r$ and $r+1$. We stress that the singularity of the differential equation system for $r=-1$ does not belong to the Riemann sheet of the physical solution, which instead is continuous and analytic for $r=-1$. In the spirit of a proof-of-concept, we have not developed a systematic approach to 
the search for useful boundary conditions to determine the unique relevant solution 
of the system. In the case at hand, continuity in $r = -1$, uniform-weight arguments, 
and the known value of the residue of the double pole in $\epsilon$ can be used to 
recover the known solution. We find
\begin{eqnarray}
  I_{\rm box} & = & \frac{k(\epsilon)}{r} \bigg[ \frac{1}{\epsilon^2} - \frac{\log r}{2 \epsilon} -
  \frac{\pi ^2}{4} + \epsilon \, \bigg( \frac{1}{2} \, {\rm Li}_3 (- r) - \frac{1}{2} \, {\rm Li}_2 (- r) 
  \log r + \frac{1}{12} \log ^3 r \nonumber \\
  && \hspace{2cm} - \, \frac{1}{4} \log(1 + r) \left(\log ^2 r + \pi ^2 \right) 
  + \frac{1}{4} \, \pi ^2 \log r + \frac{1}{2} \zeta(3) \bigg) \, + \, {\cal O} (\epsilon^2)
  \bigg] \, ,
\end{eqnarray}
matching the result reported, for example, in Ref.~\cite{Henn:2014qga}. The one-to-one 
correspondence between the two results is found by setting the overall constant 
$k(\epsilon) \, = \, 4-\frac{\pi^2}{3}\epsilon^2 -\frac{40 \zeta(3)}{3}\epsilon^3$. This result confirms that, for one-loop integrals, a parameter-based IBP strategy
can work, and the system of differential equations can be consistently solved. At the
same time, the discussion highlights the importance of an optimal choice of the
polynomials $H_i(z)$, especially beyond one loop, where a general constructive 
procedure for the solution of the system of differential equations is not yet available. In the following section, we will focus on the role that boundary terms can play in integral reduction algorithms based on Equation \eqref{eq:dw}.

\subsubsection{One-loop massless pentagon}
\label{ssec:OlMaP}

In dimensional regularisation, it is well-known \cite{Bern:1993kr} that the one-loop massless pentagon can be expressed as a sum of one-loop boxes with one external massive leg (corresponding to the contraction to a point of one of the loop propagators), up to corrections vanishing in $d=4$. In this section, we will recover this result; from the point of view of projective forms, the analysis of this case is interesting because it introduces non-vanishing boundary terms, absent in the case of the massless box integral and in the analysis of Ref.~\cite{Barucchi:1973zm}. \\

Consider the integration by parts Equation \eqref{eq:1LIBP} for a five-parameter integral with $\nu_1 = \nu_2 = \nu_3 = \nu_4 = \nu_5 = 1$, and the exponent of the $\mathcal{U}$ polynomial equal to $2\epsilon$. As we anticipated in Section \ref{subsec:GeneralFormula}, the case $\nu_i = 1$ can be considered in our algorithm, provided we keep the boundary terms. We can see it explicitly by starting with the case $h = 1$, where we obtain the equation
\beq
\label{eq:I00000}
  \int_{S_{\left\lbrace 1,2,3,4,5 \right\rbrace}} \! d \omega_3 + s_{13} \, 
  I(1, 1, 2, 1, 1; 2 \epsilon) + s_{14} \, I(1, 1, 1, 2, 1; 2 \epsilon) 
  \, = \, \frac{2 \epsilon}{2 + \epsilon} I(1, 1, 1, 1, 1; - 1 + 2 \epsilon) \, , \qquad
\eeq
with 
\beq
  d \omega_3 \, = \, d \left[ - \, \eta_{\{2,3,4,5\}} \, 
  \frac{(z_1 + z_2 + z_3 + z_4 + z_5)^{2 \epsilon}}{(2 + \epsilon)
  \left( s_{13} z_1 z_3 + s_{14} z_1 z_4 + s_{24} z_2 z_4 + s_{25} z_2 z_5 + 
  s_{35} z_3 z_5 \right)^{2 + \epsilon}} \right] \, .
\label{eq:dw3}
\eeq
Using Stokes' theorem, and considering the only subset of the boundary of the 
five-dimensional simplex where $\eta_{\{2,3,4,5\}} \neq 0$, the boundary term 
of this equation becomes 
\beq
\label{eq:boundary_term}
  \int_{S_{\{2,3,4,5\}}} \! \eta_{\{2,3,4,5\}} \, 
  \frac{(z_2 + z_3 + z_4 + z_5)^{2 \epsilon}}{\left(s_{24} z_2 z_4 + s_{25} z_2 z_5 + 
  s_{35} z_3 z_5 \right)^{2 + \epsilon}} \, = \,  
  I_{\rm box}^{(1)} (s_{25}) \, ,
\eeq
where $I_{\rm box}^{(1)}$ is a one-loop box integral with one massive external leg, with a squared mass proportional to $s_{25}$\footnote{According to the definition of Feynman parametrised integrals \eqref{eq:FP}, actually the integral in Equation \eqref{eq:boundary_term} is strictly speaking proportional to the one-loop box integral with one massive external leg, as we are leaving out some proportionality factors.}. Effectively, the propagator with index $\nu_1$ has been contracted to a point. Note that, when applying Stokes' theorem, the integration over boundary domains corresponds to the proper integration region, needed to obtain the lower-point Feynman integral, up to a sign arising from the orientation of the boundary. Reversing this orientation, when needed, produces a sign that, for example, cancels the minus sign in \eqref{eq:dw3} -- the equation must not depend on the label of the Feynman parameter.\\

Considering, in a similar way, all the possible values for $h$, the following system 
of equations is obtained
\beq
\label{eq:pentagon system}
  (2 + \epsilon) \!
  \left( \begin{array}{c c c c c}
  0 & 0 & s_{13} & s_{14} & 0 \\
  0 & 0 & 0 & s_{24} & s_{25} \\
  s_{13} & 0 & 0 & 0 & s_{35} \\
  s_{14} & s_{24} & 0 & 0 & 0 \\
  0 & s_{25} & s_{35} & 0 & 0
  \end{array}
  \right) \! \left( \begin{array}{c}
  I(2,1,1,1,1;2\epsilon) \\
  I(1,2,1,1,1;2\epsilon) \\
  I(1,1,2,1,1;2\epsilon) \\
  I(1,1,1,2,1;2\epsilon) \\
  I(1,1,1,1,2;2\epsilon) \\
  \end{array}
  \right) \! + \!
  \left( \begin{array}{c}
  I_{\text{box}}^{(1)}(s_{25}) \\
  I_{\text{box}}^{(2)}(s_{13}) \\ 
  I_{\text{box}}^{(3)}(s_{24}) \\
  I_{\text{box}}^{(4)}(s_{35}) \\
  I_{\text{box}}^{(5)}(s_{14})
  \end{array}
  \right) \hspace{-0.2pt} = 
  2 \epsilon I(11111;-1+2\epsilon)
  \left( \begin{array}{c}
  1 \\
  1 \\
  1 \\
  1 \\
  1
  \end{array}
  \right) \! , \qquad
\eeq
where the finite integral $I(1,1,1,1,1; - 1 + 2 \epsilon)$ is proportional to the pentagon 
integral in $d = 6 - 2 \epsilon$. We can now invert the system to solve for the vector made of the five-point integrals with a raised exponent and then use Equation \eqref{eq:Uconsistency} to sum all five of them into a single integral
\beq
\label{eq:consistency_pentagon}
 \sum_{i = 1}^5 I(\{i\}_1) = I(1,1,1,1,1;1 + 2 \epsilon) \; ,
\eeq
which is the pentagon integral in $d = 4-2\epsilon$ we want to compute. The result of this procedure is
\begin{eqnarray}
\label{eq:pentagon}
  2 (2 + \epsilon) \,
  I(1, 1, 1, 1, 1;1 + 2 \epsilon) & = & 
  \Bigg\{\frac{s_{13} s_{24}-s_{13} s_{25}-s_{14} s_{25}+s_{14} s_{35}-
  s_{24} s_{35}}{s_{13} s_{14} s_{25}} \, I_{\rm box}^{(1)} \nonumber \\ 
  && - \, \frac{s_{13} s_{24}+s_{13} s_{25}-s_{14} s_{25}+s_{14} s_{35}-
  s_{24} s_{35}}{s_{13} s_{24} s_{25}} \, I_{\rm box}^{(2)} \nonumber \\ 
  && - \, \frac{s_{13} s_{24}-s_{13} s_{25}+s_{14}
   s_{25}-s_{14} s_{35}+s_{24} s_{35}}{s_{13} s_{24} s_{35}} \, I_{\rm box}^{(3)} \nonumber \\
  && + \, \frac{s_{13} s_{24}-s_{13} s_{25}+s_{14} s_{25}-s_{14} s_{35}-s_{24}
   s_{35}}{s_{14} s_{24} s_{35}} \, I_{\rm box}^{(4)} \nonumber \\ 
  && - \, \frac{s_{13} s_{24}-s_{13} s_{25}+s_{14} s_{25}+s_{14} s_{35}-
  s_{24} s_{35})}{s_{14} s_{25} s_{35}} \, I_{\rm box}^{(5)} \Bigg\} \nonumber \\ 
  && + \, 2 \epsilon \, I(1, 1, 1, 1, 1; - 1 + 2 \epsilon) \, ,
\end{eqnarray}
recovering the result of Ref.~\cite{Bern:1993kr}\footnote{The correspondence between the 
coefficients reported here and those of Ref.~\cite{Bern:1993kr} can be derived using
the definition $c_i = \sum_{j = 1}^5 S_{ij}$ in their notation.}. The most important direct consequence of \eqref{eq:pentagon} is the well-known theorem stating that the one-loop massless pentagon in four dimensions can be expressed as a sum of one-loop boxes with an external massive leg, up to $O(\epsilon)$ corrections. This last statement is due to the infrared and ultraviolet 
convergence of the $6-2\epsilon$ dimensional pentagon, which implies that the 
last line of \eqref{eq:pentagon} is $O(\epsilon)$.

\subsection{Multi-loop examples}
\label{subsec:Multi-loopEx}

\begin{figure}
\centering
\begin{tikzpicture}
            \begin{feynman}
            \vertex (a1);
            \vertex[right=1cm of a1] (a2);
            \vertex[above right= 2.12 cm of a2](a3);
            \vertex[below right= 2.12 cm of a2](a4);
            \vertex[below = 1 cm of a3] (a31);
            \vertex[above = 1 cm of a4] (a32);
            \vertex[right=3cm of a2] (a5); 
            \vertex[right=1cm of a5] (a6); 
            \diagram* {
            (a1) -- (a2)
                -- [quarter left] (a3)
                -- [quarter left] (a5)
                -- [quarter left] (a4)
                -- [quarter left] (a2),
                (a5) -- (a6),
                (a2) -- (a5)
            };
            \end{feynman}
            \draw (a2) to[out=30,in=150] (a5);
            \draw (a2) to[out=-30,in=-150] (a5);
            \draw (a2) to[out=80,in=100] (a5);
            \draw (a2) to[out=-80,in=-100] (a5);
            \path
            (2.5,1.75) node {$z_1$}
            (2.5,1.1) node {$z_2$}
            (2.5,0.25) node {.}
            (2.5,0.10) node {.}
            (2.5,-0.25) node {.}
            (2.5,-0.10) node {.}
            (2.5,-1.75) node {$z_{n+1}$};
            \end{tikzpicture}
\caption{\label{fig:banana} Banana diagram with $n$-loops}
\end{figure}
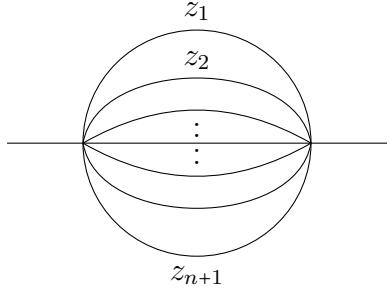

The first Symanzik polynomial for $l$-loop Feynman integrals, with $l > 1$, displays
a much more varied and intricate structure compared to the one-loop case, corresponding
to the factorial growing variety of graph topologies that can be constructed. Some 
classes of diagrams can still be described for all orders: a natural example is given by
the so-called $l$-loop banana graphs, depicted in Fig.~\ref{fig:banana}, contributing to 
two-point functions and involving $(l+1)$ propagators. The monodromy ring for these 
graphs was identified in Ref.~\cite{Ponzano:1969tk}, but this result was not (at the 
time) translated into a systematic method to construct differential equations. The two-loop sunrise graph with massive propagators 
is the simplest Feynman integral involving elliptic curves, and has been extensively 
studied both in the equal-mass case and with different internal 
masses~\cite{Broadhurst:1993mw,Muller-Stach:2011qkg,Adams:2013nia,
Adams:2014vja,Bloch:2013tra,Adams:2015gva,Adams:2015ydq,Bloch:2016izu,
Kalmykov:2016lxx,Ablinger:2017bjx,Broedel:2018qkq,Bogner:2019lfa}; furthermore, 
sunrise diagrams with massive propagators at higher loops provide early examples of 
integrals involving higher-dimensional varieties, notably Calabi-Yau manifolds~\cite{Bourjaily:2018yfy,
Broedel:2019kmn,Broedel:2021zij,Bonisch:2021yfw,Bourjaily:2022bwx}.
In this section, we will present the implementation of a reduction algorithm for banana graphs up to four loops (for equal internal masses) and two loops (for unequal internal masses). The simple algorithm consists of the generation of a large number of integration by parts relations in a format that is readable by the software \texttt{Kira} \cite{Maierhofer:2017gsa,Klappert:2020nbg,Lange:2025fba} that then provides the solution to the system. More refined possibilities for generating minimal systems (similar to what was considered in momentum space in \cite{Wu:2025aeg}) will be discussed later in the chapter.\\

As for the one-loop examples, we start the study of multi-loop banana graphs by presenting a version of Equation \eqref{eq:dw} that is specific to the case we are considering. Let us consider the equal mass case: at $l$-loop, banana graphs are characterised by the Symanzik polynomials
\begin{eqnarray}
\mathcal{U}_l \,& = &\, \sum_{i = 1}^{l+1} z_1 \ldots \hat{z_i} \ldots z_{l+1} \, , \label{eq:banfirstsym}\\
\mathcal{F}_l \,& = &\, \prod_{i = 1}^{l+1} {z_i} \frac{p^2}{\mu^2} - \frac{m^2}{\mu^2} \mathcal{U}  \, \sum_{j = 1}^{l+1} {z_j},
\label{eq:bansecsym}
\end{eqnarray}
where $\hat{z_i}$ is excluded from the product. Given these two graph polynomials, we can define the family of parametric integrals associated to $l$-loop equal mass banana graphs as
\beq
\mathcal{I}\left(\nu_1,...,\nu_{l+1},\lambda\right) = \int_{S_{l}} \, \eta_{l} \left( \prod_{j = 1}^{l+1} z_j^{\nu_j -1}\right) \,
  \frac{\mathcal{U}^{\lambda}}{\mathcal{F}^{\frac{\nu + \lambda l}{l+1}}} \, 	,
\label{eq:eqmassbanana}
\eeq
where $\nu$ is as usual defined as $\nu \equiv \sum_{i = 1}^n \nu_i$. To write Equation \eqref{eq:dw} in a symbolic way that is valid for any number of loops, we introduce raising and lowering index operators as
\begin{eqnarray}
\label{eq:indexoperators}
i^{+}\mathcal{I}\left(\nu_1,...,\nu_{l+1},\lambda\right) = \mathcal{I}\left(\nu_1,...,\nu_i+1,...\nu_{l+1},\lambda\right)   \, , \\
i^{-}\mathcal{I}\left(\nu_1,...,\nu_{l+1},\lambda\right) = \mathcal{I}\left(\nu_1,...,\nu_i-1,...\nu_{l+1},\lambda\right)\, .
\end{eqnarray}
Such index operators must not be merely understood as raising (lowering) the exponent of the corresponding Feynman parameter by one: they also keep the projective nature of the integral intact by adjusting the degree of the denominator in the integral \eqref{eq:eqmassbanana}. Similarly, we can define raising and lowering operators for the exponent of the first Symanzik polynomial
\begin{eqnarray}
\label{eq:uindexoperators}
\mathcal{U}^{+}\mathcal{I}\left(\nu_1,...,\nu_{l+1},\lambda\right) = \mathcal{I}\left(\nu_1,...,\nu_{l+1},\lambda+1\right)   \, , \\
\mathcal{U}^{-}\mathcal{I}\left(\nu_1,...,\nu_{l+1},\lambda\right) = \mathcal{I}\left(\nu_1,...,\nu_{l+1},\lambda-1\right)\, .
\end{eqnarray}
As a concrete example of applications of such definitions, we can rewrite equations \eqref{eq:Uconsistency} and \eqref{eq:Fconsistency} for the specific case of equal mass banana integrals as
\beq
\label{eq:Uconsistency_banana}
 \left( \left(\sum_{i_1<...<i_{l}}\prod_{j=1}^{l}i_j^+\right) - \mathcal{U}^+ \right)\mathcal{I}(\nu_1,...,\nu_{l+1},\lambda) = 0 \, 
\eeq
and
\beq
\label{eq:Fconsistency_banana}
\left(\frac{p^2}{m^2}\prod_{j=1}^{l+1}i_j^+ - \left(\sum_{j=1}^{l+1}j^+\right)\mathcal{U}^+ -1 \right)\mathcal{I}(\nu_1,...,\nu_{l+1},\lambda) = 0 \, ,
\eeq
which already provides some symbolic relations among the parametric integrals of this family. To write the symbolic IBP identity for equal-mass banana integrals, we introduce a further notation to include also the case of boundary terms not integrating to zero. Each time one of the indices in the notation \eqref{eq:eqmassbanana} is equal to zero, the integral is to be meant as a boundary term integrated over the sub-simplex, as in
\beq
\mathcal{I}\left(0,\nu_2,...,\nu_{l+1},\lambda\right) = \int_{S_{l-1}} \, \eta_{l-1} \left.\left( \prod_{j = 1}^{l+1} z_j^{\nu_j -1} \,
  \frac{\mathcal{U}^{\lambda}}{\mathcal{F}^{\frac{\nu + \lambda l}{l+1}}}\right)\right|_{z_1=0} \, .
\eeq
Equipped with the index raising and lowering operator and this notation, it is sufficient to consider the same choice of polynomial $H_i$ as in Equation \eqref{eq:chooseHi} to write the desired symbolic IBP relation 
\begin{align}
\label{eq:All_loop_bananaIBP}
&\left( \frac{l+1}{(\nu + l \lambda_\mathcal{U}-1)}\left(-\delta_{\nu_i,1}i^- + (\nu_i-1)i^-+\lambda_\mathcal{U}\mathcal{U}^-\left(\sum_{i_1<...<i_{l-1},i_j \neq i}\prod_{j=1}^{l-1}i_j^+\right)\right)+ \right. \nonumber \\ &-\left.\left(-\mathcal{U}^++\frac{p^2}{m^2} \prod_{i_j\neq i}i_j^+-\left(\sum_{j=1}^{l+1}j^+\right)\left(\sum_{i_1<...<i_{l-1},i_j \neq i}\prod_{j=1}^{l-1}i_j^+\right)\right)\right)\mathcal{I}(\nu_1,...,\nu_{l+1},\lambda) = 0 \, .
\end{align}
As it is written, Equation \eqref{eq:All_loop_bananaIBP} still involves heavy and abstract notation and it is not immediate to read the kind of linear relations that are generated and their usage. In the following paragraphs, we will discuss the generation of systems of linear relations necessary to write a differential equation for the basis of master integrals for banana diagrams in $d = 2-2\epsilon$ up to $l = 4$. Note that in all the notation in this section, the notion of space-time dimension has been completely absorbed in the definition of the exponent $\lambda$.  As
\beq
\lambda = \nu - \frac{(l+1)d}{2}
\eeq
to work in $2-2\epsilon$ dimensions for such integrals means that for the banana integral with no raised power of propagators is $\mathcal{I}(1,...,1,(l+1)\epsilon)$. 

\subsubsection{Two-loop equal-mass banana integral}
\label{ssec:2Lbanana}
Consider the two-loop banana integral family, also known as sunrise integrals. The all-loop Equation \eqref{eq:All_loop_bananaIBP} for $l = 2$ and $i = 1$ becomes
\begin{align}
\label{eq:2l_IBP_banana}
&\left( \frac{3}{(\nu + 2 \lambda_\mathcal{U}-1)}\left(-\delta_{\nu_1,1}1^- + (\nu_1-1)1^-+\lambda_\mathcal{U}\mathcal{U}^-\left(2^++3^+\right)\right)+ \right. \nonumber \\ &-\left.\left(-\mathcal{U}^++\frac{p^2}{m^2} 2^+3^+-\left(1^++2^++3^+\right)\left(2^++3^+\right)\right)\right)\mathcal{I}(\nu_1,\nu_2,\nu_{3},\lambda) = 0 \, .
\end{align}
Given that the internal masses are all equal, there is a complete permutation symmetry between the three Feynman parameters: it is not necessary to consider the equation for the other possible values of $i$. Equations \eqref{eq:Uconsistency_banana} and \eqref{eq:Fconsistency_banana} for $ l = 2 $ become
\begin{align}
\label{eq:Other_Equations_sunrise_U}
& \left( \left(1^+2^++2^+3^++1^+3^+\right) - \mathcal{U}^+ \right)\mathcal{I}(\nu_1,\nu_2,\nu_{3},\lambda) = 0 \\
\label{eq:Other_Equations_sunrise_F}
&\left(\frac{p^2}{m^2}(1^+2^+3^+) -  \left(1^++2^++3^+\right)\mathcal{U}^+ -1 \right)\mathcal{I}(\nu_1,\nu_2,\nu_{3},\lambda) = 0 
\end{align}
As for the one-loop examples, the problem we consider is not how to choose the master integrals: rather, we want to prove that using a system made of the linear relations introduced above it is possible to write a differential equation system whose solution represents the result of the integration of the sunrise integrals without raised propagator powers. Let us then introduce a basis of three master integrals as 
\beq
\label{eq:basissunrise}
  \Big\{ \mathcal{I}(1,1, 1; 3 \epsilon),  \mathcal{I}(2, 1, 1; 1 + 3 \epsilon),
   \mathcal{I}(0,2, 2; 1 + 3 \epsilon) \Big\} \,.
\eeq
Knowing that the cardinality of the master integrals basis is three and that one of the elements is a tadpole -- the boundary integral $\mathcal{I}(0,2, 2; 1 + 3 \epsilon) $-- there is a logic behind the choice of the element of the basis $\mathcal{I}(2, 1, 1; 1 + 3 \epsilon)$. When considering the derivative of $\mathcal{I}(1, 1, 1; 3 \epsilon)$ with respect to $r = p^2/m^2$ we find
\beq
\label{eq:d_sunrise}
\partial_r \mathcal{I}(1, 1, 1; 3 \epsilon) = -(1+2\epsilon) \mathcal{I}(2, 2, 2; 3 \epsilon)
\eeq
and the integral on the right-hand side is related to the basis by the application of the equation 
\eqref{eq:Other_Equations_sunrise_F} to $\mathcal{I}(1, 1, 1; 3 \epsilon)$:
\beq
\label{eq:FI111}
r \mathcal{I}(2, 2, 2; 3 \epsilon) - 3 \mathcal{I}(2, 1, 1;1 + 3 \epsilon) = \mathcal{I}(1, 1, 1; 3 \epsilon)\, .
\eeq
We can now freely choose any of the two integrals in the left hand side of \eqref{eq:FI111} to be the other element of the basis: we choose $\mathcal{I}(2, 1, 1;1 + 3 \epsilon)$ as the numerator of its derivative concerning $r$ has lower exponents of the Feynman parameters than the derivative of $\mathcal{I}(2, 2, 2; 3 \epsilon)$.\\

To write the differential equation system for the master integral basis \eqref{eq:basissunrise} we need two elements: the integral $\mathcal{I}(3, 2, 2;1 + 3 \epsilon) = \partial_r \mathcal{I}(2, 1, 1;1 + 3 \epsilon)$ expressed as a linear combination of master integrals and the value of the boundary integral $\mathcal{I}(0,2, 2; 1 + 3 \epsilon) $. By using Cheng-Wu theorem \ref{th:ChengWu} we can set $z_3 = 1$ in the boundary integral to obtain 

\beq
\label{eq:boundary_sunrise}
\mathcal{I}(0, 2, 2; 1 + 3 \epsilon)  = \int_{0}^{+\infty} d z_2 \frac{z_2^{3\epsilon}(-z_2(1+z_2))^{-2\epsilon}}{(1+z_2)^2} = \frac{(-1)^{-2\epsilon} (\Gamma(1+\epsilon))^2}{\Gamma(2+2\epsilon)} \, .
\eeq

The integral  $\mathcal{I}(3, 2, 2;1 + 3 \epsilon) $ can be computed as a linear combination of master integrals by generating enough parameter space linear relations and solving the system via any available software. We decided to generate the equations in a format that can be read by \texttt{Kira} via a simple algorithm (we did not optimize the algorithm at this stage) generating a large number of relations using all possible combinations of parameters up to a certain range. The number of equations we generated is 303, of which only 12 are used by \texttt{Kira} to solve the system -- showing that the optimization potential is quite large. The master integral reduction of $\mathcal{I}(3, 2, 2;1 + 3 \epsilon) $  gives the differential equation system
\begin{eqnarray}
\label{eq:sunrise_system}
  \partial_r \textbf{b} \, \equiv \, \partial_r \left(
  \begin{array}{c}
  \mathcal{I}(1, 1, 1; 3 \epsilon) \\
  \mathcal{I}(2, 1, 1; 1 + 3 \epsilon) 
  \end{array}
  \right) & = & \left(
  \begin{array}{c c}
  \frac{2 \epsilon-1}{r} &  \frac{3 (2 \epsilon-1)}{r} \\
  (3 \epsilon +1) \left(-\frac{1}{4 (r-1)}+\frac{1}{3 r}-\frac{1}{12 (r-9)}\right)
   & \frac{-2 \epsilon -1}{r-9}+\frac{-2 \epsilon -1}{r-1}+\frac{3
   \epsilon +1}{r}
  \end{array}
  \right)
  \textbf{b} \, + \nonumber \\ & + &
  \left(
  \begin{array}{c}
  0\\
  \frac{2 (-1)^{-2 \epsilon} \Gamma (\epsilon+1)^2}{(r-9) (r-1) \Gamma (2
   \epsilon+2)}
   \end{array}
   \right)
\end{eqnarray}
where the non-homogeneous part comes from the integration of the boundary term. Similarly to the one-loop box integral, this system is not reduced to an $\epsilon$ form; still, we find again a correspondence between the singular points of the differential equation system and the value of $p^2$ necessary to put on shell one ($r=1$) and three ($r=9$) internal particles. These values correspond respectively to the pseudo-threshold and threshold defined in \cite{Bogner:2019lfa} for the unequal mass case. An attempt to put the differential equation system in canonical form would produce symbol letters corresponding to elliptic integration kernels, as it is well known \cite{Broadhurst:1993mw,Laporta:2004rb,Muller-Stach:2011qkg} that this two-loop diagram cannot be expressed as a linear combination of Goncharov polylogarithms (more on this kind of functions can be found in Appendix \ref{app:SymbolsAndGoncharovPolylogarithms}). In the literature the sunrise integral is often presented as the solution to a non-homogeneous second-order differential equation; using the Gaussian method to transform first-order differential equations systems to higher order differential equations, from the system \eqref{eq:sunrise_system} we obtain 
\begin{eqnarray}
 b_1''(r) 
& + &\left(\frac{1-5
   \epsilon }{r}+\frac{2 \epsilon +1}{r-9}+\frac{2 \epsilon +1}{r-1}\right) b_1'(r) + \nonumber \\ &
   + & \left(\frac{-10 \epsilon ^2-\epsilon +3}{36 (r-9)}+\frac{2 \epsilon ^2-3 \epsilon +1}{4 (r-1)}-\frac{(\epsilon -3) (2 \epsilon -1)}{9 r}\right)b_1(r)
   = \frac{6 (-1)^{-2 \epsilon } (2 \epsilon -1) \Gamma(\epsilon +1)^2}{(r-9)
   (r-1) r \Gamma (2 \epsilon +2)}\, , \nonumber \\
   \label{eq:sunrise_diff_eq}
\end{eqnarray}
corresponding to the results available in the literature \cite{Broadhurst:1993mw,Laporta:2004rb,Muller-Stach:2011qkg}. In particular, rewriting the arbitrary parameter $\epsilon$ as $1-\frac{d}{2}$ gives the differential equation for the $d$-dimensional sunrise integral \cite{Laporta:2004rb}. The connection of the Picard-Fuchs differential equation \eqref{eq:sunrise_diff_eq} to elliptic curves can be understood by viewing the Feynman
integral as a period of a variation of a mixed Hodge structure, as described in \cite{Muller-Stach:2011qkg}.
\subsubsection{Equal-mass banana integrals at higher loops}
\label{ssec:3Lbanana}
As we have previously mentioned, banana diagrams with massive propagators at higher loops are a topic of high interest in the study of Feynman integrals as they provide examples of 
integrals involving higher-dimensional varieties, notably Calabi-Yau manifolds~\cite{Bourjaily:2018yfy,
Broedel:2019kmn,Broedel:2021zij,Bonisch:2021yfw,Bourjaily:2022bwx}. As mentioned in \cite{Muller-Stach:2011qkg}, once a differential equation for the equal mass case is found, the rank of the mixed Hodge structure connected to the integral is independent of them and so the differential equation should be of the same complexity in the unequal mass case compared to the equal mass case. It is therefore of high interest to study the equal mass limit of multi-loop banana integrals to begin the understanding of the more complicated general case.
Our derivation of a differential equation system for equal-mass banana integrals at higher loops proceeds in the same way as the two-loop case just described. In this section, we will report the symbolic identities generating the system of linear relations and the result of the master integral reduction for the differential equation system. 

\paragraph{Three-loop banana integral} For $l = 3$, equations \eqref{eq:Uconsistency} and \eqref{eq:Fconsistency} become
\begin{align}
\label{eq:3l_U}
& \left( \left(1^+2^+3^++1^+3^+4^++1^+2^+4^++2^+3^+4^+\right) - \mathcal{U}^+ \right)\mathcal{I}(\nu_1,\nu_2,\nu_{3},\nu_{4},\lambda) = 0 \\
\label{eq:3l_F}
&\left(\frac{p^2}{m^2}(1^+2^+3^+4^+) -  \left(1^++2^++3^++4^+\right)\mathcal{U}^+ -1 \right)\mathcal{I}(\nu_1,\nu_2,\nu_{3},\nu_{4},\lambda) = 0 \, .
\end{align}
As for the 2-loop example, we can use the symmetry under any permutation of the parameters to conclude that the only necessary integration by parts relation is the one for $h = 1$
\begin{align}
\label{eq:3l_IBP}
&\left( \frac{4}{(\nu + 3 \lambda_\mathcal{U}-1)}\left(-\delta_{\nu_1,1}1^- + (\nu_1-1)1^-+\lambda_\mathcal{U}\mathcal{U}^-\left(2^+3^++3^+4^++2^+4^+\right)\right)+ \right. \nonumber \\ &-\left.\left(-\mathcal{U}^++\frac{p^2}{m^2} 2^+3^+4^+-\left(1^++2^++3^++4^+\right)\left(2^+3^++3^+4^++2^+4^+\right)\right)\right)\mathcal{I}(\nu_1,\nu_2,\nu_{3},\nu_{4},\lambda) = 0 \, ,
\end{align}
where we defined integrals with one index equal to zero as boundary integrals as before. The basis of the master integrals we use is made of four integrals, one of which is a boundary
\beq
  \Big\{ \mathcal{I}(1,1,1,1; 4 \epsilon),  \mathcal{I}(2, 1, 1, 1; 1 + 4 \epsilon),
  \mathcal{I}(2, 2, 1, 1; 2 + 4 \epsilon),
   \mathcal{I}(0,3, 3, 3; 1 + 4 \epsilon) \Big\} \,.
   \label{eq:basis3lbanana	}
\eeq
The reader familiar with the literature on this kind of integrals can recognize in the non-boundary elements of the basis the integrals with zero, one, and two double propagators respectively as in momentum space-based examples. We remark that at this stage the choice of the basis follows the available literature and not an optimization criterion. The boundary integral evaluates to 
\beq
\label{eq:boundary_banana}
\mathcal{I}(0, 3, 3, 3; 1 + 4 \epsilon)  = \frac{(-1)^{1-3 \epsilon } \Gamma (\epsilon +1)^3}{\Gamma (3 \epsilon +3)} .
\eeq
By using the same trivial algorithm as before to generate IBP relations we generate 969 relations to reduce the derivatives of the master integrals to linear combinations of such master integrals. The software \texttt{Kira} uses only 14 of them to produce the following differential equation system\\
\footnotesize{
\begin{eqnarray}
\label{eq:banana3l_system}
  & \partial_r \textbf{b} & \equiv  \partial_r \left(
  \begin{array}{c}
  \mathcal{I}(1, 1, 1, 1; 4 \epsilon) \\
  \mathcal{I}(2, 1, 1, 1; 1 + 4 \epsilon)\\
  \mathcal{I}(2, 2, 1, 1; 2 + 4 \epsilon) 
  \end{array}
  \right) =  \nonumber \\  & = & \left(
  \begin{array}{c c c}
  -\frac{3 \epsilon +1}{r} & -\frac{4 (3 \epsilon +1)}{r} & 0 \\
  \frac{-4 \epsilon -1}{4 (r-4)}+\frac{4 \epsilon +1}{4 r} & \frac{4 \epsilon +1}{r}-\frac{2 (3 \epsilon +1)}{r-4} & \frac{9 \epsilon +6}{4-r} \\
\frac{(2 \epsilon +1) (4 \epsilon +1)}{3\epsilon + 2}  \left( -\frac{1}{32 (r-16)}+\frac{1}{8 (r-4)}-\frac{3 }{32 r } \right) & \frac{(2 \epsilon +1) (3 \epsilon +1)}{3\epsilon + 2} \left(-\frac{3 }{8 (r-16) }+\frac{1}{(r-4)}-\frac{5 }{8 r} \right) & \frac{-\epsilon -1}{r}-\frac{3 (2 \epsilon +1)}{2 (r-16)}+\frac{3
   (2 \epsilon +1)}{2 (r-4)}
   \end{array}
  \right)
  \textbf{b} \, + \nonumber \\ 
  & +  & \left(
  \begin{array}{c }
  0\\
  0\\
  -\frac{2 (-1)^{-3 \epsilon } \Gamma (\epsilon +1)^3}{(r-16) r \Gamma (3 \epsilon +3)}
   \end{array}
   \right)
\end{eqnarray}
}
\normalsize
We observe once again the presence of a pseudo-threshold singularity ($r=4$) and a threshold singularity ($r=16$) when all internal legs are put on shell. The system \eqref{eq:banana3l_system} can be rewritten as a higher-order differential equation using Gaussian elimination and we find
\begin{eqnarray}
&&b_1(r)^{(3)} + 
\left( \frac{3 (2 \epsilon +1)}{2 (r-16)}+\frac{3 (2 \epsilon +1)}{2 (r-4)}+\frac{3}{r} \right) b_1(r)^{(2)} + \nonumber \\ & + &
\left( \frac{1-\epsilon ^2}{r^2}-\frac{3 \left(2 \epsilon ^2+5 \epsilon +2\right)}{8 r}+\frac{6 \epsilon ^2+7 \epsilon +2}{8 (r-16)}+\frac{2 \epsilon +1}{2 (r-4)} \right) b_1(r)^{(1)} + \nonumber \\ & + & 
\left((2 \epsilon +1) \left(\frac{3 \epsilon ^2-5 \epsilon -2}{32 r^2}+\frac{-3 \epsilon ^2-\epsilon }{32 (r-4)}+\frac{9 \epsilon ^2+9 \epsilon +2}{512 (r-16)}+\frac{39
   \epsilon ^2+7 \epsilon -2}{512 r}\right) \right) b_1(r) 
+ \nonumber \\ & + &\frac{24 (-1)^{-3 \epsilon } \left(9 \epsilon ^2+9 \epsilon +2\right) \Gamma (\epsilon +1)^3}{(r-16) (r-4) r^2 \Gamma (3 \epsilon +3)}=0
\end{eqnarray}
corresponding to an irreducible third-order Picard Fuchs differential equation (derived for $d = 2$ and $x = 4/r$) reported in \cite{Broedel:2019kmn}.  While in general for distinct internal
masses in dimensional regularization, the $L$-loop banana diagram involves integrals over a
family of $(L−1)$-dimensional Calabi-Yau manifolds, the $3$-loop equal mass banana integral can be evaluated in terms of elliptic polylogarithms and iterated integrals of modular forms \cite{Bourjaily:2022bwx}.\\

\paragraph{Four-loop banana integral} The very same steps can be repeated for a higher number of loops. To test the growth of the number of equations generated by our naive algorithm, we conclude this section with the derivation of the differential equation system for $l = 4$. In that case, the relations generated by the definitions of the Symanzik polynomials are 
\begin{align}
\label{eq:4l_U}
& \left( \left(2^+3^+4^+ + 2^+3^+5^+ + 2^+4^+5^+ + 3^+4^+5^+\right) - \mathcal{U}^+ \right)\mathcal{I}(\nu_1,\nu_2,\nu_{3},\nu_{4},\nu_5,\lambda) = 0 \\
\label{eq:4l_F}
&\left(\frac{p^2}{m^2}(1^+2^+3^+4^+5^+) -  \left(1^++2^++3^++4^++5^+\right)\mathcal{U}^+ -1 \right)\mathcal{I}(\nu_1,\nu_2,\nu_{3},\nu_{4},\nu_5,\lambda) = 0 
\end{align}
while the integration by parts relation becomes
\begin{align}
\label{eq:4l_IBP}
&\left( \frac{5}{(\nu + 4 \lambda_\mathcal{U}-1)}\left(-\delta_{\nu_1,1}1^- + (\nu_1-1)1^-+\lambda_\mathcal{U}\mathcal{U}^-
\left(2^+3^+4^+ + 2^+3^+5^+ + 2^+4^+5^+ + 3^+4^+5^+\right)\right)
+ \right. \nonumber \\ &-\left.\left(-\mathcal{U}^++\frac{p^2}{m^2} 2^+3^+4^+5^+-\left(1^++2^++3^++4^++5^+\right)
\left(2^+3^+4^+ + 2^+3^+5^+ + 2^+4^+5^+ + 3^+4^+5^+\right)
\right)\right) \nonumber \\&\mathcal{I}(\nu_1,\nu_2,\nu_{3},\nu_{4},\nu_5,\lambda) = 0 \, .
\end{align}
This time, we proceed differently with the choice of the basis: we choose one boundary integral, then the 4-loop banana integral with no raised propagators, and finally three integrals that are related to each other by differentiation with respect to $r$:
\begin{eqnarray}
\label{eq:basis4lbanana	}
  \Big\{ \mathcal{I}(1,1,1,1,1; 5 \epsilon),  
  &\mathcal{I}(2, 1, 1, 1,1; 1 + 5 \epsilon),&
  \mathcal{I}(3, 2, 2, 2, 2; 1 + 5 \epsilon),\nonumber \\
  &\mathcal{I}(4, 3, 3, 3, 3; 1 + 5 \epsilon)&,
  \mathcal{I}(0,2, 1, 1, 1; 5 \epsilon) \Big\} \,.
\end{eqnarray}
In order to write a differential equation system it is sufficient to reduce the derivatives of the first and the fourth element of such a basis into a linear combination of master integrals. The boundary integral can be immediately computed via the Cheng-Wu theorem as 
\beq
\label{eq:boundary4l}
\mathcal{I}(0,2, 1, 1, 1; 5 \epsilon) = -\frac{(-1)^{-4 \epsilon } \epsilon  \Gamma (\epsilon )^4}{\Gamma (4 \epsilon +1)}
\eeq
while the derivatives of the first and fourth elements of the basis are respectively
\beq
\label{eq:drI11111}
\partial_r \mathcal{I}(1,1,1,1,1,5 \epsilon) = -\frac{(4 \epsilon +1) (\mathcal{I}(1,1,1,1,1,5 \epsilon)+5 \mathcal{I}(2,1,1,1,1,1+5 \epsilon))}{r}
\eeq
and 
\small
\begin{eqnarray}
\label{eq:drI43333}
&&\partial_r \mathcal{I}(4, 3, 3, 3, 3; 1 + 5 \epsilon)  = \nonumber \\ && - \frac{(5 \epsilon +1) \left(r^2+(r-45) (3 r-10) \epsilon -57 r+180\right)}{2 (r-25) (r-9) (r-1) r^3 (4 \epsilon +3)} \mathcal{I}(1,1,1,1,1,5 \epsilon)
+ \nonumber \\ 
&& -  \frac{(3 \epsilon +1) r^3(\epsilon +1)+28 r^2 \epsilon (3 \epsilon +1)-5 (145 \epsilon +57) r(5 \epsilon +1)+450 (5 \epsilon +1) (5 \epsilon +2)}{2 (r-25) (r-9) (r-1)
   r^3(4 \epsilon +3)} \nonumber \\ && \phantom{+}\mathcal{I}(2,1,1,1,1,1+5 \epsilon) + \nonumber \\ &&
+\left(\frac{25 \epsilon ^2+15 \epsilon +2}{r^2 (4 \epsilon +3)}+\frac{-88 \epsilon ^2-34 \epsilon +5}{72 (r-9) (4 \epsilon +3)}+\frac{-62 \epsilon ^2-39 \epsilon -4}{3 (r-1)
   (4 \epsilon +3)}+\nonumber \right. \\ && \left. +\frac{424 \epsilon ^2+486 \epsilon +137}{600 (r-25) (4 \epsilon +3)}+\frac{4766 \epsilon ^2+2849 \epsilon +233}{225 r (4 \epsilon +3)}\right) \mathcal{I}(3,2,2,2,2,1+5 \epsilon) +\nonumber \\ &&
+ \left(\frac{6 \epsilon }{r}-\frac{2 (2 \epsilon +1)}{r-25}-\frac{2 (2 \epsilon +1)}{r-9}-\frac{2 (2 \epsilon +1)}{r-1}\right)  \mathcal{I}(4,3,3,3,3,1+5 \epsilon) + \nonumber \\ &&
+\frac{24 (-1)^{-4 \epsilon
   } \Gamma (\epsilon +1)^4}{(r-25) (r-9) (r-1) r \Gamma (4 \epsilon +4)} \, .
\end{eqnarray}
\normalsize
The chosen basis makes it such that all the complexity of the linear system resides in this last entry. The result in Equation \eqref{eq:drI43333} is presented such that the complicated coefficients of each master integral are gathered together or kept separated depending on which expression is more compact. The important observation here is again that the thresholds and pseudo-thresholds appear once again as singularities of the differential equation system. The number of equations generated by our naive seeding algorithm is 3750, of which only 113 are used by \texttt{Kira} to complete the reduction. After a short report of the performances of our algorithm for the unequal mass sunrise integral, we will focus in the last part of this chapter on a promising way to reduce the number of equations that are generated, based on the algebra of polynomial ideals. We will illustrate our algorithm for the 2- and 3-loop banana integral: we postpone the analysis of the (lengthy but straightforward) fourth-order differential equation for the 4-loop banana integral to further work illustrating the optimization of the algorithm for the case $l=4$ as well.

\subsubsection{Sunrise integral with three different masses}
\label{subsub:sunrise3m}
We conclude this section of the chapter by studying the differential equation system for the 2-loop sunrise integral with three different masses in $2-2\epsilon$ dimensions. When working in dimensional regularization, the system is known to generate a fourth-order differential equation (see for example Ref. \cite{Caffo:1998du} for the origninal calculation and Ref. \cite{delaCruz:2024xit} for a interesting perspective on its connection to the integer dimension case), that for $d = 2$ can be reduced to a second order differential equation \cite{Muller-Stach:2011qkg,Adams:2014vja,
Bogner:2019lfa} either by inverting two differential operators or by considering cohomology groups of elliptic curves, as originally done. In this section, we aim at finding a non-homogeneous linear differential equation system equivalent to the one in \cite{Caffo:1998du} and to the fourth-order differential equation described in \cite{delaCruz:2024xit}. In doing so, we want to show a usage of the parameter-space IBP relations similar to the momentum space one.\\

As usual, the two symbolic relations coming from the definitions of the two Symanzik polynomials are easier to write. For the case in analysis, we set $p^2 = s$ to get
\begin{align}
\label{eq:sunrise3m_U}
& \left( \left(1^+2^++2^+3^++1^+3^+\right) - \mathcal{U}^+ \right)\mathcal{I}(\nu_1,\nu_2,\nu_{3},\lambda) = 0 \\
\label{eq:sunrise3m_F}
&\left(s(1^+2^+3^+) -  \left(m_1^21^++m_2^22^++m_3^23^+\right)\mathcal{U}^+ -1 \right)\mathcal{I}(\nu_1,\nu_2,\nu_{3},\lambda) = 0 
\end{align}
where we notice that the first one is the same as in the equal mass case, given that the Symanzik polynomial is unchanged. Inspired by the equal mass case -- and by the momentum space calculations -- we select a basis of master integrals made of 3 boundary integrals and four sunrise ones
\beq
\label{eq:basissunrise3m}
  \Big\{ \mathcal{I}(1,1, 1; 3 \epsilon),\,
  \mathcal{I}(2, 1, 1; 1 + 3 \epsilon),\,
  \mathcal{I}(1, 2, 1; 1 + 3 \epsilon),\,
  \mathcal{I}(1, 1, 2; 1 + 3 \epsilon)\Big\} \,,
\eeq
that are all different given the permutation symmetry of the first three indices is no longer valid. We can use \eqref{eq:sunrise3m_F} to immediately express the integral $\mathcal{I}(2,2, 2; 3 \epsilon)$ -- proportional to the derivative with respect to $s$ of $\mathcal{I}(1,1, 1; 3 \epsilon)$ -- as linear combination of integrals in the basis \eqref{eq:basissunrise3m}:
\beq
s \mathcal{I}(2,2, 2; 3 \epsilon)  = m_1^2\mathcal{I}(2,1, 1;1+ 3 \epsilon)+m_2^2\mathcal{I}(1,2, 1; 1+3 \epsilon)+m_3^2\mathcal{I}(1,1, 2;1+ 3 \epsilon)+ \mathcal{I}(1,1, 1; 3 \epsilon)\,.
\eeq
The system of differential equations can be constructed by expressing the integral $\mathcal{I}(3,2,2;1+3 \epsilon)$ and the other two permutations of the Feynman parameter exponents $\mathcal{I}(2,3,2;1+3 \epsilon)$ and $\mathcal{I}(2,2,3;1+3 \epsilon)$ as linear combinations of master integrals. This is achieved by generating a large number of linear relations (1590) and using the software \texttt{Kira} to find the master integral reduction, and we retrieve the system presented in \cite{Caffo:1998du}. At this stage the calculation is simply a consistency check of our IBP identities; however, this example together with the explicit form of the differential equation system will be included in the upcoming publication on the application of ideal algebra to IBP reduction in parameter space. Another interesting perspective we will mention in the last section \ref{sec:Conclusions1} is to use projective space blow-ups described in \cite{Regge:1968rhi} to potentially generate new relations simplifying the fourth-order differential equation to the second-order one for $d=2$.
\section{Algebraic ideals for improving reduction algorithms}
\label{sec:Ideals}
\subsection{Polynomial ideals and Gröbner bases}
\label{subsec:Groebner}

In this section, we review the basic theory of polynomial ideals and Gröbner bases, focusing on those aspects relevant for their applications to Feynman integrals and parametric representations. Despite we do not need a complete understanding of the rich literature on the topic -- especially as for practical applications the most important algorithms are already implemented in software as \texttt{Singular} \cite{DGPS} and \texttt{Macaulay2} \cite{M2} -- we find it important to have a general overview of the mathematical concepts foundational to the reduction algorithm described in the next section. We begin with a general overview of polynomial rings and ideals, introduce the concept of a monomial ordering, and then define Gröbner bases along with their principal properties. Standard references for this material include \cite{BeWe93,Sturmfels:1996gro,Adams:1994gro,Cox:2015ode,}.\\

\begin{definition}
\label{def:Ideal}
Let $K$ be a field, which in our applications will typically be the field of rational functions in the Mandelstam variables, and let $R = K[x_1, \dots, x_n]$ be the ring of polynomials in $n$ variables with coefficients in $K$. A subset $I \subset R$ is called an \textit{ideal} if it satisfies: 
\begin{itemize}
\item[(i)]  $0 \in I$;
\item[(ii)] if $f, g \in I$, then $f + g \in I$; 
\item[(iii)] for all $f \in I$ and $r \in R$, the product $rf \in I$.
\end{itemize} 
\end{definition}

An important fact is that every ideal $I \subset R$ is finitely generated, a consequence of Hilbert’s Basis Theorem \cite{Hilbert1890}. That is, there exist polynomials $f_1, \dots, f_m \in R$ such that
\beq
\label{eq:ideal_basis}
  I = \langle f_1, \dots, f_m \rangle \, = \, \left\{ \sum_{i = 1}^m h_i f_i \, : \, h_i \in R \right\} \,.
\eeq
The connection to the Feynman parameter IBP method, explored later in section \ref{sub:BananaIdeal}, essentially consists of being able to determine whether or not the numerator of a target integral belongs to the ideal generated by the derivatives of the second Symanzik polynomial. The choice of generators in \eqref{eq:ideal_basis} is not unique; \textit{Gröbner bases} provide a distinguished generating set for an ideal, allowing for example algorithmic manipulation and solution of systems of polynomial equations. The definition of a Gröbner basis depends crucially on a total ordering of the monomials in $R$, referred to as a \textit{monomial ordering}. Such an ordering $\prec$ is a well-ordering on the set of monomials in $x_1, \dots, x_n$ that is compatible with multiplication: if $u \prec v$, then $uw \prec vw$ for any monomial $w$.\\

Given a monomial ordering, the \textit{leading term} $\mathrm{LT}(f)$ of a non-zero polynomial $f \in R$ is the largest monomial (with respect to $\prec$) appearing in $f$, multiplied by its coefficient. 
\begin{definition}
A finite set of polynomials $\mathcal{G} = \{g_1, \dots, g_t\} \subset I$ is a Gröbner basis of the ideal $I$ if for every non-zero $f \in I$, the leading term $\mathrm{LT}(f)$ is divisible by the leading term of some $g_i$. That is,
\beq
  \forall f \in I \setminus \{0\}, \quad \exists g_i \in \mathcal{G} \, : \, \mathrm{LT}(g_i) \,|\, \mathrm{LT}(f) \,.
\eeq
\end{definition}
Classical textbooks on non-linear algebra such as Ref. \cite{michalek2021invitation} provide an equivalent definition of the set $\Gm$ to then prove it represents a basis of the ideal. Grobner bases are a version
of Gaussian elimination for polynomials of degree $n>1$ and represent a fundamental tool to simplify the study of algebraic ideals. The most well-known tool for computing Gröbner bases is the so-called Buchberger’s
algorithm \cite{Cox:2015ode}. While the complexity of the algorithm can grow rapidly with the number of variables and the degree of the generators, significant improvements and heuristics exist for cases of physical interest, for example for \textit{graded ideals}, ideals where the elements are organized by their degree\footnote{This will be the case for our application, due to the homogeneity of the Symanzik polynomials.}. Although the role of Gröbner bases is not immediately transparent from the IBP reduction performed in the next section, they play a central role in computations with polynomial ideals, including elimination theory and, most importantly for us, ideal membership tests.\\

A Gröbner basis is said to be \textit{reduced} if: (i) the leading coefficient of each $g_i \in \mathcal{G}$ is 1; and (ii) no monomial in any $g_i$ is divisible by $\mathrm{LT}(g_j)$ for $j \neq i$. Every ideal has a unique reduced Gröbner basis for each fixed monomial ordering -- we refer to \cite{Cox:2015ode} for a proof of this statement. In applications to Feynman integrals, including ours, one often considers polynomial ideals to simplify integration-by-parts identities \cite{Gluza:2010ws,Wang:2023nvh,Wu:2025aeg}, and Gröbner bases are used to simplify and systematically manipulate these relations.\\

We conclude the section by mentioning one particularly important and common ordering: the \textit{lexicographic ordering}, defined by
\beq
  x_1^{a_1} x_2^{a_2} \dots x_n^{a_n} \prec_{\mathrm{lex}} x_1^{b_1} x_2^{b_2} \dots x_n^{b_n}
  \quad \Longleftrightarrow \quad \exists\, k \text{ such that } a_1 = b_1, \dots, a_{k-1} = b_{k-1}, \, a_k < b_k \,.
\eeq

This ordering is particularly well-suited for elimination problems, as it allows the computation of elimination ideals that project an algebraic variety onto a lower-dimensional subspace. However, other orderings, such as graded reverse lexicographic (grevlex), are often preferred for computational efficiency \cite{michalek2021invitation}.\\

\subsection{Equal mass banana integrals}
\label{sub:BananaIdeal}
We dedicate this section to showing how the tools coming from ideal algebra just introduced come into play when dealing with IBP relations for Feynman parametrized integrals. Coming back to Equation \eqref{eq:dw}, which we report here again for clarity
\beq
  d \omega_{n-2} \, = \, \frac{1}{(P-1) \, \big( D(z) \big)^{P - 1}} \, \, \eta_{\{z\}} \, 
  \sum_{i = 1}^n \frac{\partial H_i(z)}{\partial z_i} \, - \, \frac{\eta_{\{z\}}}{\big( D(z) \big)^P} \, 
  \sum_{i = 1}^n  H_i \, \frac{\partial D(z)}{\partial z_i} \, ,
\eeq
we can aim at setting up a reduction strategy where the sum 
\beq
\label{eq:ideal_sum}
\sum_{i = 1}^n  H_i \, \frac{\partial D(z)}{\partial z_i}
\eeq
can be set to an arbitrary target polynomial. This would result in linear relations where the integrals in the integrated version of \eqref{eq:dw} are either boundary integrals or integrals with denominator power $P-1$, except for one single integral whose nominator is given by the linear combination defined above. If we keep positive powers of the Feynman parameters and the first Symanzik polynomials we can in principle reduce the set of integrals to smaller and smaller sets, giving a preferred direction to the master integral reduction. It should be clear how the problem we are describing corresponds exactly to testing the membership of a target polynomial to the ideal generated by the derivatives of the second Symanzik polynomial -- corresponding to $D(z)$ in \eqref{eq:ideal_sum}.\\

\subsubsection{Two-loop sunrise}
\label{ssub:2lsunrise_ideal}
We illustrate how the idea we introduced above can speed up reduction algorithms by repeating the construction of the differential equation system for the equal-mass two-loop sunrise integral. First, we define the three polynomials generating the ideal by differentiating the polynomial \eqref{eq:bansecsym} for $l = 2$:
\begin{eqnarray}
\label{eq:sunrise_generators}
 \frac{\partial \mathcal{F}}{\partial z_1}& =& z_3 z_2 (r-3)-z_2^2-z_3^2-2 z_1 (z_2+z_3) \nonumber \\
 \frac{\partial \mathcal{F}}{\partial z_2} &=&z_1 (z_3 (r-3)-2 z_2)-z_1^2-z_3 (2 z_2+z_3) \nonumber \\
 \frac{\partial \mathcal{F}}{\partial z_3}& = & z_1 (z_2 (r-3)-2 z_3)-z_1^2-z_2 (z_2+2 z_3) \,.
\end{eqnarray}
As anticipated, the coefficients belong to the ring of rational functions of the Mandelstam variables. If we consider the same master integrals basis \eqref{eq:basissunrise}, we still can reduce the derivative with respect to $r$ of $\mathcal{I}(1,1,1,3\epsilon)$  to the basis via Equation \eqref{eq:FI111}. The only necessary reduction is the one for $\mathcal{I}(3,2,2,1+3\epsilon)$. Such an integral has denominator power $P = 3 + 2\epsilon$: we can reduce it to a linear combination of integrals having $P = 2 + 2\epsilon$ and positive exponents (and boundaries) if the part of the numerator $z_1^2z_2z_3$ belongs to the ideal generated by the polynomials \eqref{eq:sunrise_generators}. We are looking for some polynomials $Q_i$ such that in the sum \eqref{eq:ideal_sum} we can write $H_i = Q_i \mathcal{U}^{1+3\epsilon}$ and we get the correct numerator of $\mathcal{I}(3,2,2,1+3\epsilon)$. We can use the command \textit{lift} of the program \texttt{Singular} to find such polynomials $Q_i$. As an example of the non-trivial choice of polynomial produced by this algorithm, we report here the polynomial $Q_1$ 
%
\begin{eqnarray}
Q_1 &=& (r ((r-8) r+2)+3) z_3^2-(r-1) z_3 \left((r-9) (r-1) r z_1-2 ((r-5) r+3) z_2\right)+ \nonumber \\&&+z_2 \left((r (2 r-3)+3)
   z_2-4 (r-1) r z_1\right)/(3 (r-9) (r-1)^2 r)\,.
\end{eqnarray}
The simple application of the IBP relation \eqref{eq:dw} with the input of the decomposition into the ideal of $z_1^2z_2z_3$ leads to the equation
\beq
\mathcal{I}(3,2,2,1+3\epsilon) = \frac{(2 (r-2) \epsilon +r-3) \mathcal{I}(2,2,2,3 \epsilon )+(r \epsilon +r+\epsilon -3) \mathcal{I}(3,2,1,3 \epsilon )}{(r-9) (r-1)
   (\epsilon +1)} + \text{boundaries}
\eeq
which, together with the Symanzik polynomials relations \eqref{eq:FI111} and 
\beq
2\mathcal{I}(3,2,1,3\epsilon) = \mathcal{I}(2,1,1,1+3\epsilon) - \mathcal{I}(2,2,2,3\epsilon)
\eeq 
gives the correct reduction
\begin{eqnarray}
\mathcal{I}(3,2,2,1+3\epsilon) &=&\frac{\left(r^2 (\epsilon +1)+10 r \epsilon -9 (3 \epsilon +1)\right) \mathcal{I}(2,1,1,3 \epsilon +1)}{2 (r-9) (r-1) r (\epsilon +1)} + \nonumber \\ &+&\frac{(r-3) (3 \epsilon +1)
   \mathcal{I}(1,1,1,3 \epsilon )}{2 (r-9) (r-1) r (\epsilon +1)}
   + \nonumber \\ &+& \text{boundaries}\,.
\end{eqnarray}
We remark that only 3 equations were used to complete the reduction of all the integrals necessary to find the differential equation system for the sunrise integral, way less than the 303 generated by the naive seeding and even less than the ones used by \texttt{Kira} to solve the system. To show the application of this reduction algorithm to another case, we analyse the slightly more complicated example of the three-loop banana integral. 
\subsubsection{Three-loop banana}
\label{ssub:3lbanana_ideal}
As we did for the 2-loop example, we begin the section by defining the generators of the polynomial ideal necessary for performing the reduction. Such generators are
\begin{eqnarray}
\label{eq:banana3l_generators}
 \frac{\partial \mathcal{F}}{\partial z_1}& =&  r z_2 z_4 z_3-z_1 z_2 z_3-z_1 z_4 z_3-z_2 z_4 z_3-z_1 z_2 z_4-\left(z_1+z_2+z_3+z_4\right) \left(z_2 z_3+z_4 z_3+z_2 z_4\right)\nonumber \\
 \frac{\partial \mathcal{F}}{\partial z_2} &=& r z_1 z_4 z_3-z_1 z_2 z_3-z_1 z_4 z_3-z_2 z_4 z_3-z_1 z_2 z_4-\left(z_1+z_2+z_3+z_4\right) \left(z_1 z_3+z_4 z_3+z_1 z_4\right) \nonumber \\
 \frac{\partial \mathcal{F}}{\partial z_3}& = & r z_1 z_2 z_4-z_1 z_2 z_3-z_1 z_4 z_3-z_2 z_4 z_3-z_1 z_2 z_4-\left(z_1+z_2+z_3+z_4\right) \left(z_1 z_2+z_4 z_2+z_1 z_4\right) \nonumber \\
  \frac{\partial \mathcal{F}}{\partial z_4}& = & r z_1 z_2 z_3-z_1 z_2 z_3-z_1 z_4 z_3-z_2 z_4 z_3-z_1 z_2 z_4-\left(z_1 z_2+z_3 z_2+z_1 z_3\right) \left(z_1+z_2+z_3+z_4\right)\,. \nonumber \\ 
\end{eqnarray}
To use algebraic ideal decomposition to generate the differential equation system for the 3-loop equal-mass banana diagram, first consider the integrals of the basis \eqref{eq:basis3lbanana	}, in particular, the non-boundary ones
\beq
  \Big\{ \mathcal{I}(1,1,1,1; 4 \epsilon),  \mathcal{I}(2, 1, 1, 1; 1 + 4 \epsilon),
  \mathcal{I}(2, 2, 1, 1; 2 + 4 \epsilon)\Big\} \,.
\eeq
When differentiating with respect to the kinematic ratio $r = p^2/m^2$, we obtain a set of integrals to reduce via linear relations. Such integrals are 
\beq
  \Big\{ \mathcal{I}(2,2,2,2; 4 \epsilon),  \mathcal{I}(3, 2, 2, 2; 1 + 4 \epsilon),
  \mathcal{I}(3, 3, 2, 2; 2 + 4 \epsilon)\Big\} \,.
\eeq
We can immediately use the consistency relation coming from the definition of $\mathcal{F}$ for such a family of diagrams, that is Equation \eqref{eq:Fconsistency_banana} which applied to the integral $\mathcal{I}(1,1,1,1; 4 \epsilon)$ gives
\beq
  \mathcal{I}(1,1,1,1; 4 \epsilon) = r \mathcal{I}(2,2,2,2; 4 \epsilon) - 4 \mathcal{I}(2,1,1,1; 1 + 4 \epsilon)
\eeq
which already represents the master integral reduction of the integral $\mathcal{I}(2,2,2,2; 4 \epsilon)$. The second integral to reduce in order of power $P$ of the denominator is the integral $ \mathcal{I}(3, 2, 2, 2; 1 + 4 \epsilon)$. Before using integration by parts \eqref{eq:dw} together with ideal decomposition, let us write the consistency relations coming from the Symanzyk polynomials that the considered integral is part of:
\beq
\label{eq:system_consistency_3lbanana}
\begin{cases}
-\mathcal{I}(2,1,1,1,4 \epsilon +2)+2\mathcal{I}(3,3,2,1,4 \epsilon +1)+2 \mathcal{I}(3,2,2,2,4 \epsilon +1)=0\\
-\mathcal{I}(3,1,1,1,4 \epsilon +2)+3 \mathcal{I}(4,2,2,1,4 \epsilon +1)+\mathcal{I}(3,2,2,2,4 \epsilon
   +1)=0\\
   r \mathcal{I}(3,2,2,2,4 \epsilon +1)-\mathcal{I}(2,1,1,1,4 \epsilon +1)-\mathcal{I}(3,1,1,1,4 \epsilon +2)-3 \mathcal{I}(2,2,1,1,4 \epsilon +2)=0
\end{cases}
\eeq
In such a system, there are four integrals that do not belong to the master integral basis and only three equations. If one of such integrals could be reduced via the ideal decomposition strategy we are developing, the reduction of $ \mathcal{I}(3, 2, 2, 2; 1 + 4 \epsilon)$ would be complete. Unfortunately, this is not so immediate as none of the parameter products in the numerators of the four non-master integrals in the system \eqref{eq:system_consistency_3lbanana} belongs to the ideal generated by the derivatives of the second Symanzik polynomial. The reduction can still be performed by considering a linear combination of integrals making a numerator that belongs to the ideal. By examining the Gr\"obner basis of the considered ideal, we consider the reduction of the integral
\beq
\label{eq:new_integral_definition}
\mathcal{J} = (r-2) (-(r-8) \mathcal{I}(3,2,2,2,4 \epsilon +1)+2  \mathcal{I}(4,2,2,1,4 \epsilon +1)+6  \mathcal{I}(3,3,2,1,4 \epsilon +1)) \, .
\eeq
At this point, the reduction procedure for $ \mathcal{I}(3, 2, 2, 2; 1 + 4 \epsilon)$ proceeds in three steps: first, we replace one of the four non-master integrals in \eqref{eq:system_consistency_3lbanana} with the integral $J$ by inverting its definition; then we solve the system using $\mathcal{J}$ and the two appearing master integrals as basis of the vector space of the solution; and finally we reduce the integral $J$ via integration by parts using the ideal decomposition. The first two steps are rather trivial and lead to
\beq
 \mathcal{I}(3,2,2,2,4 \epsilon +1) = \frac{-2 (r-2)  \mathcal{I}(2,1,1,1,4 \epsilon +1)+3 (r-2)  \mathcal{I}(2,2,1,1,4 \epsilon +2)-3 \mathcal{J}}{(r-4) (r-2)}\, .
\eeq
The decomposition in integrals having a lower denominator power of $\mathcal{J}$ starts by finding the polynomials $H_i$ such that the selected integral is the ideal element appearing in Equation \eqref{eq:dw}. Using the command \textit{lift} of the program \texttt{Singular} we find that the four polynomials are 
\begin{eqnarray}
\label{eq:Hi_banana3l}
 H_1 &=& -r z_1 z_2 \mathcal{U}_3^{1+4\epsilon} \nonumber \\ 
 H_2 &=& 2 z_1 z_2 \mathcal{U}_3^{1+4\epsilon} \nonumber \\ 
 H_3 &=& 2 z_1 z_2 \mathcal{U}_3^{1+4\epsilon} \nonumber \\ 
 H_4 &=& -2 z_1 z_2 \mathcal{U}_3^{1+4\epsilon} \, ,
\end{eqnarray}
where $ \mathcal{U}_3$ is the first Symanzik polynomial \eqref{eq:banfirstsym} for the 3-loop banana integral.
Using the parameter IBP relation with these values of $H_i$ and the usual consistency relations generated by the Symanzik polynomials, we can reduce the integral $\mathcal{J}$ to a linear combination of master-integrals -- up to boundary terms which we will analyze below. The result is
\beq\label{eq:Jreduced}
\mathcal{J} = -\frac{(r-2) ((r (8 \epsilon +5)+16 \epsilon +4) \mathcal{I}(2,1,1,1,4 \epsilon +1)+(4 \epsilon +1) \mathcal{I}(1,1,1,1,4 \epsilon ))}{3 r (3 \epsilon +2)} + \text{boundary}
\eeq
The boundary terms are given by Equation \eqref{eq:w} and correspond to taking $z_i = 0 $ for each $H_i$. For this reduction, the only non-zero boundary terms are given by $H_3$ and $H_4$, and they cancel because they carry opposite signs. The boundary term is therefore $0$ and the reduction for $ \mathcal{I}(3, 2, 2, 2; 1 + 4 \epsilon)$ becomes
\begin{eqnarray}
\mathcal{I}(3, 2, 2, 2; 1 + 4 \epsilon) = &-& \frac{(2 (r+8) \epsilon +r+4) \mathcal{I}(2,1,1,1,4 \epsilon +1)}{(r-4) r } + \nonumber \\ 
&-& \frac{3 r (3 \epsilon +2) \mathcal{I}(2,2,1,1,4 \epsilon +2)}{(r-4) r } + \nonumber \\ 
&-& \frac{(4 \epsilon +1) \mathcal{I}(1,1,1,1,4
   \epsilon )}{(r-4) r }
\end{eqnarray}
which is in agreement with the result already written in Equation \eqref{eq:banana3l_system}. Note that the reduction \eqref{eq:Jreduced} implies that all the non-master integrals in the system \eqref{eq:system_consistency_3lbanana} can be expressed as a linear combination of master integrals. \\
The last integral we need to reduce in order to complete the construction of the differential equation system for the 3-loop equal-mass banana integral is $ \mathcal{I}(3, 3, 2, 2; 2 + 4 \epsilon)$. Because of what we mentioned above, it is sufficient to reduce this integral to a linear combination of any of the integrals appearing in \eqref{eq:system_consistency_3lbanana}, to then use their reduction to complete the process. The numerator of the integral $ \mathcal{I}(3, 3, 2, 2; 2 + 4 \epsilon)$ belongs to the ideal generated by the polynomials \eqref{eq:banana3l_generators} and can therefore be reduced immediately to
\begin{eqnarray}
\label{eq:reduction_3322}
 \mathcal{I}(3, 3, 2, 2; 2 + 4 \epsilon) &=& \frac{(4 \epsilon +2) \mathcal{I}(4,2,2,1,4 \epsilon +1)}{(r-16) r (\epsilon +1)}+\frac{2 ((r-4) \epsilon +r-10) \mathcal{I}(3,3,2,1,4 \epsilon +1)}{3 (r-16) r
   (\epsilon +1)}+ \nonumber \\
   &+&\frac{2 (r (5 \epsilon +3)-2 \epsilon -9) \mathcal{I}(3,2,2,2,4 \epsilon +1)}{3 (r-16) r (\epsilon +1)} + \text{boundary} \, .
\end{eqnarray}
For the sake of clarity, we did not report the lengthy expressions for the polynomial coefficients $H_i$ for this reduction; however, they all consist of polynomials of circa 20 elements and testify once more how the use of ideal algebra can find highly non-trivial IBP seedings. Given all the integrals in Equation \eqref{eq:reduction_3322} are known as linear combinations of master integrals, the reduction is complete once the boundary term is computed. Using the same principle of  taking $z_i = 0 $ for each $H_i$ we can prove that in Equation \eqref{eq:reduction_3322}
\beq
\text{boundary}\, = \, \frac{2 \mathcal{I}(0,2,2,2,2+4\epsilon)}{3(r-16)r(1+\epsilon)}
\eeq
The system of differential equations we obtain is exactly the one reported in \eqref{eq:banana3l_system}. The difference is that for obtaining such a system via naive seeding of the polynomials $H_i$ we had generated almost 1000 equations of which only 14 were necessary, while here integration by parts was applied only twice, and in total we used 5 consistency relations coming from Symanzik polynomials. These simple examples prove that, at least in principle, it is possible to use ideal algebra to find the optimal use of relation \eqref{eq:dw} and heavily reduce the number of equations that are generated to reduce the master integrals. Some future directions starting from these observations are reported in the conclusions.

\section{Summary and perspectives}
\label{sec:Conclusions1}
In this chapter, we have presented a projective framework to derive IBP identities 
and differential equations for Feynman integrals in parameter space, updating and 
extending ideas and results that first emerged half a century ago, before
modern developments. We have emphasised the significance of the early 
mathematical results reported in~\cite{Regge:1968rhi,Ponzano:1969tk,
Ponzano:1970ch,Regge:1972ns}, which resonate strikingly with contemporary 
research. These ideas from algebraic topology were turned into a concrete 
application to one-loop diagrams by Barucchi and Ponzano~\cite{Barucchi:1973zm,
Barucchi:1974bf}. To apply these results in the modern context, we 
have shown how the analysis extends naturally to dimensional regularisation,
we have generalised the results to the multi-loop level, and we have emphasised 
the role played by boundary terms in the IBP identities, noting that they
do not vanish in general, and they provide a useful tool to link 
complicated integrals to simple ones. All these developments have explicitly 
been tested on relatively simple one- and multi-loop diagrams, recovering known 
results, including the elliptic differential equation for the equal-mass banana diagram up to four loops and the unequal mass two-loop sunrise integrals. These results extend what we have previously published in \cite{Artico:2023jrc} and \cite{Artico:2023bzt} and provide the foundations for a deeper analysis of viable improvements of our reduction algorithm in the last section of the chapter. Concepts coming from the study of polynomial ideals were applied to generate equations directly reducing the target integrals to the basis integrals (and to boundary terms) in an effective way, cutting the number of generated equations by two orders of magnitude. Integration by parts identities in parameter space and their relation to polynomial ideals open the way to a number of possible future research directions, which we are describing in the following paragraphs.

\paragraph{Implementation of efficient reduction strategies} 
It is a natural question to ask how this method compares to the usual momentum-space 
approach. This question cannot be answered in detail and quantitative
computational terms at this stage, since this is just an exploratory study, while
momentum-space techniques have been honed through decades of optimisation.
We can however make a few observations already at this stage. \\

First of all, the parameter-space method offers, to say the least, a 
rather different organisation of the calculation of an integral family, as compared to 
momentum-space algorithms. This should be evident from the concrete cases 
examined in the text: for example, the integral basis arising naturally from the 
Barucchi-Ponzano theorem for the massless box is not the same as the conventional 
one, and the differential equations that emerge are different too~\cite{Henn:2014qga}. 
We note further that how the lattice of different (integer) values of the 
indices $\nu_i$ is explored in parameter space appears different from standard IBPs. 
In the absence of boundary terms, parameter-space IBPs connect integrals with a fixed 
number of external legs, but different space-time dimensions. This is not necessarily 
a positive feature, since the goal of reduction algorithms is to a large extent to 
connect complicated integrals to simpler ones. It must however be noted that, in 
standard algorithms~\cite{Laporta:2000dsw}, the goal of achieving this simplification 
is reached in a rather roundabout way, through the ordering imposed in the recursive 
exploration of the index lattice. In parameter space, this simplifying step is specifically
associated with the novel feature of non-vanishing boundary terms, which give
lower-point integrals. These terms can in principle be reached simply by 
suitably picking the initial values of the indices, as was done for the massless pentagon 
in \ref{ssec:OlMaP}. \\

Continuing with the comparison, we observe that both the momentum-space 
algorithms and the projective one have a large degree of arbitrariness in their 
initialisation, which leaves room for optimisation. In the present chapter, we explored an optimization strategy for the simple example of two- and three-loop banana integrals based on the membership test for polynomial ideals, achieving a great degree of simplification. The examples reported are far from providing a completely automated algorithm not relying on human input (e.g. to our rotation of the basis for the three-loop case). Our current studies on the two-loop unequal-mass sunrise integral and on the massless double-box integral aim at paving a way in the direction for a reliable and competitive algorithm. The grading of the polynomial ideal, for example, can be exploited to study which numerators can be reduced for each fixed total degree; these polynomials can then be used to reduce the number of equations generated and - hopefully - provide a smaller solvable linear system. In contrast to momentum-pace algorithms,
we observe that the parameter-space approach bypasses the ambiguity due
to the choice of loop-momentum routing, which can be non-negligible for complicated
diagrams; similarly, the issue of irreducible numerators is implicitly dealt with at
the momentum integration stage. These two aspects are among the consequences
of the fact that the parameter space offers a minimal representation of Feynman 
integrals, transparently related to the symmetries of the original Feynman graph. When
complex multi-scale examples of this kind become available, a more thorough
comparison of the two approaches, including computational aspects, will
become possible.

\paragraph{A deeper understanding of the monodromy group of Feynman diagrams.}
In the historical introduction of this chapter, we mentioned how the foundational study \cite{Regge:1968rhi} and the applications to respectively multi-loop self-energies, one-loop diagrams and necklace diagrams \cite{Ponzano:1969tk,Ponzano:1970ch,Regge:1972ns} were aimed at studying the monodromy ring of such integrals, with the declared aim of determining to which extent a function is determined by it. This problem for functions of a single complex variable was posed by Riemann and solved in Ref. \cite{plemelj1964problems}, thus in principle determining the solution to one-variable problems such as multi-loop banana integrals. The topic of analytic properties of Feynman integrals and, more generally, of the S-matrix, recently received a revival of interest -- see for example the lecture notes in Refs. \cite{Mizera:2023tfe,Curry:2024mua}, with application to black-hole scattering ampliture \cite{Aminov:2024mul} and some all-loop tests \cite{Bourjaily:2020wvq}. The landscape of research possibilities from the perspective of Feynman parametric integrals is still vast, considering that a second part of \cite{Regge:1968rhi} was devoted to so-called \textit{bi-projective} forms tackling integrals with singular subsimplexes (possibly outside of the physical Riemann sheet of the Mandelstam variables). Whether these bi-projective forms -- that include blow-up effects of the projective space also employed in \cite{Muller-Stach:2011qkg} -- can generate additional identities for the unequal mass sunrise integral sufficient to reduce the order of the differential equation to two is still an open problem.

\paragraph{Applications to other kinds of integrals}
The chapter we are concluding here refers to our ongoing interest in parametric integrals, started in particular with the paper \cite{Artico:2023bzt} and the proceeding \cite{Artico:2023jrc}. The fact that these two references have already been published for more than one year allows us to add to the list research directions that other authors have identified in the papers citing our original work. Many of the possible applications of the technology we are developing depend on the notion of GKZ systems \cite{Gelfand:1990bua,Hosono:1995bm,Feng:2019bdx}, a very general class of differential equations dealing with integrals of rational forms very similar to our Feynman parametrized integrals. Among the physical quantities that can be described by GKZ systems there are cosmological correlators \cite{Grimm:2024tbg,Grimm:2025zhv} and energy correlators \cite{Moult:2025nhu}. The first are challenging quantities describing correlation function in a in non-flat space-time, such as in cosmological settings that describe an expanding universe, for which usual Feynman diagrams are supplemented with with additional time integrals at each vertex; energy flow operators provide the connection between real world collider phenomenology, and the deep underlying principles of QFT. Integrals appearing in both these configurations can be expressed as integrals of rational forms for which the parameter space IBP seems a possible language to explore a master integral reduction. Energy correlators figure as the simplest case of study of form factors squared in section 4 of Ref. \cite{He:2025zbz}, where the authors also suggest prospects towards integrated results via integral reduction in parameter space.

\chapter{Bulk-defect-defect correlators}
\label{ch:BDD}
\section{Introduction and preliminaries}
\label{sec:IntroBDD}

\subsection{The interest for defect correlators and an eagle view on the chapter}
\label{subsec:EagleView}

The chapter you are about to read describes more extensively (and expands on) the results presented in \textit{Perturbative bootstrap of the Wilson-line defect CFT: Bulk-defect-defect correlators} \cite{Artico:2024wnt} by the collaboration the author is part of. We decided to start this chapter by providing an overview of its content and of the scientific context the paper refers to. Many of the concepts and quantities introduced in this preliminary section will be properly described later in the chapter, and in particular, in the second half of this introduction, we will focus on a brief discussion of conformal field theories (CFTs) and supersymmetry. This is by no means meant to be a complete review of such rich topics: the reader interested in CFTs can refer to \cite{DiFrancesco:1997nk,Blumenhagen:2009zz} and references therein; the reader interested in supersymmetry can refer to \cite{AMATI1988169,wess1992supersymmetry,Weinberg:1995mt,muller2010introduction} and references therein.\\

This and the following chapters both deal with correlators of local operators in the context of defect CFT, in other words, correlators of local operators and one extended non-local operator referred to as a defect. Defects are central to a broad range of physical theories, from condensed-matter systems to high-energy physics. This is because, despite the presence of a defect, the local structure of a theory remains intact, preserving key features that allow for instance the application of both non-perturbative and perturbative techniques. In critical systems, conformal defects retain a significant portion of the underlying conformal symmetry, enabling the use of modern tools like the conformal bootstrap to formulate constraints on observables\footnote{See the subsection on CFTs for further details.}. Although numerical studies face challenges, substantial progress has been made on the analytic side, inspired by the development of Lorentzian inversion formulas and dispersion relations in conformal field theories (CFTs) without defects \cite{Caron-Huot:2017vep}. In recent years, a wide variety of defect CFTs have been investigated, with Wilson lines emerging as crucial probes in both the AdS/CFT correspondence \cite{Maldacena:1997re,Maldacena:1998im} and the study of confinement in Yang-Mills theories \cite{Polyakov:1978vu,Witten:1998zw}. In this context, the configuration involving a Maldacena-Wilson line in four-dimensional $\Nm=4$ super Yang-Mills (sYM) theory holds a particularly prominent position.
This setup retains many key features of its parent theory, including one-dimensional conformal symmetry, supersymmetry, and integrability.\\

Recent studies have focused on two canonical configurations: the two-point functions of bulk operators in the presence of the Wilson line and multipoint correlators of defect operators. For the two-point functions, correlators involving half-BPS operators\footnote{See \ref{ssub:Half-BPS} for a definition of half-BPS operators.} have been studied at weak and strong coupling, employing analytical bootstrap methods \cite{Barrat:2021yvp,Barrat:2022psm,Bianchi:2022ppi,Meneghelli:2022gps,Gimenez-Grau:2023fcy}, perturbative techniques \cite{Barrat:2020vch}, and exact results obtained through localization in a specific kinematic regime known as topological \cite{Drukker:2007yx,Giombi:2009ds,Giombi:2009ek,Buchbinder:2012vr,Beccaria:2020ykg}. More generally, two-point functions of bulk operators in the presence of a defect have been studied in \cite{Billo:2016cpy}, with further results on $n$-point correlation functions (including bulk-defect-defect) to be found in \cite{Lauria:2020emq}.
In the context of multipoint correlators, extensive work has been conducted on four-point functions of defect half-BPS operators using modern approaches that combine numerical conformal bootstrap and integrability \cite{Cavaglia:2021bnz,Cavaglia:2022qpg,Cavaglia:2022yvv,Cavaglia:2023mmu}.
Strong-coupling results have also been derived through direct computations \cite{Giombi:2017cqn,Giombi:2023zte} or with the analytical bootstrap \cite{Liendo:2016ymz,Liendo:2018ukf,Ferrero:2021bsb,Ferrero:2023znz,Ferrero:2023gnu}. Meanwhile, exact results have been obtained with localization techniques \cite{Giombi:2018qox}. Additionally, studies of higher-point functions at weak coupling have led to conjectures about superconformal Ward identities \cite{Barrat:2021tpn,Barrat:2022eim}, which have been confirmed and expanded upon \cite{Bliard:2024und,Barrat:2024ta}, leading to new results for five- and six-point functions \cite{Bliard:2023zpe,Peveri:2023qip,Barrat:2024nod,Barrat:2024ta2,Artico:2024wut}.\\

\begin{figure}
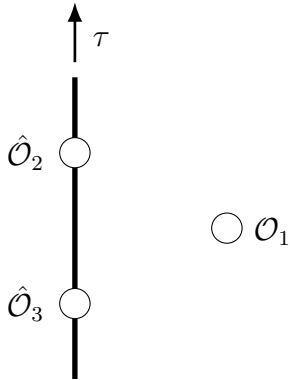

    \centering
    \IllustrationIntro
    \caption{Illustration of a bulk-defect-defect correlator $\vev{\Op_1 \Oh_2 \Oh_3}$. The two \textit{defect} operators $\Oh_{2,3}$ are representations of the one-dimensional CFT preserved by the line, while the \textit{bulk} operator $\Op_1$ on the right lives in the four-dimensional space of $\Nm=4$ sYM.}
    \label{fig:BulkDefectDefect}
\end{figure}

The configurations discussed above are generally considered the simplest correlators with non-trivial kinematics in defect CFTs. However, there is another canonical setup that has so far received little attention: correlators of one bulk and two defect operators. The setup is represented in Figure \ref{fig:BulkDefectDefect} and it is the main focus of this chapter.
This configuration is particularly interesting in the context of the Wilson-line defect CFT, as it depends on just one spacetime cross-ratio and one $R$-symmetry variable for half-BPS operators.
In contrast, two-point functions of bulk operators rely on two spacetime cross-ratios and one $R$-symmetry variable, while four-point functions of defect operators depend on one spacetime cross-ratio and two $R$-symmetry variables. Bulk-defect-defect correlators thus appear to be the simplest non-fixed correlators in this defect CFT.
To date, bulk-defect-defect correlators have been explored primarily in scalar theories with line defects \cite{Lauria:2020emq}, with some insights drawn from locality constraints \cite{Levine:2023ywq,Levine:2024wqn} and conformal block expansion \cite{Buric:2020zea,Okuyama:2024tpg}.\footnote{See also \cite{Gimenez-Grau:2020jvf,Chen:2023oax} for analyses involving boundaries.} The kinematics studied in \cite{Levine:2023ywq,Levine:2024wqn} is analogous to one of the two-point functions in CFT on real projective space studied in \cite{Giombi:2020xah,Zhou:2024ekb}. In the context of the Wilson-line defect CFT, the localization machinery has been developed in \cite{Giombi:2018hsx} for calculating these correlators exactly in a special kinematic limit.\\

In reference \cite{Artico:2024wnt} and in this chapter, we study bulk-defect-defect correlators in the Wilson-line defect CFT, focusing on the simplest case where both the bulk and defect operators are half-BPS. We do this by using a so-called \textit{perturbative bootstrap} approach, meaning that a set of non-perturbative constraints is used to reduce to the minimum the amount of perturbative information (\textit{i.e.} Feynman diagrams) necessary to determine the correlation function.
For the case of bulk-defect-defect correlators, we derive differential constraints (that we interpret as superconformal Ward identities) through a heuristic approach, and demonstrate their equivalence to the existence of a topological sector. These constraints prove useful in systematically eliminating one $R$-symmetry channel. We also examine specific limits of these correlators, where they reduce either to bulk-defect two-point functions or to the product of bulk one-point and defect two-point functions. We extend the study of the kinematic limit presented in \cite{Giombi:2018hsx} to NLO, introducing its use in this context as a tool to fix constants and to perform checks. We then study the weak and strong coupling expansions of bulk-defect-defect correlators. At weak coupling, we show that the number of $R$-symmetry channels increases order by order in perturbation theory; at next-to-leading order, we fully determine the correlators by focusing on the simplest $R$-symmetry channel, which avoids bulk vertices.
Notably, the results contain no transcendental functions, despite their potential presence in individual diagrams and conformal blocks. At strong coupling, we focus on one particular correlator and study it up to partial next-to-next-to-leading order results by using superconformal constraints.\\

The structure of the chapter is as follows. In subsections \ref{sub:CFT} and \ref{sub:SUSY} we introduce concepts in conformal field theory and supersymmetry that are particularly relevant for the study of the correlators in analysis. Then in section  \ref{sec:TheWilsonLineDefectCFT}, we establish the foundational elements necessary for the computations in this work.
Section \ref{sec:BDDNonPerturbativeConstraints} compiles the non-perturbative constraints that govern the correlators.
Perturbative results in the weak- and strong-coupling regimes are presented in section \ref{sec:PerturbativeResults}. In section \ref{sec:ConclusionsBDD}, we summarize the main findings of the chapter and describe future directions. Appendix \ref{app:Integrals} refers to both this and the next chapter and provides the integrals required for the computations discussed in the main text, as well as details on the point-splitting regularization \cite{Hagiwara:1980ys,Barci:2000cr,Beisert:2003tq,Beisert:2015uda} chosen for divergent integrals.

\subsection{Conformal field theory and defects}
\label{sub:CFT}
\subsubsection{The conformal algebra and its representations}
\label{ssub:ConfAlgebra}
We begin the introduction of the key concepts for understanding the framework of the supersymmetric Wilson-line conformal defect by introducing the concept of conformal symmetry and its representations. Conformal symmetry in $d$ dimensions is a symmetry preserving the angles of the original theory, or equivalently preserving the metric tensor up to local rescaling
\beq
\eta'_{\mu\nu}(x') = \kappa(x) \eta_{\mu\nu}(x)\,.
\label{eq:MetricConfTransf}
\eeq
Throughout this and the next chapter, we will consider Euclidean conformal field theory, meaning the metric tensor $\eta{\mu\nu}(x) = \delta_{\mu\nu}$. 
It is an immediate consequence of this definition that the \textit{Poincaré group} of special relativity is a subgroup of the conformal symmetry group. The generators of the Poincaré group are therefore also generators of the conformal group consisting of $d$ translation generators $P_\mu$ and $d(d-1)/2$ antisymmetric generators of rotations in $d$-dimensional space $M_{\mu\nu}$. We can define the action of the conformal generators on scalar fields $\Op(x)$. Working with the usual Cartesian coordinates, 
\begin{eqnarray}
& P_\mu & =  -i \partial_\mu \,,\label{eq:P}\\
& M_{\mu\nu} & =  i(x_\mu \partial_\nu - x_\nu \partial_\mu) \,.
\label{eq:M}
\end{eqnarray}
To these $d(d+1)/2$ generators in total, we can add $d+1$ additional ones generating transformations that do not belong to the conformal group while not being part of the Poincaré group. These generators correspond to one dilatation
\beq
x_\mu \rightarrow \lambda x_\mu 
\label{eq:Dilatations}
\eeq
and to $d$ special conformal transformations
\beq
x_\mu \rightarrow \frac{x_\mu - b_\mu x^2}{1-2b\cdot x+b^2 x^2}\,.
\label{eq:SpecConf}
\eeq
When applied to scalar operators $\Op(x)$, the generators of dilatations and special conformal transformations can be written as 
\begin{eqnarray}
& D & = -i x^\mu \partial_\mu,\label{eq:D}\\
& K_\mu & = -i \left( 2x_\mu x^\nu \partial_\nu - x^2 \partial_\mu \right) \,.
\label{eq:K}
\end{eqnarray}
Including spin would require adding spin terms to the rotation generator $M_{\mu\nu}$ and modifying the action of special conformal transformations $K_{\mu}$ and dilatations $D$ \cite{DiFrancesco:1997nk}. The generators of the conformal algebra are in total $(d+1)(d+2)/2$ and obey the following commutation relations\footnote{To avoid confusion, we mention explicitly these rules are valid for any explicit representation of the generators, regardless of the spin of the operator they are applied on.}
\begin{equation}
\begin{aligned}
  [D, P_\mu] &= i P_\mu, \\
  [D, K_\mu] &= -i K_\mu, \\
  [D, M_{\mu\nu}] &= 0, \\
  [P_\mu, K_\nu] &= 2i \left( \delta_{\mu\nu} D - M_{\mu\nu} \right), \\
  [P_\mu, P_\nu] &= 0, \\
  [K_\mu, K_\nu] &= 0, \\
  [M_{\mu\nu}, P_\rho] &= i (\delta_{\nu\rho} P_\mu - \delta_{\mu\rho} P_\nu), \\
  [M_{\mu\nu}, K_\rho] &= i (\delta_{\nu\rho} K_\mu - \delta_{\mu\rho} K_\nu), \\
  [M_{\mu\nu}, M_{\rho\sigma}] &= i (\delta_{\nu\rho} M_{\mu\sigma} - \delta_{\mu\rho} M_{\nu\sigma}
  + \delta_{\mu\sigma} M_{\nu\rho} - \delta_{\nu\sigma} M_{\mu\rho}).
\end{aligned}
\label{eq:ConfAlgebra}
\end{equation}
The presence of the $\delta_{\mu\nu}$ is due to the aforementioned Euclidean signature. It is possible to rewrite the conformal group transformation on $\mathbb{R}^d$ as linear group transformation on a $d$ dimensional null cone embedded in $\mathbb{R}^{d+2}$ \cite{OsbornCFTLectures}; this approach is worth mentioning but it will not be discussed further in this thesis.\\

The representations of the conformal algebra described above must also be representations of the Lorentz group, as the generators of this last one are also generators of the conformal group. We classify the representations of the Lorentz group according to their transformation properties under $SO(d)$, described by its irreducible representations. When specializing to $d=4$ Euclidean conformal field theory we can use the spin quantum numbers $j_L$ and $j_R$ of the representations of the Lorentz group $SU(2)\times SU(2)$ \cite{Peskin:1995ev,muller2010introduction}. In this thesis, we specialize to fields that are scalar, spinor, or vector -- respectively total spin 0, $1/2$, and 1. The representations of the conformal group are also characterized by a second quantum number called \textit{scaling dimension}, defined for scalar fields by the action of a scale transformation
\beq
\phi(\lambda x) = \lambda^{-\Delta} \phi(x)
\label{eq:ScaleTransf}
\eeq
and connected to the dilatation operator, as
\beq
D\phi(x) = i(x_\mu \partial^\mu + \Delta) \phi(x)\,.
\eeq
All together, we summarize in table \ref{tab:CFT_Operators} the classification of the CFT representations presented so far, to which we add for the sake of completeness representations of higher spin.

\begin{table}[h!]
\centering
\footnotesize 
\renewcommand{\arraystretch}{1.1}
\setlength{\tabcolsep}{6pt}
\begin{tabular}{|c|c|c|c|}
\hline
\textbf{Operator} & \textbf{symbol} & \textbf{Quantum Numbers $(\Delta; j_L, j_R)$} & \textbf{Example} \\
\hline
Scalar & $\phi(x)$ & $(\Delta; 0, 0)$ & Elementary/composite scalar \\
Weyl spinor (L) & $\psi_\alpha(x)$ & $(\Delta; \tfrac{1}{2}, 0)$ & Chiral fermion \\
Weyl spinor (R) & $\bar{\psi}^{\dot{\alpha}}(x)$ & $(\Delta; 0, \tfrac{1}{2})$ & Anti-chiral fermion \\
Vector & $A^\mu(x)$ & $(\Delta; \tfrac{1}{2}, \tfrac{1}{2})$ & Gauge field \\
Field strength & $F_{\mu\nu}(x)$ & $(\Delta; 1, 0)$ or $(\Delta; 0, 1)$ & Self-dual / anti-self-dual parts \\
Stress tensor & $T^{\mu\nu}(x)$ & $(\Delta; 1, 1)$ & Energy-momentum tensor \\
\hline
\end{tabular}
\caption{Classification of local operators in a 4D Euclidean conformal field theory by their quantum numbers $(\Delta; j_L, j_R)$, corresponding to scaling dimension and Lorentz spin.}
\label{tab:CFT_Operators}
\end{table}
%
At the classical level, scaling dimensions are commonly integer-valued. At the quantum level, however, they are subjected to corrections coming from the renormalization procedure: these corrections are referred to as anomalous dimensions and can be perturbatively expanded both at weak and strong coupling.\\
Applying the translation generator $P^\mu$ on an operator produces a new operator of increased scaling dimension; on the other hand, applying the generator $K^\mu$ generates operators with lower scaling dimensions. For \textit{unitary} conformal field theories there is a lower bound for the scaling dimensions of operators generated in this way: fields that saturate this lower boundary are the ones satisfying
\beq
\left[ K_\mu, \phi(0)\right] = 0\,.
\label{eq:ConfPrimaries}
\eeq
These fields are called \textit{conformal primaries} and generate a \textit{conformal multiplet} consisting of the primary and all the \textit{descendant} fields obtained by acting on the primary with the translation generator an arbitrary number of times \cite{DiFrancesco:1997nk,Gaberdiel:1999mc}.
\subsubsection{Correlation functions and operator product expansion}
\label{ssub:CorrAndOPE}
The natural subject of study in a conformal field theory is correlation functions of (primary) fields, where by correlation function (or Green's function) of quantum fields $\phi(x_i)$ we mean the vacuum expectation value of the time-ordered product of fields
\begin{equation}
G^{(N)}(x_1, x_2, \dots, x_N) = \langle 0 | \mathcal{T} \left\{ \phi(x_1) \phi(x_2) \dots \phi(x_N) \right\} | 0 \rangle
\label{eq:GreenFunct}
\end{equation}
where \( \mathcal{T} \) is the time-ordering operator, and \( |0\rangle \) denotes the vacuum state. In the path integral formalism, correlation functions are expressed as:
\begin{equation}
G^{(N)}(x_1, \dots, x_N) = 
\frac{\displaystyle \int \mathcal{D}\phi\, \phi(x_1)\dots\phi(x_N) e^{i S[\phi]}}
     {\displaystyle \int \mathcal{D}\phi\, e^{i S[\phi]}}\,.
\label{eq:PathIntCorrelator}
\end{equation}
Correlation functions of the kind \eqref{eq:GreenFunct} are strongly constrained by conformal symmetry, which for example fully determines the form of two- and three-point functions in terms of constants called \textit{conformal data} -- one-point functions are trivially zero due to translation invariance. We will describe these and more constraints when introducing more extensively such correlators in section \ref{sub:Correlators}. In this introduction, we want to describe another remarkable property of correlators in a conformal field theory: the \textit{operator product expansion} (OPE), first introduced in \cite{Wilson:1972ee}. The OPE allows for the expansion of a product of two local operators in terms of all the possible local operators in the theory, and it takes the form\cite{DiFrancesco:1997nk}
\beq
\Op_{\Delta_1}(x_1)\Op_{\Delta_2}(x_2) \sim \sum_{\Op_\Delta \, \text{primary}} \lambda_{\Delta_1\Delta_2,\Delta} C_{\Delta_1,\Delta_2,\Delta}(x_{12},\partial_2)\Op_{\Delta}(x_2)
\label{eq:OPEGeneral}
\eeq
where the sum runs over all the primaries of the theory, $C_{\Delta_1,\Delta_2,\Delta}(x_{12},\partial_2)$ is a differential operator encoding the construction of the descendants, and  $\lambda_{\Delta_1\Delta_2,\Delta}$ is a three-point function described later in equation \eqref{eq:BulkThreePointFunctions}. The OPE is not unique to conformal field theory; however, conformal symmetry guarantees the expansion converges with a non-zero radius \cite{Pappadopulo:2012jk,Rychkov:2015lca}. Its prominent position among the techniques available to study correlators in CFT is due to OPE being at the origin of the \textit{conformal block} expansion and the \textit{crossing symmetry} for multipoint correlators. Let's illustrate this idea with a four-point correlator. Conformal symmetry ensures that the correlator among four operators in four dimensions can depend on two space-time ratios we identify with $u$ and $v$. By considering the OPE of $\Op_{\Delta_1}\Op_{\Delta_2}$ and $\Op_{\Delta_3}\Op_{\Delta_4}$ we can express the correlator as a sum over the exchanged conformal primaries $\Op$. The result is a sum over functions called conformal blocks fully characterized by the symmetry group \cite{Dolan:2000ut,Dolan:2003hv}
\begin{equation}
\langle \mathcal{O}_{\Delta_1}(x_1) \mathcal{O}_{\Delta_2}(x_2) \mathcal{O}_{\Delta_3}(x_3) \mathcal{O}_{\Delta_4}(x_4) \rangle 
\sim \sum_{\mathcal{O}} \lambda_{12\mathcal{O}} \lambda_{34\mathcal{O}}\, G^{12,34}_{\Delta,\ell}(u,v)\,,
\label{eq:ConfBlockExp}
\end{equation}
where $G_{\Delta,\ell}(u,v)$ is the conformal block for a primary operator of scaling dimension $\Delta$ and spin $\ell$, and $\lambda_{ij\mathcal{O}}$ are the OPE coefficients\footnote{For the sake of completeness, we used both quantum number $\Delta$ and $\ell$ of the operators in this notation}. This same idea can be repeated considering the OPE expansion of $\Op_{\Delta_1}\Op_{\Delta_4}$ and $\Op_{\Delta_2}\Op_{\Delta_3}$ giving rise to a consistency relation known as crossing symmetry
\beq
\sum_{\mathcal{O}} \lambda_{12\mathcal{O}} \lambda_{34\mathcal{O}}\, G^{12,34}_{\Delta,\ell}(u,v)\, =
\sum_{\mathcal{O'}} \lambda_{14\mathcal{O'}} \lambda_{23\mathcal{O'}}\, G^{14,23}_{\Delta,\ell}(u,v)\,
\label{eq:CrossingTheory}
\eeq
playing a crucial role in the numerical conformal bootstrap~\cite{Rattazzi:2008pe,Poland:2018epd}.
\subsubsection{Conformal defects and defect operators}
\label{ssub:ConfDef}
\begin{figure}
\centering
\includegraphics[scale=0.6]{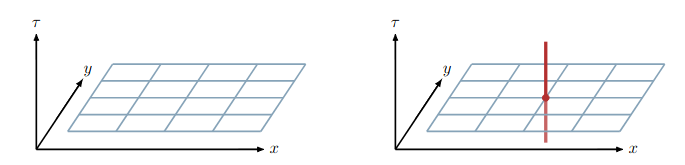}
\caption{An example of a line defect extending in the time direction in a (2 + 1)-dimensional spacetime. The grid in both figures represents the two-dimensional space. The left figure shows the bulk theory without any defects, while the right figure depicts the introduction of a line defect. Reproduced from \textit{Line Defect in Conformal Field Theory} \cite{Barrat:2024nod} by Julien Barrat.}
\label{fig:Defect}
\end{figure}

We introduced how correlation functions in conformal field theories are remarkably constrained by conformal symmetry, making their study a natural testing ground for techniques aimed at studying these functions beyond perturbation theory. Defects represent a way to introduce new physical phenomena beyond these strong constraints, while preserving part of the original symmetries. Conformal defects are non-local extensions of the bulk theory that break the conformal symmetry in a controlled way \cite{McAvity:1995zd,Billo:2016cpy,Liendo:2012hy}. For $ p$-dimensional defects embedded in a $ d$-dimensional space, we can write this symmetry breaking as 
\beq
SO(d+1,1) \longrightarrow SO(p+1,1) \otimes SO(d-p)
\label{eq:DefectsYMmBreak}
\eeq
meaning that the original conformal group is broken into a conformal symmetry on the defect and the rotation group around the defect. The value $d-p$ is referred to as the codimension of the defect. This thesis and the papers it is based on \cite{Artico:2024wnt,Artico:2024wut} deals with the case $d=4$ and \textit{line defects} having $p=1$, which is also the easiest to visualize: the presence of a line (extending into the time direction) breaks the bulk conformal symmetry into the rotation group around the line times the defect conformal symmetry. An analogous configuration with $d = 3$ and $p=1$ is reproduced in figure \ref{fig:Defect}. 
The mirror configuration of co-dimension 1 deals with defects called \textit{boundaries}, on which a wide literature is available~\cite{McAvity:1995zd,Liendo:2012hy,Bissi:2022mrs}. Surface defects having dimension 2 have also been the subject of recent studies~\cite{Gaiotto:2009fs,Holguin:2025bfe,Bissi:2022mrs,Chalabi:2025nbg}.\\

The defect CFT we will focus on in this thesis, the case of the supersymmetric Maldacena-Wilson loop in $\Nm = 4$ super Yang-Mills, is one of the instances of a QFT that can be described by an action -- in this case, the bulk action plus a defect action including fields coupling to the defect. This action and the properties of this defect theory will be further described in section \ref{sub:Wilson-line}. In this introduction, we emphasize some of the features that the presence of a defect brings into a conformal field theory, referring to works on the subject for the interested reader.

\paragraph{Defect representations and correlators}
\label{par:DefectRepr}
We can consider operator living on the defect -- local excitations on the defect -- as representations of the residual symmetry group  $SO(p+1,1) \otimes SO(d-p)$. These local operators for $p=1$ are characterized by two quantum numbers: the scaling dimension $\Dh$, the quantum number connected to the dilatation operator in the one-dimensional defect CFT; and the transverse spin $s \in \mathbb{N}/2$, connected to the rotation group $SO(3)$ \cite{Lemos:2017vnx,Giombi:2022vnz}. Using this new set of local operators it is possible to build a new kind of correlators, as introduced in section \ref{sub:Correlators}. Mixed correlators among bulk and defect operators are the primary focus of this chapter, while the next one deals with correlators of only defect operators.

\paragraph{New conformal data and defect OPE}
The presence of more possibilities of building correlators among operators in the presence of a defect and its local excitations results in the emergence of new conformal data. We will introduce all the kinematically fixed correlators in section \ref{sec:TheWilsonLineDefectCFT}, for now, we remark that in the presence of a defect, it is not true anymore that one-point functions of bulk operators are zero, because the Lorentz group is now broken by the presence of the defect. The constants appearing in the one-point functions of operators  $\Op_\Delta$ will be denoted by $a_\Delta$, while two-point functions between bulk and defect correlators are determined by the constants $b_{\Delta\Dh}$. These last constants emerge when considering a new kind of OPE, involving a bulk operator and the defect itself (see for example references \cite{Lemos:2017vnx,Billo:2013jda}). When considering the correlator among two operators in the presence of a defect (depending on two space-time ratios $r$ and $w$), it is possible to expand it in defect blocks, corresponding to the exchange of a defect operator when expanding both bulk operators with the defect. The result is 
\beq
F(r,w) = \sum_{\Dh,s} b_{\Delta_1\Dh,s}b_{\Delta_2\Dh,s} \hat{f}_{\Dh,s}(r,w)
\label{eq:DefectBlockExp}
\eeq
where the defect blocks $\hat{f}_{\Dh,s}(r,w)$ are again fixed by the symmetry. We refer the interested reader to the rich literature on the two-point bulk correlator \cite{Billo:2016cpy,Lauria:2017wav,Lauria:2018klo,Herzog:2020bqw,Herzog:2020bqw,Barrat:2020vch,Barrat:2021yvp} for more insights into this defect OPE.

\subsection{The supersymmetric extension of the Poincar\'e algebra}
\label{sub:SUSY}
Now that we have introduced conformal symmetry, its representations, and the symmetry breaking induced by conformal defects, we turn our attention to the other symmetry we will encounter and discuss throughout the next two chapters: supersymmetry. The main idea is to extend the symmetry content of a theory with Poincar\'e invariance -- whose Lie algebra we denote by $\mathcal{P}$ -- two operations are allowed:
\begin{itemize}
\item adding a local transformation acting on some internal degrees of freedom of our model -- this is the case of \textit{gauge transformations} with gauge group $\mathcal{G}$,
\item transforming the Poincar\'e Lie algebra into a graded algebra $\mathcal{P} \oplus \mathcal{L}^{(1)}$ by adding to the algebra a set of spinor charges.
\end{itemize}
In order to clarify this statement, we will now introduce the relevant theorems following the exposition of reference \cite{muller2010introduction}, while to clarify the role of conformal symmetry in the picture we refer to Appendix B of \cite{Weinberg:2000cr}.
\subsubsection{The theorems  of Coleman-Mandula and Haag, Łopusza\'nski, Sohnius}
We report here without proof the two crucial theorems marking the birth of the concept of supersymmetry. The first theorem by Coleman and Mandula \cite{Coleman:1967ad} answers the question discussing how much more symmetry we can add to a QFT beyond the Poincar\'e algebra. Under five specific assumptions, the theorem establishes that 
\begin{thm}[Coleman-Mandula theorem]
The connected symmetry group $F$ of the $S$-matrix is locally isomorphic to the direct product of a compact symmetry group $G$ and the Poincar\'e group $P$.
\label{th:Coleman-Mandula}
\end{thm}
The five assumptions are listed here:
\begin{itemize}
\item[(i)] \textit{Lorentz invariance:} the group $F$ contains a subgroup locally isomorphic to the Poincar\'e group $P$.
\item[(ii)] \textit{Finite number of particle types:} for any finite mass $M > 0$, there are only finitely many particle types (species) with mass less than or equal to that value.
\item[(iii)] \textit{Nontrivial scattering:} the $S$-matrix is not the identity; there exist interactions causing scattering between different particle states.
\item[(iv)] \textit{Analyticity:} the $S$-matrix is an analytic function of the scattering angles and energies (except possibly at poles corresponding to physical bound states).
\item[(v)]  \textit{Technical assumption:} the generators of $F$, written as integral operators in momentum space, have distributions for their kernels.
\end{itemize}
The last assumption is a technical assumption necessary to ensure the manipulations involved in the proof are well justified from a mathematical point of view. The theorem of Coleman and Mandula translates into the fact that the symmetry group of a QFT equipped with an 
$S$-matrix is isomorphic to the direct product of the Poincaré group and an internal symmetry group, thus limiting the allowed space-time symmetries of the theory. However, there are some important details hidden in the assumptions we listed that are worth discussing before introducing the role played by graded Lie algebras (or \textit{supersymmetry}) in the picture. First of all, the theorem applies to quantum field theories allowing the existence of a $S$-matrix and a mass gap: in presence of untrapped massless particles the $S$-matrix becomes infrared divergent and the assumptions of the theory are no longer satisfied, as explained in Appendix B of \cite{Weinberg:2000cr}. However, in theories with only massless particles the proof of theorem \ref{th:Coleman-Mandula} still holds \cite{Weinberg:2000cr} if we add to the Poincar\'e symmetry algebra the missing generators of the conformal group $D$ and $K^\mu$ defined in \eqref{eq:ConfAlgebra}. This remark is important as this thesis deals with supersymmetric extensions of conformal field theories\footnote{For a discussion on the $S$-matrix of superconformal theories we refer to the introduction of \cite{Beisert:2006ez}}.  \\

The extension of the group $F$ to the direct product of the conformal group with the gauge group is one important step to avoid the restrictions of the Coleman-Mandula theorem. However, the most important relaxation of the five listed assumptions is represented by the generalization of the notion of a Lie algebra to graded Lie algebras where defining relations involve also anticommutators, done by  Haag, Łopusza\'nski and Sohnius in \cite{Haag:1974qh}. Remarkably, this paper followed the publication of the first concrete supersymmetric model by Wess and Zumino \cite{Wess:1973kz} and -- but this is less surprising -- the pioneer study from Berenzin and Kac of groups with anticommuting and commuting generators \cite{Berezin:1970}. The theorem of Haag, Łopusza\'nski, and Sohnius classifies the possible algebraic structures obtained when adding anti-commuting fermionic charges to the generators of the symmetries, obtaining graded Lie algebras. For a complete discussion of the topic we refer to \cite{muller2010introduction} and references therein: in this introduction it is sufficient to mention that a graded algebra is a vector space $L$ which is a direct sum of two subspaces $L_0$ and $L_1$ equipped with a product $\circ$ such that if $u_k \in L_k$, 
\beq
u_k \circ u_j \in L_{j+k\;\text{mod}\,2} \,.
\eeq
In the context of quantum field theory, the graded algebra is formed by the Poincar\'e algebra as $L_0$ and Weyl spinor charges $Q_A^\alpha$, $\bar{Q}_{\dot{A}}^\alpha$ belonging to $L_1$. Here, the index $A$ refers to the spinor index while the index $\alpha = 1,..,\Nm$ labels different supersymmetry generators. All together, the relations we have to add to \eqref{eq:ConfAlgebra} are \cite{muller2010introduction}
\begin{equation}
\begin{aligned}
\{ Q_A^\alpha, \bar{Q}_{\dot{B}}^\beta \} &= 2 \sigma^\mu_{A\dot{B}} P_\mu \delta^{\alpha\beta} \,, \\
\{ Q_A^\alpha, Q_B^\beta \} &= 0 \,, \\
\{ \bar{Q}_{\dot{A}}^\alpha, \bar{Q}_{\dot{B}}^\beta \} &= 0 \,, \\
[P_\mu, Q_A^\alpha] &= 0 \,, \\
[P_\mu, \bar{Q}_{\dot{A}}^\alpha] &= 0 \,, \\
[M_{\mu\nu}, Q_A^\alpha] &= i(\sigma^{\mu\nu})_A{}^B Q_B^\alpha \
\end{aligned}
\label{eq:SUSYalgebra}
\end{equation}
where we set to zero the central charges $Z^{\alpha\beta}$ appearing in anti-commutation relations between spinor charges $Q_A^\alpha$ and $Q_B^\beta$. The generators of the possible gauge transformations are here left out of these relations, as well as the generators of the internal symmetry given by the rotation of the supercharges (\textit{i.e.} the rotations connected to the index $\alpha$ in $Q_A^\alpha$). Once introduced the generators of the super-Poincar\'e algebra, we can summarize its representation theory.
\subsubsection{Representations of the super-Poincar\'e algebra and superconformal primaries}
To study the representations of the super-Poincar\'e algebra we can consider the two Casimir operators whose eigenvalues label the irreducible representations: the operators $P^2$ and $C^2$. While the definition of the first Casimir operator is immediate, the second one is defined by the contraction 
\beq
C^2 = C^{\mu\nu}C_{\mu\nu}
\eeq
of the operator defined as
\beq
C^{\mu\nu} = B^\mu P^\nu - P^\mu B^\nu
\eeq
where
\beq
B^\mu = W^\mu +\frac{1}{8}\left(\bar{Q}\gamma^\mu \gamma^5 Q \right)
\eeq
and $W^\mu$ is the Pauli-Ljubanski polarization vector whose square is a Casimir operator of the Poincar\'e algebra (see for example \cite{muller2010introduction} for the definition of the polarization vector and a proof of all commutation relations necessary to prove the statements in this section). In the rest frame, it is possible to write the operator $C^2$ as the square of an operator $J_k$ whose commutation relations are the same as those of an angular momentum
\beq
C^2 = 2 m^4 J_k J^k
\eeq
with $k = 1,2,3$. A representation of the super-Poincar\'e algebra can therefore be represented by the eigenvalues $(m,j,j_3)$ corresponding to the operators $P^2,\,C^2$ and $J_3$. Although it might be tempting to interpret the value $j_3$ as the spin of the representation, this correspondence is not valid for the supersymmetric case. The construction of the different states is discussed in chapter 4 of reference \cite{muller2010introduction} together with the concept of Clifford vacuum: here we simply mention that for every value of $(m,j,j_3)$ we have (for the case with only one supersymmetric charge) four states
\beq
\left.|\Omega\right\rangle\,,\;
\bar{Q}^{\dot{1}}\left.|\Omega\right\rangle\,,\;
\bar{Q}^{\dot{2}}\left.|\Omega\right\rangle\,,\;
\bar{Q}^{\dot{1}}\bar{Q}^{\dot{2}}\left.|\Omega\right\rangle
\eeq
having respectively spin $j_3,\,(j_3+1/2),\,(j_3-1/2)$ and again $j_3$ and different parity properties.\\

In the presence of conformal symmetry and multiple supersymmetric charges, the super-Poincaré algebra is extended to the full superconformal algebra -- which includes, in addition to translations and supersymmetry transformations, also special conformal transformations and dilatations -- and the symmetry formed by the rotation of the supersymmetry label indices is called $R$-symmetry. Studying the representations of this larger algebra requires an extension of the labeling scheme beyond just $(m, j, j_3)$, as states are now organized according to their behavior under dilatations and $R$-symmetry transformations, in addition to Lorentz and supersymmetry. \\

Elaborating on the same classification given for the conformal symmetry, a key concept in this framework is that of \textit{superconformal primary operators}. These are local operators that are annihilated by the generators of special conformal transformations $K_\mu$ and by the superconformal charges  $Q^{A}_\alpha$. They form the highest-weight states of superconformal multiplets, from which all other operators (called descendants) are generated via the action of the translation generators $P^\mu$ and the supercharges $\bar{Q}^{\dot{A}}_\alpha$. Just as conformal primaries are classified by their scaling dimension $\Delta$ and Lorentz spin, superconformal primaries are labeled by their scaling dimension, spin, and $R$-symmetry quantum numbers. Certain representations are called \textit{short} or BPS-protected, meaning they satisfy shortening conditions that reduce the number of independent descendant states. These shortened multiplets are of central importance in many applications, including the study of protected operators and non-renormalization theorems. For a detailed treatment of superconformal representation theory in four dimensions, see e.g. \cite{Dolan:2002zh,Minwalla:1997ka,Cordova:2016emh}.\\

We can now conclude this introduction that listed the main concept at the foundations of our topic of study and we can start discussing the supersymmetric model we are interested in.
\section{The Wilson-line defect CFT}
\label{sec:TheWilsonLineDefectCFT}
\subsection{The bulk theory: $\mathcal{N}=4$ super Yang-Mills}
\label{subsec:N=4sYM}

The theory we study in this thesis is a supersymmetric extension of a gauge theory, whose details of representation theory go beyond the scope of this thesis. In the simplest form of supersymmetry for a gauge theory ($\mathcal{N}=1$), to each gluon $A^a_\mu$ corresponds a spin 1/2 gluino super-partner $\psi^a_\alpha$, connected to the original gluon via the Grassman odd supersymmetry generator extending the original Poincar\'e algebra (see references \cite{AMATI1988169,wess1992supersymmetry,Weinberg:1995mt,muller2010introduction} and references therein for an historical and comprehensive review). It is possible to add up to a maximum of four copies of such supersymmetry generators to remain in the realm of renormalizable gauge quantum field theories, obtaining the  $\mathcal{N}=4$ supersymmetric Yang-Mills theory, or  $\mathcal{N}=4$ sYM \cite{Brink:1976bc}. To close, the supersymmetry algebra also requires six real scalar fields obeying
\beq
\phi^{aAB} = -\phi^{aBA}
\eeq
where $A$ and $B$ label the supersymmetry generators from 1 to 4. As we described above, the field content of the $\mathcal{N} = 4$ sYM theory is
\beq
\textbf{1}\; \text{gluon} \oplus \; \textbf{4}\; \text{Weyl fermions} \oplus \; \textbf{6}\; \text{scalars} 
\eeq
and as required by supersymmetry, the bosonic and fermionic degrees of freedom are equal: eight each. It is important to stress that all fields in $\mathcal{N} = 4$ sYM belong to the adjoint representation of the group $SU(N)$, thus distinguishing gluinos from quarks (that belong instead to the fundamental representation of the gauge group). This means that the index $a$ for any field goes from 1 to $N^2-1$. The $SU(N)$ algebra generators obey the commutation relation
\beq
\left[ T^a,T^b\right] = i f^{abc}T^c
\label{eq:TCommut}
\eeq
and the trace relation
\beq
\tr T^aT^b = \frac{\delta^{ab}}{2}
\label{eq:TrT}
\eeq
where the factor $1/2$ in \eqref{eq:TrT} is a widespread convention we also adopt.\\
 
The most compact form of the Lagrangian density of the $\mathcal{N} = 4$ sYM theory can be written in 10 dimensional notation \cite{Brink:1976bc},
\beq
\mathcal{L} = \frac{1}{g^2} \tr \left( \frac{1}{2} F_{MN}F_{MN} + i \bar{\Psi}\Gamma_{N}D_{N}\Psi \right)
\eeq
where we adopt the convention that repeated indices are summed with an Euclidean metric and the matrices $\Gamma_{N}$ are 32 by 32 matrices. The four-dimensional action can be obtained by dimensional reduction: by compactification of six dimensions on the higher dimensional torus $T^6$, 6 components of the vector $A^M$ become the six real scalars $\phi$ and the other four give origin to the gluon field $A^\mu$.  In this way, one obtains a four-dimensional Lagrangian that contains gluons coupled canonically to the scalars and four complex fermions:
\begin{equation}
    \begin{split}
    S
    =\ &
    \frac{1}{g^2} \tr \int d^4 x\, \biggl( 
    \frac{1}{2} F_{\mu\nu} F_{\mu\nu}
    + D_\mu \phi^I D_\mu \phi^I
    - \frac{1}{2} [ \phi^I, \phi^J ] [ \phi^I, \phi^J ] \\
    &+
    i\, \psib \slashed{D} \psi
    + \psib \Gamma^I [ \phi^I\,, \psi ]
    + \pd_\mu \cb D_\mu c
    + \xi ( \pd_\mu A_\mu )^2
    \biggr)\,,
    \end{split}
    \label{eq:BulkAction}
\end{equation}
where $\mu = 0, . . . , 3$ are the spacetime directions and $I = 1, . . . , 6 $ are the indices of the six scalars, giving origin to a global $SO(6)$ symmetry called R-symmetry. This notation introduces a single 16-component Majorana fermion composed of the 4 Weyl fermions mentioned above. The term proportional to $\xi$ is a gauge-fixing term and we work in the Feynman gauge where $\xi = 1$.\\

The four-dimensional $\mathcal{N}=4$ sYM theory not only represents the maximally supersymmetric renormalizable gauge theory, it also preserves conformal symmetry at the quantum level \cite{Sohnius:1981sn,Mandelstam:1982cb}. This means that the theory is invariant under Poincar\'e transformations, dilatiations and special conformal transformations of the space-time; and that the $\beta$-function of the coupling $g$ is 0, meaning that the conformal symmetry is not broken by quantum fluctuations\footnote{In all this thesis, unless specified, we will refer to the \textit{superconformal phase} of $\mathcal{N}=4$ sYM where all expectations values of then scalar fields vanish. Other possibilities include the \textit{Coulomb phase} \cite{Kraus:1998hv}, which is not discussed in this thesis.}. In this case, the super-Poincar\'e algebra extends to the superconformal algebra $\mathfrak{psu}(2,2|4)$, which together with the gauge group $SU(N)$ forms the complete symmetry group of $\mathcal{N}=4$ sYM \cite{Sohnius:1981sn,Beisert:2003jj,Beisert:2003te}. It is important to stress that (super-)conformal symmetry does not prevent divergences from arising: loop corrections to scattering amplitudes can present infrared divergences to be canceled with soft and collinear particle emissions, while composite gauge invariant operators such as $\text{tr}(\phi^I\phi^I)$ are renormalized to cancel their inherent short-distance divergences and acquire therefore an anomalous dimension. We anticipate that operators whose anomalous dimension is zero are called \textit{protected} operators and represent cases for which UV divergences cancel without the need of renormalization.\\

The conformal nature of the theory $\Nm=4$ sYM makes it one of the most studied examples for the so-called AdS/CFT duality. This duality, first introduced in \cite{Maldacena:1997re}, describes a correspondence between a string theory living in an anti-de Sitter space (AdS) and a quantum field theory defined on its boundary (the Minkowski space-time). It is based on the holographic principle, which states that a gravitational theory in a $d+1$ dimensional volume $V$ can be described by a theory living on the $d$-dimensional boundary of $V$. For the $4$-dimensional theory $\Nm = 4$ sYM with gauge group $SU(N)$ and large $N$, the corresponding $5$-dimensional theory presenting the same super-conformal symmetry is a type II B string theory on $AdS_5 \times S^5$ \cite{Haag:1974qh}. The correspondence is a strong-weak correspondence in the following sense: the trustworthiness of perturbative calculations in the supersymmetric Yang-Mills theory is lost for a large \textit{effective 't Hooft coupling} $\lambda = g^2N$. However, a large effective coupling corresponds to the realm of validity of the super-gravity solution for $AdS_5 \times S^5$. The strong-coupling limit is studied perturbatively through \textit{Witten diagrams}: in this thesis, we mention them as they will appear in this chapter; however, we will not need to compute them and therefore we do not discuss them further. Examples of use of the Ads/CFT correspondence include the computation of planar gluon scattering amplitudes at strong coupling, equivalent to a minimal surface
area calculation in $AdS_5$ space, with the surface ending on a polygon formed by the gluon
momenta\cite{Alday:2007hr}. \\

For the sake of completeness, it is worth mentioning that $\Nm = 4$ sYM is invariant under the $S$-duality group leading to an interesting weak/strong-coupling duality of the electric/magnetic type known as Montonen–Olive duality \cite{Montonen:1977sn}. This duality will not be explored in this thesis and we will shortly refer to this later in this chapter when defining the Wilson line. The theory $\Nm = 4$ sYM for large $N$ is also an example of integrable field theory, meaning that its spectrum can be determined exactly for any value of the coupling by recasting the problem in terms of a set of equations that in principle determine the answer \cite{Beisert:2010jr,Gromov:2013pga}.

\subsubsection{Half-BPS operators}
\label{ssub:Half-BPS}

Local operators defined from the fields in the action \eqref{eq:BulkAction} are representations of the superconformal algebra $\mathfrak{psu}(2,2|4)$ and can therefore be classified via their quantum numbers. These numbers are the scaling dimension $\Delta$, the spin $\ell$ and the $SO(6)_R$ R-symmetry
charge $k$ \cite{Beisert:2003te,Kinney:2005ej,Dobrev:2016gqa}. When considering correlation functions in the next chapters, we focus on a subset of operators known as \textit{half-BPS} operators, from the name of Bogomol’nyi-Prasad-Sommerfield condition they satisfy  \cite{Bogomolny:1975de,Prasad:1975kr,Dolan:2001tt,Lee:1998bxa}, that is that such operators commute (or anti-commute, depending on their bosonic or fermionic nature) with half of the supersymmetry charges $Q_{\alpha}^A$:
\beq
\left[ Q_{\alpha}^A,\Op_\Delta(x)\right]_\pm = 0 \, ,
\eeq
where the subscript specifies that the parentheses can identify a commutator or anti-commutator as mentioned above. Among the half-BPS operators, we are particularly interested in scalar operators, having spin $\ell = 0$, scaling dimension $\Delta$, and the R-symmetry Dynkin labels being $[0, \Delta, 0]$. It is clear that to obtain a scalar from fields in the adjoint representation of $SU(N)$ it is necessary to consider traces of products of such fields; in our work we will consider correlation functions of \textit{single-trace} operators, referring to scalar operators defined as 
%
\begin{equation}
    \Op_{\Delta} (u,x)
    =
    \frac{1}{\sqrt{n_{\Delta}}} \tr \left( u \cdot \phi (x) \right)^\Delta\,.
    \label{eq:SingleTraceHalfBPSOperators_Bulk}
\end{equation}
Such operators are made of $k = \Delta$ scalar fields whose $R$-symmetry indices are contracted with the vector $u$ \cite{Maldacena:1997re,Dolan:2001tt,Beisert:2003te}. To keep the half-BPS property valid, the vector $u$ must satisfy the condition
\begin{equation}
    u^2
    =
    0\,.
    \label{eq:u_Bulk}
\end{equation}
These operators have integer scaling dimensions $\Delta = k$, the number of elementary fields in the operator, and are protected: the scaling dimension does not get quantum corrections coming from short distance (or ultra-violet) divergences, as the condition \eqref{eq:u_Bulk} sets the short distance limit of the two point function to zero. The normalization constants can be evaluated in the large $N$ limit to be \cite{Semenoff:2002kk}
\begin{equation}
    n_{\Delta}
    =
    \frac{\Delta \lambda^\Delta}{2^{3\Delta} \pi^{2\Delta}}\,,
    \label{eq:NormalizationConstantHalfBPS_Bulk}
\end{equation}
where we defined the 't Hooft coupling before $\lambda$ as 
\beq
\label{eq:tHooft_coupling}
\lambda = g^2 N \, .
\eeq
The restriction to single-trace operators is here purely technical: most of the content of the following chapters can be adapted to multi-trace operators of the form
\begin{equation}
    \Op_{\vec{\Delta}} (u,x)
    =
    \frac{1}{\sqrt{n_{\vec{\Delta}}}} \tr \left( u \cdot \phi (x) \right)^{\Delta_1}...\left( u \cdot \phi (x) \right)^{\Delta_n}\,
    \label{eq:MultiTraceHalfBPSOperators_Bulk}
\end{equation}
with $\vec{\Delta} = \left( \Delta_1,...,\Delta_n\right)$, with the limitation that normalization coefficients and conformal data are possibly less studied in the literature  (see \cite{Witten:2001ua,Giusto:2024trt} and references therein for studies on multi-trace operators).\\

We conclude this section by highlighting once more that half-BPS operators belong to a subclass of operators that is not at all complete when considering OPE limits: while the external operators can be chosen to be half-BPS, the exchanged operators in OPE channels can be of any kind. The \textit{Konishi} operator  \cite{Konishi:1983hf,Beisert:2003te} is an example of operator that is not half-BPS
\beq
\mathcal{K}(x) = \text{tr}(\phi^I\phi^I) \,.
\eeq
We already encountered this operator when discussing the appearance of short-distance divergences for local operators in $\Nm = 4$ sYM: indeed the Konishi operator has the same classical dimension as $\Op_2 (u,x)$, but it has a non-vanishing anomalous dimension due to the renormalization procedure.

\subsubsection{Bulk Feynman rules}
\label{ssub:BulkFR}
In this section, we present the propagators and vertices that the action \eqref{eq:BulkAction} gives origin to, and provide some of the formulas that are used throughout the thesis to compute correlation functions among operators. As a first step, we introduce the position space Feynman rules for the propagators of the fields in $\Nm = 4$ sYM, namely the already mentioned scalars, gluons, and fermions plus the ghosts generated by the Faddeev–Popov quantization procedure for gauge theories. Such propagators are:
\begin{equation}
    \begin{split}
    \text{Scalars:} \qquad 
    & \ScalarPropagator = g^2 \delta^{IJ} \delta^{ab}\, I_{12}\,, \\
    \text{Gluons:} \qquad 
    & \GluonPropagator = g^2 \delta_{\mu\nu} \delta^{ab}\, I_{12}\,, \\
    \text{Fermions:} \qquad 
    & \FermionPropagator = i g^2 \delta^{ab} \slashed{\pd}_1 I_{12}\,, \\
    \text{Ghosts:} \qquad 
    & \GhostPropagator = g^2 \delta^{ab} I_{12}\,,
    \end{split}
    \label{eq:Propagators}
\end{equation}
where we defined the four-dimensional scalar propagator as
\begin{equation}
    I_{ij} = \frac{1}{4\pi^2 x_{ij}^2}\,.
    \label{eq:PropagatorFunction4d}
\end{equation}
It is important to note that in the normalization we have chosen for the action \eqref{eq:BulkAction}, all propagators carry a power of 2 of the coupling constant $g$. In this normalization, each vertex generated by the action carries a power $g^{-2}$; however, due to the coupling constants power in the propagators it is often common to present the vertices in the form of insertion rules that have been given, for example, in \cite{Beisert:2002bb,Drukker:2009sf}. The insertion rules involving the vertices we are interested in for this chapter are
\begin{align}
    \VertexScalarScalarGluon
    &=
    - g^4 f^{abc} \delta^{IJ} (\pd_1 - \pd_2)_\mu Y_{123}\,, \label{eq:VertexScalarScalarGluon} \\
    \VertexFourScalars
    &=
    - g^6
    \left\lbrace f^{abe}f^{cde} \left( \delta^{IK}\delta^{JL}
    -
    \delta^{IL}\delta^{JK} \right)
    +
    f^{ace}f^{bde} \left( \delta^{IJ}\delta^{KL}
    -
    \delta^{IL}\delta^{JK} \right) \right. \notag \\[-1.5em]
    &\phantom{=\ }
    \left. + f^{ade}f^{bce} \left( \delta^{IJ}\delta^{KL}
    -
    \delta^{IK}\delta^{JL} \right) \right\rbrace X_{1234}\,,
    \label{eq:VertexFourScalars}
\end{align}
where the integrals $Y_{123}$ and $X_{1234}$ are given in the Appendix \ref{app:Integrals}. The other vertices we can get from the action \eqref{eq:BulkAction} are:
\begin{equation}
    \VertexFermionFermionScalar \quad
    \VertexGluonGluonGluon \quad
    \VertexFermionFermionGluon \quad
    \VertexGhostGhostGluon\ \quad
    \VertexScalarScalarGluonGluon \quad
    \VertexGluonGluonGluonGluon\ .
    \label{eq:MoreVertices}
\end{equation}
For future purposes, we also give here the insertion rule for the one-loop correction of the scalar propagator \cite{Erickson:2000af}:
\begin{equation}
    \begin{split}
    \SelfEnergyNoText &=
    \SelfEnergyDiagramOne
    +
    \SelfEnergyDiagramTwo
    +
    \SelfEnergyDiagramThree
    +
    \SelfEnergyDiagramFour \\
    &=
    - 2 g^4 N \delta^{ab} \delta^{IJ}\, Y_{112}\,.
    \end{split}
    \label{eq:SelfEnergy}
\end{equation}
with the integral $Y_{112}$ log-divergent and given in Appendix \ref{app:Integrals}.

\subsection{The Wilson-line defect CFT}
\label{sub:Wilson-line}
In this section, we introduce the extended operator known as \textit{Maldacena-Wilson loop}. This operator plays the role of a defect in the bulk theory $\Nm = 4$ sYM, breaking the superconformal symmetry in a controlled way. We will define local operators living on the defect, the defect CFT, and the defect Feynman rules necessary to compute correlation functions of bulk and defect operators. A useful starting point is the Wilson line in pure Yang-Mills theory \cite{Peskin:1995ev,Weinberg:1996kr,Beccaria:2017rbe}, a non-local operator converting the gauge transformation law of a fermion at point $x$ to that at point $y$
\begin{equation}
    \mathcal{W_B}
    =
    \frac{1}{N}
    \tr \Pm \exp \int_{\mathcal{B}} dx_\mu\,
    i A_\mu \,,
    \label{eq:YMWilsonLine}
\end{equation}
where the integration is taken along any path $\mathcal{B}$ running from $x$ to $y$. The symbol $\Pm$ indicates \textit{path ordering} of the fields in the definition of the exponential to address the non-commuting nature of algebra-valued fields in the non-abelian case. This path ordering prescription defines the correct order in which fields on the line have to be considered: by parametrising the path with a parameter $s$, fields are ordered such that
\begin{equation}
\Pm (\phi(s_1)\phi(s_2) ) = \left\lbrace
\begin{array}{c}
\phi(s_1)\phi(s_2) \quad \text{if} \; s_1>s_2\; ,\\
\phi(s_2)\phi(s_1) \quad \text{if} \; s_2>s_1\; .\\
\end{array}
\right.
\label{eq:PathOrdering}
\end{equation}
The value of Wilson line \eqref{eq:YMWilsonLine} depends on the integration path $\mathcal{B}$; the value can be thought as the phase accumulated by a particle carrying a quantum number associated with a
gauge field, such as an electric charge (Abelian case) or a color charge (non-Abelian), when it is transported around the path. When the path $\mathcal{B}$ is a closed path we obtain a \textit{Wilson loop}
\begin{equation}
    \mathcal{W_B}
    =
    \frac{1}{N}
    \tr \Pm \exp \oint_{\mathcal{B}} dx_\mu\,
    i A_\mu \,,
    \label{eq:YMWilsonLoop}
\end{equation}
which is a non-trivial gauge invariant function of $A_\mu$ -- this can be seen by rewriting the Wilson loop as a function of the field strength $F_{\mu\nu}$ using Stokes theorem \cite{Peskin:1995ev}.The value of the Wilson loop for particular closed paths $\mathcal{B}$ can be connected to physically measurable effects such as the \textit{Aharonov–Bohm effect} in quantum mechanics, where a charged particle experiences a phase shift in its wave function when it is transported around a closed path that encloses the magnetic flux, even when the transport happens in an area where the magnetic field is zero \cite{Aharonov:1959fk,Peshkin:1989zz}. In the context of pure Yang-Mills theory, the expectation value of the Wilson loop is connected to the potential energy between a fermion and an antifermion, and can provide an order parameter for the confinement and deconfinement phases \cite{Rothe:1992nt,Greensite:2011zz,Makeenko:2002uj}.\\

The Wilson line \eqref{eq:YMWilsonLine} and the Wilson loop \eqref{eq:YMWilsonLoop} break all the supersymmetry of $\Nm = 4$ sYM, as it can be easily seen by applying supersymmetry generators on their expressions. It is possible to define a supersymmetric version of the Wilson line by considering also scalar field couplings to the line \cite{Maldacena:1998im,Rey:1998ik}. The supersymmetric Maldacena-Wilson line extended in the time direction is
\begin{equation}
    \Wl
    =
    \frac{1}{N}
    \tr \Pm \exp \int_{-\infty}^\infty d\tau\,
    (i A_0 (\tau) + \theta \cdot \phi (\tau))\,,
    \label{eq:WilsonLine}
\end{equation}
and preserves half of the supersymmetric charges provided that the $SO(6)$ vector $\theta$ has the property $\theta^2 = 1$. Without loss of generality, we can set
\begin{equation}
	\theta
	=
	(0,0,0,0,0,1)\,.
	\label{eq:theta}
\end{equation}
which means only the field $\phi^6$ couples to the line. The two lines \eqref{eq:YMWilsonLine} and \eqref{eq:WilsonLine} are the two endpoints of the renormalization group flow of the line operator where the coupling to the scalar field happens with an additional coupling constant \cite{Beccaria:2021rmj,Beccaria:2022bcr}. Upon the already mentioned \textit{S-duality}, the supersymmetric Wilson line transforms into a 't Hooft line representing a supersymmetric magnetic monopole -- see references \cite{Kapustin:2005py,Gomis:2009xg,Kristjansen:2023ysz} for defect-CFT inspired studies of the topic.\\

The presence of the line extended in the time direction breaks the four-dimensional conformal symmetry into a one-dimensional conformal symmetry on the line \cite{Liendo:2018ukf}
\begin{equation}
    SO (5,1) \to SO (2,1) \times SO (3)\,,
    \label{eq:symmetryBreaking}
\end{equation}
with operators having their scaling dimension $\hat{\Delta}$ as quantum number, and the rotational symmetry around the defect with quantum number $s$, the transverse spin.
The defect also breaks the R-symmetry group $SO(6)_R$ down to $SO(5)_R$, with $k$ denoting the
corresponding quantum number. Altogether, the full supersymmetric algebra $\mathfrak{psu}(2,2|4)$ of $\Nm=4$ sYM is broken down to the defect algebra $\mathfrak{osp}(4^*|4)$  \cite{Liendo:2018ukf,Barrat:2021tpn}. 
For the supersymmetric Wilson line and the circular Wilson loop (in the case representing a quark in the fundamental representation), the expectation values can be computed exactly both at finite and large $N$. This result was first computed in \cite{Erickson:2000af} by considering that only the so-called rainbow graphs contribute to any fixed order and then confirmed rigorously using supersymmetric localization \cite{Pestun:2009nn}. At large $N$, the expectation values read
\begin{eqnarray}
&&\vev{\Wl} = 1 \nonumber \\
&&\vev{\mathcal{W_C}} = \frac{2}{\sqrt{\lambda}}I_1(\sqrt{\lambda})
\end{eqnarray}
where we used the modified Bessel function of the first kind
\beq
I_{\alpha}(x) = \sum_{k=0}^{\infty} \frac{1}{k! \, \Gamma(\alpha+k+1)} \left(\frac{x}{2}\right)^{2k+\alpha}\, .
\eeq
It is noteworthy that despite the Wilson line and the Wilson loop can be mapped into each other via conformal transformation, such a transformation generates quantum anomalies responsible for the non-trivial expectation value of the Wilson loop. This explains the discrepancy in the expectation values between the two trajectories \cite{Drukker:2000rr}. 
%
\subsubsection{Half-BPS defect operators}
Among the representations of the defect algebra, a special class consists of the scalar half-BPS operators $\Oh_\Delta$, which have protected scaling dimensions $ \Dh $ equal to the $R$-symmetry charge $k$ and $s=0$ \cite{Lemos:2017vnx,Liendo:2018ukf,Ferrero:2021bsb}.
We focus on the operators that are inserted along the trace of the Wilson line. In principle, higher-trace operators exist. However, they do not play a significant role in the study of the defect CFT in the limit of large $N$. They do, however, become relevant in the study of correlation functions that involve bulk operators in the presence of a defect, as discussed later in the chapter. Once more, we similarly define these operators to \eqref{eq:SingleTraceHalfBPSOperators_Bulk}:
\begin{equation}
    \Oh_\Dh (u,\tau)
    =
    \frac{1}{\sqrt{n_\Dh}}
    \Wl [ (u \cdot \phi(\tau))^\Dh ]\,,
    \label{eq:HalfBPSDefOperator}
\end{equation}
where $u^2 = 0$ and $u \cdot \theta = 0$ to ensure that the representation remains half-BPS, symmetric, traceless, and decoupled from the field $\phi^6$ present in the Wilson line \cite{Maldacena:1998im,Erickson:2000af}.
Here, $\Wl[\ldots]$ indicates that the fields are \textit{inserted} along the Wilson line, meaning
\begin{equation}
    \Wl [ \Oh ]
    =
    \frac{1}{N} \tr
    \Pm [\, \Oh\, \exp  \int_{-\infty}^\infty  d\tau\,
    (i A_0 (\tau) + \phi^6 (\tau)) ]\,,
    \label{eq:DefInsertion}
\end{equation}
and their presence inside the path ordering operators means that in perturbative calculations the line integrals coming from the exponential are separated in practice by the presence of the operator. Once again, these operators have integer scaling dimensions $\Dh = k$, the number of elementary fields in the operator, and their scaling dimension does not get quantum corrections coming from short-distance divergences \cite{Giombi:2017cqn,Cooke:2017qgm}. The normalization constants $n_\Dh$ depend on the coupling $\lambda$ only and for half-BPS operators, they can be computed using the methods outlined in \cite{Giombi:2018qox}.
Below are the expressions for the operators relevant to this thesis:
\begin{align}
    \nh_1
    &=
    \frac{\sqrt{\lambda}}{2\pi^2} \frac{\Ids_1}{\Ids_2}\,, \label{eq:n1} \\
    \nh_2
    &=
    \frac{1}{4\pi^4} (3 \lambda - (\Ids_1 - 2)(\Ids_1 + 10))\,, \label{eq:n2} \\
    \nh_3
    &=
    \frac{3}{8 \pi^6} \biggl(
    \frac{(5 \lambda + 72) \Ids_1 \Ids_2}{\sqrt{\lambda}}
    - \frac{\lambda (26 \Ids_1 + 3\lambda - 32) + 288(\Ids_1 - 1)}{\Ids_1 -2}
    \biggr)\,,
    \label{eq:n3}
\end{align}
where we define the function
\begin{equation}
    \Ids_a
    =
    \frac{\sqrt{\lambda}\, I_0 (\sqrt{\lambda})}{I_a (\sqrt{\lambda})}\,.
    \label{eq:Ids}
\end{equation}
Note that $n_1$ has a direct \textit{physical} interpretation, being related to the Bremsstrahlung function, as it describes the emission of soft particles from the line defect \cite{Alday:2007hr}. Similarly to the case of bulk operators, we remark that multi-trace operators on the line exist as well and are particularly relevant when considering correlators between bulk and defect operators. When studying the CFT on the line and purely defect correlators, however, we can limit our focus to single-trace operators. Note that there are line operators that are single trace and not protected, such as the insertion of $\phi^6$ or $\phi^I\phi^I$ \cite{Konishi:1983hf,Barrat:2020vch,Barrat:2022eim}. 

\subsubsection{Defect Feynman rules}
\label{ssub:DefectFR}
The presence of the Maldacena-Wilson line introduces new vertices into the theory that were not described in section \ref{subsec:N=4sYM}. One crucial vertex arises from the coupling of the Wilson line to the gluon field, expressed as
\begin{align}
    \DefectVertexOnePointGluon
    \sim
    i \delta^{\mu 0} \int_{\tau_2}^{\tau_3} d\tau_4\, I_{14}\,,
    \label{eq:DefectVertexOnePointGluon}
\end{align}
Here, the contribution of the generators of the gauge group depends on the number of insertions, which is determined by the structure of the bulk action and the correlator of interest.
Additionally, a scalar vertex exists, whose insertion rule form is:
\begin{equation}
    \DefectVertexOnePointScalar
    \sim
    \delta^{I6} \int_{\tau_2}^{\tau_3} d\tau_4\, I_{14}\,,
    \label{eq:DefectVertexOnePointScalar} 
\end{equation}
and once again the contribution from the gauge group depends on the number of insertions. Note that this scalar vertex is essential for ensuring that the Wilson line operator \eqref{eq:WilsonLine} maintains a finite expectation value without requiring renormalization \cite{Guralnik:2004yc}. Note that the vertices arising from the Maldacena-Wilson line do not carry any power of the coupling constant in our normalization (the propagators connecting to the line still carry a factor $g^2$). 
\subsection{Correlation functions of half-BPS operators}
\label{sub:Correlators}
We now turn our attention to the functions that will be studied in this chapter and the next one: correlation functions among half-BPS operators. Thanks to the OPE \eqref{eq:OPEGeneral}, correlation functions in a conformal field theory are an excellent tool to compute conformal data, such as scaling dimensions and three-point functions involving operators that otherwise could involve complicated perturbative calculations, such as non-protected operators. In this thesis, we consider correlators involving both bulk and defect scalar operators: we introduce now the shorthand notation that we will use in the following chapters to denote such operators
\begin{equation}
    \vev{\Delta_1 \ldots \Delta_m \Dh_1 \ldots \Dh_n}
    =
    \vev{\Op_{\Delta_1} (u_1, x_1) \ldots \Op_{\Delta_m} (u_m, x_m) \Wl [\Oh_{\Dh_1} (\uh_1, \tau_1) \ldots \Oh_{\Dh_n} (\uh_n, \tau_n)] }\,.
    \label{eq:Correlators_ShorthandNotation}
\end{equation}
It is clear that the defect operators in \eqref{eq:Correlators_ShorthandNotation} follow the definition for single-trace defect operators given in \eqref{eq:HalfBPSDefOperator}, while bulk operators carry one trace each, thus ensuring gauge invariance. The most studied cases in the literature are the one with $(m,n) = (2,0)$ \cite{Barrat:2021yvp,Barrat:2022psm,Bianchi:2022ppi,Meneghelli:2022gps,Gimenez-Grau:2023fcy,Barrat:2020vch,Drukker:2007yx,Giombi:2009ds,Giombi:2009ek,Buchbinder:2012vr,Beccaria:2020ykg,Billo:2016cpy} and the one with $(m,n) = (0,n)$ \cite{Cavaglia:2021bnz,Cavaglia:2022qpg,Cavaglia:2022yvv,Cavaglia:2023mmu,Giombi:2017cqn,Giombi:2023zte,Liendo:2016ymz,Liendo:2018ukf,Ferrero:2021bsb,Ferrero:2023znz,Ferrero:2023gnu,Giombi:2018qox,Barrat:2021tpn,Barrat:2022eim,Bliard:2024und,Barrat:2024ta,Bliard:2023zpe,Peveri:2023qip,Barrat:2024nod,Barrat:2024ta2,Artico:2024wut} to which the next chapter is dedicated. Conformal field theory and supersymmetry represent strong constraints on the correlation functions of the kind \eqref{eq:Correlators_ShorthandNotation}: in particular, some correlators are completely fixed up to constants. We introduce in the following paragraphs the kinematically fixed correlators of the bulk theory and of the theory in the presence of the supersymmetric Wilson-line defect. 
\subsubsection{Kinematically fixed correlators}
\label{subsubsec:FixedCorrelators}

We begin the analysis of kinematically trivial correlators by discussing the correlators of half-BPS scalar bulk operators without the line defect. These correlation functions are not included in the shorthand notation \eqref{eq:Correlators_ShorthandNotation}, since even the case $n = 0$ involves the presence of a Wilson line without insertions. We use  -- only in this section -- double brackets to identify correlators that do not involve the Wilson-line defect.

\paragraph{Two-point bulk functions.}
Two-point functions are fixed by the $4d$ conformal symmetry to be
\begin{equation}
    \llangle \Delta_1 \Delta_2 \rrangle 
    =
    \delta_{\Delta_1 \Delta_2} (12)^{\Delta_1}\,,
    \label{eq:BulkTwoPointFunctions}
\end{equation}
where we use the shorthand notation
\begin{equation}
    (ij)
    =
    \frac{u_i \cdot u_j}{x_{ij}^2}\,.
    \label{eq:4dSuperPropagator}
\end{equation}
The contraction of the vectors $u_1$ and $u_2$ is the result of supersymmetry (the correlator must not transform under $SO(6)$). It is also important to notice that these kinematically fixed correlators are called \textit{unit normalized} \cite{Semenoff:2002kk} as the normalization factor has been taken care of in the definition of the operators \eqref{eq:SingleTraceHalfBPSOperators_Bulk}.
\paragraph{Three-point bulk functions.}
As mentioned in the introduction in section \ref{sub:CFT}, conformal symmetry fixed the shape of three-point correlators as well. For scalar operators, they read
\begin{equation}
    \llangle \Delta_1 \Delta_2 \Delta_3 \rrangle
    =
    \lambda_{\Delta_1 \Delta_2 \Delta_3}
    (12)^{\Delta_{123}} (23)^{\Delta_{231}} (31)^{\Delta_{312}}\,,
    \label{eq:BulkThreePointFunctions}
\end{equation}
where we have defined
\begin{equation}
    \Delta_{ijk}
    =
    \frac{1}{2}(\Delta_i + \Delta_j - \Delta_k)\,.
    \label{eq:ShorthandThreePoint}
\end{equation}
We can note that again the $R$-symmetry vector contractions are fixed by supersymmetry and the power of the propagators is such that there is always a number $\Delta_i$ of $u_i$ vectors, as it should be to respect the $R$-symmetry structure. In the large $N$ limit, three-point functions of half-BPS single-trace operators are known to be exactly given by \cite{Lee:1998bxa}
\begin{equation}
    \lambda_{\Delta_1 \Delta_2 \Delta_3}
    =
    \frac{\sqrt{\Delta_1 \Delta_2 \Delta_3}}{N}\,.
    \label{eq:ThreePointBulk_Exact}
\end{equation}
The dimensions of the operators $\Delta_i$ and the three-point coefficients $\lambda_{\Delta_i \Delta_j \Delta_k}$ (for all the operators in $N=4$ sYM) represent the conformal data of the theory, as using the operator product expansion it is possible to retrieve any $n$-point correlation function from these data plus the conformal blocks (see for example equation \eqref{eq:ConfBlockExp}).
\paragraph{One-point bulk functions in presence of the defect} The case $m=1\,,\,n=0$ in \eqref{eq:Correlators_ShorthandNotation} represents the first example of a kinematically fixed correlation function in presence of a defect. All one-point correlation functions of local operators in a conformal field theory are vanishing to ensure Lorentz invariance of the vacuum; this statement however is only true if no defects are present in the theory. When a line defect is introduced, then 
\begin{equation}
    \vev{\Delta_1 }
    =
    a_{\Delta_1 } (1 \theta)^{\Delta_1 },
    \label{eq:OnePoint_KinematicStructure}
\end{equation}
with
\begin{equation}
    (i\theta)
    =
    \frac{u_i \cdot \theta}{|\vec{x}_i|}\,.
    \label{eq:MixedSuperPropagatorsITh}
\end{equation}
The variable $\vec{x}_i$ represents the distance vector of the point $x_1$ from the defect line. The coefficients $a_\Delta$ are new conformal data and are functions of the coupling independent of kinematic data, similarly to  $\lambda_{\Delta_i \Delta_j \Delta_k}$ \cite{Giombi:2018hsx}. 
\begin{equation}
    a_{\Delta}
    =
    \frac{\sqrt{\lambda} \sqrt{\Delta}}{2^{\Delta/2+1} N} \frac{\Ids_1}{\Ids_{\Delta}}\,.
    \label{eq:OnePoint}
\end{equation}
Since the normalization of the operators is already fixed, the numbers $a_\Delta$ contain actual new information on the defect theory \footnote{An interesting case of one-point correlation functions playing an important physical role is the one of the finite tempterature bootstrap, where defect and bulk one and two-point correlators can be written as functions
of zero-temperature data and thermal one-point functions \cite{Marchetto:2023xap,Barrat:2024aoa,Barrat:2024fwq}. }. From equation \eqref{eq:OnePoint_KinematicStructure} it is also possible to extract information on the bulk theory, in particular the scaling dimensions of the bulk fields. \\

We now shift our attention to the correlators having several bulk operators $m = 0$, representing correlation functions that involve defect operators only. Since the one-dimensional theory on the defect is still conformal, it is possible to use such symmetry to fix the two- and three-point functions as in the bulk case.

\paragraph{Two-point defect functions.}
Similarly to \eqref{eq:BulkTwoPointFunctions}, defect two-point functions are fixed by the one-dimensional conformal symmetry to be
\begin{equation}
    \vev{\Dh_1 \Dh_2}
    =
    \delta_{\Dh_1 \Dh_2} (\hat{1}\hat{2})^{\Dh_1}\,,
    \label{eq:DefectTwoPoint}
\end{equation}
where we used the notation
\begin{equation}
    (\hat{i}\hat{j})
    =
    \frac{\uh_i \cdot \uh_j}{\tau_{ij}^2}\,.
    \label{eq:1dSuperPropagator}
\end{equation}
The $R$-symmetry indices are again contracted to result in an operator that does not transform under the remaining $SO(5)$ symmetry. We highlight that the operators are unit-normalized and therefore the two-point functions do not have coupling-dependent prefactors.
\paragraph{Three-point defect functions.}
The three-point functions of defect scalar operators are once again fixed by the one-dimensional conformal symmetry and they are given by
\begin{equation}
    \vev{\Dh_1 \Dh_2 \Dh_3}
    =
    \lambdah_{\Dh_1 \Dh_2 \Dh_3}
    (\hat{1}\hat{2})^{\Dh_{123}} (\hat{2}\hat{3})^{\Dh_{231}} (\hat{3}\hat{1})^{\Dh_{312}}\,,
    \label{eq:DefectThreePointFunctions}
\end{equation}
with $\Dh_{ijk}$ defined in the same way as in \eqref{eq:ShorthandThreePoint}.

As opposed to the bulk case, the OPE coefficients $\lambdah_{\Dh_1 \Dh_2 \Dh_3}$ are not currently known in an analytical form, even in the large $N$ limit.
They can however be calculated case by case using the integrability techniques of \cite{Giombi:2018qox}.
We give explicitly here the expression for $\lambdah_{\hat{1}\hat{1} \hat{2}}$, as it is useful for our later calculations:
\begin{equation}
    \lambdah_{\hat{1}\hat{1}\hat{2}}
    =
    \frac{-2 \sqrt{\lambda}
    \Ids_1 (7 + \Ids_1) \Ids_2 \Ids_3 - (-32 + 14 \Ids_1 + \Ids_1^2) \Ids_2^2 \Ids_3 + \lambda (3 \Ids_1 \Ids_2^2 - \Ids_1^2 \Ids_3 + 9 \Ids_2^2 \Ids_3)}{4 \Ids_1 \sqrt{\lambda (3 \lambda - (-2 + \Ids_1) (10 + \Ids_1))}
    \Ids_2 \Ids_3}\,,
    \label{eq:lambdah112}
\end{equation}
with $\Ids_k$ defined in \eqref{eq:Ids}. The coefficients $\Dh_i$ and $\lambdah_{\Dh_i \Dh_j \Dh_k}$ are new conformal data -- meaning that they carry information that is not present in the original bulk theory and they appear in defect OPE -- and together with $a_\Delta$ and their bulk counterparts belong to the set of conformal data of the defect CFT. There is still one kinematically fixed correlator to analyze, carrying its independent conformal data.

\paragraph{Bulk-defect two-point functions.}
Two-point bulk-defect correlators are also fixed kinematically by conformal symmetry. 
\begin{equation}
    \vev{\Delta_1 \Dh_2}
    =
    b_{\Delta_1 \Dh_2} (1 \hat{2})^{\Dh_2} (1 \theta)^{\Delta_1 - \Dh_2},
    \label{eq:BulkDefect_KinematicStructure}
\end{equation}
with
\begin{equation}
    (i\hat{j})
    =
    \frac{u_i \cdot \hat{u}_j}{x_{ij}^2}\,.
    \qquad
    (i\theta)
    =
    \frac{u_i \cdot \theta}{|\vec{x}_i|}\,.
    \label{eq:MixedSuperPropagatorsIJ}
\end{equation}
The normalization of the operators is now fixed and it does not leave the possibility of fixing the coefficients $b_{\Delta \Dh}$ to unity: the bulk-defect coefficients are indeed an additional set of conformal data, and at weak coupling they can be expressed as \cite{Giombi:2018hsx,Barrat:2024nod}
\begin{equation}
    b_{\Delta \Dh} = \frac{\sqrt{\Delta}}{(\Delta - \Dh)!} \frac{\lambda^{(\Delta - \Dh)/2}}{2^{3(\Delta - \Dh)/2} N}
    \biggl(
    1
    + \frac{\lambda}{48} \frac{2 - 5 (\Delta - \Dh)}{2 + \Delta - \Dh}
    + \Om(\lambda^2)
    \biggr)\,.
    \label{eq:BulkDefect_WeakCoupling}
\end{equation}
Note that we consider $\Delta \geq \Dh$ in this formula,
since correlators with $\Delta < \Dh$ vanish.

In principle these OPE coefficients can be calculated exactly using either localization techniques \cite{Giombi:2018hsx} or microbootstrap.
The latter was used in \cite{Barrat:2021un,Barrat:2024nod} to efficiently produce a large number of expressions, and to find a closed form for relevant special cases such as:
\begin{align}
    b_{\Delta \hat{1}}^2
    &=
    \frac{\sqrt{\lambda} \Delta^3}{2^{\Delta + 1} N^2} \frac{\Ids_1 \Ids_2}{\Ids^2_\Delta}\,, \label{eq:BulkDefect_Exact_Delta1} \\
    b_{\Delta \hat{2}}^2
    &=
    \frac{\sqrt{\lambda} \Delta}{2^{\Delta} N^2}
    \frac{
    \left( 
    \sqrt{\lambda}I_3(\sqrt{\lambda})I_{\Delta}(\sqrt{\lambda})
    -
    I_1(\sqrt{\lambda})
      	\left(
      	2(\Delta-1)I_{\Delta-1}(\sqrt{\lambda})+
      	\sqrt{\lambda}I_{\Delta+2}(\sqrt{\lambda})
        \right)
    \right)^2}{I_1(\sqrt{\lambda})
    \left(
    I_1(\sqrt{\lambda})
       \left(
       4I_2(\sqrt{\lambda})+\sqrt{\lambda}I_5(\sqrt{\lambda})
       \right)
    \right)
    -
    \sqrt{\lambda}I_3^2(\sqrt{\lambda})
    }\,,
    \label{eq:BulkDefect_Exact_Delta2}
\end{align}
where the functions $I_a(\sqrt{\lambda})$ are the already introduced modified Bessel functions of the first kind.
When the defect operator is the identity, the two-point coefficients become the one-point coefficients of the bulk operators introduced before in equation \eqref{eq:OnePoint}. We conclude the list of the kinematically fixed correlators with a table summarizing all the correlators discussed and their associated conformal data, referencing to formulas and useful literature.
\begin{table}[h]
    \centering
    \renewcommand{\arraystretch}{1.4}
    \begin{tabular}{|c|c|c|}
        \hline
        \textbf{Correlator Type} & \textbf{Conformal Data} & \textbf{Reference} \\
        \hline
        \multicolumn{3}{|c|}{\textbf{Without Defect}} \\
        \hline
        Bulk 2-point & Operator dimensions $\Delta$ & Eq.~\eqref{eq:BulkTwoPointFunctions} \\
        Bulk 3-point & OPE coefficients $\lambda_{\Delta_1 \Delta_2 \Delta_3}$ & Eqs.~\eqref{eq:BulkThreePointFunctions}, \eqref{eq:ThreePointBulk_Exact}, \cite{Lee:1998bxa} \\
        \hline\hline
        \multicolumn{3}{|c|}{\textbf{With Defect}} \\
        \hline
        Bulk 1-point & One-point coefficients $a_\Delta$ & Eqs.~\eqref{eq:OnePoint_KinematicStructure}, \eqref{eq:OnePoint}, \cite{Giombi:2018hsx} \\
        Defect 2-point & Defect dimensions $\hat{\Delta}$ & Eq.~\eqref{eq:DefectTwoPoint} \\
        Defect 3-point & OPE coefficients $\hat{\lambda}_{\hat{\Delta}_1 \hat{\Delta}_2 \hat{\Delta}_3}$ & Eqs.~\eqref{eq:DefectThreePointFunctions}, \eqref{eq:lambdah112}, \cite{Giombi:2018qox} \\
        Bulk-defect 2-point & Bulk-defect OPE coefficients $b_{\Delta \hat{\Delta}}$ & Eqs.~\eqref{eq:BulkDefect_KinematicStructure}, \eqref{eq:BulkDefect_WeakCoupling}, \cite{Giombi:2018hsx,Barrat:2024nod} \\
        \hline
    \end{tabular}
    \caption{Kinematically fixed correlators and associated conformal data for half-BPS operators.}
    \label{tab:FixedCorrelators}
\end{table}

\subsubsection{Bulk-defect-defect correlators}
Beyond the correlation functions discussed in the previous section, there are no further correlators that are completely fixed kinematically by conformal symmetry. It is still true, though, that conformal symmetry and supersymmetry strongly constrain the correlation functions that, once stripped of a prefactor encoding the transformation properties under the generators of the conformal group, can only depend on conformally invariant ratios of variables. Higher $n$-point correlation functions in the absence of any defect, for example, depend on $n(n-3)/2$ spacetime cross-ratios, where $n < d + 1$. For $n \geq d+1$ the number is $nd-(d+1)(d+2)/2$, where the first term corresponds to the naive number of degrees of freedom, while the second term is the number of generators.\footnote{These correlators have been studied at weak and strong-coupling using a variety of techniques \cite{Drukker:2008pi,Drukker:2009sf,Bercini:2024pya}.} The same is true for pure defect correlators, as they form a $1d$-CFT, with the remark that as $n\geq d+1$ is always valid the number of conformally invariant variables is $n-3$.\\

The case of multipoint defect correlators -- $m=0$ in \eqref{eq:Correlators_ShorthandNotation} -- and the case of two-point functions in presence of the defect are the two most studied kind of correlators in the literature (see references \cite{Barrat:2021yvp,Barrat:2022psm,Bianchi:2022ppi,Meneghelli:2022gps,Gimenez-Grau:2023fcy,Barrat:2020vch,Drukker:2007yx,Giombi:2009ds,Giombi:2009ek,Buchbinder:2012vr,Beccaria:2020ykg,Billo:2016cpy,Cavaglia:2021bnz,Cavaglia:2022qpg,Cavaglia:2022yvv,Cavaglia:2023mmu,Giombi:2017cqn,Giombi:2023zte,Liendo:2016ymz,Liendo:2018ukf,Ferrero:2021bsb,Ferrero:2023znz,Ferrero:2023gnu,Giombi:2018qox,Barrat:2021tpn,Barrat:2022eim,Bliard:2024und,Barrat:2024ta,Bliard:2023zpe,Peveri:2023qip,Barrat:2024nod,Barrat:2024ta2,Artico:2024wut} and the vast literature quoted therein). In this chapter, we focus on the largely overlooked bulk-defect-defect correlators having $m = 1$ and $n = 2$ and representing a particularly simple kinematically non-fixed correlator in the defect CFT. These correlators can be expressed as
\begin{equation}
    \vev{ \Delta_1 \Dh_2 \Dh_3 }
    =
    \Km_{\Delta_1 \Dh_2 \Dh_3} \Am_{\Delta_1 \Dh_2 \Dh_3} (\zeta ; x)\,,
    \label{eq:BDD_Kinematics}
\end{equation}
where $\Am_{\Delta_1 \Dh_2 \Dh_3} (\zeta ; x)$ is a reduced correlator, and $\Km_{\Delta_1 \Dh_2 \Dh_3}$ is a superconformal prefactor encoding the transformation properties of the correlator under conformal symmetry and $R$-symmetry. The conformal cross-ratios are defined as
\begin{equation}
    x
    =
    \frac{|\vec{x}|_{1}^2 \tau_{23}^2}{(\vec{x}_{1}^2+\tau_{12}^2)(\vec{x}_{1}^2+\tau_{13}^2)}\,,
    \label{eq:SpacetimeCrossRatio}
\end{equation}
for the spacetime variables, and 
\begin{equation}
    \zeta
    =
    - \frac{1}{2} \frac{(u_1 \cdot \theta)^2 (\uh_2 \cdot \uh_3)}{(u_1 \cdot \uh_2)(u_1 \cdot \uh_3)}\,.
    \label{eq:RsymmetryCrossRatio}
\end{equation}
The $-1/2$ factor in our definition of $\zeta$ is chosen such that the kinematic limit studied in \cite{Giombi:2018hsx}, which is known as the topological sector and will be introduced later in the chapter, corresponds to the sum of all the $R$ symmetry channels defined below. The prefactor is chosen such that the leading order at weak coupling comes without a factor $\zeta$.
For instance, in the case $\Delta_1 \geq \Dh_2 + \Dh_3$, this results in
\begin{equation}
    \Km_{\Delta_1 \Dh_2 \Dh_3}
    =
    \frac{(1\hat{2})^{\Dh_2} (1\hat{3})^{\Dh_3}}{(1 \theta)^{\Delta_{231}}}\,.
    \label{eq:BDD_PrefactorExample}
\end{equation}
In our conventions, the reduced correlator is a polynomial in the $R$-symmetry variable $\zeta$.
Concretely,
\begin{equation}
    \Am_{\Delta_1 \Dh_2 \Dh_3} (\zeta ; x)
    =
    \sum_{j=1}^r
    \left(\frac{\zeta}{x} \right)^{j-1} F_j (x)\,.
    \label{eq:BulkDefectDefect_RsymmetryChannels}
\end{equation}
We call the functions $F_j$ $R$-symmetry channels. Each further power of $\zeta$ corresponds to two fewer $u_1$ vector contractions with the defect vectors $\hat{u}$ and two more scalar interactions between the line and the bulk operator. The dependency on $x$ in the prefactors is chosen such that they become $1$ in the topological limit of \cite{Giombi:2018hsx}.
The number of channels $r = r(\Delta_1, \Dh_2, \Dh_3)$ for a specific configuration is given by
\begin{equation}
    r
    =
    \left\lfloor \Delta_{1\hat{3}\hat{2}} \right\rfloor + 1 \,,
    \label{eq:NumberOfRsymmetryChannels1}
\end{equation}
for $\Delta_1 < \Dh_2+\Dh_3$ and 
\begin{equation}
    r
    =
    \Dh_3 + 1 \,,
    \label{eq:NumberOfRsymmetryChannels2}
\end{equation}
for $\Delta_1 \geq \Dh_2+\Dh_3$.\footnote{We always assume $\Dh_2 \geq \Dh_3$ without loss of generality.}. To clarify these formulas, consider an example having fixed values $\Dh_2 = 4$ and $\Dh_3 = 2$. For $\Delta_1 = 2$ and $\Delta_1 = 3$ the number of channels is only $1$. Then, for  $\Delta_1 = 4,\,5$ two $R$-symmetry channels are allowed. Finally, for $\Delta_1 \geq 6$ there are 3 different $R$-symmetry channels and their number stops growing. \\

The operator product expansion (OPE) can be used to expand the correlator \eqref{eq:BDD_Kinematics} into conformal blocks. One can either start with an OPE of the two defect operators $\Oh_{\Dh_2} \times \Oh_{\Dh_3}$, or with the OPE of the bulk operator with the line defect, $\Op_{\Delta_1} \times \Wl$.
These two operations lead to the same expression, leading to the conformal block expansion
\begin{equation}
    F_j (x)
    =
    x^{-\Delta_1}
    \sum_{\Dh\: \mathrm{prim.}}
    b_{\Delta_1 \Dh}^{(j)} \lambdah_{\Dh_2 \Dh_3 \Dh}^{(j)}
    g_{\Dh} (x)\,,
    \label{eq:ConformalBlockExpansion}
\end{equation}
where the sum runs over all the primaries, including superdescendants. Because of the equivalence of the OPE expansions, there exist no crossing relations connecting different expansions of the same correlators. Although not specified explicitly here to avoid cluttering the notation, the blocks $g_{\Dh} (x)$ depend on the scaling dimensions $\Dh_2$ and $\Dh_3$ (but not $\Delta_1$).
They have been determined in \cite{Buric:2020zea,Okuyama:2024tpg} to be
\begin{equation}
    g_{\Dh} (x)
    =
    x^{\Dh/2}\,
    _2F_1
    \left(
    (\Dh +\Dh_2-\Dh_3)/2,
    (\Dh -\Dh_2+\Dh_3)/2;
    \Dh+1/2;
    x
    \right)\,.
    \label{eq:ConformalBlocks}
\end{equation}

\begin{figure}[h]
    \centering
    \OPE
    \caption{Illustration of the operator product expansion for the bulk-defect-defect correlators.
    The result is an expansion in conformal blocks, with the OPE coefficients being given by the product of defect three-point and bulk-defect two-point coefficients.}
    \label{fig:OPE}
\end{figure}

The OPE is illustrated in Figure \ref{fig:OPE}. Representations of the correlators different from the sum \eqref{eq:ConformalBlockExpansion} can also be obtained: in particular, it is possible to express the sum as over superprimaries instead of primaries, by determining the superconformal blocks. Further in the chapter we provide the superblock expansion for the case $\vev{2 \hat{1} \hat{1}}$. An alternative expansion of bulk-defect-defect was derived in \cite{Levine:2023ywq,Levine:2024wqn} by using locality properties of the correlator: this block expansion is known as the local block expansion.\\

In the rest of the chapter, we will focus on the computation of bulk-defect-defect correlators using a mix of non-perturbative and perturbative techniques (mostly for a small 't Hooft coupling \eqref{eq:tHooft_coupling}). Before presenting the results up to NLO in the perturbative expansion, we introduce the non-perturbative constraints on the correlators which we use to reduce the perturbative information needed to complete the calculation.

\section{Non-perturbative constraints}
\label{sec:BDDNonPerturbativeConstraints}
We start the presentation of non-perturbative constraints by showing how superconformal symmetry can be used to trade a function of the spacetime cross-ratio for a constant. 
This constant, called the topological sector, can be evaluated using localization techniques \cite{Pestun:2007rz,Pestun:2016zxk,Preti:2016hhk,Giombi:2018hsx}.
We then present the superconformal block expansion -- which will be used later in section \ref{subsec:211} to study the case $\vev{2 \hat{1} \hat{1}}$ at strong coupling.
Finally, we briefly discuss locality properties of the block expansion introduced in \cite{Levine:2023ywq,Levine:2024wqn} and pinching and splitting limits, which relate parts of the bulk-defect-defect correlators to lower-point functions.

\subsection{Topological sector}
\label{subsec:Topological}
Superconformal symmetry can typically be encoded in the form of superconformal Ward identities (SCWI). 
They are powerful tools for constraining correlators of half-BPS operators, both in $\Nm=4$ sYM \cite{Dolan:2001tt,Dolan:2004iy,Dolan:2004mu} and in the Wilson-line defect CFT \cite{Liendo:2016ymz,Liendo:2018ukf,Barrat:2021tpn,Bliard:2024und,Barrat:2024ta,Artico:2024wut}.
For bulk-defect-defect correlators, SCWIs are currently not known and a rigorous derivation has yet to be done. The attempt of SCWI guessing done in \cite{Artico:2024wnt} has led to a differential constraint that is fully encoded by the existence of the \textit{topological sector}.\\
Our starting point is the topological sector calculated in \cite{Giombi:2018hsx}. This sector corresponds to a specific $R$-symmetry polarization of a bulk-defect-defect correlator, \textit{i.e.} a specific choice of vectors $u_1$, $\uh_2$, and $\uh_3$ such that $\zeta = x$. The powerful idea explored in \cite{Giombi:2018hsx} is the correspondence of the correlators in this subsector with the bosonic two-dimensional Yang-
Mills theory on $S^2$. The computation in the two-dimensional Yang-Mills
theory can then be further reduced to simple Gaussian (multi-)matrix models \cite{Giombi:2009ms,Pestun:2007rz,Giombi:2018qox}, and the
results depend only on the area of the region inside the Wilson loop. The topological sector of bulk-defect-defect correlators is related to a small deformation of the Wilson loop and can be computed by acting with differential operators on one-point correlators in the presence of the defect. The result has no kinematic dependence; in our conventions, the topological sector can be expressed as the sum of the $R$-symmetry channels of \eqref{eq:BulkDefectDefect_RsymmetryChannels}:
\begin{equation}
    \sum_{j=1}^{r} F_j (x)
    =
    \Fds_{\Delta_1 \Dh_2 \Dh_3}\,.
    \label{eq:TopologicalSector}
\end{equation}
The right-hand side $\Fds_{\Delta_1 \Dh_2 \Dh_3}$ is a constant, in the sense that it depends neither on spacetime nor $R$-symmetry variables.
It is however a function of the coupling $\lambda$, exactly as scaling dimensions and three-operator interaction constants $\lambda_{\Delta_1\Delta_2\Delta_3}$.
Equation \eqref{eq:TopologicalSector} can immediately be used to eliminate one $R$-symmetry channel in \eqref{eq:BulkDefectDefect_RsymmetryChannels}:
\begin{equation}
    F_r (x)
    =
    \Fds_{\Delta_1 \Dh_2 \Dh_3}
    -
    \sum_{j=1}^{r-1} F_j (x)\,.
    \label{eq:SolutionWI}
\end{equation}
In fact, in \cite{Artico:2024wnt}, we formulated an Ansatz for the Ward identities based on known examples  \cite{Liendo:2016ymz,Liendo:2018ukf,Barrat:2021tpn,Bliard:2024und,Barrat:2024ta,Artico:2024wut} and used the results of the correlators up to NLO at weak coupling to constrain the numerical coefficients.
We found the differential constraint
\begin{equation}
    \left.
    \bigl(
    \pd_x + \pd_\zeta
    \bigr)
    \Am_{\Delta_1 \Dh_2 \Dh_3} (\zeta; x)
    \right|_{\zeta = x}
    =
    0\,.
    \label{eq:SCWI}
\end{equation}
Unfortunately, this equation can be shown not to encode more constraints than \eqref{eq:TopologicalSector}.
It is possible that the fact that the correlator depends on only one spacetime and one $R$-symmetry cross-ratio does not leave much room for supersymmetry to constrain its form any further.
It is also possible that differential constraints not taking the form \eqref{eq:SCWI} are being missed by this strategy.
For instance, in five-point functions of half-BPS operators in $\Nm=4$ sYM, this Ansatz approach proved unsuccessful, although it is known that differential constraints exist \cite{Meneghelli:2024ta}.
For the time being, we use the constraint \eqref{eq:SolutionWI} to study the correlators at leading and next-to-leading orders at weak coupling in the rest of the chapter. Although we know that the value of the topological sector \eqref{eq:TopologicalSector} only depends on the 't Hooft coupling, to give precise results for the correlators it is important to fix the value of this constant to give correct results for bulk-defect-defect correlators. The next section deals with the expansion of localization results from \cite{Giombi:2018hsx} up to NLO at weak coupling.

\subsubsection{Localization results}
\label{subsubsec:LocalizationResults}

The topological sector can be evaluated for arbitrary $\Delta_1$, $\Dh_2$, $\Dh_3$ using the localization techniques of \cite{Giombi:2018hsx}. The derivatives with respect to the area of the Wilson-loop of the one-point bulk correlation functions lead to angular integrals of rational functions of trigonometric functions. Consider for example the case $\Delta_1\geq\Dh_2+\Dh_3$: the arguments in \cite{Giombi:2018hsx} express the localization result as
\beq
\mathbb{G}_{\Delta_1\Dh_2\Dh_3} = (- i g)^{\Dh_2+\Dh_3} 2^{-\Delta_1/2}\frac{\sqrt{\Delta_1}}{N}  \int_0^{2\pi} d\mu B(\Delta_1)  U_C(\Dh_2, \cos \theta) U_C(\Dh_3, \cos \theta)
\eeq
where we used a different letter as this result is derived for a circular Wilson line and under a different normalization\footnote{Note that this only results in the need to multiply the localization result from  \cite{Giombi:2018hsx} by a correct normalization factor to get the results reported in the following section.}. The functions $U_C(\Dh, x)$ are Chebyshev polynomials in the variable $x$ and the measure $d\mu B(\Delta_1)$ can be written as
\beq
d\mu B(\Delta_1) = \frac{ e^{i\Delta_1\theta} }{2\pi i^{\Delta_1}}  \exp (4\pi g \sin \theta) d\theta
\label{eq:LocalizationMeasure}
\eeq
In principle, the localization result is exact and the exponential in \eqref{eq:LocalizationMeasure} can be expanded to any order in perturbation theory; in practice, the integrals appearing may become complicated and prevent an analitical expression\footnote{To be completely precise, we should mention that for higher orders in perturbation theory the product of Chebyshev polynomials becomes the product of two sums of Chebyshev polynomials.}. Here we expand the localization results to next-to-leading order, obtaining explicit formulas for all cases.

\begin{figure}
\centering
 \includegraphics[width=.7\linewidth]{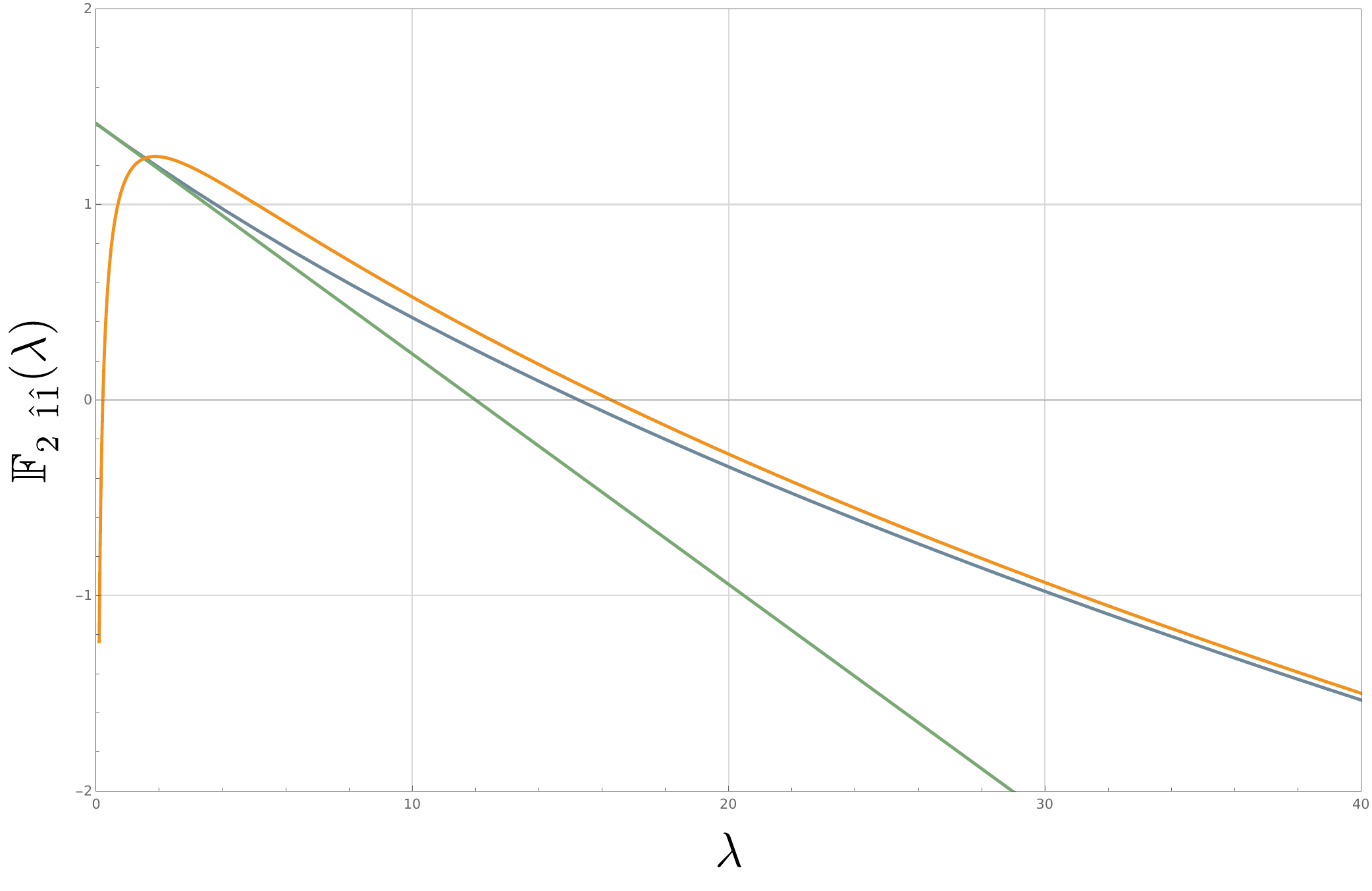}
\caption{The figure displays the topological sector for $\vev{2 \hat{1} \hat{1}}$.
The blue line is the exact expression, presented in \eqref{eq:Fds211_Exact}.
The green and orange lines correspond, respectively, to the weak- and strong-coupling expansions up to next-to-leading order (weak coupling) and next-to-next-to-leading order (strong coupling), derived in \cite{Giombi:2018hsx} and reproduced here in \eqref{eq:Fds_weak}, \eqref{eq:Fds1_weak}, and \eqref{eq:Fds211_strong}.
}
\label{fig:TopologicalSector211}
\end{figure}

\paragraph{Weak coupling.}

At weak coupling, the topological sector follows the perturbative structure
\begin{equation}
    \Fds_{\Delta_1 \Dh_2 \Dh_3}
    =
    \frac{\lambda^{a/2}}{N}
    \biggl(
    \Fds_{\Delta_1 \Dh_2 \Dh_3}^{(0)}
    +
    \lambda
    \Fds_{\Delta_1 \Dh_2 \Dh_3}^{(1)}
    + \ldots
    \biggr)\,,
    \label{eq:Fds_PerturbativeStructure}
\end{equation}
where $a = \text{min} (a_4)$ is introduced in \eqref{eq:1dSuperPropagator} and corresponds to the minimum number of propagators connecting the operator $\Op_{\Delta_1}$ and the Wilson line. For a given configuration $\vev{\Delta_1 \Dh_2 \Dh_3}$, the leading order is given by
\begin{equation}
    \begin{split}
   \Fds_{\Delta_1 \Dh_2 \Dh_3}^{(0)}
   &=
   \frac{2^{\frac{1}{2} (2\Delta_{\hat{2}\hat{3}1}-2)} i^{a-2\Delta_{\hat{2}\hat{3}1}} \sqrt{\Delta_1} }{\Gamma (a+1)} 
   \left(
   (2-2\Delta_{\hat{2}\hat{3}1}) \Theta (1-a)
   +
   2^{1-a} \Theta (a)\right)\,,
   \end{split}
   \label{eq:Fds_weak}
\end{equation}
where we use the following conventions for the Heaviside step function:
\begin{equation}
    \Theta (x)
    =
    \begin{cases}
        0\,, &\text{ if } x \leq 0\,, \\
        +1\,, &\text{ if } x > 0\,.
    \end{cases}
    \label{eq:Heaviside}
\end{equation}
This formula is the compact version of the results presented in Appendix B.1 of \cite{Giombi:2018hsx}.\\
The next-to-leading order contribution to $\Fds_{\Delta_1 \Dh_2 \Dh_3}$ is 0 for $\Delta_1 \leq \Dh_2-\Dh_3$ and $\Delta_1 < \Dh_2 + \Dh_3$ with $ \Dh_2 + \Dh_3 - \Delta_1$ odd. For the other cases ($\Delta_1 \geq \Dh_2 + \Dh_3$ or $\Delta_1 < \Dh_2 + \Dh_3$ with $ \Dh_2 + \Dh_3 - \Delta_1$ even), we find
\begin{equation}
    \begin{split}
   \Fds_{\Delta_1 \Dh_2 \Dh_3}^{(1)}
   =&
   -\frac{2^{-1+3\Delta_{\hat{2}\hat{3}1}} \sqrt{\Delta_1} \Delta_{\hat{2}\hat{3}1} \lambda^{a/2} }{N}
   \bigg[
   \frac{(-1)^{\Delta_{\hat{2}\hat{3}1}}}{2^{2\Delta_{\hat{2}\hat{3}1}}\Gamma(4+a)}
   \Theta (\Delta_{\hat{2}\hat{3}1})
     \\ 
   &+\frac{1}{\Gamma(3-2\Delta_{\hat{2}\hat{3}1})}
   \left(1-\Theta (\Delta_{\hat{2}\hat{3}1})\right)
   \bigg] -\frac{1}{12}\Fds_{\Delta_1 \Dh_2 \Dh_3}^{(0)}\,,
   \end{split}
   \label{eq:Fds1_weak}
\end{equation}
where the Heaviside step function is defined as in \eqref{eq:Heaviside}.
\paragraph{Strong coupling.}
The strong-coupling limit of the topological sector can be evaluated using the already mentioned tools of \cite{Giombi:2018hsx} for large 't Hooft coupling.
We focus here only on the case $\vev{2\hat{1}\hat{1}}$, since it is the only one that we consider at strong coupling.
It reads
\begin{equation}
    \Fds_{2\hat{1}\hat{1}}
    =
    \frac{1}{N}
    \biggl(
    -
    \frac{\sqrt{\lambda}}{\sqrt{2}}
    +
    \frac{9}{2\sqrt{2}}
    -
    \frac{15}{8\sqrt{2}\sqrt{\lambda}}
	+
    ...
    \biggr)
    +
    \ldots\,.
    \label{eq:Fds211_strong}
\end{equation}
In the above formula, the expansion within the brackets is in powers of $1/\sqrt{\lambda}$ while the dots outside of the brackets represent the expansion in $1/N$. 
The topological sector at weak and strong coupling is plotted in Figure \ref{fig:TopologicalSector211} for $\vev{2 \hat{1} \hat{1}}$, together with its exact expression introduced in the next section.

\subsection{Superblock expansion}
\label{subsec:SuperblockExpansion}

The constraints from superconformal symmetry discussed above can be used to derive an expansion in superconformal blocks for the bulk-defect-defect correlators.
Generically, the expansion in superblocks re-sums the contributions of the conformal primaries that are not superconformal in the conformal block expansion \eqref{eq:ConformalBlockExpansion}. The superconformal block expansion gives the expansion for the full correlator (as opposed to the conformal block expansion that is valid for each $R$-symmetry channel separately) and it takes the form
\begin{equation}
    \Am_{\Delta_1 \Dh_2 \Dh_3} (\zeta ; x)
    =
    \sum_{\Oh} b_{\Delta_1 \Oh} \lambdah_{\Dh_1 \Dh_2 \Oh} \Gm_{\Oh} (\zeta ; x)\,,
    \label{eq:GeneralCase_SuperblockExpansion}
\end{equation}
where $\Oh$ designates the superprimary operators described in section \ref{sub:SUSY} appearing in the OPE $\Oh_{\Dh_2} \times \Oh_{\Dh_3}$.
For the case of the Wilson-line CFT, this OPE is given, e.g., in \cite{Ferrero:2023znz} and reads
\begin{align}
    \Oh_{\Dh_2} \times \Oh_{\Dh_3}
    \longrightarrow
    \sum_{m = \Dh_{23} \text{ step }2}^{\Dh_2 + \Dh_3} \Bm_m
    +
    \sum_{i=0}^{\Dh_2 - 1} \sum_{j=0}^i \sum_{\Dh^{(0)} > 2i + \Dh_{23} + 1} \Lm_{[2i -2j, 2j + \Dh_{23}], 0}^{\Dh}\,.
    \label{eq:OPE}
\end{align}
In the case $\Dh_2 = \Dh_3$, the leading contribution corresponds to the identity through the identification $\Bm_0 = \hat{\mathbf{1}}$.
The operators $\Bm_m$ are half-BPS and therefore protected from corrections to their scaling dimensions, while the so-called \textit{long operators} $\Lm_{[a,b],0}^{\Dh}$ are unprotected and have a non-vanishing anomalous dimension.
The lowest-lying non-protected operator (both at weak and strong coupling) is $\phi^6$, which has dimension \cite{Alday:2007hr,Grabner:2020nis,Cavaglia:2021bnz}
\begin{align}
    \Delta_{\phi^6}
    &\overset{\lambda \sim 0}{=}
    1 + \frac{\lambda}{4 \pi^2} + \ldots\,, \label{eq:Delta_phi6_weak} \\
    \Delta_{\phi^6}
    &\overset{\lambda \gg 1}{=}
    2 - \frac{5}{\sqrt{\lambda}} + \ldots\,. \label{eq:Delta_phi6_strong}
\end{align}
In this section, we will focus on the bulk-defect-defect correlator of the lowest dimensional protected operators: $\vev{2 \hat{1} \hat{1}}$. For this correlator, the content of the OPE is
\begin{equation}
    \Oh_1 \times \Oh_1 \longrightarrow \hat{\mathbf{1}} + \Bm_2 + \sum_{\Dh^{0}\geq 1} \Lm_{[0,0],0}^\Dh\,.
    \label{eq:OPE_Oh1}
\end{equation}
We now discuss how to derive the explicit form of the superblock expansion presented in \cite{Artico:2024wnt}\footnote{We thank Gabriel Bliard and Philine van Vliet for sharing their preliminary results for these superblocks and useful conversations around this topic.}. First of all, it is necessary to remark that in a perturbative setting, operators can become degenerate \cite{Liendo:2018ukf,Lauria:2020emq}.
The degeneracy grows with the tree-level scaling dimension $\Delta^{(0)}$.
The large $N$ limit helps to reduce the degeneracy, but there are still many operators with the same tree-level scaling dimensions.\footnote{A counting of the number of operators can be found in \cite{Ferrero:2023znz}.}

From \eqref{eq:OPE_Oh1}, we can read the expansion of $\vev{2 \hat{1} \hat{1}}$ in superblocks:
\begin{equation}
    \Am_{2\hat{1}\hat{1}} (\zeta; x)
    =
    a_2 \Gm_{\hat{\mathbf{1}}} (\zeta ; x) + b_{2\hat{2}} \lambdah_{\hat{1}\hat{1}\hat{2}} \Gm_{\hat{2}} (\zeta ; x) + \sum_{\Dh} b_{2 \Dh} \lambdah_{\hat{1} \hat{1} \Dh} \Gm_{\Dh} (\zeta ; x)\,,
    \label{eq:211_SuperblockExpansion}
\end{equation}
where $\Dh$ in the last sum refers to the operators of $\Lm_{[0,0],0}^{\Dh}$.

Each superblock can be decomposed into an $R$-symmetry and a spacetime part through the following sum over the Dynkin labels of the $R$-symmetry and the scaling dimensions, here labelled as $r$:
\begin{equation}
    \Gm_{\Oh} (\zeta ; x)
    =
    \sum_{a,b,r}
    \alpha_{[a,b],r} h_{[a,b]}(\zeta) g_{r} (x)\,,
    \label{eq:Superblock_Decomposition}
\end{equation}
where $g_{\Dh} (x)$ refers to the bosonic blocks already introduced in \eqref{eq:ConformalBlocks}.
The range of the sums is determined by considering the content of the supermultiplets.
This analysis was done in \cite{Ferrero:2023znz} for the case of the Wilson-line defect CFT.
We then apply the superconformal constraints \eqref{eq:SCWI} on the blocks (or equivalently the topological sector constraint \eqref{eq:SolutionWI} to the correlator) to fix the open coefficients $\alpha_{[a,b],r}$.
This amounts to requiring that individual blocks satisfy the symmetries of the correlator.

The identity block is simply given by
\begin{equation}
    \Gm_{\hat{\mathbf{1}}} (\zeta ; x)
    =
    - \frac{2 \zeta}{x}\,,
    \label{eq:Superblock_Identity}
\end{equation}
where the factor $2$ is there to account for the one present in the definition \eqref{eq:RsymmetryCrossRatio}.

For the operator $\Oh_2$, we see in (B.4) of \cite{Ferrero:2023znz} that the block should take the form
\begin{equation}
    \Gm_{\hat{2}} (\zeta; x)
    =
    \alpha_{[0,2],2} h_{[0,2]} (\zeta) x^{-1} g_2 (x)
    +
    \alpha_{[2,0],3} h_{[2,0]} (\zeta) x^{-1} g_3 (x)
    +
    \alpha_{[0,0],4} h_{[0,0]} (\zeta) x^{-1} g_4 (x)\,.
    \label{eq:Superblock_B2_Form}
\end{equation}
Note that we have selected scalar contributions in the supermultiplet.
Generically, $R$-symmetry blocks depend on the external operators, and for $\vev{2\hat{1}\hat{1}}$ they are expected to take the form
\begin{equation}
    h_{[a,b]} (\zeta)
    =
    \beta_{[a,b]}^{(0)}
    +
    \beta_{[a,b]}^{(1)} \zeta\,,
    \label{eq:RsymmetryBlock_Ansatz}
\end{equation}
where the coefficients $\beta_{[a,b]}^{(0)}$ are so far unfixed.
Applying the SCWI \eqref{eq:SCWI} and choosing the normalization according to the highest-weight channel, we finally obtain
\begin{equation}
    \Gm_{\hat{2}} (\zeta; x)
    =
    \biggl( 1 - \frac{2}{5} \zeta \biggr) x^{-1} g_2 (x) - \frac{12}{175} \frac{\zeta}{x} g_4 (x)\,,
    \label{eq:Superblock_B2}
\end{equation}
The same method can be applied to the superblocks of the long operators, yielding
\begin{equation}
    \begin{split}
        \Gm_{\Dh} (\zeta ; x)
        &=
        -\frac{\zeta}{x} g_\Dh (x)
        +
        \biggl( 1-\frac{2 \left(\Delta ^2+3 \Delta +1\right) \zeta }{4 \Delta ^2+12 \Delta +5} \biggr) x^{-1} g_{\Dh+2} (x) \\
        &\phantom{=\ }
        -
        \frac{(\Delta +2)^2 (\Delta +3)^2 }{(2 \Delta +3) (2 \Delta +5)^2 (2 \Delta +7)} \frac{\zeta}{x} g_{\Dh+4} (x)\,.
    \end{split}
    \label{eq:SuperblockLong}
\end{equation}

The expansion in superblocks can be expanded in the coupling constant, through
\begin{equation}
    \Delta
    =
    \Delta^{(0)} + \lambda \gamma^{(1)}_{\Delta^{(0)}} + \ldots\,,
    \label{eq:Delta_Expansion}
\end{equation}
and similarly for OPE coefficients.
This allows a perturbative analysis, which will play an important role further in the chapter.
It is important however to notice that, as mentioned above, OPE coefficients and scaling dimensions become degenerate in a perturbative setting.
Instead of individual coefficients, we are therefore forced to consider their \textit{average} only.
For instance,
\begin{equation}
    \vev{ b_{2 \Dh}^{(\ell)} \lambdah_{\hat{1} \hat{1} \Dh}^{(\ell)} }
    =
    \sum_{\text{deg}} b_{2 \Dh}^{(\ell)} \lambdah_{\hat{1} \hat{1} \Dh}^{(\ell)}\,,
    \label{eq:AverageOPE_Definition}
\end{equation}
where the sum is over all the operators that have the same tree-level scaling dimension $\Delta^{(0)}$.
Note that we are using here the notation of \cite{Ferrero:2023znz}.
Moreover, the same analysis can be performed at strong coupling, and it is known that in this regime the degeneracy is lifted slower than at weak coupling \cite{Liendo:2018ukf,Ferrero:2021bsb}.\\

The superblock expansion can be studied in the topological sector as well.
It is well-known that only half-BPS contributions survive in this kinematic limit \cite{Giombi:2018hsx}.
For the case of $\vev{2 \hat{1} \hat{1}}$, we have
\begin{equation}
    \Fds_{2 \hat{1} \hat{1}}
    =
    -2 a_2 + \vev{b_{2\hat{2}} \lambdah_{\hat{1} \hat{1} \hat{2}}}\,.
    \label{eq:Fds211_AlmostExact}
\end{equation}
This provides an exact formula for the topological sector discussed in Section \ref{subsubsec:LocalizationResults}.
The average above corresponds to the contributions from the two operators $\Oh_{2}$ and $\Oh_{(0|2)}$ -- an operator defined from multi-trace operators via a Gram-Schmidt procedure \cite{Giombi:2018hsx} to obtain orthogonal operators to the single trace ones\footnote{In reference \cite{Artico:2024wnt} we define the non orthogonal multi-trace operators employed in this procedure.}.
This degeneracy is in fact absent, as it can be seen in the following way.
From the Gram-Schmidt procedure mentioned above, the good operator $\Oh_{(0|2)}$ takes the form
\begin{equation}
    \Oh_{(0|2)}
    \sim
    \bigl(
    \hat{\mathrm{O}}_{(0|2)}
    -
    \vev{\Oh_2 \hat{\mathrm{O}}_{(0|2)}} \Oh_2
    \bigr)\,.
    \label{eq:DoubleTrace_GramSchmidt}
\end{equation}
The corresponding OPE coefficient $b_{2(0|2)}$ is of order $\Om(N^0)$.
However the three-point function $\lambdah_{\hat{1} \hat{1} (0|\hat{2})}$ vanishes as a consequence of \eqref{eq:DoubleTrace_GramSchmidt}.
The topological sector is therefore simply given by
\begin{equation}
    \Fds_{2 \hat{1} \hat{1}}
    =
    -2 a_2 + b_{2 \hat{2}} \lambdah_{\hat{1} \hat{1} \hat{2}}\,.
    \label{eq:Fds211_Exact}
\end{equation}
This expression is plotted in figure \ref{fig:TopologicalSector211} from weak to strong coupling, using the analytic results \eqref{eq:OnePoint}, \eqref{eq:BulkDefect_Exact_Delta2} and \eqref{eq:lambdah112}, and is shown to agree well with the perturbative results of \cite{Giombi:2018hsx}.

\subsection{Pinching and splitting}
\label{subsec:PinchingAndSplittingLimits}

In this section, we present two kinematic limits corresponding to the reduction of some bulk-defect-defect correlators to kinematically fixed functions, in particular to two-point bulk-defect functions (pinching limit) and products of one-point bulk correlators and two-point defect correlators (splitting limit). In our pre-turbative bootstrap framework introduced in \cite{Artico:2024wnt,Artico:2024wut}, lower-point functions serve as further non-perturbative information constraining the correlators under study and can serve to fix overall constraints in the $R$-symmetry channels.

\subsubsection{Pinching limit}
\label{subsubsec:PinchingLimits}

The pinching of the two defect operators into one having scaling dimension being the sum of the original two, i.e., the limit of the correlator $\tau_3 \rightarrow \tau_2$ and  $\uh_3 \rightarrow \uh_2$,  results in the OPE coefficient $b_{\Delta_1 \Dh}$ after accounting for the different normalization constants:
\begin{align}
    \vev{ \Delta_1 \Dh }
    =
    \frac{\sqrt{\nh_{\Dh_2} \nh_{\Dh_3}}}{\sqrt{ \nh_{\Dh}} }
    \lim\limits_{3 \to 2}\,
    \vev{ \Delta_1 \Dh_2 \Dh_3 }
    =
    b_{\Delta_1 \Dh}\,
     (12)^{\Dh} (1 \theta)^{\Delta_1 - \Dh} \,,
    \label{eq:PinchingBulkDefectDefect}
\end{align}
with $\Dh = \Dh_2 + \Dh_3$.
This provides a valuable limit of the bulk-defect-defect correlator, since the coefficients $b_{\Delta_1 \Dh}$ can be evaluated (see \eqref{eq:BulkDefect_WeakCoupling}). For this limit to be possible, it is necessary that at least one channel without contractions $(\hat{2}\hat{3})$ is present, as it becomes zero in this limit. This condition corresponds to $\Delta_1\geq\Dh_2+\Dh_3$.
If one inserts the expansion in channels \eqref{eq:BulkDefectDefect_RsymmetryChannels} in \eqref{eq:PinchingBulkDefectDefect}, only the channel $F_1$ computed at $x = 0$ survives, as $\zeta = 0$
\begin{align}
   \vev{ \Delta_1 \Dh }
    &=  (12)^{\Dh_2 + \Dh_3} (1\theta)^{-2\Delta_{\hat{2}\hat{3}1}}
    \biggl(
    F_1^{(0)} (0)
    + \lambda
    \biggl(
    F_1^{(1)} (0) - \frac{F_1^{(0))} (0)}{48}
    \biggr)
    + \Om(\lambda^2)
    \biggr)\,,
    \label{eq:BulkDefectInF0}
\end{align}
from which it is possible to extract the pinching constraints at leading and next-to-leading orders:
\begin{align}
    F_1^{(0)} (0) &=
    \frac{\sqrt{\Delta_1}}{(-2\Delta_{\hat{2}\hat{3}1})!} \frac{\lambda^{-\Delta_{\hat{2}\hat{3}1}}}{2^{-3\Delta_{\hat{2}\hat{3}1}} N}\,, \label{eq:PinchingConstraint_LO} \\
    F_1^{(1)} (0) &=
    \frac{\sqrt{\Delta_1}}{(-2\Delta_{\hat{2}\hat{3}1})!} \frac{\lambda^{-\Delta_{\hat{2}\hat{3}1}}}{2^{(-3\Delta_{\hat{2}\hat{3}1}+2)} N} 
    \frac{(1+2\Delta_{\hat{2}\hat{3}1})}{3 (2 - 2\Delta_{\hat{2}\hat{3}1})}\,.
    \label{eq:PinchingConstraint_NLO}
\end{align}

It might be tempting to consider another pinching limit, which consists of bringing the bulk operator in a third position on the line in order to generate a double-trace defect operator.
This operator would not be orthonormal to the single-trace operators and will be discussed later in the chapter.

\subsubsection{Splitting limit}
\label{subsubsec:SplittingLimits}

A different limit comes from considering a large separation between the bulk operator and the defect ones, i.e., $|\vec{x}_1| \to \infty$.
In the correlator expressed in $R-$symmetry channels \eqref{eq:BulkDefectDefect_RsymmetryChannels}, this corresponds to the limit $x \rightarrow 0$ of the whole correlator without setting $\zeta = 0$ (which would correspond to the previous pinching as mentioned above).
In this case, the bulk-defect-defect correlator factorizes into a product of a two-point defect correlator and a one-point function of a bulk operator. Note that since the operators do not change, there is no role played by the normalization factors. The correlator thus becomes  
\begin{equation}
    \lim\limits_{\chi \to 0}\,
    \vev{ \Delta_1 \Dh_2 \Dh_3 } = \vev{\Delta_1} \vev{\Dh_2 \Dh_3}\,,
    \label{eq:SplittingADelta}
\end{equation}
from which \eqref{eq:BulkDefectDefect_RsymmetryChannels} simplifies to 
\begin{equation}
        \lim\limits_{x \to 0}\,
    \vev{ \Delta_1 \Dh_2 \Dh_3 } =  a_{\Delta_1} \Km_{\Delta_1\Dh_2\Dh_3} \left(-2\frac{\zeta}{x}\right)^{\Dh_2} \delta_{\Dh_2\Dh_3}\,,
    \label{eq:Splitting}
\end{equation}
where the coefficient $a_{\Delta_1}$ is given in \eqref{eq:OnePoint}. Note that the factor $(-2)$ in \eqref{eq:Splitting} is there because of the definition we have chosen of the $R$-symmetry cross ratio \eqref{eq:RsymmetryCrossRatio}, which simplifies the formulation of the topological sector.
The splitting limit can serve as a constraint or consistency check for the case $\Dh_2 = \Dh_3$: Equation \eqref{eq:Splitting} combined with \eqref{eq:BulkDefectDefect_RsymmetryChannels} gives
\begin{equation}
        F_{\Dh_2 +1}(0) =  \left( - 2\right)^{\Dh_2} a_{\Delta_1} \delta_{\Dh_2\Dh_3}\,,
    \label{eq:SplittingCheck}
\end{equation}
which is valid non-perturbatively. Note that \eqref{eq:SplittingCheck} is consistent with the non-perturbative result for the topological sector of $\vev{2\hat{1}\hat{1}}$ in \eqref{eq:Fds211_AlmostExact}.
\subsection{Locality}
\label{sub:Locality}
In this section, we review the principles behind the local block expansion introduced for bulk-defect-defect correlators in \cite{Levine:2023ywq,Levine:2024wqn}. We decide to keep this analysis at the level of the superblock expansion, but we stress that the original paper does not consider supersymmetry: here we simply mean that we apply the analysis to all $R$-symmetry channels, potentially considering them separately. We begin by considering the analyticity property of the correlator of not having a discontinuity for $z\geq 1$\footnote{We define the discontinuity operator $\mathcal{I}_z$ acting on $F(z)$ as $\mathcal{I}_z  F(z) = \lim_{\epsilon \to 0} \left( F(z+i\epsilon)-F(z-i\epsilon)\right)/2i$}, as this point does not correspond to any of the operators crossing each other positions. 
\beq
\mathcal{I}_z \Am_{\Delta_1\Dh_2\Dh_3} = 0 \qquad \text{for $z \geq 1$}
\label{eq:DiscAm}
\eeq
However, at this point the conformal blocks appearing in the (super)-conformal block expansion \eqref{eq:211_SuperblockExpansion} are discontinuous -- as all conformal blocks involved for any dimensions of the external operators\cite{Buric:2020zea,Okuyama:2024tpg} are. By pluggin the expansion \eqref{eq:211_SuperblockExpansion} in \eqref{eq:DiscAm} we get the following infinite sum
\beq
   \mathcal{I}_z \left( \sum_{\Oh} b_{\Delta_1 \Oh} \lambdah_{\Dh_1 \Dh_2 \Oh} \Gm_{\Oh} (\zeta ; x) \right) = 0
   \label{eq:DiscAmSum}
\eeq
running over the superconformal primaries. Given that the sum is an infinite sum, it is not justified to swap the sum and the discontinuity operator\footnote{When the sum runs over a finite number of fields we can consider the sum of the discontinuities of the blocks.} In order to de-clutter the notation, we redefine the coefficients of the superblocks as
\beq
c_\Dh = b_{\Delta_1 \Oh} \lambdah_{\Dh_1 \Dh_2 \Oh}
\eeq
where, for practical applications in perturbative frameworks, degenerate operators will be considered together in the same coefficient. To make use of the condition on the OPE coefficients expressed in the sum in \eqref{eq:DiscAmSum}, we want to write a dispersion relation for the correlator in the analysis. Given that we are interested in manipulating the contour around $z = 0$ -- where the correlator branch cut for $z < 0$ starts -- we consider a modified correlator 
\beq
\bar{\Am} = \Am_{\Delta_1\Dh_2\Dh_3} - \sum_{0<\Dh<2\tilde{\alpha}} c_\Dh \Lm_{\Dh}^{\tilde{\alpha}} (\zeta ; x)
\eeq
where
\beq
\Lm_{\Dh}^{\tilde{\alpha}} (\zeta ; x) = \Gm_{\Oh} (\zeta ; x) + O(z^{\tilde{\alpha} + \epsilon})\,.
\eeq
\begin{figure}
\centering
\includegraphics[width=.7\linewidth]{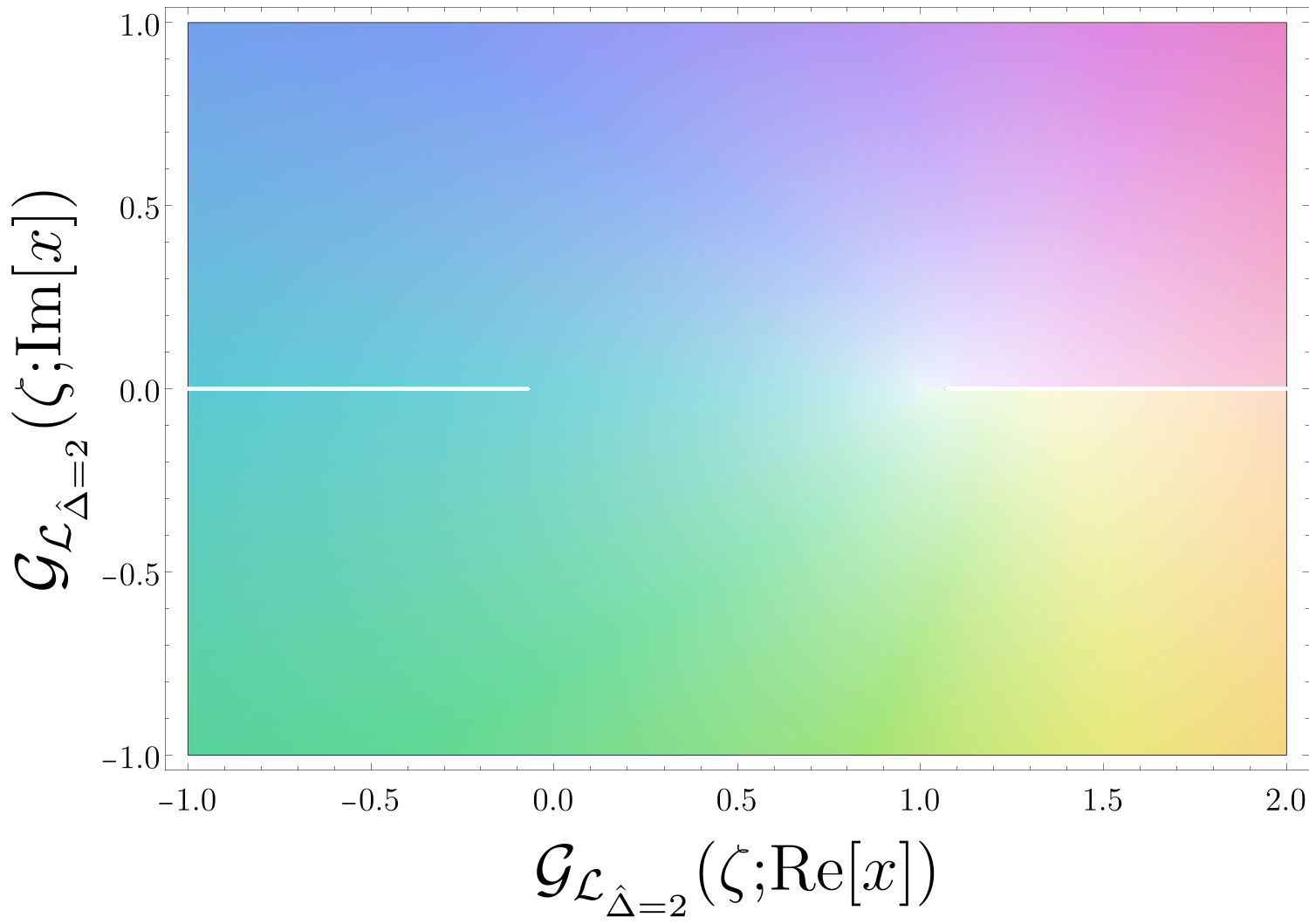}
\caption{The figure displays the analytic structure of the superblock $\Gm_{\Dh =2} (\zeta;x)$. We see that they have a discontinuity at $x \geq 1$, which is incompatible with the locality requirement discussed here. Note that the superblocks $\Gm_{\Dh > 2} (\zeta;x)$ exhibit the same behavior.
}
\label{fig:SuperBlockAnalytic}
\end{figure}
%
The Cauchy formula
\beq
\bar{\Am}(\zeta ; w) = \oint \frac{dz}{2\pi i} \left( \frac{w}{z} \right)^{\tilde{\alpha}+1} \frac{\bar{\Am} (\zeta ; z)}{z-w}
\eeq
can then be manipulated to become a dispersion relation for the modified correlator, \textit{i.e.} an integral involving its discontinuity, which is non-zero only for $z<0$.
\beq
\bar{\Am}(\zeta ; w) = \int_{-\infty}^0 \frac{dz}{\pi } \left( \frac{w}{z} \right)^{\tilde{\alpha}+1} \frac{\Im_z\bar{\Am} (\zeta ; z)}{z-w}
\eeq
we can then define the local blocks as the same dispersion formulas for the superblocks and then find
\beq
 \Lm_{\Dh}^{\tilde{\alpha}} (\zeta ; w) = \Gm_{\Oh} (\zeta ; w) - \int_1^{+\infty} \frac{dz}{\pi} \left( \frac{w}{z} \right)^{\tilde{\alpha}+1} \frac{\Im_z \Gm_{\Oh} (\zeta ; z)}{z-w}\,,
\eeq
which then gives a dispersion formula for the bulk-defect-defect correlator which can be proven \cite{Levine:2023ywq,Levine:2024wqn} to be equivalent to the condition
\beq
\sum_{\Oh} c_\Dh \Gm_{\Oh} (\zeta ; x) = \sum_{\Oh} c_\Dh \Lm_{\Dh}^{\tilde{\alpha}} (\zeta ; w)\,.
\eeq
This result is particularly interesting because, as we have mentioned above, there is no crossing relation for bulk-defect-defect correlators as both OPE limits give the same expression. Locality conditions give a similar constraint on the correlator with respect to crossing, and the condition translates into a sum rule for the OPE coefficients
\beq
\sum_\Dh c_\Dh \theta_n^{\tilde{\alpha}}\left(\Dh\right) = 0
\label{eq:SumRulesLocality}
\eeq 
where the functions $\theta_n^{\tilde{\alpha}}\left(\Dh\right)$ are given in equation (3.27) of \cite{Levine:2023ywq}. Despite being a relevant non-perturbative constraint for bulk-defect-defect correlators, locality at the moment does not play a role in our perturbative bootstrap approach; however, they represent a relevant consistency check and can be employed in the future for the determination of strong-coupling correlators at NNLO (see section \ref{subsub:StrongCoupling}).
\section{Perturbative results}
\label{sec:PerturbativeResults}
In this part of the chapter, we present explicit results for the bulk-defect-defect correlators introduced in section \ref{sub:Correlators}, and for which we listed non-perturbative constraints in the previous section \ref{sec:BDDNonPerturbativeConstraints}.
We begin by determining the contributions to the correlator $\vev{2 \hat{1} \hat{1}}$ up to next-to-leading order at weak and strong coupling, and we extract the corresponding CFT data using the superblocks presented in Section \ref{subsec:SuperblockExpansion}.
At strong coupling, we also present partial results for the next-to-next-to-leading order, for which one rational function remains unfixed but is shown to be strongly constrained by the collection of sum rules \ref{eq:SumRulesLocality}. We then generalize our results by computing all correlators up to next-to-leading order at weak coupling. We observe that transcendental terms are systematically absent at this order, which results in relations for the OPE coefficients.
\subsection{The correlator $\vev{2 \hat{1} \hat{1}}$ }
\label{subsec:211}
The correlation function between one protected bulk operator of dimension two and two protected defect operators of dimension one represents the easiest example of bulk-defect-defect correlator, since the trace in the definition of the half-BPS bulk operators \eqref{eq:SingleTraceHalfBPSOperators_Bulk} implies that the dimension one bulk operator of this kind is trivially zero. Because of its simplicity, the correlator $\vev{2 \hat{1} \hat{1}}$ is the ideal case to test how the combination of non-perturbative and perturbative information can determine bulk-defect-defect correlators at weak coupling. Many of the properties of this initial correlator are indeed maintained in correlators of operators of higher scaling dimension: let us then start the analysis of $\vev{2 \hat{1} \hat{1}}$ by establishing the notation and examining the leading order contribution.
\subsubsection{Weak coupling}
\label{subsub:WeakCoupling}

\paragraph{Perturbative structure}
\label{par:PerturbativeStructure_weak}

At weak coupling, the bulk-defect-defect correlator $\vev{2 \hat{1} \hat{1}}$ has the following perturbative structure at large $N$:
\begin{equation}
    \Am_{2\hat{1}\hat{1}} (\zeta;x)
    =
    \frac{1}{N}
    \bigl(
    \Am_{2\hat{1}\hat{1}}^{(0)} (\zeta;x)
    +
    \lambda\, \Am_{2\hat{1}\hat{1}}^{(1)} (\zeta;x)
    +
    \ldots
    \bigr)
    + \ldots\,,
    \label{eq:PerturbativeStructure_211_weak}
\end{equation}
where as usual the $\ldots$ inside the brackets refer to higher powers of $\lambda$, while the $\ldots$ at the end of the expression refer to corrections in $N$.
As explained in \eqref{eq:BulkDefectDefect_RsymmetryChannels}, the correlator $\vev{2\hat{1}\hat{1}}$ consists of \textit{two} $R$-symmetry channels:
\begin{equation}
    \Am_{2\hat{1}\hat{1}}^{(\ell)} (\zeta ; x)
    =
    F_1^{(\ell)} (x) + \frac{\zeta}{x} F_2^{(\ell)} (x)\,.
    \label{eq:RsymmetryChannels_211}
\end{equation}
From the number of $R$-symmetry channels and the solution to the topological constraint \eqref{eq:SolutionWI} it immediately follows that one channel is sufficient to completely determine the whole correlator. This makes the case in analysis ideal to test the validity of our approach by computing both channels up to NLO and testing how the perturbative bootstrap approach can simplify the computation of bulk-defect-defect correlators compared to a pure perturbative approach.
\paragraph{Leading order}
\label{par:LeadingOrder_weak}

We first determine the leading order for the correlator $\vev{2\hat{1}\hat{1}}$.
At this order, the correlator consists of a single Feynman diagram, which can be represented as
\begin{equation}
    \Am_{2 \hat{1} \hat{1}}^{(0)} (\zeta ; x)
    =
    \BulkDefectDefectLO\,,
    \label{eq:LO_FeynmanDiagram_211}
\end{equation}
and only consists of the free propagators defined in \eqref{eq:Propagators}.
It is easy to see that the $R$-symmetry channels take the form
\begin{equation}
    F_1^{(0)} (x)
    =
    c_0\,,
    \qquad
    F_2^{(0)} (x)
    =
    0\,,
    \label{eq:LO_Fixed_211}
\end{equation}
where $c_0$ is a constant that can be fixed either through direct calculation, from the pinching constraint of \eqref{eq:PinchingConstraint_LO}, or the topological sector \eqref{eq:Fds_weak}.
At this order, we find
\begin{equation}
    c_0
    =
    b_{2\hat{2}}^{(0)} \lambdah_{\hat{1}\hat{1}\hat{2}}^{(0)}
    =
    \sqrt{2}\,.
    \label{eq:LO_Constant_211}
\end{equation}
Note that the identity operator in the OPE \eqref{eq:OPE_Oh1} does not contribute to the superconformal block expansion at this order, i.e., we have
\begin{equation}
    a_2^{(0)}
    =
    0\,.
    \label{eq:LO_Identity_211}
\end{equation}

It is a trivial task to extract the CFT data for these results using the superblock expansion \eqref{eq:211_SuperblockExpansion}.
For the long operators labelled $\Lm_{[0,0],0}^\Dh$ \cite{Liendo:2018ukf}, we find the following closed form:
\begin{equation}
    \vev{ b_{2\Dh}^{(0)} \lambdah_{\hat{1}\hat{1}\Dh}^{(0)} }
    =
    \begin{cases}
        0\,, & \text{ if } \Dh^{(0)} \text{ odd}\,, \\
        \frac{(-1)^{\Dh/2} \sqrt{\pi} \Gamma (\Dh+2)}{2^{\Dh+1/2} \Gamma (\Dh+3/2)}\,, & \text{ if } \Dh^{(0)} \text{ even}\,.
    \end{cases}
    \label{eq:LO_CFTData_211}
\end{equation}
This is the supersymmetric version of the formula given in \cite{Buric:2020zea,Okuyama:2024tpg,Levine:2023ywq,Levine:2024wqn}, which corresponds to an expansion in bosonic blocks. We remark that in any perturbative expansion, operators become degenerate and therefore the CFT data are obtained as averages over operators having the same classical scaling dimension.

\paragraph{Next-to-leading order}
\label{par:NextToLeadingOrder_weak}

\begin{table}
    \centering
    \caption{Feynman diagrams contributing to the correlator $\vev{2\hat{1}\hat{1}}$ at next-to-leading order.}
    \begin{tabular}{lc}
        \hline \\[1pt]
        $F_1^{(1)} (x)$ & 
        \BulkDefectDefectNLOSEOne \;\;
        \BulkDefectDefectNLOSETwo \;\;
        \BulkDefectDefectNLOX \;\;
        \BulkDefectDefectNLOH \;\;
        \BulkDefectDefectNLOYOne \;\;
        \BulkDefectDefectNLOYTwo  \\[5ex]
        \hline \\[1pt]
        $F_2^{(1)} (x)$ & 
        \BulkDefectDefectNLOFoneOne \;\;
        \BulkDefectDefectNLOFoneTwo \;\;
        \BulkDefectDefectNLOFoneThree\\[5ex]
        \hline
    \end{tabular}
    \label{table:Diagrams211NLO}
\end{table}
At NLO, the Feynman diagrams corresponding to the correlator are listed in Table \ref{table:Diagrams211NLO}. 
Note that each open line in \eqref{eq:NLO_FeynmanDiagrams_F2_211_1}-\eqref{eq:NLO_FeynmanDiagrams_F2_211_2} can end on the orange dots placed along the defect. Due to the operator normalization (\eqref{eq:n1} and following), it is important to remember that a contribution from the lower order diagram is also present, multiplied by the NLO expansion of the normalization factor. Thanks to the non-perturbative constraints, we do not have to calculate all the diagrams in table \ref{table:Diagrams211NLO}.
In particular, we can focus on the second $R$-symmetry channel and therefore on the diagrams that do not contain bulk vertices. The integrals necessary to compute this channel arise from the coupling of $\phi^6$ with the Wilson line and are
\begin{equation}
  I(a) =  \int_{-\infty}^a d\tau_1 I_{1\tau_1}  ... \int_{-\infty}^{\tau_{n-1}} d\tau_{\Delta_1-(\Dh_2 + \Dh_3)} I_{1\tau_{\Delta_1-(\Dh_2 + \Dh_3)}} = \frac{1}{n!}\left(\int_{-\infty}^a d\tau I_{1\tau}\right)^{n}\,,
  \label{eq:IntegratePhi6OnTheLine}
\end{equation}
and similar from a point $a$ to $+\infty$.In this case, the number of nested integrals $n$ is two.  
Thanks to their relation to the one-dimensional integral of a single propagator, the integrals of the kind \eqref{eq:IntegratePhi6OnTheLine} are easy to compute using the results  
\begin{equation}
    \int_{-\infty}^a d\tau I_{1\tau} = \frac{\pi + 2 \arctan\left( \frac{a}{|x_\perp|}\right) }{2 |x_\perp|}\,,
    \label{eq:MasterIntegralDefet1}
\end{equation}
and
\begin{equation}
    \int_a^{+\infty} d\tau I_{1\tau} = \frac{\pi - 2 \arctan\left( \frac{a}{|x_\perp|}\right) }{2 |x_\perp|}\,.
    \label{eq:MasterIntegralDefet2}
\end{equation}
From the relations listed above, we obtain the results for the two diagrams contributing to the channel we focus on
\begin{align}
    \BulkDefectDefectNLOFoneOne
    &= - \mathcal{K}_{2\hat{1}\hat{1}} \frac{1}{16 \sqrt{2} \pi^2} \frac{\zeta}{x} \biggl( \pi + 2 \text{arctan} \biggl( \sqrt{\frac{1-x}{x}} \biggr) \biggr)^2\,, \label{eq:NLO_FeynmanDiagrams_F2_211_1} \\[1.5ex]
    \BulkDefectDefectNLOFoneTwo
    &= - \mathcal{K}_{2\hat{1}\hat{1}} \frac{1}{16 \sqrt{2} \pi^2} \frac{\zeta}{x} \biggl( \pi - 2 \text{arctan} \biggl( \sqrt{\frac{1-x}{x}} \biggr) \biggr)^2\,,
    \label{eq:NLO_FeynmanDiagrams_F2_211_2}
\end{align}
where the prefactors result from symmetry factors and trace contributions. Putting the two diagrams together, we obtain for the $R$-symmetry channel $F_2$ the result
\begin{align}
    F_2^{(1)} (x)
    &=
    - \frac{1}{8 \sqrt{2} \pi^2}
    (\pi^2 + 4 \text{arccos}^2 \sqrt{x})\,.
    \label{eq:NLO_ResultF2_211}
\end{align}
As a consistency test, we can check that this expression is compatible with the splitting constraint of Section \ref{subsec:PinchingAndSplittingLimits}:
\begin{equation}
    F_2^{(1)} (0)
    =
    -2 a_2^{(1)}
    =
    - \frac{1}{4 \sqrt{2}}\,.
    \label{eq:NLO_Identity_211}
\end{equation}


The channel $F_1$ can now be obtained through the supersymmetry constraints of Section \ref{subsec:Topological}.
The solution \eqref{eq:SolutionWI} to the Ward identity implies
\begin{equation}
    F_1^{(1)} (x)
    =
    \Fds_{2\hat{1}\hat{1}}^{(1)}
    -
    F_2^{(1)} (x)\,.
    \label{eq:NLO_AfterWI_211}
\end{equation}
Using the results of \cite{Giombi:2018hsx} (summarized in Section \ref{subsubsec:LocalizationResults}), we can calculate the topological sector, and we get
\begin{equation}
    \Fds_{2\hat{1}\hat{1}}^{(1)}
    =
    - \frac{1}{6 \sqrt{2}}\,.
    \label{eq:NLO_Topological_211}
\end{equation}
The correlator is now completely fixed.
Notice that this result was obtained without considering the first line of Table \ref{table:Diagrams211NLO}, where the Feynman diagrams consist of bulk vertices and are significantly harder than the first line. This strategy will allow us in the next sections to provide results for any correlator up to NLO in the weak coupling perturbative expansion. We decide therefore to show the validity and the advantages of this approach by explicitly computing the diagrams contributing to the channel $F_1^{(1)} (x)$. The results are obtained using the Feynman rules of Section \ref{ssub:BulkFR} and \ref{ssub:DefectFR} and the integrals of Appendix \ref{app:Integrals}, where the regularization for UV divergent integrals is also described.\\

We begin with the self-energy diagrams.
Using the insertion rule \eqref{eq:SelfEnergy}, it is easy to find that
\begin{equation}
    \BulkDefectDefectNLOSEOne
    = \frac{\log (x)+2 \log (\epsilon )-2}{4 \sqrt{2} \pi ^2}\,,
    \label{eq:DiagramSEOne}
\end{equation}
while
\begin{equation}
    \BulkDefectDefectNLOSETwo
    = \frac{-\log \left(\tau _3\right)+\log (\epsilon )-1}{2 \sqrt{2} \pi ^2}\,.
    \label{eq:DiagramSETwo}
\end{equation}
Note that these results are given in the conformal limit $\tau_3 \to \infty$.
In other words, each diagram contains corrections in $1/\tau_3$ and divergences proportional to $\log \tau_3$, however since we know that the correlator depends on a single cross-ratio $x$, they are expected to cancel once added up. As expected from the final result \eqref{eq:NLO_AfterWI_211}, we will indeed observe a cancellation of the $\log \tau_3$ terms when summing all diagrams contributing to the $R$-symmetry channel $F_1^{(1)} (x)$. Next, we consider the two diagrams that involve bulk vertices but no integration along the Wilson line.
The first one contains the X-integral defined in appendix \ref{app:Integrals} and pinched to.
Taking into account all the prefactors, it reads
\begin{equation}
    \BulkDefectDefectNLOX
    = -\frac{\log (x)+2 \log (\epsilon )-2}{8 \sqrt{2} \pi ^2}\,,
    \label{eq:DiagramX}
\end{equation}
The second diagram involves the pinched F-integral (defined in appendix \ref{app:Integrals}), from which we obtain\footnote{We thank Julius Julius and Philine van Vliet for pointing out corrections to this expression.}
\begin{equation}
    \BulkDefectDefectNLOH
    = -\frac{\log \left(x\right)+2 \log (\epsilon )}{8 \sqrt{2} \pi ^2}\,.
    \label{eq:DiagramF}
\end{equation}
Finally, two diagrams involve an integral along the Wilson line, as well as a bulk vertex.
These integrals are more challenging but can be performed analytically in the conformal frame $\tau_3 \to \infty$. For example, the integrals can be manipulated to extract their divergences and then evaluated numerically to be matched to an ansatz made of the expected functions appearing in the result.
The first integral gives
\begin{equation}
    \BulkDefectDefectNLOYOne
    = -\frac{3 \log (x)-12 \left[\arctan\left(\sqrt{\frac{1}{x}-1}\right)\right]^2+6 \log (\epsilon )+\pi ^2-6}{24 \sqrt{2} \pi ^2}\,,
    \label{eq:DiagramYOne}
\end{equation}
where it should be understood that the gluon propagator can connect to each orange dot on the line.
The second one yields
\begin{equation}
    \BulkDefectDefectNLOYTwo
    = \frac{3\log \left( x \right) -2\left(-6 \log \left(\tau _3\right)+3 \log (\epsilon )+\pi ^2-6\right)}{24 \sqrt{2} \pi ^2}\,.
    \label{eq:DiagramYTwo}
\end{equation}
When adding all the diagrams, we find that the divergences cancel as well as the $\log \tau_3$ terms.
The result is in perfect agreement with \eqref{eq:NLO_AfterWI_211}; the functions appearing in the single integrals and the integration complexity however increase in this $R$-symmetry channel, and the combination of non-perturbative techniques with perturbative computations represents an advantage. In the following sections, we will prove how this advantage allows the computation of correlators of half-BPS scalar operators having unspecified scaling dimensions. Beyond providing an explicit check of the validity of the perturbative bootstrap approach, the integrals reported in this section can also be used to compute correlation functions of non-primary operators obtained by acting with differential operators on $\Op_2(x_1)$. 

\paragraph{CFT data.}
The results above can be used for determining the CFT data at next-to-leading order.
Notice that the expressions \eqref{eq:NLO_ResultF2_211} and \eqref{eq:NLO_AfterWI_211} are entirely \textit{rational}, i.e., no transcendental functions are appearing at this order.
This is surprising, since we know that the operators exchanged in the OPE generally have anomalous dimensions at one loop, while we have seen in \eqref{eq:LO_CFTData_211} that operators with even $\Dh$ have a tree-level OPE coefficient.
This observation can be summarized in the following equation:
\begin{equation}
    \vev{ b_{2 \Dh}^{(0)} \lambdah_{\hat{1}\hat{1}\Dh}^{(0)} \gamma_\Dh^{(1)} }
    =
    0\,,
    \label{eq:NLO_NoLog_211}
\end{equation}
where the brackets $\vev{ \ldots }$ should be understood in the average sense explained in \eqref{eq:AverageOPE_Definition} and below.
We comment on the absence of logarithmic terms in Section \ref{subsubsec:OnTheAbsenceOfTranscendentalFunctions}.

In \eqref{eq:LO_CFTData_211}, we saw that operators with odd tree-level scaling dimension have vanishing OPE coefficients at order $\Om(\lambda^0/N)$.
They start to contribute at one loop, and we find the following closed form for their average:
\begin{equation}
    \vev{ b_{2 \Dh}^{(1)} \lambdah_{\hat{1}\hat{1}\Dh}^{(1)} }_{\Dh \text{ odd}}
    =
    \frac{(-1)^{\Dh/2} i\, \Gamma(\Dh+2)}{2^{\Dh+5/2} \sqrt{\pi} \Gamma(\Dh+3/2)} H_{\Dh+1}\,,
    \label{eq:NLO_OPECoeffsOdd_211}
\end{equation}
with $H_n$ the harmonic numbers defined through
\begin{equation}
    H_n
    =
    \sum_{k=1}^n \frac{1}{k}\,.
    \label{eq:HarmonicNumbers}
\end{equation}
The lowest operator described by \eqref{eq:NLO_OPECoeffsOdd_211} corresponds to $\phi^6$ and is not degenerate.
Its corresponding OPE coefficient is given by
\begin{equation}
    b_{2 {\phi^6}}^{(1)} \lambdah_{\hat{1}\hat{1}\phi^6}^{(1)}
    =
    - \frac{1}{2\sqrt{2} \pi}\,.
    \label{eq:NLO_OPECoeffsphi^6_211}
\end{equation}

Long operators with an even tree-level scaling dimension receive corrections to their tree-level value given in \eqref{eq:LO_CFTData_211}.
It is difficult to find an analytic closed form, however, they can be expressed in the form of a recursion relation:
\begin{align}
    \vev{ b_{2 \Dh}^{(1)} \lambdah_{\hat{1}\hat{1}\Dh}^{(1)} }_{\Dh \text{ even}}
    &=
    - \frac{(1)_{\Dh/2}}{12 \sqrt{2} (5/2)_{\Dh/2}}
    \biggl( 1 - \frac{4}{\pi} \frac{(\Dh+1)(\Dh+3)}{\Dh^2} \biggr) \notag \\
    &\phantom{=\ }
    - \frac{\Dh/2+1}{\Gamma(-\Dh/2)}
    \sum_{h=2 \text{ step } 2}^{\Dh-2}
    \frac{i^h \Gamma (h+5/2) \Gamma((h-\Dh)/2)}{\Gamma^2(h/2+1) \Gamma((h+\Dh+5)/2)}
    \vev{ b_{2 \Dh}^{(1)} \lambdah_{\hat{1}\hat{1}\Dh}^{(1)} }\,.
    \label{eq:NLO_OPECoeffsEven_211}
\end{align}
Although this is not a closed form, note that recursion relations are extremely efficient, even for high $\Dh$.
In the formula above, we used the Pochhammer symbols, defined through
\begin{equation}
    (a)_n
    =
    \frac{a!}{(a-n)!}\,.
    \label{eq:Pochhammer}
\end{equation}

\subsubsection{Strong coupling}
\label{subsub:StrongCoupling}
\begin{figure}
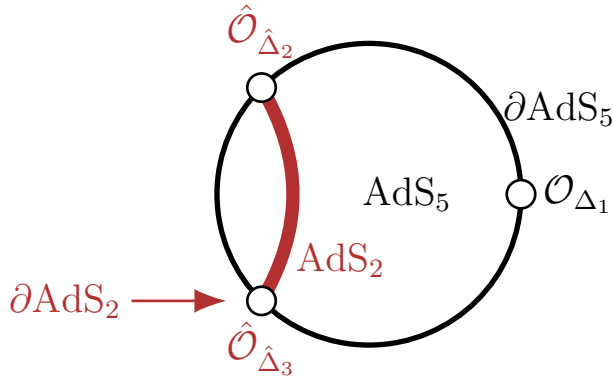

    \centering
    \WittenDiagrams
    \caption{Illustration of the Witten diagrams for the bulk-defect-defect correlators at strong coupling.
    As labelled on the figure, the inside of the circle represents the AdS$_5$ spacetime, while its boundary corresponds to the dual CFT$_4$ (in this case, $\Nm=4$ sYM).
    The bold red line represents the string worldsheet that is dual to the Wilson line.
    The worldsheet and the boundary CFT coincide on a line, which is the Maldacena-Wilson line \eqref{eq:WilsonLine}.
    The bulk operator $\Op_{\Delta_1}$ lives in the CFT$_4$, while the operators $\Oh_{\Dh_2}$ and $\Oh_{\Dh_3}$ are representations of the CFT$_1$ dual to the AdS$_2$ surface.
    Note that the fact that the worldsheet and the CFT$_4$ coincide at two points only in the figure is an artefact of the representation.
    }
    \label{fig:WittenDiagrams}
\end{figure}

We now consider the bulk-defect-defect correlator  $\vev{2\hat{1}\hat{1}}$ in the strong-coupling regime, providing the leading and next-to-leading orders with the corresponding CFT data\footnote{After this thesis was reviewed by the referees, the strong coupling calculation has been updated by the authors of \cite{Artico:2024wnt} -- independently from the content of the referee report. For this reason, we refer the reader interested in the strong coupling part of the calculation to the updated version of reference \cite{Artico:2024wnt} available on arXiv.}. At next-to-next-to-leading order, we determine through an Ansatz the transcendental part of the correlator, and comment on the fact that the rational part is constrained by the sum rules \eqref{eq:SumRulesLocality} arising from locality. We study the strong-coupling regime through a perturbative expansion at large $N$ of the form
\begin{equation}
    \Am_{2 \hat{1} \hat{1}} (\zeta;x)
    =
    \frac{\sqrt{\lambda}}{N}
    \biggl(
    \Am_{2 \hat{1} \hat{1}}^{(0)} (\zeta;x)
    +
    \frac{1}{\sqrt{\lambda}} \Am_{2 \hat{1} \hat{1}}^{(1)} (\zeta;x)
    +
    \frac{1}{\lambda} \Am_{2 \hat{1} \hat{1}}^{(2)} (\zeta;x)
    +
    \ldots
    \biggr)
    + \ldots\,.
    \label{eq:StrongCoupling_PerturbativeStructure}
\end{equation}
As in \eqref{eq:PerturbativeStructure_211_weak}, the dots refer to corrections in $1/\sqrt{\lambda}$ inside the brackets and in $1/N$ outside the brackets.
The relevant Witten diagrams are listed in Table \ref{table:WittenDiagrams211NLO}, following the conventions of Figure \ref{fig:WittenDiagrams}. In this section, we will not compute them and we mostly use them as guides for the discussion; the interested reader can refer to \cite{Witten:1998qj,DHoker:2002nbb} for an introduction to the topic of Witten diagrams and holography.

\paragraph{Leading order}
\label{par:LeadingOrder_strong}

At leading order, there is only one disconnected Witten diagram, which corresponds to the contribution of the identity operator.
We have seen in Section \ref{subsec:SuperblockExpansion} that the identity only appears in $F_2$ (in the form of a constant), and thus we have
\begin{equation}
    F_2^{(0)} (x)
    =
    - 2 a_2^{(0)}
    =
    - \frac{1}{\sqrt{2}}\,,
    \qquad
    F_1^{(0)}
    =
    0\,.
    \label{eq:StrongCoupling_Results_LO}
\end{equation}
The CFT data for the long operators is easy to extract and it is
\begin{equation}
    \vev{ b_{2 \Dh}^{(0)} \lambdah_{\hat{1}\hat{1}\Dh}^{(0)} }
    =
    0\,.
    \label{eq:StrongCoupling_CFTData_LO}
\end{equation}
This expression is conservative, as it represents an average over operators having the same classical dimension and does not give information on each operator. Nevertheless, it is easy to convince oneself that individual operators also have vanishing OPE coefficients as at leading order either they do not couple with the bulk operator or with the two defect ones:
\begin{equation}
    b_{2 \Dh}^{(0)} \lambdah_{\hat{1}\hat{1}\Dh}^{(0)}
    =
    0\,.
    \label{eq:StrongCoupling_CFTDataDisentangled_LO}
\end{equation}

\paragraph{Next-to-leading order}
\label{par:NextToLeadingOrder_strong}

\begin{table}
    \centering
    \caption{Witten diagrams contributing to the bulk-defect-defect correlators in the strong-coupling limit (on the CFT side).
    The first two diagrams are disconnected and correspond to the factorized limit $\vev{\Delta_1} \vev{\Dh_2 \Dh_3}$, if non-vanishing.
    The third one is connected and was considered in \cite{Levine:2023ywq}.
    In our setup, it describes the pinching limit $\vev{\Delta_1 (\Dh_2 + \Dh_3)}$.
    The fourth diagram is non-trivial and starts contributing at order $\Om(\sqrt{\lambda}/N)$.
    Note that other diagrams that we do not show certainly also contribute to this order.\\}
    \begin{tabular}{rc}
        \hline \\[.1pt]
        $\vev{\Delta_1 \Dh_2 \Dh_3} =$
        &
        \WittenDiagramOne \;\; $+$ \;\;
        \WittenDiagramTwo \;\; $+$ \;\;
        \WittenDiagramThree \;\; $+$ \;\;
        \WittenDiagramFour \;\; $+ \ldots$ \\[6ex]
        \hline
    \end{tabular}
    \label{table:WittenDiagrams211NLO}
\end{table}
At next-to-leading order, there are two Witten diagrams contributing to the correlator, namely the second and third ones in Table \ref{table:WittenDiagrams211NLO}.
The first contribution is disconnected and corresponds to $a_2^{(1)}$.\footnote{Note that the disconnected part can be evaluated at all orders in $\lambda$ using the protected CFT data.}
The second one is the leading connected term, which consists only of free propagators.
We conclude then that the $R$-symmetry channels are again constant.
Moreover, this diagram corresponds, up to a prefactor, to the leading term of the bulk-defect correlator $\vev{2 \hat{2}}$.
This results in
\begin{equation}
    F_1^{(1)} (x)
    =
    b_{2\hat{2}} \lambda_{\hat{1} \hat{1} \hat{2}}^{(1)}
    =
    \frac{3}{\sqrt{2}}\,.
    \label{eq:StrongCoupling_F2_NLO}
\end{equation}
Meanwhile, the disconnected term gives
\begin{equation}
    F_2^{(1)} (x)
    =
    \frac{3}{2\sqrt{2}}\,.
    \label{eq:StrongCoupling_Disc_NLO}
\end{equation}

To extract the conformal data at this order, we first observe that the spectrum consists of defect operators with \textit{even} scaling dimensions at tree level, in agreement with expectations from the AdS/CFT correspondence \cite{Liendo:2016ymz,Ferrero:2021bsb}.
For instance, the operator $\phi^6$ flows from $\Dh=1$ at weak coupling to a two-particle state with $\Dh=2$ at strong coupling \cite{Alday:2007hr}.
Although the interpretation differs from weak coupling, the CFT data at the next-to-leading order take the same form up to an overall constant:
\begin{equation}
    \vev{ b_{2 \Dh}^{(1)} \lambdah_{\hat{1}\hat{1}\Dh}^{(1)} }
    =
    \frac{(-1)^{\Dh/2} 3 \sqrt{\pi} \Gamma(\Dh+2)}{2^{\Dh+3/2} \Gamma(\Dh+3/2)}\,.
    \label{eq:StrongCoupling_CFTData_NLO}
\end{equation}
We remark that since the leading order contributions are zero, there is no term involving anomalous dimensions.

\paragraph{Partial results at next-to-next-to-leading order}
\label{par:PartialResultsAtNextToNextToLeadingOrder}

\begin{figure}
\centering
  \includegraphics[width=.65\linewidth]{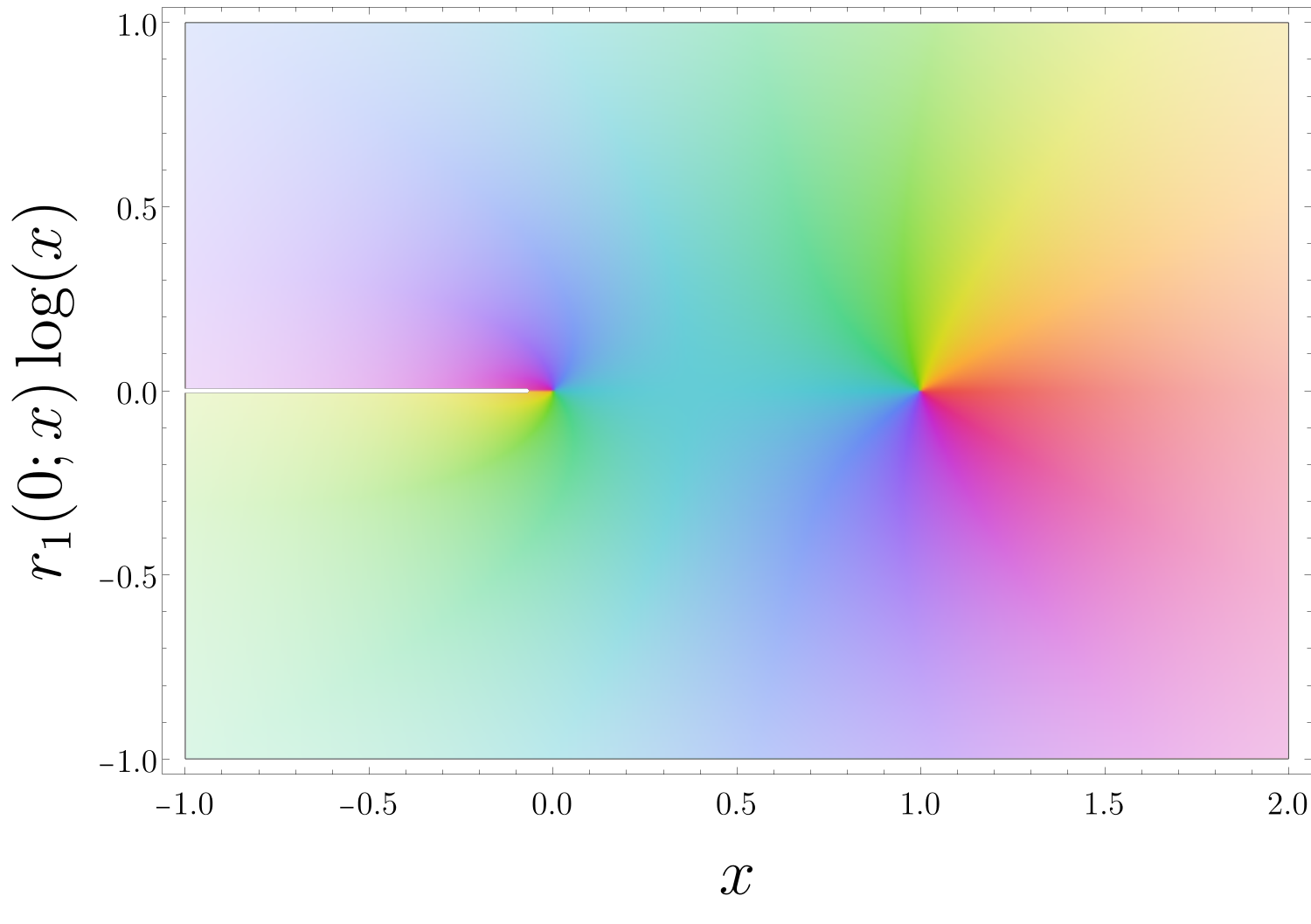}
\caption{Analytic structure of the $\log$ term at next-to-next-to-leading order at strong coupling.
Although individual blocks are discontinuous at $x \geq 1$, the resummed term obeys the locality condition of \cite{Levine:2023ywq}.
}
\label{fig:ContourPlots_LogTermsStrong}
\end{figure}

We now discuss partial results for the next-to-next-to-leading order, in which we expect several Witten diagrams to appear.
One example is the fourth diagram of Table \ref{table:WittenDiagrams211NLO}.
We give here the transcendental part of the $\vev{2 \hat{1} \hat{1}}$ correlator, and discuss constraints on the rational part.\\

We start by formulating an Ansatz based on the expected structure of the correlator.
In the strong-coupling regime, the transcendentality weight typically increases in steps of $1$ with powers of $1/\sqrt{\lambda}$ \cite{Beccaria:2019dws,Ferrero:2021bsb}.
Moreover, since we are dealing with a function of a single spacetime cross-ratio, this means that the correlator takes the following form at transcendentality one:
\begin{equation}
    \Am_{2\hat{1}\hat{1}}^{(2)} (\zeta ; x)
    =
    r_0 (\zeta ; x) + r_1 (\zeta ; x) \log x\,,
    \label{eq:NNLO_Ansatz}
\end{equation}
where the functions $r_j (\zeta;x)$ are rational functions (both at $x \sim 0$ and $x \sim 1$) as well as polynomials of degree one in $\zeta$ in order to reflect the structure presented in \eqref{eq:RsymmetryChannels_211}. We can therefore write the functions $r_j (\zeta ; x)$ as
\begin{equation}
    r_j (\zeta ; x)
    =
    s_i^{(0)} + \zeta s_i^{(1)}\,.
    \label{eq:NNLO_Ansatz2}
\end{equation}

We can study the logarithmic terms by using the fact that degeneracies are not lifted at next-to-leading order in the scaling dimensions.\footnote{See, e.g., \cite{Ferrero:2021bsb} for an explanation and an application in four-point functions of defect operators.}
This means that we can extract the anomalous dimensions from the averages, as it is the same for all degenerate operators:
\begin{equation}
    \vev{ b_{2 \Dh}^{(1)} \lambdah_{\hat{1} \hat{1} \Dh}^{(1)} \gamma_\Dh^{(1)} }
    =
    \vev{ b_{2 \Dh}^{(1)} \lambdah_{\hat{1} \hat{1} \Dh}^{(1)} } \gamma_\Dh^{(1)}\,.
    \label{eq:NNLO_Degeneracies}
\end{equation}
Note that this statement is assuming that the operators $\Oh_{\Dh}$ are single-trace.
This assumption is justified, as if we denote an arbitrary multi-trace operator of dimension $\Dh$ by $\Oh_{(\mathbf{\Dh})}$, we can remark that
\begin{equation}
    b_{2 \mathbf{\Dh}}
    \sim
    \begin{cases}
        \Om(1)\,, &\text{ if } \Oh_{(\mathbf{\Dh})} = \Oh_{(0|2)}\,, \\
        \Om(1/N^2)\,, &\text{ otherwise}\,.
    \end{cases}
    \label{eq:NNLO_DoubleTraceStory}
\end{equation}
Thus, at large $N$, we only need to worry about $\Oh_{(0|2)}$, since three-point functions of defect multi-trace operators are suppressed.
In fact, the coefficients $b_{2 \hat{2}} \lambdah_{\hat{1}\hat{1}\hat{2}}$ and $b_{2 (\hat{0}|\hat{2})} \lambdah_{\hat{1}\hat{1}(\hat{0}|\hat{2})}$ are seen to scale as $1/N$ by noting that
\begin{equation}
    \lambdah_{\hat{1}\hat{1}(\hat{0}|\hat{2})}
    \sim
    \frac{1}{N}\,.
    \label{eq:DefectThreePoint_ScalingN}
\end{equation}
This scaling is true if the three-point function $\lambdah_{\hat{1}\hat{1}(\hat{0}|\hat{2})}$ is non-vanishing. We are now in a similar situation to the one described below equation \eqref{eq:DoubleTrace_GramSchmidt}. 
In fact, by requiring that $\Oh_{(0|2)}$ is orthogonal to $\Oh_{2}$, we are setting at the same time $\lambdah_{\hat{1}\hat{1}(\hat{0}|\hat{2})}=0$, similarly to \eqref{eq:DoubleTrace_GramSchmidt}.
To summarize, only single-trace operators appear on the left-hand side of \eqref{eq:NNLO_Degeneracies}.
We believe this to be true at all orders in the coupling constant.

The anomalous dimensions for the long operators $\Oh_{\Lm_{[0,0],0}^\Dh}$ are known \cite{Liendo:2018ukf,Ferrero:2021bsb} and they are given by the quadratic Casimir eigenvalue for singlets of $R$-symmetry and transverse spin:
\begin{equation}
    \gamma_{\Dh}^{(1)}
    =
    - \frac{\Dh (\Dh + 3)}{2}\,.
    \label{eq:NNLO_gammaDh}
\end{equation}
Plugging these values inside the superblock expansion and comparing to the Ansatz \eqref{eq:NNLO_Ansatz}, we find the rational function corresponding to the log terms. The result is
\begin{equation}
    r_1 (\zeta ; x)
    =
    \frac{3}{\sqrt{2}} (x - \zeta)\,.
    \label{eq:NNLO_LogTerms}
\end{equation}

We can further constrain the correlator by using the superconformal Ward identity \eqref{eq:SCWI} to remove one of the remaining two rational functions.
From the expression \eqref{eq:SolutionWI}, we get
\begin{equation}
    s_1^{(0)} (x)
    =
    \frac{21}{8 \sqrt{2}} \frac{1}{x} - \frac{s_0^{(0)} (x)}{x}\,,
    \label{eq:NNLO_SolutionWI}
\end{equation}
and the correlator at next-to-next-to-leading order is now fully fixed up to the rational function $s_0^{(0)} (x)$.

\paragraph{Sum rules.}
We conclude this section on strong coupling by mentioning that the remaining rational function can in principle be further constrained from consistency conditions.
The open rational function follows the superblock expansion
\begin{align}
    s_0^{(0)} (x)
    &=
    - \frac{9}{4 \sqrt{2}}
    +
    \biggl(
    \vev{ b_{2 \Dh}^{(2)} \lambdah_{\hat{1} \hat{1} \Dh}^{(2)} }_{\Dh^{(0)} = 2}
    -
    \frac{9}{10 \sqrt{2}}
    \biggr) x \notag \\
    &\phantom{=\ }
    +
    \biggl(
    \vev{ b_{2 \Dh}^{(2)} \lambdah_{\hat{1} \hat{1} \Dh}^{(2)} }_{\Dh^{(0)} = 4}
    +
    \frac{8}{9} \vev{ b_{2 \Dh}^{(2)} \lambdah_{\hat{1} \hat{1} \Dh}^{(2)} }_{\Dh^{(0)} = 2}
    +
    \frac{457 \sqrt{2}}{945}
    \biggr) x^2
    + \ldots\,.
    \label{eq:NNLO_RationalLeftover}
\end{align}
Strong constraints on these OPE coefficients can be derived by demanding locality, as it was shown in \cite{Levine:2023ywq}.
Since we are interested only in the rational terms, and the logarithmic terms are local, the coefficients of a corresponding bosonic block expansion are constrained by the locality sum rules \eqref{eq:SumRulesLocality} in the form
\begin{equation}
    c_{m+n}
    +
    \sum_{k=0}^m c_k\, \theta^{1+m}_n (2+2k)
    =
    0\,,
    \label{eq:NNLO_SumRules}
\end{equation}
where $\theta_n^m (\Delta)$ is defined in Equation (3.27) of \cite{Levine:2023ywq}.
Here, $m$ is a positive integer assumed to be small but unknown.
The sum rules alone are however insufficient for fixing all the coefficients if $m>0$.
Perhaps an input from the Witten diagram side can help to make the sum rules useful for fixing $s_0^{(0)} (x)$.

\subsection{Generalization to arbitrary $\Delta_1$, $\Dh_2$, $\Dh_3$}
\label{subsec:GeneralizationToArbitraryDelta1Dh2Dh3}
In this section, we present the analysis for the general correlator $\vev{\Delta_1 \Dh_2 \Dh_3}$ at weak coupling up to next-to-leading order. We do not extract the CFT data, although we note that it is in principle possible to do it case by case in the same way as in the sections above (see for example \ref{par:LeadingOrder_weak}). We have seen in \eqref{eq:LO_FeynmanDiagram_211} and \eqref{eq:NLO_FeynmanDiagrams_F2_211_1}-\eqref{eq:NLO_FeynmanDiagrams_F2_211_2} that the correlator $\vev{2 \hat{1} \hat{1}}$ is fixed up to next-to-leading order by Feynman diagrams that do not contain bulk vertices. As it will be further clarified below, this observation holds for the general case $\vev{\Delta_1 \Dh_2 \Dh_3}$ as well.
We are therefore interested in diagrams of the form
\begin{equation}
    \BulkDefectDefectGeneralLeadingOrder\,.
    \label{eq:GeneralFeynmanDiagrams}
\end{equation}
Here, the thick solid lines designate $a_k$ free propagators.
The coefficients $a_{k=1, \ldots, 4}$ refer to the possible allowed contractions without involving a bulk vertex.
They satisfy the consistency relations
\begin{equation}
    \begin{split}
        a_1 + a_2 + a_4 &= \Delta_1\,, \\
        a_1 + a_3 &= \Dh_2\,, \\
        a_2 + a_3 &= \Dh_3\,.
    \end{split}
    \label{eq:ConsistencyConditions_GeneralCase}
\end{equation}
It is transparent that this system of equations does not have a unique solution: there are multiple ways to realize this condition that lead to consistent diagrams. However, to a higher value of $a_4$ corresponds a diagram contributing to a higher order in the perturbative expansion\footnote{This is not true anymore at strong coupling, where as we have seen the first non zero order corresponds to the exchange of the identity operator on the defect \eqref{eq:StrongCoupling_Results_LO}.}.\\

The Feynman diagrams \eqref{eq:GeneralFeynmanDiagrams} can be used to understand the perturbative weak-coupling structure of the bulk-defect-defect correlators.
General correlators have to take the following form:
\begin{equation}
    \Am_{\Delta_1 \Dh_2 \Dh_3} (\zeta ; x)
    =
    \frac{\lambda^{a/2}}{N} \biggl(
    \Am_{\Delta_1 \Dh_2 \Dh_3}^{(0)} (\zeta ; x)
    +
    \lambda \Am_{\Delta_1 \Dh_2 \Dh_3}^{(1)} (\zeta ; x)
    + \ldots
    \biggr)
    + \ldots\,,
    \label{eq:PerturbativeStructure_weak_General}
\end{equation}
where we defined $a=\text{min}(a_4)$, i.e., the lowest number of propagators that can connect the bulk operators and the line defect for a given configuration characterized by $\Delta_1$, $\Dh_2$, $\Dh_3$, while still satisfying \eqref{eq:ConsistencyConditions_GeneralCase}. The number of $R$-symmetry channels can grow arbitrarily high, as can be seen in \eqref{eq:NumberOfRsymmetryChannels1}.
However, at low orders, most of the $R$-symmetry channels are suppressed for a given configuration, meaning that they contribute only at higher orders in the perturbative expansion. This can be understood by counting the powers of $\lambda$ we add when removing two propagators connecting $x_1$ with the defect operators and replacing them with one propagator connecting $\tau_2$ and $\tau_3$ and two propagators connecting $x_1$ with the Wilson line.
Concretely, we get
\begin{align}
    \Am_{\Delta_1 \Dh_2 \Dh_3}^{(0)}
    &=
    F_1^{(0)} (x)\,, \\
    \Am_{\Delta_1 \Dh_2 \Dh_3}^{(1)}
    &=
    F_1^{(1)} (x) + \frac{\zeta}{x} F_2^{(1)} (x)\,, \\
    &\phantom{||}\vdots
    \label{eq:RsymmetryChannels_PerturbativeStructure}
\end{align}
and a new channel appears at each order until the full number of channels has been reached.
In the following paragraphs, we study the correlators up to next-to-leading order, in which only \textit{two} $R$-symmetry channels contribute. The key point is represented by the presence of the topological sector which trades the most complicated perturbative channel with the difference between the topological constant and the simplest channel, made of diagrams of the form \eqref{eq:GeneralFeynmanDiagrams}.
\subsubsection{Leading order at weak coupling}
The leading order is given by the topological sector \eqref{eq:Fds_weak}, since only one channel contributes and it is constant.
To see that this is the case, consider the Feynman diagrams \eqref{eq:GeneralFeynmanDiagrams} for the distinct cases $\Dh_2 - \Dh_3 \leq \Delta_1 < \Dh_2 + \Dh_3$ and $\Delta_1 \geq \Dh_2 + \Dh_3$.
In the first case, no propagator is connecting to the defect when minimizing $a_4$, and thus the leading $R$-symmetry channel is constant.
The second configuration leads to considering nested defect integrals.
Using the identity \eqref{eq:IntegratePhi6OnTheLine}, we can rewrite them in terms of a single master defect integral:
\begin{equation}
    \vev{\Delta_1 \Dh_2 \Dh_3}^{(0)}
    =
    c_0 (\Delta_1, \Dh_2, \Dh_3)
    \biggl(
    \int_{-\infty}^{\infty} d\tau_4\, I_{14}
    \biggr)^{-2\Delta_{\hat{2}\hat{3}1}}
    =
    c_0 (\Delta_1, \Dh_2, \Dh_3)
    \biggl(
   	\frac{\pi}{|x_\perp|}
    \biggr)^{-2\Delta_{\hat{2}\hat{3}1}}\,.
    \label{eq:LO_Integrals_Case2}
\end{equation}
This integral is elementary and its solution is given by the sum of \eqref{eq:MasterIntegralDefet1} and \eqref{eq:MasterIntegralDefet2}.
The coefficient $c_0 (\Delta_1, \Dh_2, \Dh_3)$ encodes the symmetry factors and the contributions from the traces.
It is given by the topological sector \eqref{eq:Fds_weak} (or the pinching limit \eqref{eq:PinchingConstraint_NLO} when applicable).

\subsubsection{Next-to-leading order at weak coupling}
At next-to-leading order, we can use the supersymmetry constraint \eqref{eq:SolutionWI} and focus on calculating the channel $F_2$ only:
\begin{equation}
    F_1^{(1)} (x)
    =
    \Fds_{\Delta_1 \Dh_2 \Dh_3}^{(1)}
    -
    F_2^{(1)} (x)\,.
    \label{eq:211_SolutionWI}
\end{equation}
We only need to solve the generalized version of \eqref{eq:NLO_FeynmanDiagrams_F2_211_1}-\eqref{eq:NLO_FeynmanDiagrams_F2_211_2} to obtain all the correlators. We can group the correlators into two different groups:
\begin{itemize}
    \item $\Dh_2-\Dh_3 \leq \Delta_1 \leq \Dh_2 - \Dh_3 + 1$:
    These configurations are the ones where only one $R$-symmetry channel exists.
    In this case, the correlator is topological and $\Am_{\Delta_1 \Dh_2 \Dh_3}$ is equal to the topological sector $\Fds_{\Delta_1 \Dh_2 \Dh_3}$.
    \item $\Dh_2-\Dh_3 + 1  < \Delta_1$:
    In the conformal frame $\tau_3 \to \infty$, the only surviving diagrams are the ones depicted in \eqref{eq:GeneralFeynmanDiagrams}, for which the value of $a_4$ is now raised by $2$.
    The correlator can then be determined from the function
    \begin{equation}
        F_2^{(1)} (x)
        =
        c_1 (\Delta_1, \Dh_2, \Dh_3)
        \sum_{\pm} \biggl(
        \pi \pm 2 \text{arctan} \biggl( \sqrt{\frac{1-x}{x}} \biggr)
        \biggr)^{a+2}\,,
        \label{eq:F2_NLO_Generalization}
    \end{equation}
    which is the generalization of \eqref{eq:NLO_FeynmanDiagrams_F2_211_1}-\eqref{eq:NLO_FeynmanDiagrams_F2_211_2}. The sum over the signs should be understood as the presence of two summands with opposite signs.
    The constant $c_1$ is then fixed by the topological sector, pinching/splitting limits of Section \ref{subsec:PinchingAndSplittingLimits} when possible and elementary Wick contractions otherwise.
\end{itemize}
\subsubsection{On the absence of transcendental functions}
\label{subsubsec:OnTheAbsenceOfTranscendentalFunctions}

We now comment on the absence of transcendental functions observed at next-to-leading order.
In this setup, supersymmetry forces the combination of the Feynman diagrams gathered in section \ref{subsec:211} to form a rational function. From the (super-)conformal block expansion \eqref{eq:211_SuperblockExpansion}, however, this cancellation is not immediately visible and results in constraints over CFT data. It is unclear at present whether we should expect this property to appear in other theories: for example, the result is still valid for the correlator $\vev{2\hat{1}\hat{1}}$ also for the non-supersymmetric Wilson line (as the diagrams in the first channel are the same).
We provide in this section a few hints about this curious cancellation for the correlators analyzed in the previous sections.

The first thing to notice is that, since we are dealing with OPE coefficients at leading order, we can restrict our operators to \textit{effective} ones, i.e., operators that have spin $s=0$ and non-vanishing bulk-defect and three-point functions.
Such operators are of the form
\begin{equation}
    \Oh_{\Dh}^{\text{eff}} \sim \Tds^{i_1 \ldots i_\Delta} \Wl [ \phi^{i_1} \ldots \phi^{i_\Delta} (\phi^6)^\Delta] + \ldots\,,
    \label{eq:Oheff}
\end{equation}
with $\Tds^{i_1 \ldots i_\Delta}$ a tensor structure 
This is required for the tree-level three-point functions to be potentially non-vanishing.
The terms hidden in the dots might contribute however to $b_{\Delta_1 \Dh}$.

\paragraph{Example: $\Dh=2$ for $\vev{2 \hat{1} \hat{1}}$.}
Let us illustrate Equation \eqref{eq:NLO_NoLog_211} for the OPE coefficient $\vev{b_{2 \Dh}^{(0)} \lambdah_{\hat{1}\hat{1}\Dh}^{(0)}}|_{\Dh=2}$, to convince ourselves that it is not a trivial relation.
From the correlator $\vev{2 \hat{1} \hat{1}}$, we expect through \eqref{eq:LO_CFTData_211} that
\begin{equation}
    \vev{b_{2 \Dh}^{(0)} \lambdah_{\hat{1}\hat{1}\Dh}^{(0)} \gamma_\Dh^{(1)}}_{\Dh^{(0)} = 2} = 0\,.
    \label{eq:211_CFTDataCancellation_2}
\end{equation}
At $\Dh^{(0)} = 2$, there are two operators contributing in the OPE $\Oh_1 \times \Oh_1$ which have the form \eqref{eq:Oheff}.
At one loop, they are defined as \cite{Correa:2018fgz}
\begin{equation}
    \Oh_{\pm} (x)
    =
    \phi^i \phi^i
    \pm
    \sqrt{5} \phi^6 \phi^6\,,
    \label{eq:Opm}
\end{equation}
through orthogonalization of the anomalous dimension matrix, and with $i \in \left\lbrace 1,...,5\right\rbrace$. The one-loop anomalous dimensions are given by \cite{Barrat:2024nod}
\begin{equation}
    \gamma_{\pm}^{(1)}
    =
    \frac{5 \pm \sqrt{5}}{16 \pi^2}\,.
    \label{eq:gammaOpm}
\end{equation}
From our results at leading order (see \eqref{eq:LO_CFTData_211}) we observe that
\begin{equation}
    \vev{ b_{2 \Dh}^{(0)} \lambdah_{\hat{1}\hat{1}\Dh}^{(0)} }_{\Dh^{(0)}=2}
    =
    b_{2 +}^{(0)} \lambdah_{\hat{1}\hat{1} +}^{(0)}
    +
    b_{2 -}^{(0)} \lambdah_{\hat{1}\hat{1} -}^{(0)}
    =
    - \frac{2 \sqrt{2}}{5}\,,
    \label{eq:Cancellation_Step}
\end{equation}
from which we conclude that
\begin{equation}
    b_{2 \pm}^{(0)} \lambdah_{\hat{1}\hat{1} \pm}^{(0)}
    \neq
    0\,.
    \label{eq:Cancellation_OPECoefficientsNonVanishing}
\end{equation}
Moreover, we have argued in \eqref{eq:Oheff} that the two three-point function coefficients are equal.
The equation \eqref{eq:211_CFTDataCancellation_2} can then be written as
\begin{equation}
    b_{2 +}^{(0)} \gamma_+^{(1)}
    +
    b_{2 -}^{(0)} \gamma_-^{(1)}
    =
    0\,,
    \label{eq:Cancellation_RelationOPECoefficients}
\end{equation}
from which we can extract the disentangled data
\begin{equation}
    b_{2 \pm}^{(0)} \lambdah_{\hat{1}\hat{1} \pm}^{(0)}
    =
    \frac{\sqrt{2}}{5} (1 \pm \sqrt{5})\,.
    \label{eq:Cancellation_DisentangledData}
\end{equation}
The absence of transcendental terms appears then to be the consequence of non-trivial relations among CFT data and a satisfying theoretical explanation is still missing.
It would be therefore interesting to understand how this expression generalizes for higher-length operators, and if it persists for models without supersymmetry.
\section{Summary and perspectives}
\label{sec:ConclusionsBDD}
In this chapter -- based on reference \cite{Artico:2024wnt} -- we have introduced and studied the correlation functions involving one bulk and two defect half-BPS operators in the context of the Wilson-line defect CFT in $\Nm=4$ sYM.
This setup for a supersymmetric defect preserves many properties of $\Nm=4$ sYM without defects, such as supersymmetry, integrability, and a large subset of conformal symmetry. These properties are used to strongly constrain correlators of local bulk operators and defect excitations: our focus is in particular on scalar operators whose scaling dimension is protected via supersymmetry.
Through a combination of non-perturbative constraints and perturbative insights, we derived novel results for external operators with arbitrary scaling dimensions at weak coupling. The non-perturbative constraints were presented in Section \ref{sec:BDDNonPerturbativeConstraints} and included in particular the results coming from supersymmetric localization, as well as remarks on the analyticity properties of the correlators.  We presented the perturbative results in Section \ref{sec:PerturbativeResults}: at weak coupling, we have shown that up to next-to-leading order the correlators are rational functions of the cross-ratios and we observe a cancellation of the expected transcendental terms.
This is explained through intricate relations between the OPE coefficients and the scaling dimensions, which follow from supersymmetry: it is unclear what the fate of this observation is in other models.
In the strong-coupling regime, we provided results for the correlator $\vev{2 \hat{1} \hat{1}}$ up to next-to-leading order, and presented partial results for the next-to-next-to-leading order contribution.
In this case, the correlator is known up to a rational function of the spacetime cross-ratio, which is further constrained through sum rules in the gist of \cite{Levine:2023ywq}.
\begin{figure}
\centering
\begin{subfigure}{.65\textwidth}
  \includegraphics[width=1\linewidth]{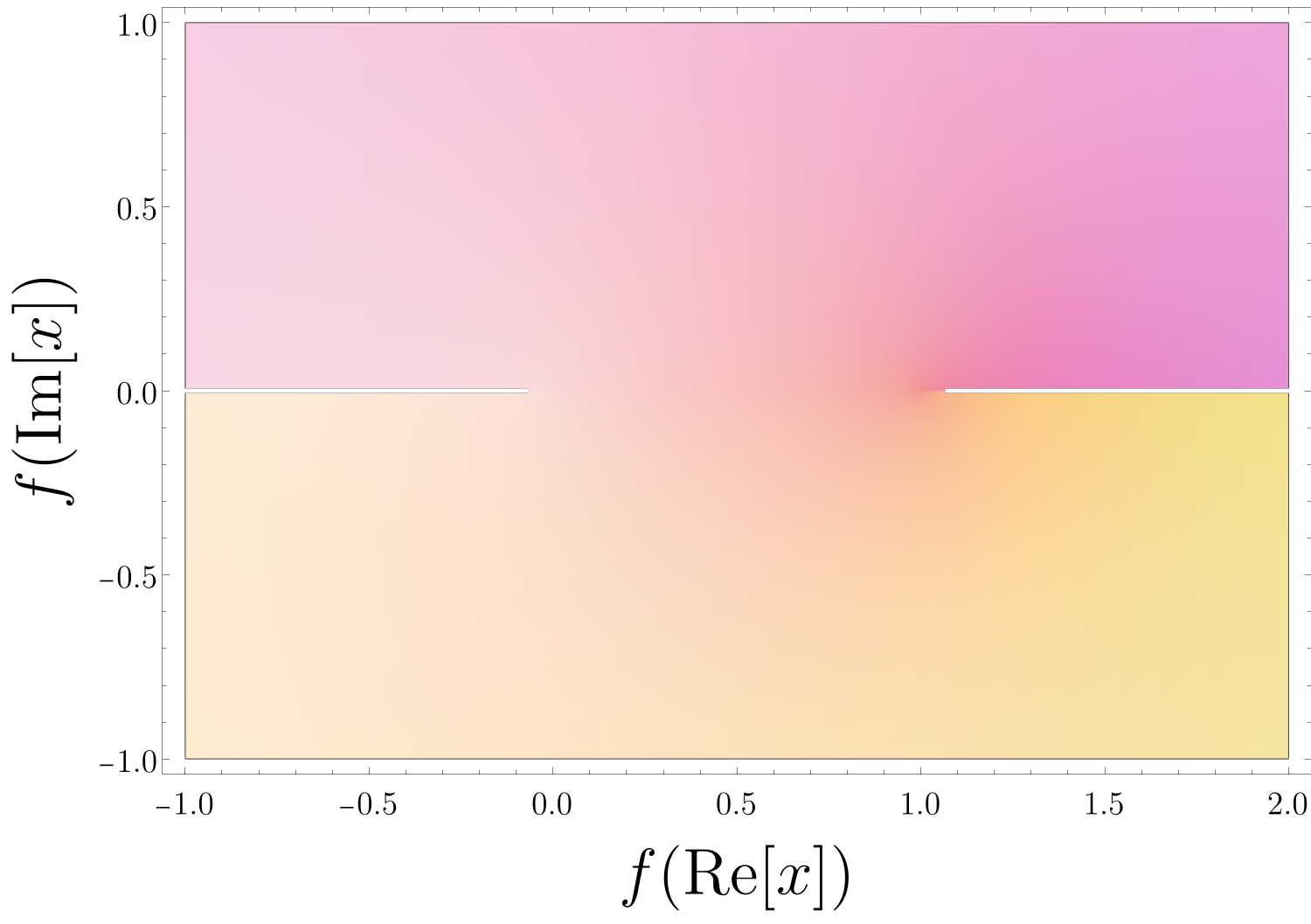}
\end{subfigure}%
\caption{Complex plot of the Feynman diagram given in Equation \eqref{eq:NLO_FeynmanDiagrams_F2_211_1}
Here, we defined $f(x) = \bigl(\pi + \arctan (\sqrt{(1-x)/x}) \bigr)^2$.
We observe a discontinuity at $x \geq 0$, incompatible with the locality constraint.
}
\label{fig:ComplexPlot_ArcTan}
\end{figure}
This work opens the path to several interesting research directions that can be pursued. In the following paragraphs, we list some of the most relevant ones.
\paragraph{Functional space.} 
In the weak-coupling regime, little is known about the functional space of bulk-defect-defect correlators. In the case of the Wilson line in $\Nm=4$ sYM, we encounter trigonometric functions of the spacetime cross-ratio $x$, which are rational for the whole range $x \in [0,1]$. Given that the variable $x$ is directly related to the angle formed by the two defect operators while keeping the bulk operator $\Op_{\Delta_1}$ fixed, a hypothesis worth testing is whether an angular variable makes the NNLO integrals more accessible. In other words, it would be interesting to study the integrals encountered at next-to-next-to-leading order using the variable $\phi = \arccos x$, and see if it leads to simplifications. Once the space of functions is understood, a valuable tool to further constrain the result might be locality, in the spirit of the discontinuity analysis presented in section \ref{sub:Locality}. Consider the plot presented in figure \ref{fig:ComplexPlot_ArcTan}: the functions appearing as the result of individual Feynman diagrams can have non-physical branch cuts; if the space of functions is known, applying locality constraints can partially fix the contribution from other diagrams, as non-physical discontinuities must cancel.
\paragraph{Loss of transcendentality and supersymmetric Ward identities.}
As stressed in section \ref{subsubsec:OnTheAbsenceOfTranscendentalFunctions}, at NLO we observe a surprising cancellation of the transcendental terms. Logarithmic terms are expected to appear from the conformal block expansion, and their absence can be explained through relations between the OPE coefficients. However, it would be valuable to gain a deeper insight into the reasons that prevent logarithms from appearing, in particular to see if supersymmetry is ultimately responsible for this cancellation.
This cancellation is not unique to supersymemtric correlators: the bulk-defect-defect correlator of the lowest-lying operators in the $O(N)$ model, using the $\veps$-expansion at the Wilson-Fisher fixed point (see \cite{Cuomo:2021kfm,Gimenez-Grau:2022czc,Gimenez-Grau:2022ebb,Bianchi:2022sbz,Pannell:2023pwz,Giombi:2023dqs} for related works) has a similar property. At next-to-leading order, a single diagram contributes:
\begin{equation}
\vev{\phi(x_1) \hat{\phi}(\tau_2) \hat{\phi}(\tau_3)} = \ONDiagramNLO \sim \pi^2-4\arccos \left(\sqrt{x}\right)
\label{eq:ONModel}
\end{equation}
We observe in this case as well an absence of transcendental terms, although this may be a mere coincidence as other bulk-defect-defect correlators in the $O(N)$ model are transcendental at next-to-leading order\footnote{We thank Miguel Paulos, Philine van Vliet and Julius Julius for valuable discussions on the topic of transcendentality in the $O(N)$ model.}. Future studies could investigate this phenomenon for more models (e.g. fishnet field theory in the presence of a defect \cite{Wu:2020nis,Gromov:2021ahm} or fermionic defect CFT \cite{Giombi:2022vnz, Barrat:2023ivo,Pannell:2023pwz})  and different external operators.\\
If the origin of this cancellation resides in supersymmetry, it might be the case that the simple guess of superconformal Ward identities in \eqref{eq:SCWI} cannot encode the constraints acting on the correlator and that a more careful derivation in the gist of \cite{Liendo:2016ymz} is necessary. Ongoing efforts are being made to approach the problem using the defect superspace formalism; it could be also interesting to study whether the phenomenon is present also in the case of the non-supersymmetric Wilson-line defect, where the diagrams are similar. This might clarify the role of the bulk and defect supersymmetry in this cancellation phenomenon.
\begin{figure}
\centering
\begin{subfigure}{.65\textwidth}
  \includegraphics[width=1\linewidth]{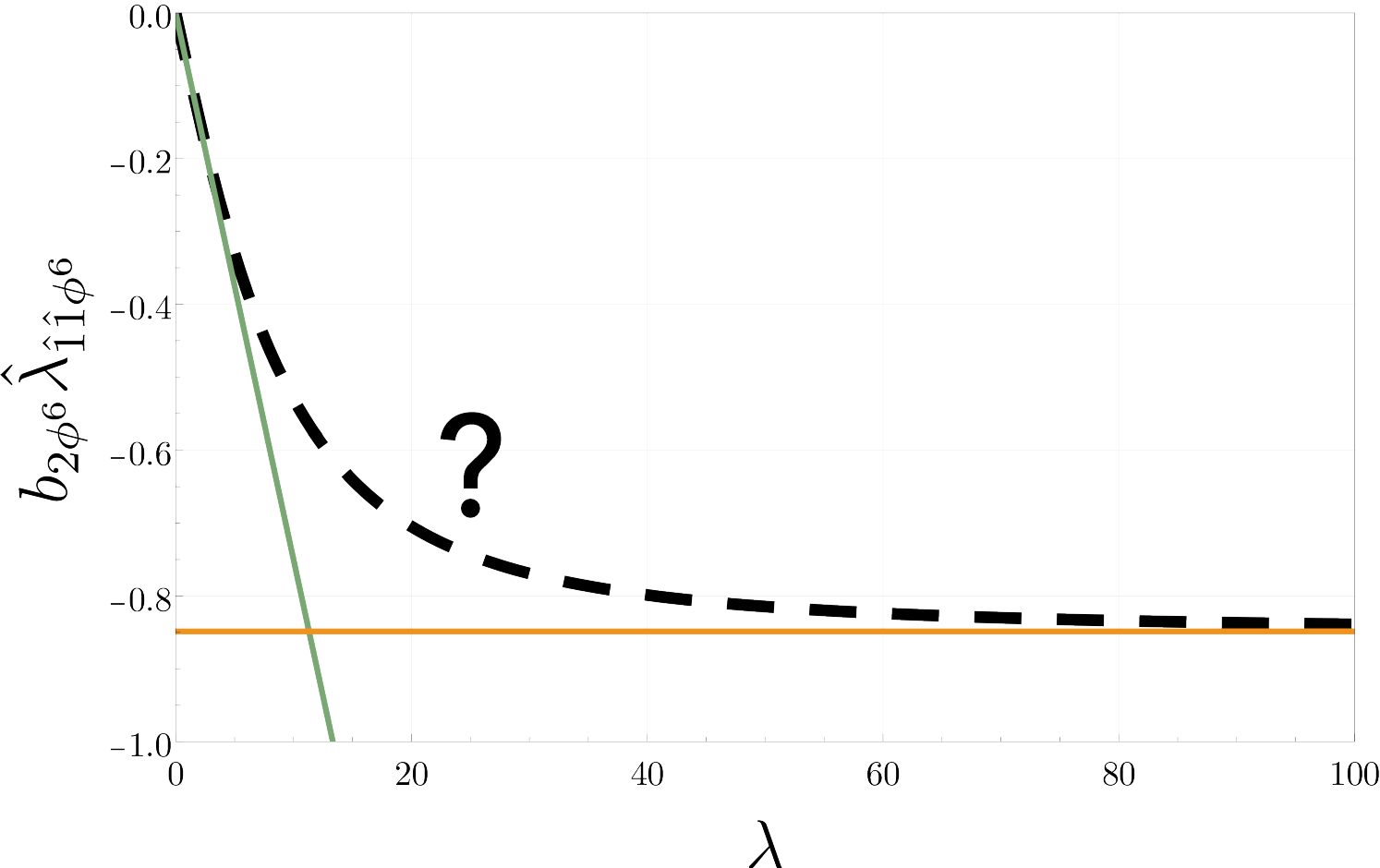}
\end{subfigure}%
\caption{Plot of the OPE coefficient $b_{2\phi^6} \lambdah_{\hat{1}\hat{1}\phi^6}$ as a function of the coupling constant.
The green line corresponds to the weak-coupling value, extracted from $\vev{2 \hat{1} \hat{1}}$, while the orange line represents the strong-coupling regime.
The black dashed line is a two-point Pad{\'e} approximation between these two regimes.
An interesting goal would be to use numerical bootstrap methods for studying this intermediate regime.
}
\label{fig:PlotConclusions}
\end{figure}
\paragraph{Bootstrap directions.}
The locality sum rules \eqref{eq:SumRulesLocality} are amenable to a numerical study, especially in this case where the spectrum is known through integrability across the conformal manifold \cite{Cavaglia:2021bnz,Cavaglia:2022qpg,Cavaglia:2022yvv}. However, typical bootstrap strategies are suffering from the fact that the OPE coefficients are not positive, as is the case for four-point functions of identical operators. An alternative method would be to truncate the sum \textit{à la} Gliozzi \cite{Gliozzi:2016cmg} and use an adapted version of the Tauberian theorem to estimate the tail, similarly to what was done in thermal cases \cite{Qiao:2017xif,Marchetto:2023xap,Barrat:2024aoa}.
It would also be interesting to see a full-fledged bootstrap study of a defect CFT that involves the four-point functions of defect operators $\vev{\Dh_1 \Dh_2 \Dh_3 \Dh_4}$, the two-point functions of bulk operators (in the presence of the defect) $\vev{\Delta_1 \Delta_2}$, together with the bulk-defect-defect correlators $\vev{\Delta_1 \Dh_2 \Dh_3}$. This should result in intertwining relations that might provide interesting relations for lifting degeneracies in perturbative settings. A good candidate of an observable to study numerically and that appears in $\vev{2 \hat{1} \hat{1}}$ is the combination $b_{2\phi^6} \lambdah_{\hat{1}\hat{1} \phi^6}$ (see Figure \ref{fig:PlotConclusions}). The three-point function $\lambda_{\hat{1}\hat{1} \phi^6}$ has been studied extensively and is known precisely from weak to strong coupling, while presently very little is known about $b_{\hat{1}\phi^6}$.
The numerical approach would also profit from integrated correlator relations, such as the ones derived for $\Nm=4$ SYM in \cite{Dorigoni:2022zcr,Alday:2023pet} and for line defects in \cite{Drukker:2022pxk,Cavaglia:2022qpg,Cavaglia:2022yvv,Pufu:2023vwo,Billo:2023ncz,Billo:2024kri,Dempsey:2024vkf}. A study of bulk-defect-defect integrated correlators is to this day not available in the literature and would be a good candidate for a deeper understanding of integrated correlators in the presence of a defect.

\chapter{Multipoint correlators}
\label{chapter:MP}
\section{Introduction and preliminaries}
\label{sec:PreliminariesMulti}

\subsection{The interest for multipoint correlators and the structure of this chapter}
This chapter completes the previous one with the analysis of correlators among multiple defect excitations in the context of the supersymmetric Wilson-line defect conformal field theory, and it is largely based on the paper \textit{Perturbative bootstrap of the Wilson-line defect CFT: Multipoint correlators} \cite{Artico:2024wut} by the author of this thesis and other collaborators. The aim of this third chapter is to apply the principles of the perturbative bootstrap to challenging multipoint defect correlators that involve multiple space-time and $R$-symmetry ratios, building on the principles described in the previous chapter. In this introduction, we will re-introduce some of the key concepts of defect CFT study and of the perturbative bootstrap, describing the importance of multipoint correlators and the approach we follow for computing them at weak-coupling up to next-to-next-to-leading order. The second part of this introduction focuses on the description of the vast literature on the topic of multipoint correlators and the different approaches applied to their study, which hopefully can clarify our contribution to this interesting and challenging topic.\\

Defects play a crucial role as observables in physics, with applications spanning from condensed-matter systems to high-energy physics.
In condensed matter, line defects typically correspond to point-like impurities in atomic lattices, while in pure Yang-Mills gauge theories, Wilson lines are essential in probing confinement \cite{Polyakov:1978vu,Witten:1998zw}.
In the previous chapter we have discussed how in conformal field theories, there exists a specialized class of defects, known as conformal defects, which break conformal symmetry in a controlled manner \cite{Billo:2016cpy,Lauria:2020emq}.
For line defects, this symmetry-breaking preserves a one-dimensional CFT.
Conformal line defects have been studied in a wide range of critical systems, from the $\mathrm{O}(N)$ model (and related theories) in various dimensions \cite{Cuomo:2021kfm,Gimenez-Grau:2022czc,Gimenez-Grau:2022ebb,Bianchi:2022sbz,Aharony:2023amq,Cuomo:2024psk} to supersymmetric theories \cite{Drukker:1999zq,Drukker:2000rr,Semenoff:2001xp,Drukker:2011za,Bianchi:2020hsz}, employing powerful techniques such as the conformal bootstrap \cite{Liendo:2016ymz,Liendo:2018ukf,Ferrero:2021bsb,Ferrero:2023znz,Ferrero:2023gnu,Barrat:2021yvp,Barrat:2022psm,Bianchi:2022ppi,Meneghelli:2022gps,Gimenez-Grau:2023fcy,Barrat:2020vch,Barrat:2021tpn,Barrat:2022eim,Bliard:2024und,Barrat:2024ta,Bliard:2023zpe,Peveri:2023qip,Barrat:2024nod,Barrat:2024ta2,Artico:2024wut}, integrability \cite{Cavaglia:2021bnz,Cavaglia:2022qpg,Cavaglia:2022yvv,Cavaglia:2023mmu}, and supersymmetric localization \cite{Drukker:2007yx,Giombi:2009ds,Giombi:2009ek,Buchbinder:2012vr,Beccaria:2020ykg,Giombi:2018qox}.
As introduced in section \ref{sub:SUSY} of last chapter, four-dimensional $\Nm=4$ Super Yang-Mills occupies a special position in the space of quantum field theories, due to its rich structure: it is a conformal field theory, believed to be integrable \cite{Beisert:2003tq,Beisert:2010jr}, and it has a well-studied holographic dual in the context of the AdS/CFT correspondence \cite{Maldacena:1997re,Witten:1998qj}.
A particularly notable conformal defect in this theory is the supersymmetric Maldacena-Wilson loop \cite{Maldacena:1998im}, whose definition in Euclidean space along a path $\Cm$ we repeat here:
\begin{equation}
    \Wm_{\Cm} = \frac{1}{N} \tr \Pm \exp \oint_{\Cm} d\tau (i \dot{x}_\mu A_\mu (\tau) + |\dot{x}|\, \theta^I \phi^I (\tau))\,,
    \label{eq:MaldacenaWilsonLoop}
\end{equation}
where $\theta^{I=1, \ldots, 6}$ is a polarization vector that defines which of the $\mathfrak{so}(6)_R$ scalar fields couple to the defect.
For a circular geometry, the operator becomes half-BPS, and the exact expectation value of this operator is given by a Bessel function \cite{Erickson:2000af,Drukker:2000rr,Pestun:2007rz,Pestun:2009nn}.
For an infinite straight line, it reduces to the simple value of $1$.
In recent years, the Wilson-line defect CFT in $\Nm=4$ sYM has garnered significant attention and it has been explored through approaches such as the conformal bootstrap \cite{Liendo:2016ymz,Liendo:2018ukf,Ferrero:2021bsb,Barrat:2021yvp,Barrat:2022psm,Ferrero:2023znz,Ferrero:2023gnu,Bonomi:2024lky}, integrability \cite{Giombi:2009ds,Giombi:2018qox,Giombi:2018hsx}, and a combination of both techniques, known as bootstrability \cite{Cavaglia:2021bnz,Cavaglia:2022qpg,Cavaglia:2022yvv,Cavaglia:2023mmu}.
Perturbative calculations, both at weak \cite{Barrat:2021tpn,Barrat:2022eim,Bianchi:2022ppi,Artico:2024wnt} and strong coupling \cite{Giombi:2017cqn,Gimenez-Grau:2023fcy,Giombi:2023zte,Artico:2024wnt}, have yielded a remarkable amount of data.\\
\begin{figure}
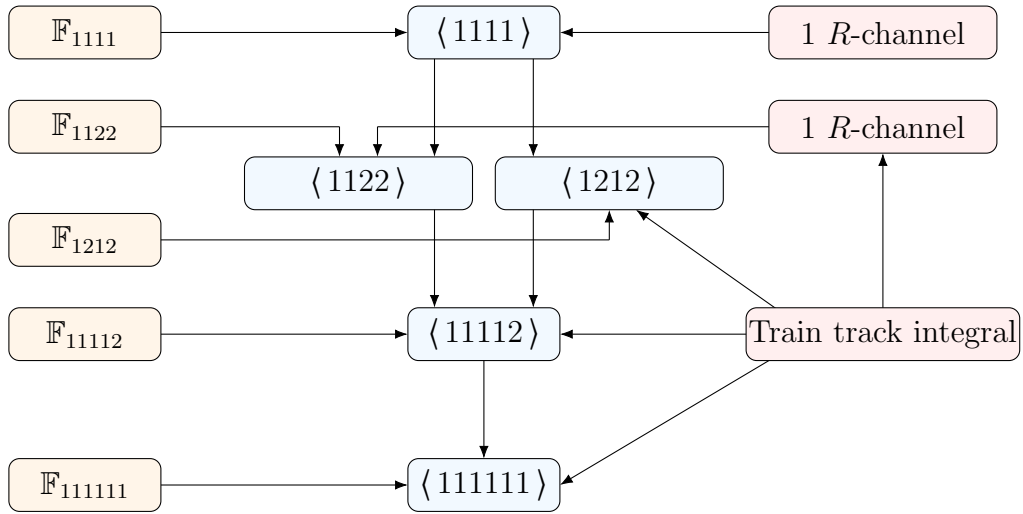

    \centering
    \DiagramIntro
    \caption{Illustration of the strategy used for calculating multipoint correlation functions. The center column consists of the correlators that are computed throughout this paper. The left column refers to the data accessible non-perturbatively, while the right column refers to the perturbative input. The correlators $\vev{1111}$ and $\vev{1122}$ require the input of one $R$-channel each, while the other correlators necessitate the topological data $\Fds$ and a single six-point integral, known as the train track.}
    \label{fig:DiagramIntro}
\end{figure}
%

In this chapter, following reference \cite{Artico:2024wut}, we focus on multipoint correlation functions of defect half-BPS operators. Several motivations drive this direction of study.
Higher-point correlation functions are a central focus of conformal bootstrap studies, as they encode vast amounts of CFT data and provide an alternative framework to the conventional analysis of multiple four-point functions.
Early progress in this area has been made, e.g., with studies of five-point functions in the Ising model \cite{Poland:2023vpn,Poland:2023bny} and six-point functions in one-dimensional systems \cite{Antunes:2023kyz,Harris:2024nmr}.\footnote{See \cite{Bercini:2020msp,Antunes:2021kmm,Buric:2021ywo,Buric:2021ttm,Buric:2021kgy,Kaviraj:2022wbw,Bargheer:2024hfx} for various works in this direction.}
The Wilson-line defect CFT is particularly well-suited for advancing these techniques, and a fascinating perspective is to merge the techniques of \cite{Antunes:2023kyz} and \cite{Cavaglia:2021bnz} to do a multipoint bootstrability study.
We aim to demonstrate how symmetry considerations, combined with appropriately constructed Ans\"atze, can lead to novel analytical results in the weak-coupling regime. In order to complete this task, multipoint superconformal Ward identities have been conjectured in \cite{Barrat:2021tpn}, before being confirmed and extended in \cite{Bliard:2024und,Barrat:2024ta}. The solution to these SCWI for multipoint correlators was presented in \cite{Artico:2024wut} and will be discussed later in the chapter.
In this chapter, we introduce a bootstrap approach for the perturbative analysis of multipoint correlators at weak coupling, leading to new results for the five-point function $\vev{11112}$ and the six-point function $\vev{111111}$.
This perturbative bootstrap method leverages non-perturbative constraints such as superconformal symmetry, crossing symmetry, and the pinching behavior of correlators to lower-point functions.
Remarkably, the correlators we will present in this chapter are all governed only by either one or two functions of the spacetime cross-ratios, significantly simplifying the computations.
This reduction allows us to focus on a selected class of diagrams, from which we can determine the correlators at next-to-next-to-leading order; moreover, we specifically demonstrate that the six-point train track integral in the collinear limit controls the correlators.
By inputting the result for this integral \cite{Rodrigues:2024znq} along with the non-perturbative constraints, we fully determine the correlators, providing new analytical expressions.
Each correlator is systematically constructed by building upon lower-point results, as illustrated in Figure \ref{fig:DiagramIntro}.
In the diagram in the figure, we identify the non-perturbative information with the symbol we use for the topological sector and a light orange color, while all necessary perturbative information comes in red on the left. The pinching to lower point operators corresponds to moving vertically among correlators having the sum of scaling dimensions of external operators equal to six.\\

The structure of the chapter is as follows.
In the rest of the section \ref{sec:PreliminariesMulti}, we introduce our notation on multipoint correlators and then provide a brief overview of the literature on the Wilson-line defect one-dimensional CFT, exploring the different approaches that have been employed for this challenge.
Section \ref{sec:NonPerturbativeConstraints} outlines the non-perturbative constraints that form the foundation for our calculations, while the perturbative results are reported in section \ref{sec:PertResultsMultip}. In particular, given the dependence of our results on lower-point functions, section \ref{sub:FourPointFunctions} presents explicit calculations for the correlators $\vev{1111}$, $\vev{1122}$, and $\vev{1212}$ up to next-to-next-to-leading order.
Our main results — the correlators $\vev{11112}$ and $\vev{111111}$ at next-to-next-to-leading order — are detailed in section \ref{sub:HigherPointFunctions}. Section \ref{sec:ConclusionsMulti} summarizes the chapter and proposes directions for future exploration. This chapter is completed by four appendices: appendix \ref{app:SymbolsAndGoncharovPolylogarithms} reviews symbols and Goncharov polylogarithms; appendix \ref{app:Integrals} provides the integrals necessary throughout this work; and appendices \ref{app:FeynmanDiagramsOf1111} and \ref{app:FeynmanDiagramsOf1122} summarize the Feynman diagram computations for the correlators $\vev{1111}$ and $\vev{1122}$, respectively.
\subsection{Multipoint defect correlators: definitions and conventions}
\label{sub:MultipointDefinition}
This section provides the foundational material essential for the discussion in the rest of the chapter.
An overview of the Wilson-line defect CFT was already included in section \ref{sec:TheWilsonLineDefectCFT} of the previous chapter, therefore we refer to that section for the definition of the bulk and defect theories and operators and to references \cite{Maldacena:1998im,Rey:1998ik} as fundamental papers introducing the supersymmetric Wilson-line. Here, we present the correlation functions that are the focus of subsequent sections and we discuss a special conformal integral (the six-point train track) that turns out to be essential for the calculation of multipoint correlators. The relevant Feynman rules necessary for the computation of the bulk and defect integrals discussed in this chapter are reported in appendix \ref{app:Integrals}.\\

For our calculations, we will make use of the following OPE coefficients regarding defect operators:
\begin{align}
    \lambda_{112} &= \frac{-2 \sqrt{\lambda}
   \Ids_1 (7 + \Ids_1) \Ids_2 \Ids_3 - (-32 + 14 \Ids_1 + \Ids_1^2) \Ids_2^2 \Ids_3 + \lambda (3 \Ids_1 \Ids_2^2 - \Ids_1^2 \Ids_3 + 9 \Ids_2^2 \Ids_3)}{4 \Ids_1 \sqrt{\lambda (3 \lambda - (-2 + \Ids_1) (10 + \Ids_1))}
  \Ids_2 \Ids_3} \,, \label{eq:lambda112} \\
    \lambda_{123} &= \frac{\sqrt{3} \Ids_2 \sqrt{\sqrt{\lambda} (\lambda  (-26 \Ids_1-3 \lambda +32)-288 (\Ids_1-1))+(\Ids_1-2) \Ids_1 \Ids_2 (5
   \lambda +72)}}{\sqrt{\lambda } \sqrt{(\Ids_1-2) \Ids_1 \Ids_2 (3 \lambda -(\Ids_1-2) (\Ids_1+10))}} \,, \label{eq:lambda123} \\
    \lambda_{222} &= \frac{6 (\Ids_1-17) \lambda +2 (\Ids_1-2) (\Ids_1 (\Ids_1+14)+148)}{(3 \lambda -(\Ids_1-2) (\Ids_1+10))^{3/2}}\,,\label{eq:lambda222}
\end{align}
where the functions $\Ids_a$ are defined in \eqref{eq:Ids}.

\subsubsection{Four-point functions}
\label{subsubsec:Correlators_FourPointFunctions}

We start our analysis with the simplest correlators of defect scalar half-BPS operators (defined in \eqref{eq:HalfBPSDefOperator}) with non-trivial kinematics: the four-point functions.
A reduced correlator can be defined as
\begin{equation}
    \vev{\Delta_1 \Delta_2 \Delta_3 \Delta_4}
    =
    \Km_{\Delta_1 \Delta_2 \Delta_3 \Delta_4} \Am_{\Delta_1 \Delta_2 \Delta_3 \Delta_4} (x; r,s)\,,
    \label{eq:4Pt_Correlator}
\end{equation}
where $\Km_{\Delta_1 \Delta_2 \Delta_3 \Delta_4}$ is a (super)conformal prefactor, and $\Am_{\Delta_1 \Delta_2 \Delta_3 \Delta_4} (x; r,s)$ is a dimensionless function depending on one spacetime cross-ratio $x$ and two $R$-symmetry variables $r$ and $s$. This can be done (with suitable differences to account for the multi-variate case) for any correlation function of half-BPS scalar operators, regardless of the scaling dimension.
The spacetime cross-ratio is defined as
\begin{equation}
    x
    =
    \frac{\tau_{12} \tau_{34}}{\tau_{13} \tau_{24}}\,,
    \label{eq:4Pt_CrossRatio_x}
\end{equation}
while the $R$-symmetry cross-ratios are given by
\begin{equation}
    r
    =
    \frac{(u_1 \cdot u_2)(u_3 \cdot u_4)}{(u_1 \cdot u_3)(u_2 \cdot u_4)}\,,
    \qquad
    s
    =
    \frac{(u_1 \cdot u_4)(u_2 \cdot u_3)}{(u_1 \cdot u_3)(u_2 \cdot u_4)}\,.
    \label{eq:4Pt_CrossRatio_rs}
\end{equation}

The reduced correlator can be decomposed into $R$-symmetry channels (or $R$-channels):
\begin{equation}
    \Am_{\Delta_1 \Delta_2 \Delta_3 \Delta_4}
    =
    \sum_{i=1}^{r} R_i\, F_i (x)\,,
    \label{eq:4Pt_RSymmetryChannels}
\end{equation}
where $r = r(\Delta_1, \ldots, \Delta_4)$ is the number of channels, and $R_i$ represents the basis elements, which remain unspecified for now but are linearly independent.
Beyond four-point functions, the number of channels $r$\footnote{Not to be confused with the $R$-symmetry cross ratio, but the meaning will always be clear from the context.} can be determined using a recursion relation for $n$-point functions independently of the chosen basis: 
\begin{equation}
    r (\Delta_1, \ldots, \Delta_n)
    =
    r (\Delta_1 - 1, \Delta_2 - 1, \ldots, \Delta_n)
    + \ldots
    +
    r (\Delta_1 - 1, \Delta_2, \ldots, \Delta_n - 1)\,,
    \label{eq:Recursion_RSymmetryChannels}
\end{equation}
with the initial conditions
\begin{equation}
    \begin{split}
        r (\Delta_1, \ldots, \Delta_i, 0, \Delta_{i+2}, \ldots, \Delta_n) &= r (\Delta_1, \ldots, \Delta_i, \Delta_{i+2}, \ldots, \Delta_n)\,, \\
        r (\Delta) &= 0\,, \\
        r (\Delta_1, \Delta_2) &= \delta_{\Delta_1 \Delta_2}\,.
    \end{split}
    \label{eq:Recursion_StartingValues}
\end{equation}
For external operators all having dimension $\Delta = 1$, the number of channels is given by the closed-form expression
\begin{equation}
    r(1,1, \ldots, 1)
    =
    (n-1)!!\,.
    \label{eq:Recursion_SpecialCase}
\end{equation}

In this chapter, we focus on three specific cases, all of which involve exactly three $R$-symmetry channels:\footnote{It is straightforward to extend this analysis to the more general correlators $\vev{11kk}$ and $\vev{1k1k}$, which also contain three $R$-symmetry channels.}
\begin{align}
        \vev{1111} &= (13)(24) \Am_{1111} (x;r,s)\,, \label{eq:4Pt_1111} \\
        \vev{1122} &= (13)(24)(34) \Am_{1122} (x;r,s)\,, \label{eq:4Pt_1122} \\
        \vev{1212} &= (13)(24)^2 \Am_{1212} (x;r,s)\,. \label{eq:4Pt_1212}
\end{align}
The shorthand notation for propagators equipped with an $R$-symmetry vector product was already introduced in \eqref{eq:1dSuperPropagator}.
Note that with the convention for the prefactors we introduced above, it is possible to express the reduced correlator $\Am_{\Delta_1 \Delta_2 \Delta_3 \Delta_4}$ as a linear function of the two $R$-symmetry ratios. A natural choice for the basis of $R$-symmetry channels is
\begin{equation}
    \Am (x;r,s)
    =
    F_1 (x)
    +
    \frac{r}{x^2} F_2 (x)
    +
    \frac{s}{(1-x)^2} F_3 (x)\,,
    \label{eq:4Pt_NaturalBasis}
\end{equation}
where the prefactors are chosen such to make the expression of the topological sector equal to the sum of the channels $F_j(x)$, as it will be clear in later sections.
The functions $F_j (x)$ are different for each of the correlators $\vev{1111}$, $\vev{1122}$, and $\vev{1212}$, but we omit additional subscripts to streamline the notation.
The context will clarify which correlator the respective $F_j$ functions refer to.

\subsubsection{Five-point functions}
\label{subsubsec:Correlators_FivePointFunctions}

We now focus on five-point functions, specifically on the case of $\vev{11112}$.
While more general configurations can be studied using the techniques presented here, we restrict ourselves to this case for clarity.
The reduced correlator is defined as
\begin{equation}
    \vev{11112}
    =
    \Km_{11112} \Am_{11112} ( \lbrace x; r, s, t \rbrace )\,,
    \label{eq:11112_Correlator}
\end{equation}
where the kinematic dependence should be understood as
\begin{equation}
    \lbrace x; r,s,t \rbrace
    =
    ( x_1, x_2; r_1, s_1, r_2, s_2, t_{12} )\,.
    \label{eq:11112_Shorthand}
\end{equation}
Indeed, all five-point functions depend on two spacetime cross-ratios:
\begin{equation}
    x_1
    =
    \frac{\tau_{12} \tau_{45}}{\tau_{14} \tau_{25}}\,,
    \qquad
    x_2
    =
    \frac{\tau_{13} \tau_{45}}{\tau_{14} \tau_{35}}\,,
    \label{eq:11112_CrossRatios_x}
\end{equation}
and five $R$-symmetry variables:
\begin{gather}
        r_1
        =
        \frac{(u_1 \cdot u_2) (u_4 \cdot u_5)}{(u_1 \cdot u_4) (u_2 \cdot u_5)}\,,
        \qquad
        s_1
        =
        \frac{(u_1 \cdot u_5) (u_2 \cdot u_4)}{(u_1 \cdot u_4) (u_2 \cdot u_5)}\,, \notag \\
        r_2
        =
        \frac{(u_1 \cdot u_3) (u_4 \cdot u_5)}{(u_1 \cdot u_4) (u_3 \cdot u_5)}\,,
        \qquad
        s_2
        =
        \frac{(u_1 \cdot u_5) (u_3 \cdot u_4)}{(u_1 \cdot u_4) (u_3 \cdot u_5)}\,, \label{eq:11112_CrossRatio_rst} \\
        t_{12}
        =
        \frac{(u_1 \cdot u_5) (u_2 \cdot u_3) (u_4 \cdot u_5)}{(u_1 \cdot u_4) (u_2 \cdot u_5) (u_3 \cdot u_5)}\,. \notag
\end{gather}
The (super)conformal prefactor $\Km_{11112}$ is chosen as
\begin{equation}
    \Km_{11112}
    =
    (14)(25)(35)\,.
    \label{eq:11112_Prefactor}
\end{equation}
Following the recursion relation \eqref{eq:Recursion_RSymmetryChannels}, the reduced correlator can be decomposed into six $R$-symmetry channels:
\begin{equation}
    \Am_{11112}
    =
    \sum_{i=1}^6 R_i F_i (x_1, x_2)\,,
    \label{eq:11112_RSymmetryChannels}
\end{equation}
for which the natural choice of basis is
\begin{equation}
    \lbrace R_i \rbrace
    =
    \biggl\lbrace
    1, \frac{r_1}{x_1^2}, \frac{s_1}{(1-x_1)^2}, \frac{r_2}{x_2^2}, \frac{s_2}{(1-x_2)^2}, \frac{t_{12}}{x_{12}^2}
    \biggr\rbrace\,.
    \label{eq:11112_NaturalBasis}
\end{equation}
Again, we see how this choice of the $R$-symmetry ratio allows for the reduced correlator $\Am_{11112}$ to be linear in the $R$-symmetry ratios, and for the topological sector to be the sum of the six channels.
\subsubsection{Six-point functions}
\label{subsubsec:Correlators_SixPointFunctions}

We now examine the six-point function of elementary insertions $\Op_1$.
Analogous to the lower-point functions discussed previously, this correlator can be expressed as
\begin{equation}
    \vev{111111}
    =
    \Km_{111111} \Am_{111111} ( \lbrace x; r, s, t \rbrace )\,.
    \label{eq:111111_Correlator}
\end{equation}
The function $\Am_{111111}$ depends on three spacetime cross-ratios and nine $R$-symmetry variables.\footnote{In fact, it depends on eight variables only.
For convenience, we keep the ninth variable throughout the paper, though it should be kept in mind that it can be expressed in terms of the other eight cross-ratios.
We thank Pietro Ferrero for bringing this to our attention. This does not modify the validity of the solution to the SCWI later presented.}
The spacetime cross-ratios are defined as
\begin{equation}
    x_1
    =
    \frac{\tau_{12}\tau_{56}}{\tau_{15}\tau_{26}}\,,
    \quad
    x_2\
    =
    \frac{\tau_{13}\tau_{56}}{\tau_{15}\tau_{36}}\,,
    \quad
    x_3\
    =
    \frac{\tau_{14}\tau_{56}}{\tau_{15}\tau_{46}}\,,
    \label{eq:111111_CrossRatio_x}
\end{equation}
while the $R$-symmetry variables are
\begin{gather}
    r_1
    =
    \frac{(u_1 \cdot u_2)(u_5 \cdot u_6)}{(u_1 \cdot u_5)(u_2 \cdot u_6)}\,,
    \quad
    r_2
    =
    \frac{(u_1 \cdot u_3)(u_5 \cdot u_6)}{(u_1 \cdot u_5)(u_3 \cdot u_6)}\,,
    \quad
    r_3
    =
    \frac{(u_1 \cdot u_4)(u_5 \cdot u_6)}{(u_1 \cdot u_5)(u_4 \cdot u_6)}\,, \notag \\
    s_1
    =
    \frac{(u_1 \cdot u_6)(u_2 \cdot u_5)}{(u_1 \cdot u_5)(u_2 \cdot u_6)}\,,
    \quad
    s_2
    =
    \frac{(u_1 \cdot u_6)(u_3 \cdot u_5)}{(u_1 \cdot u_5)(u_3 \cdot u_6)}\,,
    \quad
    s_3
    =
    \frac{(u_1 \cdot u_6)(u_4 \cdot u_5)}{(u_1 \cdot u_5)(u_4 \cdot u_6)}\,, \notag \\
    t_{12}
    =
    \frac{(u_1 \cdot u_6)(u_2 \cdot u_3)(u_5 \cdot u_6)}{(u_1 \cdot u_5)(u_2 \cdot u_6)(u_3 \cdot u_6)}\,,
    \quad
    t_{13}
    =
    \frac{(u_1 \cdot u_6)(u_2 \cdot u_4)(u_5 \cdot u_6)}{(u_1 \cdot u_5)(u_2 \cdot u_6)(u_4 \cdot u_6)}\,, \notag \\
    t_{23}
    =
    \frac{(u_1 \cdot u_6)(u_3 \cdot u_4)(u_5 \cdot u_6)}{(u_1 \cdot u_5)(u_3 \cdot u_6)(u_4 \cdot u_6)}\,.
    \label{eq:111111_CrossRatio_rst}
\end{gather}
The (super)conformal prefactor is chosen as
\begin{equation}
    \Km_{111111}
    =
    \frac{(15)^2 (26) (36) (46)}{(16) (56)}\,.
    \label{eq:111111_Prefactor}
\end{equation}
This choice is motivated by the natural basis of $R$-symmetry channels which, with this choice of prefactor, is
\begin{equation}
    \begin{split}
    \lbrace R_i \rbrace
    &=
    \biggl\lbrace
    \frac{t_{23}}{x_{23}^2}, \frac{t_{23} r_1}{x_{23}^2 x_1^2}, \frac{t_{23} s_1}{x_{23}^2 (1-x_1)^2},
    \frac{t_{13}}{x_{13}^2}, \frac{t_{13} r_2}{x_{13}^2 x_2^2}, \frac{t_{13} s_2}{x_{13}^2 (1-x_2)^2},
    \frac{t_{12}}{x_{12}^2}, \frac{t_{12} r_3}{x_{12}^2 x_3^2}, \frac{t_{12} s_3}{x_{12}^2 (1-x_3)^2}, \\
    & \phantom{=\ }
    \phantom{\biggl\lbrace}
    \frac{r_1 s_2}{x_1^2 (1-x_2)^2}, \frac{r_1 s_3}{x_1^2 (1-x_3)^2}, \frac{r_2 s_1}{x_2^2 (1-x_1)^2}, \frac{r_2 s_3}{x_2^2 (1-x_3)^2}, \frac{r_3 s_1}{x_3^2 (1-x_1)^2}, \frac{r_3 s_2}{x_3^2 (1-x_2)^2}
    \biggr\rbrace\,,
    \end{split}
    \label{eq:111111_NaturalBasis}
\end{equation}
i.e., this time $\Am_{111111}$ is a polynomial in the $R$-symmetry variables \eqref{eq:111111_CrossRatio_rst}, as it is not possible to have a linear function of such variables saturating the number of $R$-symmetry channels. The prefactors are as always chosen such that the topological sector is the sum of the $R$-symmetry channels.

\subsubsection{One integral to rule them all}
\label{subsub:OneIntegralToRuleThemAll}

A central part of the perturbative bootstrap approach to multipoint correlators is to demonstrate that higher-point correlators at next-to-next-to-leading order can be computed by imposing symmetry constraints, provided we know \textit{one} integral.
This section introduces this integral -- the six-point train track -- and discusses some important limits used throughout the chapter.\\

The six-point train track integral is a conformal two-loop integral defined as
\begin{equation}
    B_{123,456}
    =
    \TrainTrack\
    =
    \int d^4 x_7\, I_{15} I_{25} I_{35} X_{4567}\,,
    \label{eq:TrainTrack_Definition}
\end{equation}
which is expected to be elliptic when the external points are four-dimensional \cite{Bourjaily:2017bsb,Bourjaily:2018ycu,Ananthanarayan:2020ncn,Loebbert:2020glj,Kristensson:2021ani,Morales:2022csr,McLeod:2023qdf}.
The integral $X_{4567}$ is defined in appendix \ref{app:Integrals}.
This integral can be expressed in terms of polylogarithms in the collinear limit, i.e., when all external points are aligned.
This configuration was studied in \cite{Rodrigues:2024znq}, and the result can be expressed in terms of Goncharov polylogarithms as
\begin{equation}
    B_{123,456}
    =
    \frac{I_{15} I_{24} I_{36}}{128 \pi^{4}} b_{123,456} (x_1, x_2, x_3) \,,
    \label{eq:TrainTrack_Result}
\end{equation}
where
\begingroup
\allowdisplaybreaks
\begin{align}
    b_{123,456}
    &=
    \frac{x_{13}^2}{x_1 x_2 (1-x_3) x_{12}}
    \bigl(
    -G(1,x_1) G(1,x_2) G(1,x_3)+G(1,x_2) G(x_3,x_1) G(1,x_3) \notag \\
    &\phantom{=\ } -G(1,x_1) G(x_3,x_2)
    G(1,x_3)+G(x_3,x_1) G(x_3,x_2) G(1,x_3) \notag \\
    &\phantom{=\ }-2 G(1,0,x_1) G(1,x_3)+2 G(1,x_2,x_1)
     G(1,x_3)+2 G(x_3,0,x_1) G(1,x_3) \notag \\
    &\phantom{=\ }-2 G(x_3,x_2,x_1) G(1,x_3)+G(x_3,x_1)
    G(0,1,x_2)-G(1,x_1) G(0,x_3,x_2) \notag \\
    &\phantom{=\ }+G(x_3,x_1) G(1,0,x_2)+G(0,x_3) (-G(1,x_2)
    G(x_3,x_1)+G(1,x_1) (G(1,x_2) \notag \\
    &\phantom{=\ }+G(x_3,x_2))+2 G(1,0,x_1)-2 G(1,x_2,x_1))-G(x_3,x_2)
    G(1,x_2,x_1) \notag \\
    &\phantom{=\ }+G(x_3,x_2) G(1,x_3,x_1)+G(1,x_1) G(1,x_3,x_2)-G(x_3,x_1)
    G(1,x_3,x_2) \notag \\
    &\phantom{=\ }-G(1,x_1) G(x_3,0,x_2)-G(1,x_2) G(x_3,1,x_1)+G(0,x_2) (-2 G(0,x_3)
    G(1,x_1) \notag \\
    &\phantom{=\ }+2 G(1,x_3) G(1,x_1)-2 G(1,x_3)
    G(x_3,x_1)-G(1,x_3,x_1)+G(x_3,1,x_1)) \notag \\
    &\phantom{=\ }+G(1,x_1) G(x_3,1,x_2)-G(x_3,x_1)
    G(x_3,1,x_2)+G(1,x_2)
    G(x_3,x_2,x_1) \notag \\
    &\phantom{=\ }+G(1,0,x_3,x_1)-G(1,x_2,x_3,x_1)+G(1,x_3,0,x_1)-G(1,x_3,x_2,x_1) \notag \\
    &\phantom{=\ }-
    G(x_3,0,1,x_1)-G(x_3,1,0,x_1)+G(x_3,1,x_2,x_1)+G(x_3,x_2,1,x_1)
    \bigr)\,.
\end{align}
\endgroup
The variables $x_1, x_2, x_3$ correspond to the six-point spacetime cross-ratios defined in \eqref{eq:111111_CrossRatio_x}.
Note that, in the collinear limit, the result of the integral depends on the ordering of the external points, i.e., the subscripts on $B$ are not generally commutative.
The results for different orderings can be found in \cite{Rodrigues:2024znq} and are related via analytic continuation (that is non-trivial for multivariate Goncharov polylogarithms).
We provide a review of Goncharov polylogarithms in Appendix \ref{app:SymbolsAndGoncharovPolylogarithms}.

Interestingly, the well-known conformal kite integral \cite{Usyukina:1994iw,Drummond:2006rz} can be obtained as a special pinching limit of the train track:
\begin{align}
    K_{13,24}
    =
    \KiteIntegral
    =
    I_{13}^{-1} \lim\limits_{6 \to 1, 5 \to 3}
    B_{123,546}
    =
    \frac{I_{13} I_{24}}{512 \pi^{6}} \Phi^{(2)} (x)\,,
    \label{eq:KiteIntegral}
\end{align}
where
\begin{equation}
    \begin{split}
    \Phi^{(2)} (x) &=
    \frac{1}{x (1-x)}
    \bigl(
    G(0,0,1,x)+G(1,1,0,x)
    +G(0,1,0,x)+G(1,0,1,x) \\
    &\phantom{=\ }
    -2 (G(1,0,0,x) + G(0,1,1,x))
    \bigr)\,,
    \end{split}
\end{equation}
with $x$ being the four-point cross-ratio defined in \eqref{eq:4Pt_CrossRatio_x}. The train-track integral will play an important role in the upcoming sections.

\paragraph{Non-conformal integrals from conformal integrals}
\label{par:NonConformalIntegralsFromConformalIntegrals}

A worth-mentioning feature of the train track integral is that it can also be used to access \textit{non-conformal} integrals previously not known.
For example, the $H$-integral, defined as
\begin{equation}
    H_{12,34}
    =
    \Hintegral\
    =
    \int d^4 x_5\, I_{15} I_{25}\, Y_{345}\,,
    \label{eq:H1234_Definition}
\end{equation}
is non-conformal and remains unsolved to the best of our knowledge, but whose derivatives are well known and play a crucial role in the calculation of defect correlators at NLO in \cite{Barrat:2021tpn}, for example. It can be derived from the conformal train track integral as follows:
\begin{equation}
    H_{12,34}
    =
    \lim\limits_{\tau_5 \to \infty} I_{25}^{-1} I_{35}^{-1} B_{125,345}\,.
    \label{eq:H_From_B}
\end{equation}
This expression is valid even when the points are not aligned, suggesting that the $H$-integral is likely elliptic in the general case.
In the collinear limit, the $H$-integral can be expressed in Goncharov polylogarithms and depends on two variables (up to a prefactor): 
\begin{equation}
    H_{12,34} =
    \frac{h(u,v)}{8192 \pi^{10} \tau_{12} \tau_{34}}\,,
   \label{eq:H1234_1d}
\end{equation}
with
\begin{equation}
    \begin{split}
    h (u,v) &=
    G (\ub,v)
    \left( G (0,1,u)+G (1,0,u) \right)
    +
    2 (G (0,u) G (1,0,v) - G (1,u) G (1,0,v)) \\
    &\phantom{=\ }
    - G(1,v) \left(G (0,1,u)+G (1,0,u)-2 G (1,1,u) \right)
    +
    G (1,u) G (1,1-u,v) \\
    &\phantom{=\ }
    -
    2 G (0,u) G (1-u,0,v)-G (1,0,1-u,v) - G (1,1-u,0,v) \\
    &\phantom{=\ }
    +
    G (1-u,0,1,v)+G (1-u,1,0,v)\,,
   \end{split}
   \label{eq:H1234_1d_Result}
\end{equation}
where we defined the variables
\begin{equation}
    u = \frac{\tau_{12}}{\tau_{14}}\,, \qquad
    v = \frac{\tau_{34}}{\tau_{14}}\,.
    \label{eq:H_Variables}
\end{equation}

\subsection{A short review of the literature}
\label{sub:Literature}
In this section, we want to provide a short but comprehensive overview of the literature on the topic of multipoint correlators of half-BPS operators belonging to the one-dimensional defect CFT represented by the supersymmetric Wilson line. By multipoint correlators in this chapter, we refer to any correlator of more than three external operators, including the case of four-point correlators in the definition. This choice is slightly non-standard, as the literature on four-point correlators is so vast that they can be considered a topic on their own; however, they do not represent a special case in this chapter, as the tools we use to study four-point correlators are the same we use for higher-point ones. We start our exposition of the available techniques with an overview of the literature on four-point correlators.

\subsubsection{Four-point correlators}
\label{subsub:FourPointLiterature}
The four-point correlators \eqref{eq:4Pt_Correlator} have been a longstanding focus of any study involving operators in a CFT, as these correlators represent a simple yet interesting case of functions that are not completely fixed by conformal symmetry. Thanks to the OPE introduced already in \eqref{eq:OPEGeneral}, four-point correlators contain a vast amount of CFT data (actually infinite, although difficult to disentangle) and can be the subject of a numerical study in special cases, thanks to crossing and positivity constraints. Numerous techniques have been applied to the study of four-point functions, including conformal bootstrap, dispersion relations, supersymmetric localization, integrability, and the perturbative bootstrap explored in the rest of the chapter.

\paragraph{The conformal bootstrap.} 
To illustrate the application of the conformal bootstrap to the study of the four-point correlation function in the Wilson line-defect CFT we refer to \cite{Liendo:2016ymz} by Pedro Liendo and Carlo Meneghelli, to \cite{Liendo:2018ukf} by Liendo, Meneghelli, and Mitev, and to \cite{Ferrero:2021bsb,Ferrero:2023znz,Ferrero:2023gnu} by Ferrero and Meneghelli. The conformal bootstrap of four-point correlators performed in these papers introduces some of the ingredients that we are also going to use in our perturbative bootstrap: the superconformal Ward identities, the crossing equations, and the Ansatz of harmonic polylogarithms (HPLs). The study of the invariance of the correlators under the superalgebra $\mathfrak{osp}(4^*|4)$ leads to superconformal Ward identities strongly constraining the $R$-symmetry channels. The solution to such SCWI reduces the study of the correlator of the lowest-dimensional half-BPS scalar operators to the study of one function which is once again constrained by the crossing symmetry. The decomposition of correlators in superconformal blocks -- also respecting the SCWI -- according to selection rules of the form
\begin{equation}
    \Oh_1 \times \Oh_1 \longrightarrow \hat{\mathbf{1}} + \Bm_2 + \sum_{\Dh^{0}\geq 1} \Lm_{[0,0],0}^\Dh\,,
    \label{eq:SelRules_Oh1}
\end{equation}
together with the definition of a suitable Ansatz of HPLs allows for fixing the correlator $\vev{1111}$ up to fourth order ($N^3LO$ at strong coupling in $AdS_2$) from the knowledge of the unperturbed theory, symmetries, and consistency conditions. 
\paragraph{Dispersion relations.} In reference \cite{Bonomi:2024lky}, Bonomi and Forini reproduced the same impressive strong coupling calculation without making use of any Ansatz of HPLs by using a dispersion relation for the four-point correlator derived directly from the Lorentzian inversion formula \cite{Simmons-Duffin:2017nub}. Such a dispersion formula takes the form
\beq
\Am(z) = 	\int_0^1 dw w^{-2} \text{dDisc}\left[ \Am(z) \right] K(z,w)
\eeq
where $K(z,w)$ is an integration kernel and dDisc is the double discontinuity operator. Such an integral can be solved by plugging in the (super)conformal block expansion order by order in perturbation theory and using physical model assumptions that are similar to the ones used in the perturbative bootstrap; however, no Ansatz is needed to complete the calculation. The integration kernel and the integrals are non-trivial, and they were solved up to fourth order for strong 't Hooft coupling leading to a result consistent with \cite{Ferrero:2023gnu}. Note that some of the assumptions on the degeneracy of the spectrum and the so-called \textit{braiding symmetry}\footnote{The interested reader can refer to \cite{Ferrero:2023gnu} for a deeper understanding of this symmetry} are not valid anymore for weak coupling, thus making it impossible a straightforward application of the same bootstrap and dispersion relation techniques to weakly coupled correlators. 
\paragraph{Supersymmetric localization and integrability.}
An extremely relevant contribution to the computation of a particular polarization of multipoint correlator (called the topological sector) using supersymmetric localization is \cite{Giombi:2018qox} by Simone Giombi and Shota Komatsu. The localization technique \cite{Pestun:2007rz} consists in transforming the infinite-dimensional path integral of the Wilson-loop action into an integral defined over a finite-dimensional locus, and originally connected the exact expectation value of the Wilson-loop in terms of a matrix-model. Reference \cite{Giombi:2018qox} expands the localization result to multipoint correlators of insertions of generic scaling dimension, showing the correct Gram-Schmidt procedure to obtain the topological sector in the presence of operators of dimension bigger than one. The result expresses the topological correlator as a differential operator acting on the expectation value of a Wilson-loop 
\beq
\vev{\Dh_1...\Dh_n} = \prod_{k=1}^n F_{\Dh_k}(A') \vev{\Wm_\Cm(A')}|_{A'=A}\,.
\eeq
The connection to integrability techniques\footnote{By integrability here we mean that the theory has an infinite number of conserved quantities that constrain the dynamics and allow for the exact computation of observables at any value of the coupling constant.} emerges for large $N$. In the planar limit, the topological sector can be rewritten as
\beq
\vev{\Dh_1...\Dh_n} = \oint d\mu \prod_{k=1}^n Q_{\Dh_k}(x)\,,
\eeq
which is an integral over a measure that is not relevant to our discussion and that most importantly involves the functions $Q_{\Dh_k}(x)$ directly connected to the Quantum Spectral Curve (QSC) \cite{Gromov:2013pga,Gromov:2014caa} introduced in this context by Nikolay Gromov, Vladimir Kazakov, S\'ebastien Leurent, and Dmytro Volin. The QCS was originally developed as a reformulation of the spectral problem of $\Nm = 4$ sYM as a set of nonlinear Riemann–Hilbert-type equations for a set of analytic functions. In the context of supersymmetric localization for multipoint correlation functions, the importance of the connection to integrability is that the QSC is inherently a non-perturbative tool, thus allowing not only the computation of the topological sector for weak and strong coupling, but also the numerical calculation of exact results. 
\paragraph{The bootstrability approach.} A set of four papers by Cavaglià, Gromov, Julius, and Preti \cite{Cavaglia:2021bnz,Cavaglia:2022qpg,Cavaglia:2022yvv,Cavaglia:2023mmu} published between 2021 and 2023 marked an important unification of the integrability-based and the bootstrap-based approaches to compute four-point correlators. This remarkable approach starts from the idea that the Ansatz of HPLs used in the conformal bootstrap approach can be further constrained by the knowledge of the spectrum of the theory, computed for any value of the coupling constant by the Quantum Spectral Curve. The knowledge of the spectrum from QSC -- together with known results for integrated correlators for integrability -- allowed for the first determination of the correlator $\vev{1111}$ analytically at weak coupling up to next-to-next-to-leading order, which due to the absence of the braiding simmetry was not considered in the conformal bootstrap of \cite{Ferrero:2021bsb,Ferrero:2023znz,Ferrero:2023gnu}. The approach based on integrability and bootstrap, called \textit{bootstrability}, also allows for the first determination of four-point functions at finite coupling, thus leading to precise numerical constraints for some previously unknown conformal data.
\subsubsection{The perturbative calculation of higher-point correlators}
The combination of perturbative and non-perturbative techniques that form the starting point of the perturbative bootstrap approach was introduced in references \cite{Barrat:2021tpn,Barrat:2022eim} by Julien Barrat, Pedro Liendo, Giulia Peveri, and Jan Plefka. Despite being developed before bootstrability, we present it now as it represents the bridge between the study of four-point functions and the study of multipoint correlators having more than four external operators. In these two papers, the authors first developed a weak-coupling recursion relation that captures correlators at next-to-leading
order involving an arbitrary number of the elementary scalar fields -- protected and non-protected. Then, from the NLO correlators, they observed that all the correlators are annihilated by a special class of differential operators, representing the first draft of the multipoint superconformal Ward identities then formally derived and extended in \cite{Bliard:2024und}. The NLO recursion relation can be used to compute any correlator of protected external operators at weak coupling up to NLO, thus setting the starting point for the NNLO study performed in \cite{Artico:2024wut} and in this chapter.
\begin{figure}
\centering
\includegraphics[scale=0.35]{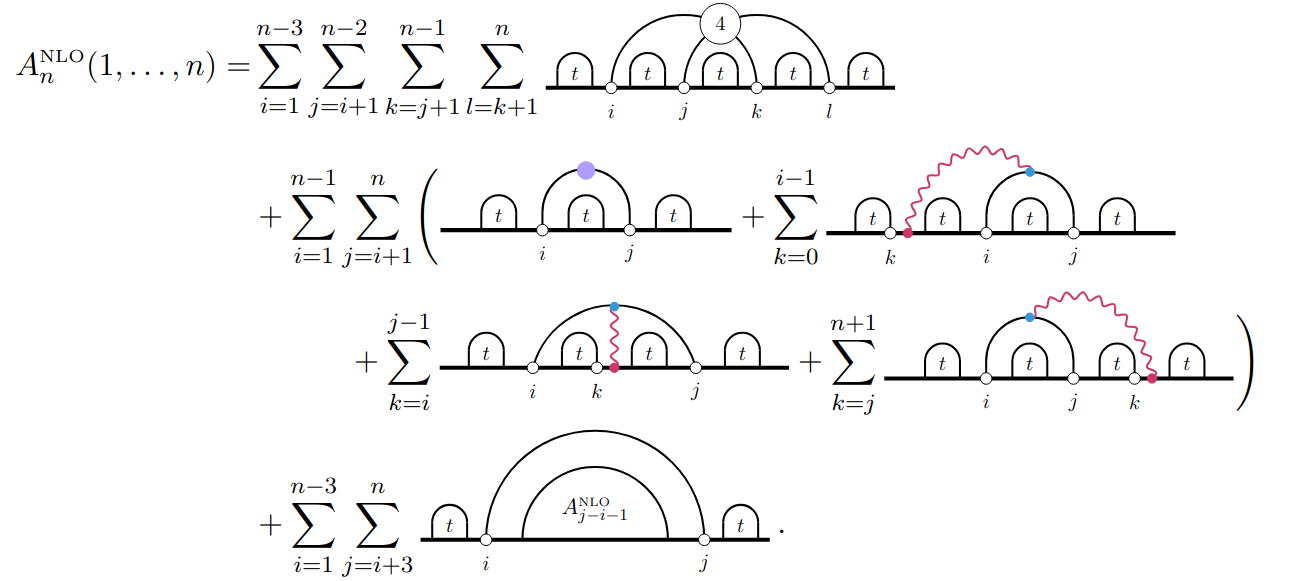}
\caption{Graphical representation of the $n$-point recursion relation from \cite{Barrat:2021tpn}. This diagrammatic formula produces all the relevant
Feynman diagrams for an arbitrary n-point function of $\Oh_1$ operators.}
\label{fig:RecRelation}
\end{figure}
Figure \ref{fig:RecRelation} shows the NLO recursion relation developed in \cite{Barrat:2021tpn,Barrat:2022eim}. This recursion relation makes it possible to compute all NLO correlators of protected scalar operators, however, this requires solving the integrals that appear in the relation. This task can be completed for NLO, however, a similar computation at NNLO is probably out of reach and involves a lot of unnecessary information coming from different $R$-symmetry channels that are related through crossing or superconformal symmetry. Our perturbative bootstrap for multipoint correlators starts from the non-perturbative superconformal Ward identities they obey in order to reduce the number of integrals to perform. Then, an Ansatz of HPLs is written in the spirit of the conformal bootstrap and the coefficients are fixed with all perturbative and non-perturbative constraints available. \\

Multipoint correlators are gaining relevance in the CFT community, as the growing literature shows (beyond \cite{Artico:2024wut}, \cite{Poland:2023vpn,Poland:2023bny,Antunes:2023kyz,Harris:2024nmr,Bercini:2020msp,Antunes:2021kmm,Buric:2021ywo,Buric:2021ttm,Buric:2021kgy,Kaviraj:2022wbw,Bargheer:2024hfx} represent different research directions around this topic). Among the papers we listed, reference \cite{Poland:2023vpn} on five-point correlators and \cite{Antunes:2023kyz} on six-point correlators represent two relevant examples to understand how higher-point correlators can play a role in the future of the conformal bootstrap. Since the conformal bootstrap has mostly been applied to four-point correlation functions with external scalar operators, most of the known OPE coefficients involve two scalars and one spinning operator.  OPE coefficients involving multiple spinning operators are typically not known, but they can indeed be studied by imposing consistency conditions on higher-point correlation functions with external scalar operators. Higher-point bootstrap represents a qualitatively novel and promising direction as it naturally combines the constraints from infinitely many four-point crossing equations. With this in mind, we conclude our summary of the available techniques for multipoint correlators in CFT (and especially on the supersymmetric Wilson-line defect) with table \ref{tab:summary_multipoint} summarizing the available results in the literature, and we proceed with listing the on-perturbative constraints multipoint correlators obey.
\begin{table}[ht]
\centering
\scriptsize
\begin{tabularx}{\textwidth}{@{}l X X X@{}}
\toprule
\textbf{Technique} & \textbf{Core Features} & \textbf{References} & \textbf{Main Results } \\
\midrule
\textbf{Conformal Bootstrap} & Crossing symmetry, SCWI, and HPL Ansatz constrained by $\mathfrak{osp}(4^*|4)$ superalgebra and selection rules & \cite{Liendo:2016ymz,Liendo:2018ukf,Ferrero:2021bsb,Ferrero:2023znz,Ferrero:2023gnu} & $\vev{1111}$ computed up to $N^3$LO using bootstrap constraints \textbf{(strong coupling)} \\
\addlinespace[0.4em]
\textbf{Dispersion Relation} & Lorentzian inversion formula, no functional Ansatz; integral kernel acts on double discontinuity & \cite{Bonomi:2024lky,Simmons-Duffin:2017nub} & $\vev{1111}$ computed up to $N^3$LO from integral representation \textbf{(strong coupling)} \\
\addlinespace[0.4em]
\textbf{Supersymmetric Localization} & Path integral reduces to matrix model; topological sector via differential operators; valid for all $\lambda$ & \cite{Giombi:2018qox,Pestun:2007rz} & Topological $n$-point correlators obtained analytically \textbf{(all couplings -- also valid at finite $N$)} \\
\addlinespace[0.4em]
\textbf{Integrability (QSC)} & Non-perturbative spectral problem solved; expectation values via Q-functions & \cite{Gromov:2013pga,Gromov:2014caa,Giombi:2018qox} & Spectrum of the theory determined exactly \textbf{(all couplings)} \\
\addlinespace[0.4em]
\textbf{Bootstrability} & Combines QSC spectrum with bootstrap Ansatz; integrated data fixes correlator form; includes numerics & \cite{Cavaglia:2021bnz,Cavaglia:2022qpg,Cavaglia:2022yvv,Cavaglia:2023mmu} & $\vev{1111}$ computed analytically to NNLO \textbf{(weak coupling)}; numerical data at \textbf{finite coupling} \\
\addlinespace[0.4em]
\textbf{Perturbative Techniques} & Weak-coupling recursion; SCWI reduces complexity & \cite{Barrat:2021tpn,Bliard:2024und} & $n$-point correlators at NLO computed recursively \textbf{(weak coupling)} \\
\bottomrule
\end{tabularx}
\caption{Summary of techniques and results for multipoint correlators on the supersymmetric Wilson line before reference \cite{Artico:2024wut} was published.}
\label{tab:summary_multipoint}
\end{table}

\section{Non-perturbative constraints}
\label{sec:NonPerturbativeConstraints}

In this section, we present the non-perturbative constraints that are instrumental for deriving the correlators in later sections \ref{sub:FourPointFunctions} and \ref{sub:HigherPointFunctions}. As in the previous chapter \ref{ch:BDD}, the goal is to reduce to the minimum the amount of perturbative information needed to compute defect correlators (multipoint correlators, in this chapter), therefore making the calculation possible even in presence of Feynman diagrams in other channels that have not been computed yet.
We begin the section by discussing in \ref{subsec:SuperconformalSymmetry} the constraints imposed by superconformal symmetry in the form of superconformal Ward identities, whose solutions are then further refined by the requirement of crossing symmetry in section \ref{subsec:CrossingSymmetry}.
A remarkable conclusion of this analysis is that all the correlators under consideration depend on a maximum of two functions of the spacetime cross-ratios.
We then explore in section \ref{subsec:Pinching} additional constraints on these functions emerging from the pinching limits of the correlators, relating limits of $n$-point functions to lower point functions. The analysis in each section is presented according to the number of the external operators considered.

\subsection{Superconformal symmetry}
\label{subsec:SuperconformalSymmetry}

As studied in section \ref{subsec:Topological} of the previous chapter, the constraints arising from superconformal symmetry can be expressed in the form of superconformal Ward identities (SCWI). In the case of the analysis of multipoint defect correlators, it is no longer true that all known SCWI are equivalent to the presence of a topological sector (whose existence can nevertheless be derived from the SCWI).
We present in this section how these constraints can be applied to the correlators presented in section \ref{sub:MultipointDefinition}.\footnote{Note that some of the content presented in this section overlaps with \cite{Barrat:2024ta}.}
To begin, we review the case of four-point functions, which was previously addressed in \cite{Liendo:2016ymz,Liendo:2018ukf}.
Building on this approach, we then extend the analysis to higher-point functions and provide new solutions to the SCWI for the correlators $\vev{11112}$ and $\vev{111111}$ that can be used to compute such correlators at weak coupling up to NNLO.

\subsubsection{Four-point functions}
\label{subsubsec:WI_FourPointFunctions}
Superconformal symmetry imposes stringent constraints on the four-point correlators \eqref{eq:4Pt_Correlator} under the form of a superconformal Ward identity.
These constraints are embodied in the differential constraint
\begin{equation}
    \left.
    \biggl(
    \frac{1}{2} \pd_x
    +
    \alpha \pd_r
    - (1-\alpha) \pd_s
    \biggr)
    \Am_{\Delta_1 \Delta_2 \Delta_3 \Delta_4} (x;r,s)
    \right|_{r \to \alpha x, s \to (1-\alpha) (1-x)}
    =
    0\,,
    \label{eq:4Pt_SCWI}
\end{equation}
for $\alpha \in \mathbb{R}$. These constraints have been originally derived and solved in \cite{Liendo:2016ymz} using superspace techniques.

As we have already mentioned, it follows from \eqref{eq:4Pt_SCWI} that all four-point functions exhibit a \textit{topological sector}, meaning that the dependence on kinematic variables disappears when the $R$-symmetry variables are aligned with the spacetime ones.
Specifically,
\begin{equation}
    \Am_{\Delta_1 \Delta_2 \Delta_3 \Delta_4} (x;x^2,(1-x)^2)
    =
    \Fds_{\Delta_1 \Delta_2 \Delta_3 \Delta_4}\,,
    \label{eq:4Pt_Topological}
\end{equation}
where $\Fds_{\Delta_1 \Delta_2 \Delta_3 \Delta_4}$ is a function of the coupling $\lambda$ alone that can be determined using localization techniques \cite{Giombi:2018qox}.
For our cases of interest, the topological sectors are
\begin{align}
    \Fds_{1111} &= \frac{3 \Ids_2^2}{\lambda \Ids_1^2} (\lambda + 8 - 4 \Ids_1)\,, \label{eq:1111_Fds} \\
    \Fds_{1122} &= \frac{1}{{8 \Ids_1 \Ids_2^2 \Ids_3 \Ids_4 \sqrt{\lambda } (3 \lambda -(\Ids_1-2)
   (\Ids_1+10))}} \bigl(
   15 \Ids_1 \Ids_3 \Ids_2^3 \lambda ^{3/2} \notag \\
   &\phantom{=\ }
   +
   \Ids_4 \bigl(\Ids_3 \bigl(\Ids_1^3 \lambda ^{3/2}+3 (\Ids_1+10)
   \Ids_2 \Ids_1^2 \lambda +3 \Ids_2^2 \Ids_1 \sqrt{\lambda } (\Ids_1 (\Ids_1+20)+14 \lambda +256) \notag \\
   &\phantom{=\ }
   +\Ids_2^3 (9 (3
   \Ids_1-56) \lambda +(\Ids_1-2) (\Ids_1 (\Ids_1+32)+832))\bigr) \notag \\
   &\phantom{=\ }
   -
   6 \Ids_1 \Ids_2^2 \lambda  \left(\Ids_1 \sqrt{\lambda
   }+(\Ids_1+28) \Ids_2\right)\bigr)
   \bigr)\,.
   \label{eq:1122And1212_Fds} \\
   \Fds_{1212} &= \Fds_{1122}
\end{align}

As discussed in section \ref{subsubsec:Correlators_FourPointFunctions}, the correlators $\vev{1111}$, $\vev{1122}$, and $\vev{1212}$ depend on three distinct $R$-symmetry channels.
Applying the superconformal Ward identities to the natural basis \eqref{eq:4Pt_NaturalBasis} leads to differential relations between these channels:
\begin{equation}
    \begin{split}
        F_1' (x) = - \frac{1}{1-x} F_3'(x)\,, \\
        F_2' (x) = - \frac{x}{1-x} F_3'(x)\,,
    \end{split}
    \label{eq:4Pt_NaiveSCWI}
\end{equation}
where $F'(x)$ denotes the derivative with respect to $x$.
This system of equations, however, is cumbersome as it requires integrating one channel to obtain another.
This issue becomes even more challenging for higher-point correlators, where the $R$-symmetry channels depend on multiple spacetime cross-ratios.
We expect however that there exists a basis in which two functions can be eliminated, reducing the solution of the Ward identity to a dependence on a single function, following the methods of \cite{Liendo:2016ymz}.
Below, we outline how to construct such a change of basis for the four-point functions we discuss.

We first define a new $R$-symmetry basis
\begin{equation}
    \Am (x;r,s)
    =
    \sum_{i=1}^{3} \tilde{R}_i (x;r,s) G_i(x)\,,
    \label{eq:4Pt_NewBasis}
\end{equation}
where $\Am$ may refer to $\Am_{1111}$, $\Am_{1212}$ or $\Am_{1122}$.
To circumvent the issue mentioned below \eqref{eq:4Pt_NaiveSCWI}, we impose the conditions as follows. The case for four external operators serves as a useful starting point for solving the SCWI for multipoint correlators as the steps to follow remain the same for higher values of $n$.
\begin{enumerate}
    \item One of the functions, $G_1(x)$, corresponds to the topological sector:
    \begin{equation}
        G_1 (x) = \Fds\,,
        \label{eq:4Pt_TopologicalCondition}
    \end{equation}
    where $\Fds$ represents $\Fds_{1111}$, $\Fds_{1212}$ or $\Fds_{1122}$.
    \item The derivative of $G_2(x)$ should not appear in the Ward identities, which leads to the condition
    \begin{equation}
        \left. \tilde{R}_2 \right|_{r \to \alpha x, s \to (1-\alpha) (1-x)}
        =
        0\,.
        \label{eq:4Pt_DerivativeCondition}
    \end{equation}
    \item We demand that $G_3 (x)$ does not appear when applying the Ward identity, but only its derivative.
    This can be obtained by demanding that the corresponding basis element $\tilde{R}_3(x)$ is an invariant of the Ward identity:
    \begin{equation}
        \left.
        \biggl(
        \frac{1}{2} \pd_x
        +
        \alpha \pd_r
        - (1-\alpha) \pd_s
        \biggr)
        \tilde{R}_3
        \right|_{r \to \alpha x, s \to (1-\alpha) (1-x)}
        =
        0\,.
        \label{eq:4Pt_SCWICondition}
    \end{equation}
\end{enumerate}
The remaining coefficients are arbitrary as long as the basis elements are linearly independent.
We normalize the solution by imposing the following additional conditions:
\begin{enumerate}
    \item[4.] Since at weak coupling and large $N$ the natural basis \eqref{eq:4Pt_NaturalBasis}, $F_1(x)$ is simpler than $F_2(x)$ and $F_3(x)$, we relate $G_2(x)$ to $F_1(x)$ directly:
    \begin{equation}
        G_2 (x) \sim F_1 (x)\,,
        \label{eq:4Pt_PlanarityCondition}
    \end{equation}
    allowing a proportionality function of $x$.
    \item[5.] Anticipating the results of section \ref{subsec:CrossingSymmetry}, we demand that $G_2(x)$ is anti-self-crossing for $\vev{1111}$ and $\vev{1212}$:
    \begin{equation}
        G_2 (x) + G_2 (1-x)
        =
        0\,,
        \label{eq:4Pt_CrossingCondition}
    \end{equation}
    which implies that $G_2(x)$ vanishes at leading order:
    \begin{equation}
        G_2 (x)
        =
        0 + \Op(\lambda)\,.
        \label{eq:4Pt_LeadingOrderCondition}
    \end{equation}
    Note that the condition \eqref{eq:4Pt_CrossingCondition} does not hold for $\vev{1122}$.
    This is due to the fact that this correlator does not cross to itself (except in a trivial way), as commented in Section \ref{subsubsec:Crossing_FourPointFunctions}.
\end{enumerate}

Based on these conditions, we can define an appropriate change of basis between the natural basis \eqref{eq:4Pt_NaturalBasis} and $\tilde{R}_j$ to apply the Ward identities.
For instance, the following dictionary provides a convenient change of basis:
\begin{equation}
    \begin{split}
        G_1 (x) &= F_1 (x) + F_2 (x) + F_3 (x) = \Fds\,, \\
        G_2 (x) &= F_1 (x)\,, \\
        G_3 (x) &= \frac{1}{2} ((2x-1) F_1 (x)+F_2 (x)-F_3 (x))\,.
    \end{split}
    \label{eq:4Pt_ChangeOfBasis}
\end{equation}
Finally, relabeling $G_3(x)$ as $f(x)$, the solution can be written succinctly as
\begin{equation}
    \Am (x;r,s)
    =
    \frac{1}{2} \biggl( \frac{r}{x^2} + \frac{s}{(1-x)^2} \biggr) \Fds
    +
    \pd_x ( \xi f(x) )\,,
    \label{eq:4Pt_SolutionSCWI}
\end{equation}
where the auxiliary function $\xi$ is defined as
\begin{equation}
    \xi
    =
    1 - \frac{r}{x} - \frac{s}{1-x}\,.
    \label{eq:4Pt_HelpFunction}
\end{equation}
The function $f(x)$ is closely related to the solution to the SCWI in \cite{Liendo:2018ukf}, which is obtained using a different change of $R-$symmetry basis. The choice here was made by requesting \eqref{eq:4Pt_CrossingCondition} for $\vev{1111}$ and $\vev{1212}$, which disallows the presence of a constant term in $f(x)$.\footnote{The results presented later in section \ref{sub:FourPointFunctions} seem to include a constant, but this is an artefact of the chosen fibration basis.
A series expansions around $x \sim 0$ or $x \sim 1$ shows no constant is present.}
The following relation is useful to keep in mind:
\begin{equation}
    f'(x)
    =
    F_1(x)\,,
    \label{eq:fFromF1}
\end{equation}
which follows directly from the choice of basis above.
In section \ref{sub:FourPointFunctions}, we will provide the function $f(x)$ for the correlators $\vev{1111}$, $\vev{1122}$, and $\vev{1212}$ up to next-to-next-to-leading order.
In the following sections, we extend this method to solve the SCWI for higher-point functions, such as $\vev{11112}$ and $\vev{111111}$.

\subsubsection{Five-point functions}
\label{subsubsec:WI_FivePointFunctions}

The superconformal Ward identities discussed for four-point functions in \eqref{eq:4Pt_SCWI} have a natural extension to multipoint correlators.
It was conjectured in \cite{Barrat:2021tpn} that for five-point functions, the SCWI takes the following form:
\begin{equation}
    \sum_{i = 1,2}
    \biggl(
    \frac{1}{2} \partial_{x_i}
    +
    \alpha_i \partial_{r_i}
    -
    (1-\alpha)_i \partial_{s_i}
    \biggr)
    \Am_{\Delta_1 \ldots \Delta_5} \bigr|_{r_i \to \alpha_i x_i, s_i \to (1-\alpha)_i (1-x_i), t_{ij} \to \alpha_{ij} x_{ij}} = 0\,,
    \label{eq:5Pt_SCWI_Conjecture}
\end{equation}
where $\alpha_i \in \mathbb{R}$.
Later work \cite{Bliard:2024und,Barrat:2024ta} found that these constraints are part of a more general set of Ward identities:
\begin{equation}
    \sum_{i \neq j}^2 \beta_i
    \biggl(
    \frac{1}{2} \partial_{x_i}
    +
    \alpha_i \partial_{r_i}
    -
    (1-\alpha)_i \partial_{s_i}
    +
    \alpha_{ij} \partial_{t_{12}}
    \biggr)
    \Am_{\Delta_1 \ldots \Delta_5} \bigr|_{r_i \to \alpha_i x_i, s_i \to (1-\alpha)_i (1-x_i), t_{ij} \to \alpha_{ij} x_{ij}} = 0\,,
    \label{eq:5Pt_SCWI}
\end{equation}
with $\alpha_i, \beta_i \in \mathbb{R}$.
This equation encodes the constraints imposed by superconformal symmetry that are known so far.

Similarly to the case of four-point functions, the SCWIs imply the existence of a topological subsector for five-point functions that can be computed using supersymmetric localization  \cite{Giombi:2018qox}.
For instance, in the case of $\vev{11112}$, the exact evaluation of the topological sector yields
\begin{equation}
    \Fds_{11112}
    =
    \frac{6 \Ids_2^2}{\lambda \Ids_1^2} \frac{2(\Ids_1 - 2)(\Ids_1 + 28) + \lambda (2 \Ids_1 - 19)}{\sqrt{3 \lambda - (\Ids_1 - 2)(\Ids_1 + 10)}}\,.
    \label{eq:11112_Fds}
\end{equation}

The differential constraints in \eqref{eq:5Pt_SCWI} in the natural basis lead once again to a differential equation system of difficult solutions; however, they can be solved using the same principles discussed for four-point functions.
Below, we outline the steps for transforming the natural basis of $\vev{11112}$ into a form analogous to \eqref{eq:4Pt_SolutionSCWI}.

We introduce a new basis for the correlator:
\begin{equation}
    \Am_{11112}
    =
    \sum_{j=1}^6 \tilde{R}_j G_j (x_1, x_2)\,.
    \label{eq:11112_NewBasis}
\end{equation}
In analogy to the four-point case, we impose the following conditions on the functions $G_j$:
\begin{enumerate}
    \item The topological limit is given by
    \begin{equation}
        G_1 (x_1, x_2)
        =
        \Fds_{11112}\,,
        \label{eq:11112_TopologicalCondition}
    \end{equation}
    meaning that all other basis functions vanish when $r_i \to x_i^2, s_i \to (1-x_i)^2, t_{ij} \to x_{ij}^2$.
    \item The derivatives of $G_{2,3,4} (x_1, x_2)$ do not appear after applying the Ward identities,\footnote{The number of functions that can be eliminated in this way is not arbitrary.
    Starting with the Ansatz
    \begin{equation}
        \tilde{R}_{i=2, \ldots, i_\text{max}+1}
        =
        a_1^{(i)} + a_2^{(i)} \frac{r_1}{x_1} + a_3^{(i)} \frac{s_1}{1-x_1} + a_4^{(i)} \frac{r_2}{x_2} + a_5^{(i)} \frac{s_2}{1-x_2} + a_6^{(i)} \frac{t_{12}}{x_{12}}\,,
        \label{eq:11112_DerivativeCondition}
    \end{equation}
    we find that $i_\text{max}=3$ to ensure linear independence.} leading to the condition
    \begin{equation}
        \left. \tilde{R}_{2,3,4} \right|_{r_i \to \alpha_i x_i, s_i \to (1-\alpha)_i (1-x_i), t_{ij} \to \alpha_{ij} x_{ij}}
        =
        0\,.
        \label{eq:11112_DerivativeConditionExplained}
    \end{equation}
    \item The remaining basis elements, $\tilde{R}_{j=5,6}$, are chosen to satisfy the Ward identities \eqref{eq:5Pt_SCWI}.
    \item We demand that the functions $G_{j=2,3,4} (x_1, x_2)$ are directly related to the simplest channels at weak coupling, i.e.,
    \begin{equation}
        \begin{split}
        G_2 (x_1, x_2) &\sim F_1 (x_1, x_2)\,, \\
        G_3 (x_1, x_2) &\sim F_3 (x_1, x_2)\,, \\
        G_4 (x_1, x_2) &\sim F_4 (x_1, x_2)\,.
        \end{split}
        \label{eq:11112_PlanarityCondition}
    \end{equation}
    \item Anticipating Section \ref{subsec:CrossingSymmetry}, we select the functions $G_{j=5,6}$ such that they are related by crossing symmetry, i.e.,
    \begin{equation}
        G_5 (x_1,x_2)
        =
        G_6 (1-x_2,1-x_1)\,.
        \label{eq:11112_CrossingCondition}
    \end{equation}
\end{enumerate}

The transformation between the natural and new basis is not unique, but one possible choice that satisfies these conditions is
\begin{equation}
    \begin{split}
    G_1 (x_1, x_2) &= \sum_{i=1}^6 F_i (x_1, x_2) = \Fds_{11112}\,, \\
    G_2 (x_1, x_2) &= F_1 (x_1, x_2)\,, \\
    G_3 (x_1, x_2) &= \frac{F_3 (x_1, x_2)}{1-x_1}\,, \\
    G_4 (x_1, x_2) &= \frac{F_4 (x_1, x_2)}{x_2}\,, \\
    G_5 (x_1, x_2) &= (1-x_2) F_1 (x_1,x_2) + \frac{(1-x_2) F_3 (x_1,x_2)}{1-x_1} + F_5 (x_1,x_2)\,, \\
    G_6 (x_1, x_2) &= x_1 F_1 (x_1,x_2)+F_2 (x_1,x_2)+\frac{x_1 F_4 (x_1,x_2)}{x_2}\,.
    \end{split}
    \label{eq:11112_ChangeOfBasis}
\end{equation}
After applying the Ward identities and eliminating $G_{1,\ldots,4}$, the five-point function solution takes the elegant form
\begin{equation}
    \begin{split}
        \Am_{11112}
        &=
        \frac{t_{12}}{x_{12}^2} \Fds_{11112}
        + \rho f_1
        + \pd_{x_1} (\xi_1 f_1)
        + \pd_{x_2} (\xi_2 f_1)
        + \pd_{x_2} (\eta f_2)\,,
    \end{split}
    \label{eq:11112_SolutionSCWI}
\end{equation}
where we have relabeled $G_5 \to f_1$ and $G_6 \to f_2$.
The auxiliary functions are given by
\begin{align}
    \xi_1 &= \frac{1-x_1}{1-x_2} \biggl( 1 - \frac{r_1}{x_1} - \frac{s_1}{1-x_1} \biggr)\,, \label{eq:11112_HelpFunction1} \\
    \xi_2 &= 1 - \frac{r_1}{x_1} - \frac{s_2}{1-x_2} + \frac{t_{12}}{x_{12}}\,, \label{eq:11112_HelpFunction2} \\
    \eta &= \frac{x_2}{x_1} \biggl( \frac{r_1}{x_1} - \frac{r_2}{x_2} - \frac{t_{12}}{x_{12}} \biggr)\,, \label{eq:11112_HelpFunction3} \\
    \rho&= - \frac{1}{1-x_2}\biggl( 1 - \frac{r_1}{x_1^2} \biggr)\,. \label{eq:11112_HelpFunction4}
\end{align}
It is however important to note that the solution \eqref{eq:11112_SolutionSCWI} does not satisfy the Ward identities on its own the Ward identities, as the crossing condition we have introduced in \eqref{eq:11112_CrossingCondition} is still not manifest.
There exists indeed one additional constraint on the derivatives of $f_1$ and $f_2$, which can be expressed as
\begin{equation}
    \biggl(
    \pd_{x_1} + \frac{x_2}{x_1} \pd_{x_2}
    \biggr) f_2
    +
    \biggl(
    \frac{1-x_1}{1-x_2} \pd_{x_1} + \pd_{x_2}
    \biggr) f_1
    =
    0\,.
    \label{eq:11112_ExtraConstraint}
\end{equation}
This constraint is not crucial for our purposes, as in the next section, we show that crossing symmetry allows us to eliminate $f_2$ altogether.
However, it should be taken into account when deriving superconformal blocks in the gist of \cite{Liendo:2016ymz}.

\subsubsection{Six-point functions}
\label{subsubsec:WI_SixPointFunctions}
As described in  \cite{Bliard:2024und,Barrat:2024ta}, the superconformal Ward identities for six-point functions can be written as
\begin{equation}
    \sum_{i \neq j}^3 \beta_i
    \biggl(
    \frac{1}{2} \partial_{x_i}
    +
    \alpha_i \partial_{r_i}
    -
    (1-\alpha)_i \partial_{s_i}
    +
    \alpha_{ij} \partial_{t_{ij}}
    \biggr)
    \Am_{\Delta_1 \ldots \Delta_6} \bigr|_{r_i \to \alpha_i x_i, s_i \to (1-\alpha)_i (1-x_i), t_{ij} \to \alpha_{ij} x_{ij}} = 0\,.
    \label{eq:6Pt_SCWI}
\end{equation}
As with lower-point functions, \eqref{eq:6Pt_SCWI} implies the existence of a topological sector.
The topological limit of the six-point function $\vev{111111}$ is given by
\begin{equation}
    \Fds_{111111}
    =
    \frac{15 \Ids_2^3}{\lambda^{3/2} \Ids_1^3} ((\lambda + 24) \Ids_1 - 8 ( \lambda + 6))\,.
    \label{eq:111111_Fds}
\end{equation}
The Ward identities can be solved for $\vev{111111}$ in the same way as for the other correlators, using the principles described in the previous sections.
However, the solution is too lengthy to be displayed explicitly; the complete expression is provided in the ancillary \textsc{Mathematica} notebook attached to reference \cite{Artico:2024wut}.

Importantly, the Ward identities allow us to eliminate ten functions from the general solution.
Additionally, the remaining four functions are subject to non-trivial differential constraints, similar to those seen for the five-point correlator, and are related by crossing.
While solving these constraints in a closed form is not critical for computing the correlator, it is a relevant research direction the exploration of their direct solutions in terms of a lower number of functions.

\subsection{Crossing symmetry}
\label{subsec:CrossingSymmetry}

We now turn our attention to the constraints imposed by crossing symmetry -- playing a fundamental role in \cite{Liendo:2018ukf}, see also the explanation of crossing relations in \cite{CFTdefectsNotesHerzog} -- on the multipoint correlators discussed in this chapter. As mentioned above, they are particularly relevant in the solution to the SCWI because they further reduce the number of functions that are necessary to describe the correlator. In practice, this means that an Ansatz of HPLs for one of the functions also provides the Ansatz for another function related by crossing. The two Ans\"atze can then be brought into the same fibration basis of HPLs.

\subsubsection{Four-point functions}
\label{subsubsec:Crossing_FourPointFunctions}

For four-point functions, crossing symmetry imposes the relation
\begin{equation}
    \vev{\Op_{\Delta_1} (1) \Op_{\Delta_2} (2) \Op_{\Delta_3} (3) \Op_{\Delta_4} (4)}
    =
    \vev{\Op_{\Delta_1} (1) \Op_{\Delta_4} (4) \Op_{\Delta_3} (3) \Op_{\Delta_2} (2)}\,,
    \label{eq:4Pt_CrossingSymmetry}
\end{equation}
with $(i) = (u_i, \tau_i)$.
The implications of crossing symmetry for specific four-point correlators, such as $\vev{1111}$, $\vev{1122}$, and $\vev{1212}$, were investigated in \cite{Liendo:2018ukf}.
We review here those results.
For the correlators $\vev{1111}$ and $\vev{1212}$, the crossing relation \eqref{eq:4Pt_CrossingSymmetry} leads to the following conditions:
\begin{align}
    F_1 (x) &= F_1 (1-x)\,, \label{eq:4Pt_F1_Crossing} \\
    F_2 (x) &= F_3 (1-x)\,. \label{eq:4Pt_F2F3_Crossing}
\end{align}
These relations imply that one channel can be eliminated by crossing symmetry, while another channel is found to be self-crossing.
As mentioned in Section \ref{subsubsec:WI_FourPointFunctions}, this results in the function $f(x)$ of \eqref{eq:4Pt_SolutionSCWI} satisfying the following anti-self-crossing condition:
\begin{equation}
    f(x)
    =
    - f(1-x)\,.
    \label{eq:4Pt_AntiSelfCrossing}
\end{equation}
It is crucial to point out that the correlator $\vev{1122}$ is related to $\vev{1221}$ via crossing symmetry.
In this case, the channels themselves are not constrained, and the associated function $f(x)$ does \textit{not} obey the anti-self-crossing relation \eqref{eq:4Pt_AntiSelfCrossing}.

\subsubsection{Five-point functions}
\label{subsubsec:Crossing_FivePointFunctions}
The crossing symmetry relation \eqref{eq:4Pt_CrossingSymmetry} can be naturally extended to higher-point functions, using relations coming from the possible OPE expansions (see for example figure 5 in \cite{CFTdefectsNotesHerzog}).
For the five-point correlator considered in \eqref{eq:11112_Correlator}, crossing symmetry takes the form
\begin{equation}
    \vev{\Op_{1} (1) \Op_{1} (2) \Op_{1} (3) \Op_{1} (4) \Op_{2} (5)}
    =
    \vev{\Op_{1} (4) \Op_{1} (3) \Op_{1} (2) \Op_{1} (1) \Op_{2} (5)}\,.
    \label{eq:5Pt_CrossingSymmetry}
\end{equation}
This crossing relation imposes constraints on the $R$-channels introduced in \eqref{eq:11112_NaturalBasis}.
Specifically, the following relations hold:
\begin{align}
    F_1 (x_1, x_2) = F_1 (1-x_2, 1-x_1)\,, \label{eq:11112_F1_Crossing} \\
    F_2 (x_1, x_2) = F_5 (1-x_2, 1-x_1)\,, \label{eq:11112_F2F5_Crossing} \\
    F_3 (x_1, x_2) = F_4 (1-x_2, 1-x_1)\,, \label{eq:11112_F3F4_Crossing} \\
    F_6 (x_1, x_2) = F_6 (1-x_2, 1-x_1)\,. \label{eq:11112_F6_Crossing}
\end{align}
The choice of basis made in Section \ref{subsubsec:WI_FivePointFunctions} for solving the Ward identities results in the simple crossing relation
\begin{equation}
    f_1 (x_1, x_2)
    =
    f_2 (1-x_2, 1-x_1)\,,
    \label{eq:11112_f1f2_Crossing}
\end{equation}
which implies that the correlator ultimately depends on a \textit{single function}, denoted in subsequent sections by $f(x_1,x_2) := f_1(x_1, x_2)$.
Determining this function at weak coupling is the subject of section \ref{subsub:11112}.

\subsubsection{Six-point functions}
\label{subsubsec:Crossing_SixPointFunctions}

Next, we explore the crossing symmetry constraints for six-point functions, focusing on the correlator $\vev{111111}$.
Since all the operators are identical, the reduced correlators exhibit numerous relations.
These relations are summarized in Table \ref{tab:SixPoint_Crossing}, where redundant relations are omitted.
The choice of basis for solving the Ward identities can be made such that, after imposing crossing symmetry, the correlator depends on two functions $f_1 (x_1,x_2,x_3)$ and $f_2(x_1,x_2,x_3)$.
As we will see in Section \ref{subsub:111111}, our choice of basis results in $f_2 (x_1,x_2,x_3) =0$ up to next-to-next-to-leading order (meaning that for $N^3LO$ the function is no longer $0$).
We thus focus on determining $f_1 (x_1,x_2,x_3)$ in subsequent section \ref{sub:HigherPointFunctions}.
The explicit relations used to relate $G_j$ to $f_{1,2}$ can be found in the ancillary notebook of reference \cite{Artico:2024wut} where these results were first described.

\begin{table}[]
    \centering
    \begin{tabular}{|c|ccccc|}
        \hline
        $(x_1, x_2, x_3)$ & $F_1$ & $F_2$ & $F_3$ & $F_4$ & $F_{14}$ \\[.25em]
        $(- \frac{x_{12}}{x_2}, - \frac{x_{13}}{x_3}, 1-x_1)$ & $F_7$ & $F_9$ & $F_8$ & $F_{12}$ & $F_{14}$ \\[.25em]
        $(\frac{x_1 x_{23}}{x_2 x_{13}}, \frac{x_1 (1-x_2)}{x_2 (1-x_1)}, \frac{x_1}{x_2})$ & $F_{10}$ & $F_2$ & $F_{11}$ & $F_{15}$ & $F_{14}$ \\[.25em]
        $(\frac{x_{12} (1-x_3)}{x_{13} (1-x_2)}, \frac{x_{12}}{x_{13}}, \frac{x_{12} x_3}{x_{13} x_2})$ & $F_6$ & $F_9$ & $F_3$ & $F_4$ & $F_{14}$ \\[.25em]
        $(- \frac{x_{23}}{1-x_2}, - \frac{x_{23}}{x_3 (1-x_2)}, \frac{x_{23} (1-x_1)}{x_{13} (1-x_2)})$ & $F_5$ & $F_2$ & $F_8$ & $F_{12}$ & $F_{14}$ \\
        $(1-x_3, \frac{1-x_3}{1-x_1}, \frac{1-x_3}{1-x_2})$ & $F_{13}$ & $F_9$ & $F_{11}$ & $F_{15}$ & $F_{14}$ \\[.25em]
        \hline
    \end{tabular}
    \caption{Summary of the relations between the $R$-channels in the natural basis \eqref{eq:111111_NaturalBasis}, induced by the crossing symmetry of $\vev{111111}$.
    The left column represents the arguments of the functions $F_j$.
    The equalities between the $R$-channels should be understood column-wise.}
    \label{tab:SixPoint_Crossing}
\end{table}

\subsection{Pinching}
\label{subsec:Pinching}

In this section, we describe the constraints that arise from reducing multipoint correlators from lower-point functions, commonly referred to as \textit{pinching}.
This phenomenon occurs when higher-weight operators are formed by bringing together fields from distinct points.
Specifically, the operators defined via \eqref{eq:HalfBPSDefOperator} possess the interesting property that higher-weight operators can be constructed by pinching together operators of lower length.
For example,
\begin{equation}
    \Wl[\underbrace{\phi^{I_{n-1}} (\tau_{n-1}) \phi^{I_n} (\tau_n)}_{\text{two operators of length $1$}}]
    \overset{\tau_n \to \tau_{n-1}}{\longrightarrow}
    \Wl[\underbrace{\phi^{I_{n-1}} (\tau_{n-1}) \phi^{I_n} (\tau_{n-1})}_{\text{one operator of length $2$}}]\,.
    \label{eq:Pinching}
\end{equation}
This property is important because lower-point correlators can impose constraints on higher-point functions.
Note that in $\Nm=4$ SYM, the pinching of half-BPS operators does not close within the same class of operators; instead, it generates higher-trace operators; this is not true however for operators on the defect (provided the limit of the $u$ vectors is also taken accordingly). 
On a broader level, one can interpret pinching as indicating that the correlators of scalar insertions are fully determined by the chain of fundamental insertions
\begin{equation}
    \vev{\phi^{I_1} (\tau_1) \ldots \phi^{I_n} (\tau_n)}\,.
    \label{eq:TheWorldDependsOnElementaryInsertions}
\end{equation}
Below, we enumerate the specific constraints that pinching imposes on the correlators of interest in this chapter.

\subsubsection{Four-point functions}
\label{subsubsec:Pinching_FourPointFunctions}

We first discuss the constraints on four-point functions arising from pinching, where two operators are combined to yield a three-point function.

\paragraph{$\vev{1111}$.}
For the correlator of elementary operators $\Op_1$, pinching the last two operators to obtain the three-point function $\vev{112}$ gives
\begin{equation}
    \lim_{4 \to 3} \vev{1111}
    =
    (13)(23) (F_1 (0) + F_3 (0))
    =
    \frac{\sqrt{n_2}}{n_1} \vev{112}\,,
    \label{eq:1111_PinchingTo112}
\end{equation}
with $n_1$ and $n_2$ the normalization constants given in \eqref{eq:n1} and \eqref{eq:n2}, respectively.
These normalization constants are required because the pinching operation \eqref{eq:Pinching} is applied at the level of the fields rather than the unit-normalized operators.

\paragraph{$\vev{1122}$.}
The correlator $\vev{1122}$ can be pinched in two distinct ways.
First, it can be pinched such that it collapses to $\vev{222}$:
\begin{equation}
    \lim_{2 \to 1} \vev{1122}
    =
    (13)(24)(34) (F_1 (0) + F_3 (0))
    =
    \frac{\sqrt{n_2}}{n_1} \vev{222}\,.
    \label{eq:1122_PinchingTo222}
\end{equation}
Second, one can bring the middle operators together in order to obtain
\begin{equation}
    \lim_{3 \to 2} \vev{1122}
    =
    (12)(24)^2 (F_1 (1) + F_2 (1))
    =
    \frac{\sqrt{n_3}}{\sqrt{n_1 n_2}} \vev{132}\,.
    \label{eq:1122PinchingTo132}
\end{equation}

\paragraph{$\vev{1212}$.}
One can proceed similarly for $\vev{1212}$, which can be collapsed to $\vev{123}$ or $\vev{132}$, which result into two constraints:
\begin{align}
    \lim_{4 \to 3} \vev{1212}
    &=
    (13)(23)^2 (F_1 (0) + F_3 (0))
    =
    \frac{\sqrt{n_3}}{\sqrt{n_1 n_2}} \vev{123}\,, \label{eq:1212_PinchingTo123} \\
    \lim_{3 \to 2} \vev{1212}
    &=
    (12)(24)^2 (F_1 (1) + F_2 (1))
    =
    \frac{\sqrt{n_3}}{\sqrt{n_1 n_2}} \vev{132}\,.
    \label{eq:1212_PinchingTo132}
\end{align}
This concludes the study of the pinching limit for four-point functions.
\subsubsection{Five-point functions}
\label{subsubsec:Pinching_FivePointFunctions}

Five-point functions can be pinched to produce lower-point functions, such as four-point functions. The pinching to three-point functions does not add further information as four-point functions are already consistent with that limit.
For the correlator $\vev{11112}$, two particularly useful pinching limits are
\begin{align}
    \lim\limits_{4 \to 3} \vev{11112} &= (13)(25)(35) \biggl( F_1(x,1) + F_4 (x,1) + \frac{r}{x} F_2(x,1) + \frac{s}{(1-x)} (F_3 (x,1) + F_6 (x,1)) \biggr) \notag \\
    &= \frac{\sqrt{n_2}}{n_1} \vev{1122}\,, \label{eq:11112_PinchingTo1122} \\
    \lim\limits_{3 \to 2} \vev{11112} &= (14)(25)^2 \biggl( F_1(x,x) + \frac{r}{x} (F_2(x,x) + F_4(x,x)) + \frac{s}{(1-x)} (F_3 (x,x) + F_5 (x,x)) \biggr) \notag \\
    &= \frac{\sqrt{n_2}}{n_1} \vev{1212}\,. \label{eq:11112_PinchingTo1212}
\end{align}
Pinching to $\vev{1113}$ is less informative because $\vev{1113}$ is an extremal correlator, yielding only a number, as discussed in \cite{Barrat:2021tpn}. In fact, it presents only one $R$-symmetry channel and the presence of a topological sector ensures this channel is a constant \cite{Giombi:2018qox}.
For our purposes, the relations \eqref{eq:11112_PinchingTo1122} and \eqref{eq:11112_PinchingTo1212} are sufficient to determine the correlators of interest up to next-to-next-to-leading order. Note that this pinching does not fix only one constant as in the previous examples, but instead reduces a complicated two-variable function to a non-trivial function of one variable spanned by a basis of linearly independent functions, thus providing a number of linear relations among coefficients. We will further clarify this statement when discussing the solution for the function determining the correlator up to NNLO.

\subsubsection{Six-point functions}
\label{subsubsec:SixPointFunctions}

Six-point functions can be pinched to produce lower-point functions, in particular five-point functions. Again, once this more complicated pinching is done, there are no further constraints coming from the pinching to lower-point functions.
For the correlator $\vev{111111}$, a particularly useful pinching limit is
\begin{equation}
	\begin{split}
	\lim\limits_{6 \to 5} \vev{111111}
	&=
    (14)(25)(35)
    \lim_{\veps\to0} \biggl\lbrace
     \left( F_{14} (x_1\veps,x_2\veps,\veps) + F_{15}(x_1\veps,x_2\veps,\veps)\right)  \\ 
     & \phantom{(14)(25)(35) \lim_{\veps\to0} \lbrace} + \frac{r_1}{x_1}\left( F_{10} (x_1\veps,x_2\veps,\veps) + F_{11}(x_1\veps,x_2\veps,\veps)\right) \\ 
     & \phantom{(14)(25)(35) \lim_{\veps\to0} \lbrace} + \frac{s_1}{(1-x_1)} \left( F_{4} (x_1\veps,x_2\veps,\veps) + F_{6}(x_1\veps,x_2\veps,\veps)\right) \\
     & \phantom{(14)(25)(35) \lim_{\veps\to0} \lbrace} + \frac{r_2}{x_2}\left( F_{12} (x_1\veps,x_2\veps,\veps) + F_{13}(x_1\veps,x_2\veps,\veps)\right) \\ 
     & \phantom{(14)(25)(35) \lim_{\veps\to0} \lbrace} + \frac{s_2}{(1-x_2)}\left( F_{1} (x_1\veps,x_2\veps,\veps) + F_{3}(x_1\veps,x_2\veps,\veps)\right) \\
     & \phantom{(14)(25)(35) \lim_{\veps\to0} \lbrace} \left. + \frac{t_{12}}{x_{12}}\left( F_{7} (x_1\veps,x_2\veps,\veps) + F_{9}(x_1\veps,x_2\veps,\veps)\right) \right\rbrace \\
     &=
     \frac{\sqrt{n_2}}{n_1}\vev{11112} \,.
	\end{split}
	\label{eq:111111_PinchingTo11112}
\end{equation}
The limit $\veps\to 0$ means to take the limit of all three six-point cross-ratios to zero while keeping their ratios fixed to the two five-point cross-ratios. This way of taking the pinching limit is necessary as the conformal frame we have chosen does not allow taking the pinching limit in the same trivial way as before: since we set $t_5\rightarrow 1$ and $t_6\rightarrow +\infty$, the only way to retrieve the pinching to five points is to consider the limit defined above. As we will discuss later in the chapter, this limit is the only pinching necessary to completely fix the solution for the six-point correlator, as it produces a high number of linear relations among the coefficients of the functions spanning the solution. %
\section{Perturbative results}
\label{sec:PertResultsMultip}

\subsection{Four-point functions}
\label{sub:FourPointFunctions}

In this section, we now compute the four-point functions introduced in Section \ref{subsubsec:Correlators_FourPointFunctions}, employing the non-perturbative constraints given by the solution of the superconformal Ward identities \eqref{eq:4Pt_SolutionSCWI}, the topological sector \eqref{eq:1111_Fds}-\eqref{eq:1122And1212_Fds}, and the pinching relations \eqref{eq:1111_PinchingTo112}-\eqref{eq:1212_PinchingTo132}.
For all correlators, we focus on the channel $F_1$, whose diagrams are the simplest at weak coupling (due to the planarity feature of the large $N$ expansion) from which the function $f(x)$ can be deduced.
Both functions will be expanded perturbatively as follows:
\begin{align}
    F_1 (x)
    &=
    \sum_{\ell=0}^\infty \lambda^\ell F_1^{(\ell)} (x)\,, \label{eq:F1_PerturbativeExpansion} \\
    f (x)
    &=
    \sum_{\ell=0}^\infty \lambda^\ell f^{(\ell)} (x)\,.
    \label{eq:f_PerturbativeExpansion}
\end{align}
In this section, we compute $F_1$ using Feynman diagrams: they represent the perturbative input needed in the perturbative bootstrap framework after the listed non-perturbative information has been used to simplify the calculation.
These results will serve as the foundation for calculating higher-point functions in Section  \ref{sub:HigherPointFunctions}.
\subsubsection{$\vev{1111}$}
\label{subsub:1111}

We begin by revisiting the simplest four-point function, $\vev{1111}$.
This correlator was computed up to next-to-leading order in \cite{Kiryu:2018phb,Barrat:2021tpn} and extended to next-to-next-to-leading order in \cite{Cavaglia:2022qpg}.
In the latter case, a combination of the conformal bootstrap and integrability techniques was employed to derive the result, as discussed in section \ref{sub:Literature}.\\

In the following, we demonstrate how to achieve the same result using Feynman diagrams.
Interestingly, we find that only the two lowest orders in a small $x$ expansion of $F_1$ are required to determine the full correlator up to NNLO, if using a transcendentality-based Ansatz.
Consequently, numerical integration alone would have sufficed to obtain the analytical result.
We show however that all diagrams can, in fact, be computed analytically in this case. The low orders in the coupling $\lambda$ are straightforward to compute by focusing on the channel $F_1$ and the relations \eqref{eq:fFromF1} and \eqref{eq:4Pt_AntiSelfCrossing}.

\paragraph{Leading order.}
At leading order for large $N$, no Feynman diagrams contribute to the channel $F_1$, as this channel corresponds to non-planar contractions (this is the reason why we choose this channel to compute perturbatively). For this reason, we have
\begin{equation}
    F_1^{(0)} (x)
    =
    0\,.
    \label{eq:1111_F1_LO}
\end{equation}
As explained in section \ref{subsubsec:Crossing_FourPointFunctions}, the function $f(x)$ is chosen to be antisymmetric.
Therefore, at leading order, we readily obtain
\begin{equation}
    f^{(0)} (x)
    =
    0\,.
    \label{eq:1111_f_LO}
\end{equation}
as the developed solution to the SCWI \eqref{eq:fFromF1} implies that $f(x)$ is simply the antiderivative of $F_1(x)$ respecting the correct boundary conditions -- here, anti-symmetry under crossing.
\paragraph{Next-to-leading order.}
At next-to-leading order, the channel $F_1$ is determined by a single diagram, which we refer to as the X-diagram in appendix \ref{app:Integrals}. After removing the unit-normalization and the $R$-symmetry variables, this diagram evaluates to
\begin{equation}
    \NLOX
    =
    \frac{\lambda^3}{4} X_{1234}\,,
    \label{eq:1111_NLO_Diagram}
\end{equation}
where $X_{1234}$ is defined in \eqref{eq:X1234}.
The prefactor of $1/4$ comes from the symmetry factor and the trace of the diagram. It is interesting to note that the channels $F_2$ and $F_3$ are more intricate to compute. These channels involve boundary diagrams and contain functions of transcendentality weight $2$, while the channel $F_1$ has transcendentality weight $1$ only.
Therefore, it is a significant simplification to use the Ward identities from \eqref{eq:4Pt_SCWI} to derive the entire correlator from $F_1$ alone.By expressing the integral in terms of Goncharov polylogarithms, the X-diagram results in the following for the channel $F_1$:
\begin{equation}
    F_1^{(1)} (x)
    =
    - \frac{1}{8\pi^2 x (1-x)} (x G(0,x) + (1-x) G(1, x))\,.
    \label{eq:1111_F1_NLO}
\end{equation}
From this, we can extract the corresponding function $f(x)$:
\begin{equation}
    f^{(1)} (x)
    =
    \frac{1}{8 \pi^2} \biggl( \frac{\pi^2}{6} + G(1,0,x) - G(0,1,x) \biggr)\,
    \label{eq:1111_f_NLO}
\end{equation}
determining the full correlator.
\paragraph{Next-to-next-to-leading order.} At next-to-next-to-leading order, the calculation becomes more intricate, involving both bulk and boundary diagrams. Table \ref{table:Diagrams1111NNLO} presents the relevant diagrams for calculating the channel $F_1$.
The complexity of this step is evident, but it would be even more challenging if we were required to compute the channels $F_2$ and $F_3$ directly, as those would introduce diagrams with multiple integrals along the Wilson line, and two-loop self-energy corrections.

\begin{table}[t!]
    \centering
    \caption{The relevant Feynman diagrams for the computation of $F_1(x)$ at next-to-next-to-leading order.
    The horizontal double line separates {\normalfont bulk} from {\normalfont boundary} diagrams.
    In the last row, the colored dots along the Wilson line indicate where the gluon can connect.
    Explicit expressions can be found in Appendix \ref{app:FeynmanDiagramsOf1111}.}
    \begin{tabular}{lc}
        \hline
        Self-energy & \DefectSSSSTwoLoopsSelfEnergyOne\ \DefectSSSSTwoLoopsSelfEnergyTwo\ \DefectSSSSTwoLoopsSelfEnergyThree\ \DefectSSSSTwoLoopsSelfEnergyFour\ \\[2ex]
        \hline
        XX & \DefectSSSSTwoLoopsXXOne\ \DefectSSSSTwoLoopsXXTwo\ \\[2ex]
        \hline
        XH & \DefectSSSSTwoLoopsXHOne\ \DefectSSSSTwoLoopsXHTwo\ \DefectSSSSTwoLoopsXHThree\ \DefectSSSSTwoLoopsXHFour\ \\[2ex]
        \hline
        Spider & \DefectSSSSTwoLoopsSpider\ \\[2ex]
        \hline \hline
        XY & \DefectSSSSTwoLoopsXYOne\ \DefectSSSSTwoLoopsXYTwo\ \DefectSSSSTwoLoopsXYThree\ \DefectSSSSTwoLoopsXYFour\ \\[2ex]
        \hline
    \end{tabular}
    \label{table:Diagrams1111NNLO}
\end{table}

The diagrams can be computed explicitly, as described in appendix \ref{app:FeynmanDiagramsOf1111}.
The resulting expression for the channel $F_1$ at NNLO is
\begin{equation}
    \begin{split}
    F_1^{(2)} (x)
    &=
    - \frac{1}{64 \pi^4 x (1-x)}
    \biggl(
    \frac{\pi^2}{3} (x G(0,x) + G(1,x)) \\
    &\phantom{=\ }
    +
    x (G(0,0,1,x) - G(1,1,0,x) + G(1,0,1,x) - G(0,1,0,x) + 3 \zeta_3) \\
    & \phantom{=\ }
    +
    G(1,0,1,x)
    +
    G(1,1,0,x)
    -
    2 (G(0,1,1,x) + G(1,0,0,x) + G(0,1,0,x))
    \biggr)\,.
    \end{split}
    \label{eq:1111_F1_NNLO}
\end{equation}
From this expression, we can deduce the corresponding function $f(x)$, which takes the form:
    \begin{equation}
        \begin{split}
            f^{(2)} (x)
            &=
            \frac{1}{64 \pi^4}
            \biggl(
            \frac{\pi^4}{15}
            +
            3 \zeta_3 G(1,x)
            +
            \frac{\pi^2}{3} (G(1,0,x) -G(0,1,x) + G(1,1,x)) \\
            &\phantom{=\ }
            +
            2( G(1,1,0,1,x) - G(0,0,1,0,x) + G(0,0,1,1,x) - G(1,1,0,0,x) \\
            &\phantom{=\ } + G(0,1,0,0,x) - G(1,0,1,1,x) )
            +
            G(1,0,1,0,x) - G(0,1,0,1,x) \\
            &\phantom{=\ } + G(1,0,0,1,x) - G(0,1,1,0,x)
            \biggr)\,.
        \end{split}
        \label{eq:1111_f_NNLO}
    \end{equation}
Notice that all terms have homogeneous transcendentality, and that the coefficients of the Goncharov polylogarithms and zeta functions are simple rational numbers. This structure was exploited in prior works using a bootstrap approach, where an Ansatz was constructed and solved for these coefficients \cite{Cavaglia:2022qpg}.
In the present context, the Feynman diagrams serve to calculate the coefficients of the Ansatz by direct comparison.
It should be pointed out that the coefficients of such an Ansatz are fully fixed by knowing $F_1$ only up to the following order:
\begin{equation}
    F_1^{(2)} (x)
    =
    \frac{1}{64 \pi^4}
    \biggl(
    - \log^2 x - \frac{\pi^2 - 12}{3} \log x + \frac{\pi^2}{3} - 3(2 + \zeta_3) - \frac{3}{2} x \log^2 x + \Om(x \log x)
    \biggr)\,.
    \label{eq:1111_F1_FirstOrdersInx}
\end{equation}
These \textit{four} coefficients could have been determined numerically if the Feynman diagrams of Table \ref{table:Diagrams1111NNLO} had been too complicated to calculate analytically\footnote{Note that some of the diagrams in Table \ref{table:Diagrams1111NNLO} are divergent. The numerical evaluation here described can be performed after extracting these divergences via integral manipulations.}.
Understanding the necessary number of terms in the expansion for $N^3LO$ would provide insight into extending these methods further.
The Ansatz approach also plays a role in developing the perturbative bootstrap algorithm discussed in Section \ref{sub:HigherPointFunctions}, which will be used to derive higher-point results.

\subsubsection{$\vev{1122}$}
\label{subsub:1122}

We now examine the four-point functions of two elementary fields $\Op_1$ and two composite operators $\Op_2$.
It is convenient to decompose the correlator into its fully connected and factorized components, as follows:
\begin{equation}
    \Am_{1122}
    =
    \Am_{1122}^{\text{(fact)}}
    +
    \Am_{1122}^{\text{(conn)}}\,
    \label{eq:1122_DisconnectedAndConnected}
\end{equation}
where by factorized we mean diagrams that are given products of lower-point defect Feynman diagrams.
For the channel of interest $F_1$, defined via \eqref{eq:4Pt_NaturalBasis}, the factorized part corresponds to the correlator $\vev{1111}$ discussed in section \ref{subsub:1111}, though with necessary subtractions for overcounted terms:
\begin{equation}
    \left.
    \Am_{1122}^{\text{(fact)}}
    \right|_{(13)(24)}
    =
    \frac{n_1^2}{n_2}
    \left.
    \Am_{1111}
    \right|_{(13)(24)}
    -
    \text{(overcounted)}\,.
    \label{eq:1122_DisconnectedFrom1111}
\end{equation}
The overcounted terms correspond to boundary diagrams and are discussed later in this section.
Up to next-to-leading order, there is no overcounting, and thus the second term in \eqref{eq:1122_DisconnectedFrom1111} can simply be discarded.
\paragraph{Leading order.}
At leading order, there is once more no contribution to $F_1$ due to planarity.
Given our chosen normalization, we have
\begin{equation}
    f^{(0)} (x)
    =
    0\,.
    \label{eq:1122_f_LO}
\end{equation}
Note that the fact that $f(x)$ is the same function at LO as the one for the correlator $\vev{1111}$ does not mean the whole correlator is the same, due to the different normalization. 
\paragraph{Next-to-leading order.}
At next-to-leading order, there is no fully connected contribution, so the correlator is identical to $\vev{1111}$ after a proper change of normalization as described above. Given that we factorize in the definition of the challel the normalization factors, the different normalizations do not change the $R$-symmetry channel. 
The channel $F_1$ is therefore given by \eqref{eq:1111_F1_NLO}, and the solution to the Ward identities is
\begin{equation}
    f^{(1)} (x)
    =
    \frac{1}{8 \pi^2} \biggl( \frac{\pi^2}{6} + G(1,0,x) - G(0,1,x) \biggr)\,.
    \label{eq:1122_f_NLO}
\end{equation}

\paragraph{Next-to-next-to-leading order} At next-to-next-to-leading order, new diagrams contribute to $F_1$ in addition to the factorized terms corresponding to $\vev{1111}$.
These diagrams are shown in Table \ref{table:Diagrams1122NNLO}.
We have only four fully connected bulk diagrams, but we must also account for the overcounting discussed in \eqref{eq:1122_DisconnectedFrom1111}.
From table \ref{table:Diagrams1111NNLO}, it is clear that three factorized boundary diagrams need to be subtracted from $\vev{1111}$, as depicted in the third row of table \ref{table:Diagrams1122NNLO}.

\begin{table}[t!]
    \centering
    \caption{The diagrams contributing at next-to-next-to-leading order to the channel $F_1$ of the correlator $\vev{1122}$.
    Here we include only the connected contributions, which are called XX and XH, and the overcounted terms (XY) which have to be subtracted from the product of $\vev{1111}$ and $\vev{11}$ in order to obtain the factorized part.
    Explicit expressions can be found in Appendix \ref{app:FeynmanDiagramsOf1122}.}
    \begin{tabular}{lc}
        \hline
        XX & \NNLOXXOne \quad \NNLOXXTwo \\[2ex]
        \hline
        XH & \NNLOXHOne \quad \NNLOXHTwo \\[2ex]
        \hline \hline
        XY (subtractions) & \NNLOXYSubtractOne\ \NNLOSubtractFactorOne \quad \NNLOXYSubtractTwo\ \NNLOSubtractFactorOne \quad \NLOX\ \NNLOSubtractFactorTwo \\[2ex]
        \hline
    \end{tabular}
    \label{table:Diagrams1122NNLO}
\end{table}

These diagrams are straightforward to compute using standard techniques, and the results are summarized in Appendix \ref{app:FeynmanDiagramsOf1122}.
The corresponding channel $F_1$ is given by
\begin{equation}
    \begin{split}
    F_1^{(2)} (x)
    &=
    \frac{1}{64 \pi^4 x (1-x)}
    \biggl(
    - \frac{\pi^2}{6} (4 x G(0,x) + (x-2) G(1,x)) \\
    &\phantom{=\ }
    +
    \frac{x}{2} (2 G(0,1,0,x) - G(1,0,1,x) - 2 G(0,0,1,x) + G(1,1,0,x) - 3 \zeta_3)\\
    &\phantom{=\ }
    +
    2 (G(1,0,0,x) + G(0,1,1,x))
    -
    2 G(0,1,0,x) - G(1,0,1,x)
    -
    G(1,1,0,x)
    \biggr)\,.
    \end{split}
    \label{eq:1122_F1_NNLO}
\end{equation}
The solution to the Ward identities is
\begin{equation}
    \begin{split}
    f^{(2)} (x)
    &=
    \frac{1}{64 \pi^4}
    \biggl(
    \frac{17 \pi^4}{180}
    +
    \frac{3}{2} \zeta_3 G(1,x)
    +
    \frac{\pi^2}{6} ( 4 G(1,0,x) - 2 G(0,1,x) + G(1,1,x) ) \\
    &\phantom{=\ }
    -
    2 (G(1,0,1,1,x) + G(1,1,0,0,x) + G(0,0,1,0,x) - G(0,0,1,1,x) \\
    &\phantom{=\ }
    - G(0,1,0,0,x))
    -
    G(0,1,0,1,x) - G(0,1,1,0,x) + G(1,0,0,1,x) \\
    &\phantom{=\ }
    +
    G(1,0,1,0,x)
    +
    \frac{3}{2} G(1,1,0,1,x)
    +
    \frac{1}{2} G(1,1,1,0,x)
    \biggr)\,.
    \end{split}
    \label{eq:1122_f_NNLO}
\end{equation}
Note that this expression is similar to \eqref{eq:1111_f_NNLO}, in that it contains no rational functions of $x$ -- only Goncharov polylogarithms and $\zeta$ functions. Still, the different normalization and presence of fully connected diagrams distinguish the two correlators. The NNLO result can be partially verified by pinching $\vev{1122}$ to $\vev{132}$ and $\vev{222}$, as described in \eqref{eq:1122PinchingTo132}-\eqref{eq:1122_PinchingTo222}: we find a perfect match with the localization data provided in \eqref{eq:lambda123}-\eqref{eq:lambda222}. The results presented in this and in the next sections were confirmed and broadened beyond weak coupling using integrability techniques in \cite{Cavaglia:2024dkk}.

\subsubsection{$\vev{1212}$}
\label{subsec:1212}
We conclude this section on the four-point function by calculating the correlator $\vev{1212}$, the correlator involving two dimension-one operators and two dimension-two operators that we still have not examined.
This correlator is fully fixed up to next-to-next-to-leading order simply by the train track integral \eqref{eq:TrainTrack_Definition}.
\paragraph{Leading order.}
At leading order, the function $F_1$ vanishes, as with the other correlators in this section.
Thus, we have
\begin{equation}
    f^{(0)} (x)
    =
    0\,.
    \label{eq:1212_f_LO}
\end{equation}

\paragraph{Next-to-leading order.}
At next-to-leading order, the intertwining of operators $\Op_1$ and $\Op_2$ makes it impossible to draw a NLO planar diagram for the channel $F_1$.
Therefore, we obtain
\begin{equation}
    F_1^{(1)} (x)
    =
    0\,.
    \label{eq:1212_F1_NLO}
\end{equation}
This means that the solution to the Ward identities is constant also at this order.
The anti-self-crossing condition \eqref{eq:4Pt_AntiSelfCrossing} fixes it to be zero:
\begin{equation}
    f^{(1)} (x)
    =
    0\,.
    \label{eq:1212_f_NLO}
\end{equation}
\paragraph{Next-to-next-to-leading order.}

At next-to-next-to-leading order, the correlator is fully governed by a single integral, which corresponds to the pinching limit of the train track integral to the kite integral, as discussed in \eqref{eq:KiteIntegral}:
\begin{equation}
    \NNLOFourPtTrainTrack\
    =
    - \frac{\lambda^5}{8} I_{24} K_{13,24}\,.
    \label{eq:1212_NNLO_TrainTrackDiagram}
\end{equation}
Here, the numerical prefactor accounts for symmetry factors and for the traces of the operators and Wilson line.
The channel $F_1$ is then
\begin{equation}
    \begin{split}
    F_1^{(2)} (x)
    &=
    \frac{1}{64 \pi^4 x (1-x)}
    \biggl(
    - \frac{\pi^2}{6} (4 G(0,x) + (2-x) G(1,x)) \\
    &\phantom{=\ }
    +
    x \biggl(\frac{1}{2} G(1, 1, 0, x) - G(0, 0, 1, x)
    + G(0, 1, 0, x) - \frac{1}{2} G(1, 0, 1, x)
    - \frac{3}{2} \zeta_3\biggr) \\
    &\phantom{=\ }
    + 2 G(1, 0, 0, x) - G(1, 1, 0, x)
    - 2 G(0, 1, 0, x) - G(1, 0, 1, x)
    + 2 G(0, 1, 1, x)
    \biggr)\,,
    \end{split}
    \label{eq:1212_F1_NNLO}
\end{equation}
which gives the solution to the Ward identities as
\begin{equation}
    \begin{split}
    f^{(2)} (x)
    &=
    \frac{1}{64 \pi^4}
    \biggl(
    -
    \frac{7 \pi^4}{180}
    +
    \frac{1}{2}
    (G(0,0,0,1,x) + G(0,0,1,0,x) + G(0,1,0,1,x) \\
    &\phantom{=\ }+ G(0,1,1,0,x) - G(1,0,0,1,x) - G(1,0,1,0,x) - G(1,1,0,1,x) \\
    &\phantom{=\ }- G(1,1,1,0,x)
    +
    2 G(1,0,1,1,x) - 2 G(0,1,0,0,x) \\
    &\phantom{=\ }+ 2 G(1,1,0,0,x) - 2 G(0,0,1,1,x))
    \biggr)\,.
    \end{split}
    \label{eq:1212_f_NNLO}
\end{equation}
Once again, no rational functions multiplying Goncharov polylogarithms appear in this solution. 
As mentioned before, this observation plays a crucial role in the next section for defining an Ansatz for the correlators $\vev{11112}$ and $\vev{111111}$.
\subsection{Higher-point functions}
\label{sub:HigherPointFunctions}

Building upon the results of the previous section, we now introduce the perturbative bootstrap method for determining the higher-point functions $\vev{11112}$ and $\vev{111111}$ up to next-to-next-to-leading order.
Specifically, we derive these correlators with minimal reliance on Feynman diagrams, utilizing either vanishing contributions or channels composed of a single train track diagram.
The key ingredient is the construction of a suitable Ansatz for multipoint correlators, based on symbols and Goncharov polylogarithms.
The remaining input consists of protected data (and the lower point functions that have been derived in the previous sections).

\subsubsection{$\vev{11112}$}
\label{subsub:11112}

\begin{figure}
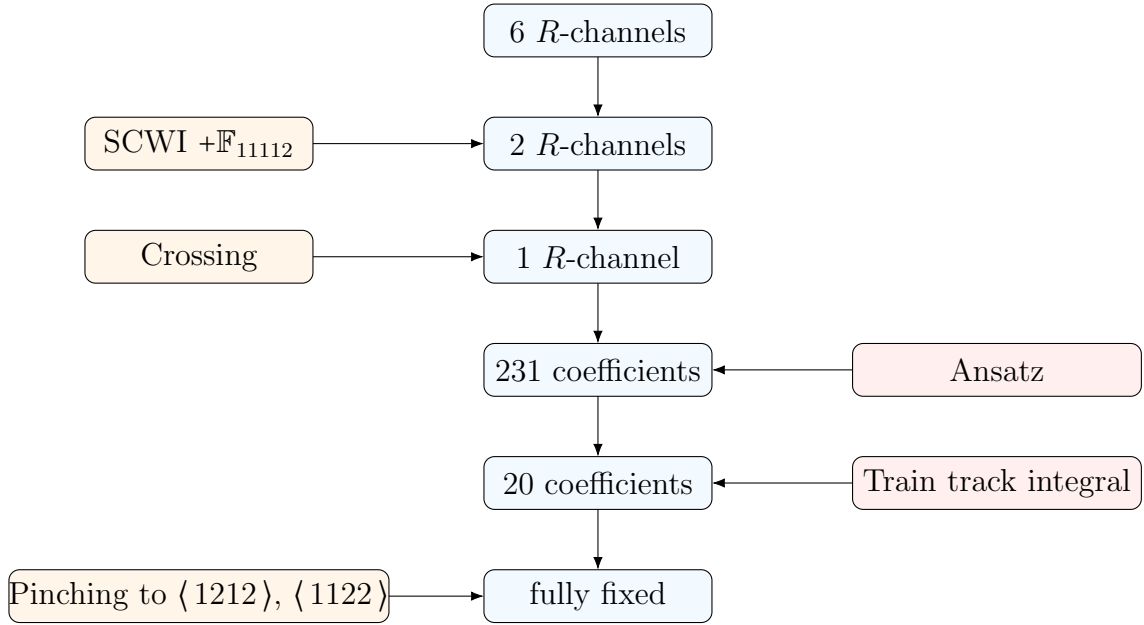

    \centering
    \MethodFivePoint
    \caption{
    Illustration of the steps involved in the bootstrap method used to derive the correlator $\vev{11112}$ at next-to-next-to-leading order.
    Superconformal Ward identities and crossing symmetry determine five of the six $R$-channels, in the basis $\tilde{R}_j$.
    An educated Ansatz is then formulated based on homogeneous transcendentality, limiting the number of open coefficients to $231$.
    By inputting the train track integral \eqref{eq:TrainTrack_Result}, which completely determines the channel $F_1$, all but $20$ coefficients are fixed.
    The remaining coefficients are resolved using the pinching constraints discussed in Section \ref{subsubsec:Pinching_FivePointFunctions}.}
    \label{fig:11112_Illustration}
\end{figure}

We start with the five-point function $\vev{11112}$, introduced earlier in Section \ref{subsubsec:Correlators_FivePointFunctions}.
Below, we outline a method to derive this correlator up to next-to-next-to-leading order without the need to explicitly compute all the relevant diagrams.
The key idea is to leverage the non-perturbative constraints discussed in Section \ref{sec:NonPerturbativeConstraints}, reducing the number of functions to determine from $6$ to $1$.
Next, we construct an Ansatz for the perturbative order of interest and incorporate the channel $F_1$ from diagram computations.
At next-to-leading order, the channel $F_1$ vanishes, while at next-to-next-to-leading order, it is determined by the train track integral.
This procedure is explained in further detail below and is summarized in Figure \ref{fig:11112_Illustration}. The numbers of coefficients reported correspond to the next-to-next-to-leading order calculation.

\paragraph{Leading order.}
At leading order, the correlator is determined up to a constant by the fact that the channel $F_1$, as defined by \eqref{eq:11112_NaturalBasis}, vanishes:
\begin{equation}
    F_1^{(0)} (x_1, x_2)
    =
    0\,.
    \label{eq:11112_F1_LO}
\end{equation}
With our chosen normalization for the solution of the Ward identities \eqref{eq:11112_SolutionSCWI}, we simply find
\begin{equation}
    f^{(0)} (x_1, x_2)
    =
    1\,.
    \label{eq:11112_f_LO}
\end{equation}

\paragraph{Next-to-leading order.}
The next-to-leading order is more interesting and, although it is already known \cite{Barrat:2021tpn}, we use it here to demonstrate our bootstrap method.

The first step is to construct a suitable perturbative Ansatz for the function $f(x_1,x_2)$, defined via \eqref{eq:11112_SolutionSCWI} and \eqref{eq:11112_f1f2_Crossing}, following a perturbative expansion akin to \eqref{eq:f_PerturbativeExpansion}.
At any given order in perturbation theory, we assume the following properties for the Ansatz:
\begin{enumerate}
    \item $f^{(\ell)} (x_1, x_2)$ contains no rational functions, consisting solely of Goncharov polylogarithms and $\zeta$-functions;\footnote{It is likely that the function space would need to be expanded beyond a certain order in perturbation theory.
    For instance, multiple $\zeta$-values are certainly expected to appear.}
    \item $f^{(\ell)} (x_1, x_2)$ is homogeneous in transcendentality;
    \item $f^{(\ell)} (x_1, x_2)$ has transcendentality weight $2\ell$.
\end{enumerate}

At next-to-leading order, this implies that the Ansatz must include a constant and Goncharov polylogarithms of at most transcendentality $2$.
The condition of homogeneous transcendentality immediately cancels functions of transcendentality $1$, as they would multiply $\zeta_1$, which diverges.
To construct the appropriate Goncharov polylogarithms, we write the Ansatz in symbols, with the elements
\begin{equation}
    \xi_i
    =
    \lbrace
    x_1, 1-x_1, x_2, 1-x_2, x_{12}
    \rbrace\,.
    \label{eq:11112_SymbolBasis}
\end{equation}
The Ansatz then takes the form
\begin{equation}
    f^{(1)} (x_1, x_2)
    =
    c_0
    +
    \sum_{i \neq j} c_{ij}\, \xi_i \otimes \xi_j\,,
    \label{eq:11112_AnsatzInSymbols}
\end{equation}
where terms with $i=j$ are excluded since they would correspond to second-order anomalous dimensions in a conformal block expansion, which are not expected to appear.
In other words, we exclude $\log^2$ terms in the OPE limits $x_i \to 0, (1-x_i) \to 0, x_{ij} \to 0$. Note that this Ansatz has to be read as the function $f^{(1)} (x_1, x_2)$ being the sum of a constant and weight two HPLs whose symbol is given by the sum in \eqref{eq:11112_AnsatzInSymbols}

The Ansatz in \eqref{eq:11112_AnsatzInSymbols} is not yet in a usable form, as it is not finite for every choice of coefficients.
Requiring finiteness (also known as the integrability condition\footnote{Note that here the term \textit{integrability} simply means that the symbol corresponds to a combination of HPLs and has nothing to do with quantum integrability.}), we fix $6$ out of the $21$ free coefficients, yielding the following Ansatz:
\begin{equation}
    \begin{split}
        16 \pi^2 f^{(1)} (x)
        &=
        c_0
        +
        c_9 G(1, x_1) G(1, x_2)
        +
        G(0, x_1) ((c_2 + c_{11}) G(0, x_2) + c_7 G(1, x_2)) \\
        &\phantom{=\ }
        +
        c_{10} G(1, x_2) G(x_2, x_1)
        +
        G(0, x_2) ((c_6 + c_{13}) G(1, x_1) + c_4 G(x_2, x_1)) \\
        &\phantom{=\ }
        +
        (c_4 + c_{12}) G(0, 0, x_2)
        +
        c_5 G(0, 1, x_1)
        +
        (c_8 + c_{10}) G(0, 1, x_2) \\
        &\phantom{=\ }
        +
        c_{11} G(0, x_2, x_1)
        +
        c_1 G(1, 0, x_1)
        +
        (c_3 + c_{14}) G(1, 0, x_2) \\
        &\phantom{=\ }
        +
        c_{13} G(1, x_2, x_1)
        +
        (-c_4 + c_{11} + c_{12}) G(x_2, 0, x_1) \\
        &\phantom{=\ }
        +
        (-c_{10} + c_{13} + c_{14}) G(x_2, 1, x_1)\,,
    \end{split}
    \label{eq:11112_Ansatz}
\end{equation}
where we have relabeled the coefficients from $c_0$ to $c_{14}$ for clarity.
A review of the integrability condition for symbols is found in Appendix \ref{app:SymbolsAndGoncharovPolylogarithms}.
With the Ansatz constructed, we can now input data from the $R$-channels to fix as many coefficients as possible.
For instance, it is straightforward to see that the channel $F_1$ remains zero at this order:
\begin{equation}
    F_1^{(1)} (x)
    =
    0\,.
    \label{eq:11112_F1_NLO}
\end{equation}
Comparing this result to \eqref{eq:11112_Ansatz} via the Ward identities fixes $13$ more coefficients.
The final two coefficients are fixed by applying the pinching conditions for $\vev{11112} \to \vev{1122}$ and $\vev{11112} \to \vev{1212}$, as given in \eqref{eq:11112_PinchingTo1122}–\eqref{eq:11112_PinchingTo1212}.
The final result is
\begin{equation}
    \begin{split}
    f^{(1)} (x_1, x_2)
    &=
    \frac{1}{8 \pi^2}
    \biggl(
    \frac{\pi^2}{2}
    +
    G(0, x_2) G(1, x_1)
    -
    G(1, x_1) G(1, x_2)
    +
    G(1, x_2) G(x_2, x_1) \\
    &\phantom{=\ }
    +
    G(0, 1, x_2)
    -
    G(1, 0, x_2)
    +
    G(1, x_2, x_1)
    -
    G(x_2, 1, x_1)
    \biggr)\,.
    \end{split}
    \label{eq:11112_f_NLO}
\end{equation}
This result can be compared with the explicit calculation performed in \cite{Barrat:2021tpn} using diagrammatic recursion relations, and we observe perfect agreement between the two methods.

\paragraph{Next-to-next-to-leading order.}
We apply the same bootstrap method to compute the five-point function $\vev{11112}$ at next-to-next-to-leading order, a result that was not present in the literature before this calculation was performed in \cite{Artico:2024wut}.
The Ansatz must now include terms of transcendentality up to weight $4$, while terms of transcendentality $3$ are excluded by the homogeneous transcendentality condition (they would multiply the divergent $\zeta_1$).
This results in an Ansatz expressed in symbols with initially $651$ coefficients. However, imposing the finiteness condition immediately fixes $420$ of these coefficients, leaving $231$ free coefficients to determine.
On the input side, it is crucial to note that the channel $F_1$ is now determined by a single diagram: the train track integral discussed in Section \ref{subsub:OneIntegralToRuleThemAll}.
This is given by the pinched six-point diagram
\begin{align}
    \NNLOFivePt
    =
    - \frac{\lambda^5}{8}
    B_{125,345}\,,
    \label{eq:11112_Diagram_NNLO}
\end{align}
where, for the sake of avoiding cluttering the notation, we do not include the $R$-symmetry variables.
The integral can be evaluated analytically, and in terms of Goncharov polylogarithms it reads
\begin{equation}
    \begin{split}
        F_1^{(2)} (x_1, x_2)
        &=
        \frac{1}{128 \pi^4 x_1 (1-x_2)}
        \bigl(
        2 (G(0, x_2) G(1, 0, x_1) - 
        G(1, x_2) G(1, 0, x_1)) \\
        &\phantom{=\ } - 
        G(x_2, x_1) (G(0, 1, x_2) + G(1, 0, x_2)) + 
        G(1, x_1) (-2 G(0, 0, x_2) \\
        &\phantom{=\ } + G(0, 1, x_2) + G(1, 0, x_2)) - 
        G(0, x_2) G(1, x_2, x_1)\\
        &\phantom{=\ } + 2 G(1, x_2) G(x_2, 0, x_1) + 
        G(1, 0, x_2, x_1) + G(1, x_2, 0, x_1) \\
        &\phantom{=\ } - G(x_2, 0, 1, x_1) - 
        G(x_2, 1, 0, x_1))
        \bigr)\,.
    \end{split}
    \label{eq:11112_F1_NNLO}
\end{equation}
This result, along with the pinching conditions, is sufficient to completely fix the Ansatz.
Although the full result for $f^{{(2)}} (x_1, x_2)$ is lengthy and is provided in the ancillary \textsc{Mathematica} notebook of \cite{Artico:2024wut}, we present a couple of terms here to remark on the structure of the Ansatz:
\begin{equation}
    f^{(2)} (x_1, x_2)
    =
    \frac{1}{64 \pi^4}
    \bigl(
    G(x_2, x_2, 1, 0, x_1)
    -
    2 G(x_2, x_2, 1, x_2, x_1)
    +
    \ldots
    \bigr)\,.
\end{equation}

\subsubsection{$\vev{111111}$}
\label{subsub:111111}

\begin{figure}
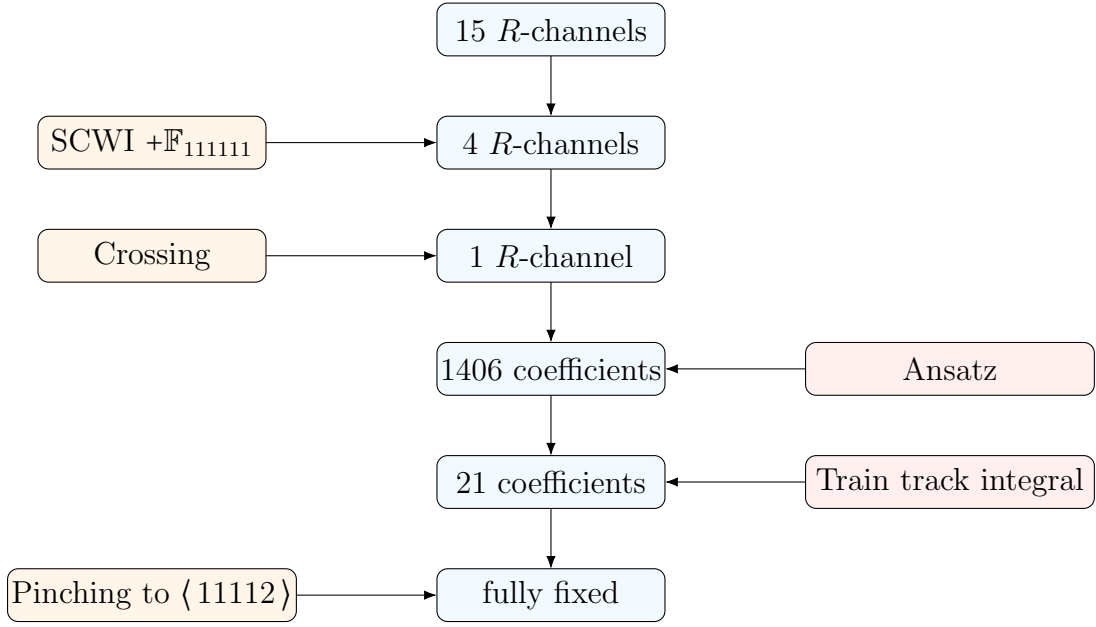

    \centering
    \MethodSixPoint
    \caption{Illustration of the method used for bootstrapping the six-point function $\vev{111111}$ at next-to-next-to-leading order.}
    \label{fig:111111_Illustration}
\end{figure}

We now compute the six-point function $\vev{111111}$ using the same method as in the previous section, but adapted to the case of three spacetime cross-ratios.
In particular, the basis of symbols is now
\begin{equation}
    \xi_i
    =
    \lbrace
    x_1, x_2, x_3,
    1-x_1, 1-x_2, 1-x_3,
    x_{12}, x_{13}, x_{23}
    \rbrace\,.
\end{equation}
In the following paragraphs, we only list the results since the steps are fully equivalent to the five-point case. The perturbative bootstrap workflow is reported here in Figure \ref{fig:111111_Illustration}, where the coefficients refer to the calculation at next-to-next-to-leading order.

\paragraph{Leading order.}
At leading order, the functions $f_{1,2}(x_1,x_2,x_3)$ are simply given by
\begin{equation}
	\begin{split}
    f_1^{(0)} (x_1,x_2,x_3)
    &=
    1\,, \\
    f_2^{(0)} (x_1,x_2,x_3)
    &=
    0\,.
    \end{split}
    \label{eq:111111_f_LO}
\end{equation}
$f_2$ has been chosen such that it is in a one-to-one correspondence with the channel $F_{14}$ of the natural basis.
This choice is convenient for our calculations, as it is easy to see that $F_{14}=0$ up to next-to-next-to-leading order.

\paragraph{Next-to-leading order.}
At next-to-leading order, we obtain
\begin{equation}
\small
    \begin{split}
        f_1^{(1)} (x_1,x_2,x_3)
        &=
        \frac{1}{12 \pi^2}
        \biggl(
        G(0,1,x_1) - G(1,0,x_1)
        +
        2\bigl(
        G(1,x_1) G(1,x_2) \\
        &\phantom{=\ }
        - G(x_2,x_1) G(1,x_2) - G(1,x_1) G(1,x_3) - G(0,x_2) (G(1,x_1)-G(x_3,x_1)) \\
        &\phantom{=\ }
        + G(1,x_3) G(x_3,x_1) + G(x_2,x_1) G(x_3,x_2) + G(1,x_3) G(x_3,x_1) \\
        &\phantom{=\ }
        + G(x_2,x_1) G(x_3,x_2)
         +G(0,x_3) ( G(1,x_1) - G(x_3,x_1) + G(x_3,x_2)) \\
        &\phantom{=\ }
        - G(x_3,x_1) G(x_3,x_2)
        - G(0,1,x_2) - G(0,1,x_3) + G(0,x_3,x_2) \\
        &\phantom{=\ }
        + G(1,0,x_2) + G(1,0,x_3) - G(1,x_2,x_1) + G(1,x_3,x_1)
        + G(x_2,1,x_1) \\
        &\phantom{=\ } - G(x_2,x_3,x_1) - G(x_3,0,x_2) - G(x_3,1,x_1) + G(x_3,x_2,x_1)
        \bigr) \\
        &\phantom{=\ }
        +
        3\bigl(
        G(1,x_3) G(1,x_2) - G(1,x_3) G(x_3,x_2) - G(0,x_3) G(1,x_2) \\
        &\phantom{=\ }
        - G(1,x_3,x_2) + G(x_3,1,x_2)
        \bigr)
        +
        \frac{2 \pi^2}{3}
        \biggr)\,, \\
        f_2^{(1)} (x_1,x_2,x_3)
        &=
        0\,,
    \end{split}
\end{equation}
which is in agreement with the results of the recursion relation in \cite{Barrat:2021tpn}.

\paragraph{Next-to-next-to-leading order.}

At next-to-next-to-leading order, the correlator is controlled by the train track integral introduced in Section \ref{subsub:OneIntegralToRuleThemAll}.
The corresponding diagram is
\begin{equation}
    \NNLOSixPt
    =
    \frac{\lambda^5}{8} B_{156,234}\,.
\end{equation}
This is the only diagram appearing in the channel $F_4$. There are two other channels that are given by only one six-point train-track diagram, but they are related to the channel $F_4$ by crossing symmetry.
The determination of $f_1$ is completely analogous to the calculation of $f$ for the correlator $\vev{11112}$, the only difference being technical since we deal with a high number of open coefficients.
Before imposing the integrability condition, the Ansatz consists of $6643$ coefficients.
Finiteness reduces this number to $1406$ unknowns.
In order to deal with such a high number of terms, we create a linear system of equations in \textsc{Mathematica} to match the Ansatz to $f_4$, that we iteratively solve by using row reduction.
Inputting the train-track fixes all the coefficients except for $21$ of them.
The pinching limit \eqref{eq:111111_PinchingTo11112} consists of six relations.
One of them suffices to fix the remaining open coefficients, while the other five serve as checks of our final result.
Since the expressions are lengthy, for the solution of the Ward identities $f_1(x_1, x_2, x_3)$ we refer to the ancillary notebook of \cite{Artico:2024wut}, where the $R-$channels in the natural basis are also given explicitly.
To illustrate the shape of the results, we show here some of the contributing terms:
\begin{equation}
	\begin{split}
		f_1^{(2)} (x_1, x_2, x_3)
		&=
		\frac{1}{64 \pi^4}
		\biggl(
		G(x_2, x_3, 1, 0, x_1)
		- \frac{8}{3}
		G(x_3, x_3, x_3, x_2, x_1)
		+
		\ldots
		\biggr)\,, \\
		f_2^{(2)} (x_1,x_2,x_3) &= 0\,.
	\end{split}
\end{equation}

\paragraph{Checks.}

\begin{figure}
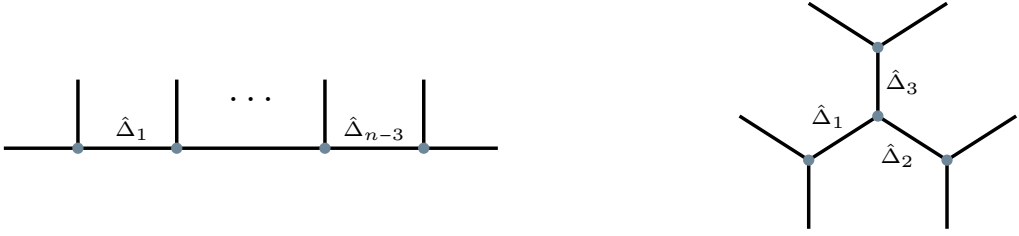

\centering
\begin{subfigure}{.55\textwidth}
  \centering
  \CombChannel
\end{subfigure}%
\begin{subfigure}{.45\textwidth}
  \centering
  \SnowflakeChannel
\end{subfigure}
\caption{The two OPE configurations used for performing checks on our result for the six-point function $\vev{111111}$ are displayed here.
The left figure presents the comb channel and the leading non-trivial exchange of operators.
The right figure shows the leading non-trivial contribution in the snowflake channel.}
\label{fig:CombsAndSnowflakes}
\end{figure}

We now proceed to some elementary checks for the results of this section.
Specifically, we wish to compare the lowest OPE coefficients for each correlator to the existing literature. The reason is that since a large amount of non-perturbative information has been used to find the correlator, we must think of independent non-trivial checks that can verify our solution is correct.

For the six-point function, we can use the two different OPE channels, namely the comb and snowflake configurations depicted in figure \ref{fig:CombsAndSnowflakes}.\footnote{We thank Andrea Cavagli{\`a} and Marco Meineri for suggesting these checks.}

In the case of the correlator $\vev{11112}$, the easiest check that one can perform is the coefficient
\begin{equation}
    F_2 (\eta_1, \eta_2)
    =
    \lambda_{112} + \Om(\eta_1, \eta_2)\,,
    \label{eq:11112_BlockExpansion}
\end{equation}
where $\eta_i$ refers to the cross-ratios of the comb channel used in Section 4.2.2 of \cite{Barrat:2024nod}.
This OPE coefficient is protected and given exactly in \eqref{eq:lambda112}.
Perturbatively, it reads
\begin{equation}
    \lambda_{112} = 1 + \frac{\lambda}{48} - \frac{29\lambda^2}{23040} + \Om(\lambda^3)\,,
    \label{eq:lambda112Pert}
\end{equation}
and we find a perfect match order by order with our correlator.
A stronger check would be to compare the OPE coefficients of unprotected operators.
The simplest case would then be the combination $\lambda_{11\phi^6}^2 \lambda_{2 \phi^6\phi^6}$.
However, to the best of our knowledge, this coefficient is presently not known up to next-to-next-to-leading order.

The check is less trivial for the six-point function $\vev{111111}$.
Using again the variables $\eta_i$ defined in \cite{Barrat:2024nod}, our correlator takes the form
\begin{equation}
    F^{(2)}_2 (\eta_1, \eta_2, \eta_3)
    =
    \frac{\pi^2 - 18 + 6 \log \eta_1}{192 \pi^4} \lambda^2\, \eta_1
    +
    \Om(\eta_1^2, \eta_2, \eta_3)\,,
    \label{eq:111111_BlockExpansion}
\end{equation}
from which we can read the OPE coefficient and the anomalous dimension of the unprotected operator $\phi^6$:
\begin{equation}
	\begin{split}
		\left. \lambda_{11\phi^6}^2 \eta_1 \right|_{\Om(\lambda^2)}
    		&=
    		\frac{\pi^2 - 18}{192 \pi^4}\,, \\
    		\gamma^{(1)}_{\phi^6}
    		&=
    		\frac{1}{4 \pi^2}\,.
	\end{split}
	\label{eq:111111_lambda116}
\end{equation}
These values match the literature (see, e.g., \cite{Cavaglia:2021bnz}).
A similar check can be performed in another OPE limit, which corresponds to the snowflake configuration.
Using the variables $z_i$ of \cite{Barrat:2022eim,Peveri:2023qip,Barrat:2024nod} and the conformal blocks of \cite{Fortin:2023xqq}, the highest-weight channel is then given by
\begin{equation}
    F^{(2)}_2 (z_1, z_2, z_3)
    =
    \frac{\pi^2 - 18 + 6 \log z_1 z_2}{192 \pi^4} \lambda^2\, z_1 z_2
    +
    \Om(z_1^2, z_2^2, z_3)\,,
    \label{eq:111111_BlockExpansion2}
\end{equation}
which again matches \eqref{eq:111111_lambda116}.

Note that an expansion in superconformal blocks would give access to the three-point function $\vev{\phi^6 \phi^6 \phi^6}$ at next-to-next-to-leading order, a quantity which is presently unknown to the best of our knowledge.
The superblocks are currently an active subject of research and they are still not completely available; therefore we postpone this analysis to future work.

\section{Summary and perspectives}
\label{sec:ConclusionsMulti}
Building on the perturbative bootstrap methods described in chapter \ref{ch:BDD} and applied to bulk-defect-defect correlators, in this chapter we explored the challenging multipoint correlators of scalar half-BPS insertions along the Maldacena-Wilson line in $\Nm=4$ sYM.
Specifically, we focused on the five-point function $\vev{11112}$ and the six-point function $\vev{111111}$, while deriving new results for $\vev{1122}$ and $\vev{1212}$ as well.
Our method combined both perturbative and non-perturbative insights into these correlators.
On the non-perturbative front, we employed the superconformal Ward identities introduced in \cite{Liendo:2016ymz,Bliard:2024und,Barrat:2024ta}, applying crossing symmetry constraints to reduce the correlators to a low number of functions of the spacetime cross-ratios.
These functions were further constrained by pinching operators down to lower-point functions. On the perturbative side, we found that, for $\vev{11112}$ and $\vev{111111}$, the correlator was governed by a single integral, referred to as the six-point train track.
This integral was evaluated in \cite{Rodrigues:2024znq} for all relevant orderings in the case of aligned external operators, while the general case still represents an active topic of research \cite{Bourjaily:2017bsb,Bourjaily:2018ycu,Ananthanarayan:2020ncn,Loebbert:2020glj,Kristensson:2021ani,Morales:2022csr,McLeod:2023qdf}.
By incorporating this result into a carefully constructed Ansatz, based on transcendentality properties, we succeeded in fully determining the correlators up to next-to-next-to-leading order.
The work presented here is currently the last step of a line of research started with references \cite{Barrat:2021tpn,Barrat:2022eim}, continued in the PhD thesis \cite{Barrat:2024nod} and \cite{Peveri:2023qip} and now reframed under the label of perturbative bootstrap in \cite{Artico:2024wut} and \cite{Artico:2024wnt}. There are several promising avenues for further exploration following this work. We will conclude this chapter with a description of possible follow-up studies, with some details regarding how we plan to approach them. \\

\paragraph{Four-point functions of higher-dimensional protected operators} While our focus in the chapter was on higher-point correlators, the techniques we developed here can also be applied to obtain four-point functions involving higher-weight operators.
A natural next step could be to construct a recursion relation, similar to the one presented in \cite{Barrat:2021tpn}, for four-point functions at next-to-next-to-leading order. This would extend the results of \cite{Kiryu:2018phb} for arbitrary correlators $\vev{\Delta_1 \Delta_2 \Delta_3 \Delta_4}$ to the next-to-next-to-leading order, which could be particularly valuable for bootstrap applications as it would provide further constraints on defect CFT data. The first step in this direction would be to compute the next correlator in terms of complexity: the correlator $\vev{2222}$. This correlator depends on one space-time cross ratio and two $R$-symmetry cross ratios and has 6 $R$-symmetry channels, making it in principle easier to bootstrap than other correlators presented in this chapter. The solution to the SCWI would be different, but the steps involved are the same: we also observe one channel to be zero up to $NNLO$ included. If the (pinched) eight-point integral representing this channel at $N^3LO$ is computed, a calculation of this correlator could be achieved at an order in the weak coupling expansion never reached before for defect correlators, thus providing a valuable input for further extending the bootstrability method \cite{Cavaglia:2021bnz,Cavaglia:2022qpg,Cavaglia:2022yvv,Cavaglia:2023mmu,Cavaglia:2024dkk}.

\paragraph{Six-point bootstrap}
Another interesting extension would be to conduct a numerical analysis of the correlator $\vev{111111}$, building upon the methods outlined in \cite{Antunes:2023kyz}.
This approach could potentially be merged with the bootstrability techniques applied so far to four-point functions \cite{Cavaglia:2021bnz,Cavaglia:2022qpg,Cavaglia:2022yvv,Cavaglia:2023mmu,Cavaglia:2024dkk}, thereby providing access to bounds on four-point functions and to a broader range of CFT data.
Such a framework would serve as an excellent testing ground for methods that may apply to higher-dimensional systems.
To carry out this analysis, it will be crucial to determine the superconformal blocks in the comb channel, a task that is currently being addressed in ongoing work \cite{Barrat:2024ta2}.
This work seeks to rederive and extend the strong-coupling results of \cite{Giombi:2023zte} for $\vev{111111}$, employing the techniques introduced in \cite{Barrat:2024ta2}.
Additionally, integrated correlators in the gist of \cite{Drukker:2022pxk,Cavaglia:2022qpg,Pufu:2023vwo,Fiol:2023cml,Billo:2023ncz,Billo:2024kri,Dempsey:2024vkf,Drukker:2024ta} were shown to significantly improve the numerical bootstrap results. Constraints coming from integrated correlators and integrability currently play no role in the perturbative bootstrap we presented; in the future, they could either serve as checks or be included in the method to further reduce the amount of perturbative information needed. The research question guiding this study is to understand how constraining the residual supersymmetry is and whether multipoint correlators at weak coupling can be bootstrapped without any diagrammatic computation as for strong-coupling \cite{Ferrero:2021bsb,Ferrero:2023znz,Ferrero:2023gnu}.
It would certainly be useful for this direction to revisit the solution of the Ward identities for six-point functions to understand if more constraints can be used to reduce the number of functions to one after imposing crossing symmetry.
An interesting quantity that could be computed is the currently unknown three-point function $\vev{\phi^6 \phi^6 \phi^6}$, thanks to the interplay between comb and snowflake channels in six-point functions.

\paragraph{Wilson-line defect in different gauge group representations.} Reference \cite{Beccaria:2022bcr} is part of a series of papers\footnote{See the bibliography of \cite{Beccaria:2022bcr} for the other papers in the series} discussing a non-local operator interpolating between the supersymmetric and non-supersymmetric Wilson-line defect. In this paper, the authors generalize previous results for the beta function and Wilson loop expectation value to the case of an arbitrary representation of the gauge group and beyond the planar limit. Vacuum expectation values of Wilson-line defects in a general representation will have coefficients that depend on color invariants \cite{Fiol:2018yuc}, and which can be connected via AdS/CFT duality to amplitudes of the open topological string of Berkovits-Vafa \cite{Fiol:2013hna}. A study of multipoint correlators for a defect in a general $SU(N)$ representation is still missing from the literature and it is definitely within reach using perturbative bootstrap techniques at weak coupling, considering that the integrals involved are the same. The SCWI applied for the computation of weak coupling correlation functions can then be applied to derive strong coupling correlators as well, generalizing the results of this chapter.

\paragraph{Finite $N$ correlators and $S$-duality.} A worth-exploring research direction is represented by the study of multipoint correlators at finite $N$, \textit{i.e.} including non-planar diagrams. Beyond their intrinsic interest, finite $N$ results for multipoint correlators at weak coupling could be mapped to strong coupling results for the 't Hooft line via the $S$-duality briefly introduced in \ref{subsec:N=4sYM}. In its simplest form, $S$-duality prescribes a change of the coupling constant as in
\beq
g\rightarrow g '=\frac{4\pi}{g} \,.
\eeq
The expansion of correlators concerning a coupling constant, however, is not well defined in the large $N$ limit: to circumvent the issue we can define a new coupling for 't Hooft lines \cite{Pufu:2023vwo} 
\beq
\lambda '\rightarrow {g '}^2N
\eeq
and multipoint correlators on a 't Hooft line can therefore be expanded at large $N$ in a strong coupling expansion in terms of inverse powers of $\lambda'$. The large $N$ strong-coupling results can then be independently calculated through analytic bootstrap building on the methods of \cite{Ferrero:2021bsb,Ferrero:2023znz,Ferrero:2023gnu}, resulting in a so-far missing description of multipoint correlators on a 't Hooft line. The 't Hooft line has recently received interest for its exciting connections to localization, holography, and integrability \cite{Kristjansen:2023ysz,Kristjansen:2024dnm,Kristjansen:2024map}. While most of the results so far deal with correlators involving bulk operators, there must be a defect tilt operator corresponding to the superprimary $\Oh_1$ for the Wilson line whose multipoint correlators we can consider. Since the ’t Hooft line in the absence of backreaction is described by a planar D1-brane embedded vertically in $AdS_5$ \cite{Kristjansen:2024map}, strong coupling correlators of defect operators can have an interpretation as weak coupling correlators in an $AdS$ background via $AdS/CFT$ duality.
\paragraph{Beyond $N=4$ sYM.} This work underscores the significant role of train track integrals in multipoint correlators of defect operators.
We anticipate that, at next-to-next-to-next-to-leading order, the eight-point function $\vev{11111111}$ will similarly be governed by lower-point functions and the integral 
\begin{equation}
    \NNNLOEightPt\,.
    \label{eq:EightPointIntegral}
\end{equation}
While obtaining the lower-point functions might pose challenges, it would still be valuable to further investigate the class of integrals \eqref{eq:EightPointIntegral}.
They have been studied in the literature and are conjectured to be integrable \cite{Chicherin:2017frs,Kazakov:2023nyu,Loebbert:2024fsj,Duhr:2024hjf}.
In cases where the external points are not aligned, elliptic behavior is expected.
Drawing from our experience with the six-point function, it is plausible that all $n$-point train track integrals, in the collinear limit, could be expressible in terms of Goncharov polylogarithms.
One promising approach to study such integrals could involve the Wilson-line defect CFT of the fishnet theory \cite{Gromov:2021ahm}, where the vertices significantly limit the number of contributing diagrams for a given configuration.
By combining techniques from the conformal bootstrap and integrability, it has been possible to derive exact correlators within this theory \cite{Grabner:2017pgm,Gromov:2018hut}.
It would be fascinating to explore to what extent these insights can be extended to the defect CFT framework.
Finally, the magnetic line defect CFT in the $\mathrm{O}(N)$ model has garnered significant interest, both from the perspective of large $N$ expansions and the $\veps$-expansion.
A similar defect CFT has been explored in Yukawa theories \cite{Giombi:2022vnz}, with correlators studied in \cite{Barrat:2023ivo}.
Yukawa CFTs are thought to exhibit emergent supersymmetry under specific configurations \cite{Fei:2016sgs}. These models hold interesting applications in condensed-matter physics, and the study of multipoint correlators in this context could yield valuable insights. In the framework of the $\veps$-expansion, many of the structures observed in this paper will likely persist, even in the absence of supersymmetry.

\chapter*{Conclusions}
\markboth{Conclusions}{}
\addcontentsline{toc}{chapter}{Conclusions}
In this final section, we provide a summary of the results presented in each chapter and some concluding remarks. Many of the possible research directions have been listed in Sections \ref{sec:Conclusions1}, \ref{sec:ConclusionsBDD}, and \ref{sec:ConclusionsMulti}; for this reason, the perspectives presented in the following paragraphs try to give a broader idea of interesting connections among the topics presented in these thesis and the existing research in theoretical physics. 
\section*{Summary of the results}
As announced in the Preface, this whole thesis has been focused on the study of the functional space of Feynman integrals and correlation functions in a defect CFT using a variety of analytical methods. This interest has produced parallel studies whose results are collected in this section. Chapter \ref{chapter:IBP} focused on developing a new class of integration-by-parts (IBP) identities for Feynman integrals formulated in parametric space and rooted in the projective geometry of the integrands. Their validity was tested through explicit applications to several one-loop and multi-loop examples, where they were shown to reproduce known results, including elliptic differential equations for the two-loop unequal-mass sunrise and up to the four-loop equal-mass banana diagram. Moving beyond the formal construction of these identities, the chapter introduced a concrete strategy for reducing integrals to a basis of master integrals. A central innovation was the translation of this reduction problem into a membership test for polynomial ideals, drawing a powerful bridge between Feynman integral reduction and non-linear algebra. This approach enabled the efficient derivation of differential equations for multi-loop integrals, with the substantial reduction of generated equations by up to two orders of magnitude for the equal mass banana integral at two and three loops. Although the resulting algorithm is not yet competitive with the most advanced momentum-space techniques, the parameter-space formulation exhibits promising structural advantages: it bypasses the ambiguity of loop-momentum routing, treats irreducible numerators implicitly, and is naturally suited to integrals with manifest graph symmetries. These features, alongside connections to GKZ systems and algebraic geometry, suggest that IBP identities in parameter space could play a central role in both simplifying and classifying Feynman integrals in future problems, also beyond particle scattering.\\

In Chapter \ref{ch:BDD} and Chapter \ref{chapter:MP} we have explored correlators in $\Nm = 4$ super Yang-Mills theory, enriched by the presence of a half-BPS Wilson line. First, we focused on correlators involving one bulk and two defect operators. These observables are situated at the intersection of two fundamental principles in quantum field theory: the powerful constraints of supersymmetry and conformal invariance, and the analytic structure of correlation functions as determined by locality and unitarity. Using such principles, we were able to prove that all bulk-defect-defect correlators up to next-to-leading order are determined by one master integral that can be expressed as a rational function of the space-time cross-ratio. The remarkable and unexpected cancellation of transcendental terms at next-to-leading order suggests the existence of further constraints on the space of functions underlying these observables — a space yet to be systematically explored. Building upon the perturbative bootstrap techniques developed in Chapter \ref{ch:BDD} for bulk-defect-defect correlators, Chapter \ref{chapter:MP} extended this framework to the significantly more intricate multipoint correlation functions of scalar half-BPS operators inserted along the Maldacena-Wilson line. The chapter focused on the five-point function $\vev{11112}$ and the six-point function $\vev{111111}$, providing new perturbative results up to NNLO and a strategy to tackle the differential constraints represented by the superconformal Ward identities. Notably, it also derived correlators such as $\vev{1122}$ and $\vev{1212}$, demonstrating the generality of the methods and providing input data for integrability-based approaches. Central to this chapter was the use of the mentioned superconformal Ward identities and crossing symmetry, which drastically reduce the number of independent functions of cross-ratios describing the correlators. These functional degrees of freedom were further constrained by “pinching” higher-point functions into known lower-point correlators. On the perturbative side, the key technical advancement was the reduction of the six-point and five-point correlators to the evaluation of a single integral known as the six-point train-track integral. Previously computed for the aligned configurations of operators, this integral provided a powerful constraint that -- when embedded into a physically motivated Ansatz based on transcendentality and symmetries -- enabled full determination of the correlators up to next-to-next-to-leading order at weak coupling. This development represents a natural culmination of a line of research now formalized under the umbrella of the perturbative bootstrap program.

\section*{Future directions}
We conclude this thesis with some final remarks on future research directions connecting the different parts of this thesis to contemporary topics in theoretical physics. The final sections of each chapter already contain numerous detailed suggestions of research projects that would represent the natural prosecution of the research presented in this thesis; relevant examples include the deeper understanding of the non-linear algebra behind master integral reduction problems for projective forms, the analysis of the constraints that can determine the loss of transcendentality observed for bulk-defect-defect correlators, and the study of defect correlators involving spinning operators via the multipoint bootstrap approach. In the following paragraphs, we list some broader perspectives that could represent valuable connections to the results and techniques presented in the three chapters of this thesis.\\

A natural question that the reader might be wondering now is whether there is a more direct connection between the IBP identities presented in the first chapter and the perturbative bootstrap introduced in the rest of the thesis than the general study of the functional space of quantum field theories. In other words, we want to establish ways in which the reduction to master integrals can have a direct impact on the perturbative bootstrap. While it is definitely true that reduction tools may be useful in tackling challenging bulk or defect integrals appearing in higher orders of the perturbative expansion, we report here a more ambitious connection, starting from the notion of cosmological correlators \cite{Maldacena:2002vr,Arkani-Hamed:2015bza,Arkani-Hamed:2017fdk}. Cosmological correlators are correlation functions of fields -- e.g. curvature perturbations -- evaluated at late times, typically on the spatial slices at the end of inflation. These quantities are directly connected to observables like the Cosmic Microwave Background anisotropies. Computing these correlators in perturbation theory requires supplementing the standard Feynman rules with additional time integrals at each vertex, since virtual particles can be created at any point in the prior evolution of the universe; consequently, even tree-level computations of cosmological correlators present significant challenges. In \cite{Grimm:2025zhv} a simple representation of tree-level cosmological correlators was investigated by viewing them as solutions to GKZ systems of differential equations, whose close connection to parameter-based IBP identities has already been mentioned in Chapter \ref{chapter:IBP}. We report here a simplified version of the example of cosmological correlator presented in equation (3.11) of Ref. \cite{Grimm:2025zhv}, clearly similar to the parametric integrals studied in the first chapter of this thesis.
\begin{equation*}
I(X,Y;\vec{\alpha}) = \int_{R_+^{N_v}}d^{N_v}x \frac{\prod_{i=1}^{N_v} x_i^{\alpha_i - 1}}{p(x_i)}
\end{equation*}
where $X$ and $Y$ are coordinates of the external points and $p(x_i)$ is a polynomial in the integration variables. The similarity with Feynman integrals is striking, although the non-homogeneity of $p(x_i)$ requires further investigations. As stated in \cite{DiPietro:2021sjt}, on a rigid de Sitter space the correlators of fields on the future boundary will be invariant under the dS isometry group $SO(1, d + 1)$, which is also the Euclidean conformal group in d dimensions. The cosmological correlators of a QFT in dS can therefore be expected to form the correlators of some (potentially non-unitary) Euclidean CFT, in an instance of the still little explored dS/CFT correspondence \cite{Strominger:2001pn,Witten:2001kn}.	The cosmological bootstrap program introduced in \cite{Arkani-Hamed:2018kmz} strives to use a simplified analytic structure for correlators in perturbation theory, together with symmetries and
singularities, to fully determine the final answer without reference to bulk time evolution: the study of cosmological correlators could represent a point of contact among the analytic perturbative techniques introduced in this thesis.\\

Another research direction of great interest because of its connections with several topics of relevance is constituted by removing the large $N$ assumption from the perturbative weak coupling computations done in the setup of the half-BPS Wilson line defect CFT. This topic has been mentioned more than once in the summary of each chapter, and in this paragraph, we are going to sketch some motivation for finite $N$ calculations -- beyond their intrinsic interest. A solid starting point is the OPE expansion for bulk-defect-defect correlators reported in Eq.~\eqref{eq:ConformalBlockExpansion}. As we have seen, both the bulk-to-defect OPE and the defect-OPE lead to a conformal block expansion of the form
\begin{equation*}
    F_j (x)
    =
    x^{-\Delta_1}
    \sum_{\Dh\: \mathrm{prim.}}
    b_{\Delta_1 \Dh}^{(j)} \lambdah_{\Dh_2 \Dh_3 \Dh}^{(j)}
    g_{\Dh} (x)\,,
\end{equation*}
where the sum runs over the conformal primaries exchanged in the OPE. At large $N$, we have seen in multiple cases that the OPE between defect operators only generated single-trace operators; however, the bulk-to-defect OPE will also bring multi-trace operators whose OPE coefficients are suppressed by inverse powers of $N$. This remark is sufficient to conclude that at finite $N$ all the correlators we have studied will witness the appearance of multi-trace operators in the OPE channels, including multipoint correlators. A large part of the conformal data of the finite $N$ defect CFT is currently unknown: a study of correlators at finite $N$, combined to a study of the line defect in different representations, will give access to a large amount of new conformal data and to a deeper understanding of the spectrum of the theory\footnote{For representations of the half-BPS Wilson line that are different from the fundamental one, there are more defect operators of dimension one beyond the single trace one and $\phi^6$, a feature so far little explored.}. We also mentioned in Chapter \ref{ch:BDD} the so-called $S$-duality, a finite $N$ correspondence mapping the Maldacena-Wilson line into a supersymmetric magnetic monopole, the 't Hooft line. Combined with results such as the ones presented in \cite{Kristjansen:2024map} for operators defined in the bulk, this non-trivial application of the $S$-duality would open some insights into the hard topic of multipoint correlators for the 't Hooft line at strong coupling. \\

A further appearance of finite $N$ data in contemporary research comes from the area of finite temperature bootstrap \cite{Iliesiu:2018fao,Marchetto:2023fcw,Marchetto:2023xap,Barrat:2024aoa,Barrat:2025wbi,Barrat:2025nvu}. We shortly referred to this topic several times in the thesis and we will dedicate this last paragraph to elaborating on the thermal bootstrap. With thermal bootstrap, we refer to the study of correlation functions on a space $S_\beta^1 \times \mathbb{R}^{d-1}$, where the time direction is compactified in a circle of radius $\beta$ given by the inverse of the temperature. The reason why the defect bootstrap and the thermal bootstrap are often presented as connected is that both are ways of breaking the conformal symmetry in a controlled way. In both cases, vacuum expectation values of operators assume a non-zero value due to the emergence of a `natural scale' of the problem: the distance from the defect in one case, the inverse temperature in the other. Local properties as the OPE are still valid at finite temperature -- they will now have a finite radius of convergence -- and given the presence of non-vanishing one-point functions the first non-trivial correlator will be the two-point function of local operators, expressed in (thermal) conformal blocks as \cite{Iliesiu:2018fao}
\begin{equation*}
G(\tau) = \langle \phi(\tau, \mathbf{0}) \, \phi(0, \mathbf{0}) \rangle_\beta
= \sum_{\mathcal{O}} a_{\mathcal{O}} \, \frac{1}{\beta^\Delta} \, \tilde{g}_{\Delta}\left( \frac{\tau}{\beta} \right),
\end{equation*}
where the connection to the zero temperature bootstrap is that the constant $a_{\mathcal{O}}$ depends on the thermal one-point functions and the zero temperature two- and three-point functions. The sum is taken on the spectrum of the theory; in the case of a thermal defect CFT (see for example \cite{Barrat:2024aoa}), we have seen that the spectrum depends on $N$\footnote{Multi-trace operators whose conformal data will be subleading at large $N$ can also be defined in the bulk. }. To take advantage of the simplicity of the large $N$ limit, we could hope that a similar simplification happens for the thermal OPE and that the sum above can be taken only on single-trace conformal primaries; this is unfortunately not possible due to the dependence on $N$ of thermal one-point functions as it is shown in Ref. \cite{Iliesiu:2018fao} for the $O(N)$ model at large $N$. It is therefore necessary to keep the whole spectrum of conformal primaries each carrying its dimension whose value must be studied at finite $N$. There are still important simplifications observed at large $N$ and large coupling $\lambda$, such as the reduction of two-point functions of half-BPS operators to the mean field theory result and stress tensor contributions which would be worth studying in a thermal defect set-up, where the stress tensor is simply absent for one-dimensional defects. The future applications of bootstrap to challenging spaces -- higher dimensional or multiple defects, cosmological and thermal backgrounds -- can profit from a deeper understanding of CFTs far from the planar large $N$ limit.


\appendix

\chapter{Integrals}
\label{app:Integrals}
\pagenumbering{roman}
This appendix is dedicated to the integrals encountered throughout this work. We also provide the regularization chosen for UV divergent integrals.
\section{Bulk Feynman integrals}
We consider here bulk integrals, i.e., integrals that do not involve the Wilson line as an internal vertex.
Integrals along the defect are one-dimensional and easy to perform with standard packages, such as \textsc{PolyLogTools}, by using the analytical expressions of this section.

The master integral for massless scalar four-point integrals at next-to-leading order is the well-known X-integral.
\begin{align}
    X_{1234}
    =
    \XIntegral\
    =
    \int d^4 x_5\, I_{15} I_{25} I_{35} I_{45}
    =
    \frac{I_{12} I_{34}}{16\pi^2}\ z \zb\, D( z, \zb)\,,
    \label{eq:X1234}
\end{align}
with the Bloch-Wigner function
\begin{equation}
    D(z, \zb)
    =
    \frac{1}{z - \zb} \left( 2 \Li_2 (z) - 2 \Li_2 (\zb) + \log z \zb\, \log \frac{1-z}{1-\zb} \right)\,,
    \label{eq:BlochWigner}
\end{equation}
where the cross-ratios $z$ and $\zb$ are defined through
\begin{equation}
    z \zb = \frac{x_{12}^2 x_{34}^2}{x_{13}^2 x_{24}^2}\,.
    \qquad
    (1-z)(1-\zb) = \frac{x_{14}^2 x_{23}^2}{x_{13}^2 x_{24}^2}\,.
    \label{eq:4dCrossRatios}
\end{equation}
Note that the Bloch-Wigner function is crossing symmetric:
\begin{equation}
    D(z, \zb) = D(1-z\,, 1-\zb)\,.
    \label{eq:BWIsCrossingSymmetric}
\end{equation}
From the X-integral, it is possible to obtain the three-point Y-integral defined as
\begin{align}
    Y_{123}
    =
    \YIntegral\
    &=
    \int d^4 x_4\, I_{14} I_{24} I_{34} \notag \\
    &=
    \lim\limits_{\tau_4 \to \infty} I_{34}^{-1} X_{1234}
    =
    \frac{I_{12}}{16\pi^2} z \zb\, D( z, \zb)\,,
    \label{eq:Y123}
\end{align}
where now the cross-ratios $z, \zb$ are defined as the limit $\tau_4 \to \infty$ of the four-point variables defined above:
\begin{equation}
    z \zb
    =
    \frac{I_{13}}{I_{12}}\,,
    \qquad
    (1-z)(1-\zb)
    =
    \frac{I_{13}}{I_{23}}\,.
    \label{eq:CrossRatiosY}
\end{equation}
The Y-integral becomes $\log$-divergent when two points are pinched together:
\begin{equation}
    Y_{112} = Y_{122} = \IntegralYOneTwoTwo\ = - \frac{I_{12}}{16\pi^2} \left( \log \frac{I_{12}}{I_{11}} - 2 \right)\,.
    \label{eq:Y112}
\end{equation} 
The pinching of two points in the X-integral can be related to Y-integrals:
\begin{equation}
    X_{1233}
    =
    \IntegralXOneTwoThreeThree\
    =
    \frac{1}{2} ( I_{13} Y_{223} + I_{23} Y_{113} ) - \frac{I_{13} I_{23}}{32 \pi^2} \log \frac{I_{13} I_{23}}{I_{12}^2}\,.
    \label{eq:X1233}
\end{equation}
For completion, we also consider the limit where the external points coincide pairwise.
This is equivalent to a two-point integral with doubled propagators. In this case, we have
\begin{equation}
    X_{1122}
    =
    \IntegralXOneOneTwoTwo\
    =
    2 I_{12} Y_{112} - \frac{I_{12}^2}{8 \pi^2}\,.
    \label{eq:X1122}
\end{equation}

We often encounter derivatives of the $Y$-integral, for which the following identities hold:
\begin{equation}
    \begin{split}
    \partial_{1 \mu} Y_{123} &= - (\partial_{2 \mu} + \partial_{3 \mu}) Y_{123}\,, \\[.6em]
    \partial_1^2\, Y_{123} &= - I_{12} I_{13}\,, \\
    \left( \partial_1 \cdot \partial_2 \right) Y_{123} &= \frac{1}{2} \left( I_{12} I_{13} + I_{12} I_{23} - I_{13} I_{23} \right)\,.
    \end{split}
    \label{eq:YIdentities}
\end{equation}
All these identities are elementary to prove, using integration by parts and the scalar Green's equation
\begin{equation}
    \pd_1^2 I_{12}
    =
    - \delta^{(4)} (x_{12})\,.
    \label{eq:GreensEquation}
\end{equation}

As mentioned in the main text, the H-integral, defined through
\begin{equation}
    H_{12,34}
    =
    \int d^4 x_5\, I_{15} I_{25}\, Y_{345}\,,
    \label{eq:H1234}
\end{equation}
is \textit{not} conformal, and as far as we can tell, no exact solution has been obtained yet.
In this work, it appears in two contexts: associated with derivatives, or in the one-dimensional limit.
In the first case, using the identities given in \eqref{eq:YIdentities}, it is straightforward to obtain the following relations:
\begin{equation}
    \begin{split}
    \partial_1^2 H_{12,34}
    &=
    - I_{12}\, Y_{134}\,, \\
    \left( \partial_1 \cdot \partial_2 \right) H_{12,34}
    &=
    \frac{1}{2} \left[ I_{12} ( Y_{134} + Y_{234} ) - X_{1234} \right]\,.
    \end{split}
    \label{eq:H_Identities}
\end{equation}
There is no identity known for $\left( \partial_1 \cdot \partial_3 \right) H_{12,34}$.
For the second case, in which the points $x_k$ are aligned, the integral $H_{12,34}$ can be obtained analytically from the train track integral, as explained in Section \ref{subsub:OneIntegralToRuleThemAll} (see in particular \eqref{eq:H_From_B} for the result itself).

We also encounter divergent H-integrals, which correspond to the case in which two points on the same side coincide.
It is easy to derive analytical results for the specific cases of interest in the conformal frame \eqref{eq:ConformalFrame}:
\begin{align}
    H_{23,11} &= \frac{1}{8 \pi^2} (1 - \log \veps) Y_{123} - \frac{1}{16 \pi^2} A_1\,, \label{eq:H2311} \\
    H_{13,22} &= \frac{1}{8 \pi^2} (1 - \log \veps) Y_{123} - \frac{1}{16 \pi^2} A_2\,, \label{eq:H1322} \\
    H_{12,33} &= \frac{1}{8 \pi^2} (1 - \log \veps) Y_{123} - \frac{1}{16 \pi^2} A_3\,, \label{eq:H1233}
\end{align}
where the integral $A_k$ is defined as
\begin{equation}
    A_k = \int d^4 x_5\, I_{15} I_{25} I_{35} \log x_{k5}^2\,.
    \label{eq:Ak}
\end{equation}
In the conformal frame \eqref{eq:ConformalFrame}, the relevant results for this integral are
\begin{align}
    A_1 &= \log x\, Y_{123}\,, \label{eq:A1} \\
    A_2 &= (\log x + \log (1-x)) Y_{123}\,, \label{eq:A2} \\
    A_3 &= \log (1-x)\, Y_{123}\,. \label{eq:A3}
\end{align}

The identities \eqref{eq:H_Identities} can be used to determine integrals that arise in the computation of Feynman diagrams.
For instance, diagrams with two scalar-scalar-gluon vertices give rise to the expression
\begin{equation}
    F_{12,34}
    =
    \frac{\pd_{12} \cdot \pd_{34}}{I_{12} I_{34}} H_{12,34}\,,
    \label{eq:F1234_Definition}
\end{equation}
where we have used the shorthand notation $\pd_{ij}^\mu := \pd_i^\mu - \pd_j^\mu$.
This integral can be elegantly expressed in terms of X- and Y-integrals as
\begin{equation}
    \begin{split}
    F_{12,34}
    =\ &
    \frac{X_{1234}}{I_{13}I_{24}} - \frac{X_{1234}}{I_{14}I_{23}} + \left( \frac{1}{I_{14}} - \frac{1}{I_{13}} \right) Y_{134} + \left( \frac{1}{I_{23}} - \frac{1}{I_{24}} \right) Y_{234} \\
    & + \left( \frac{1}{I_{23}} - \frac{1}{I_{13}} \right) Y_{123} + \left( \frac{1}{I_{14}} - \frac{1}{I_{24}} \right) Y_{124}\,.
    \end{split}
    \label{eq:F1234_Result}
\end{equation}
Note that the F-integral also reduces to a simple expression when two external points coincide:
\begin{equation}
    \begin{split}
    F_{13,23} =\ \IntegralFOneThreeTwoThree\ =\ &  \frac{1}{2}
    \left(\frac{Y_{113}}{I_{13}} + \frac{Y_{223}}{I_{23}} \right) + Y_{123} \left( \frac{1}{I_{13}} + \frac{1}{I_{23}} - \frac{2}{I_{12}} \right) \\
    &+ \frac{1}{32 \pi^2} \log \frac{I_{13} I_{23}}{I_{12}^2}\,.
    \end{split}
    \label{eq:F1323}
\end{equation}

Another important family of integrals is
\begin{equation}
    K_{ij}
    =
    \int d^4 x_5\, I_{15} I_{25} I_{35}\, Y_{ij5}\,.
    \label{eq:Kij_Definition}
\end{equation}
In the conformal frame \eqref{eq:ConformalFrame}, if one of the points $(i,j)$ is $\tau_4$, the integral gives
\begin{equation}
    K_{i4}
    =
    \frac{I_{34}}{16 \pi^2} ( (2 + \log \tau_4^2) Y_{123} - A_i)\,.
    \label{eq:Ki4}
\end{equation}
If both points are $\tau_4$, we have
\begin{equation}
    K_{44}
    =
    - \frac{1}{8 \pi^2} I_{34} Y_{123} (\log \veps - 1 - \log \tau_4)\,.
    \label{eq:K44}
\end{equation}

For fermionic integrals, there exists a special star-triangle identity:
\begin{equation}
    \VertexFermionFermionScalar
    =
    \spd_1 \spd_3\, Y_{123}
    =
    - 4 \pi^2 \sx_{12} \sx_{23} I_{12} I_{13} I_{23}\,,
    \label{eq:StarTriangle}
\end{equation}
which is used to compute integrals with a Yukawa coupling (for instance the diagram given in \eqref{eq:1111_Spider}).
\section{Point-splitting regularization}
In this section, we clarify the treatment of UV divergent integrals in chapters \ref{ch:BDD} and \ref{chapter:MP}, as it is different from the much discussed dimensional regularization of the integrals in chapter \ref{chapter:IBP}. Whenever we are in the presence of a UV-divergent integral of the kinds discussed above, we regularize it via point-splitting. This consists in introducing an ultraviolet cutoff by separating coincident points: a small parameter $\epsilon$ is imposed such that the integration domain excludes configurations where any two integration points, or an integration point and an external insertion, approach closer than a distance $\epsilon$. All the $\log(\epsilon)$ factors in the results of bulk (and defect) loop integrals come from this point splitting. \\

However, the naive definition of point-splitting regularization given above can raise questons on gauge invariance. In particular, when dealing with algebra valued fields, naive point-splitting could spoil gauge invariance as the traces involved in the definition of the operators would consider fields at different space-time points. The solution is to define a gauge invariant version of point splitting by inserting an infinitesimal Wilson line connecting the two points that have been split. Let $\Phi(x)$ denote a field in the adjoint representation of the gauge group.
A local composite operator, e.g.\ $\mathcal O(x)=\tr\!\big(\Phi^2(x)\big)$,
can be defined in a gauge-invariant way via point-splitting as
\begin{equation}
\label{eq:pointsplit_operator}
\mathcal O_\epsilon(x)
=
\tr\!\left[
\Phi\!\left(x+\tfrac{\epsilon}{2}\right)\,
U\!\left(x+\tfrac{\epsilon}{2},\,x-\tfrac{\epsilon}{2}\right)\,
\Phi\!\left(x-\tfrac{\epsilon}{2}\right)
\right],
\end{equation}
where $U(x,y)$ is the Wilson line
\begin{equation}
\label{eq:wilson_line}
U(x,y)
=
\mathcal P \exp\!\left(
ig \int_{y}^{x} \! dz^\mu \, A_\mu(z)
\right).
\end{equation}
Expanding the Wilson line for infinitesimal separation $\epsilon^\mu$,
one finds
\begin{equation}
\label{eq:wilson_expansion}
U\!\left(x+\tfrac{\epsilon}{2},\,x-\tfrac{\epsilon}{2}\right)
=
\mathbf 1
+ ig\,\epsilon^\mu A_\mu(x)
+ \mathcal O(g^2 \epsilon^2).
\end{equation}
The local operator $\mathcal O(x)$ is recovered by taking the coincident-point limit $\epsilon\to 0$ after renormalization. The contribution of the Wilson line to the point-split operator
\eqref{eq:pointsplit_operator} is then
\begin{equation}
\label{eq:wilson_correction_operator}
\delta_\text{W}\mathcal O_\epsilon(x)
=
ig\,\epsilon^\mu\,
\tr\!\Big[
\Phi\!\left(x+\tfrac{\epsilon}{2}\right)
A_\mu(x)
\Phi\!\left(x-\tfrac{\epsilon}{2}\right)
\Big]
+ \mathcal O(\epsilon^2).
\end{equation}
What we observe is that the gauge-covariant point-splitting definition differs from naive point splitting
by terms proportional to $\epsilon^\mu A_\mu$, originating from the expansion of
the infinitesimal Wilson line.
At one loop in $\mathcal N=4$ SYM, ultraviolet divergences are at most logarithmic,
and no divergence can compensate the explicit factor of $\epsilon$: consequently, all Wilson-line contributions vanish in the coincident-point limit
$\epsilon\to0$, and naive point splitting is sufficient at this order. This is the reason why when dealing with UV divergences arising from one-integrals (as all our examples), we can use the simpler naive definition of point splitting to recover gauge invariant results.

\chapter{Goncharov Polylogarithms}
\label{app:SymbolsAndGoncharovPolylogarithms}
In this appendix, we provide a brief overview of symbols and their connection to Goncharov polylogarithms, which are key elements in formulating the Ansatz in Section \ref{sub:HigherPointFunctions}.

\section{Symbols}
\label{subsec:Symbols}

Symbols are formally defined via the Hopf algebra structure of multiple polylogarithms.
Specifically, the symbol of an expression represents the highest iteration of the coproduct, modulo $i \pi$.
This feature is particularly useful for building an Ansatz, as symbols contain essential information about the arguments of the Goncharov polylogarithms.
For example, the symbol of the dilogarithm $\text{Li}_2(x)$ is given by
\begin{equation}
    \Sm ( \text{Li}_2 (x) )
    =
    - (1-x) \otimes x\,.
    \label{eq:Symbols_Example}
\end{equation}
The \textsc{Mathematica} package \textsc{PolyLogTools} is particularly effective for manipulating symbols and Goncharov polylogarithms.

Symbols obey the following properties:
\begin{align}
    A \otimes ( \pm 1 ) \otimes B &= 0\,, \label{eq:Symbols_Preperty1} \\
    A \otimes (x \cdot y) \otimes B
    &=
    A \otimes x \otimes B + A \otimes y \otimes B\,.
    \label{eq:Symbols_Property2}
\end{align}

In Section \ref{sub:HigherPointFunctions}, we demonstrate how expressing correlators through symbols facilitates the construction of an Ansatz.
However, the resulting expression may not always be finite.
Finiteness is achieved by enforcing the \textit{integrability condition}.
For an expression written as
\begin{equation}
    S
    =
    \sum c_{i_1 \ldots i_n} A_{i_1} \otimes \ldots \otimes A_{i_n}\,,
    \label{eq:Symbols_Ansatz}
\end{equation}
where the $c_{i_1 \ldots i_n}$ are numerical coefficients, $S$ is finite if it satisfies
\begin{equation}
    \sum c_{i_1 \ldots i_n} A_{i_1} \otimes \ldots \otimes A_{i_{p-1}} \otimes A_{i_{p+2}} \otimes \ldots \otimes A_{i_n} d\log A_{i_p} \wedge d\log A_{i_{p+1}} = 0\,,
    \label{eq:Symbols_IntegrabilityCondition}
\end{equation}
for all pairs $(i_p, i_{p+1})$ with $1 \leq p < n$.
This condition is efficiently implemented in \textsc{PolyLogTools}.

Additionally, symbols obey the \textit{shuffle algebra} relation:
\begin{equation}
    \Sm (x \cdot y)
    =
    \Sm (x) \shuffle\,\Sm(y)\,,
    \label{eq:Symbols_ShuffleAlgebra}
\end{equation}
which serves as a useful property for simplifying expressions.

\section{Goncharov polylogarithms}
\label{subsec:GoncharovPolylogarithms}

Goncharov polylogarithms are a crucial component of our Ansatz.
Here, we define Goncharov polylogarithms and outline their key properties.

One definition of Goncharov polylogarithms is given by the iterated integral
\begin{equation}
    G(a_1, \ldots, a_n, x)
    =
    \int_0^x \frac{dt}{t-a_1} G(a_2, \ldots, a_n, t)\,,
    \label{eq:Goncharov_Definition}
\end{equation}
with $n \geq 0$, and the initial condition
\begin{equation}
    G(x)
    =
    1\,.
    \label{eq:Goncharov_InitialCondition}
\end{equation}
The transcendentality weight of a Goncharov polylogarithm defined in this manner is $n$.
In the special case where all arguments are zero except for the last one, Goncharov polylogarithms reduce to
\begin{equation}
    G(\underbrace{0, \ldots, 0}_{n \text{ times}}, x)
    =
    \frac{1}{n!} \log^n x\,.
    \label{eq:Goncharov_SpecialCase1}
\end{equation}
Another notable special case is
\begin{equation}
    G(0, \ldots, 0, a, x)
    =
    - \text{Li}_n \biggl( \frac{x}{a} \biggr)\,,
    \label{eq:Goncharov_SpecialCase2}
\end{equation}
where $\text{Li}_n$ denotes the $n$th dilogarithm.
Goncharov polylogarithms also obey a shuffle algebra inherited from the symbols.
This relation is expressed as
\begin{equation}
    G(\vec{a}, x) G(\vec{b}, x)
    =
    \sum_{\vec{c} = \vec{a} \shuffle \vec{b}} G(\vec{c}, x)\,.
    \label{eq:Goncharov_ShuffleAlgebra}
\end{equation}

\chapter{Feynman diagrams for $\vev{1111}$}
\label{app:FeynmanDiagramsOf1111}

In this appendix, we gather the results for the diagrams relevant for the computation of the correlator $\vev{1111}$ at next-to-next-to-leading order, for which the results are presented in Section \ref{subsub:1111}.
The diagrams contributing to the channel $F_1$ are gathered in Table \ref{table:Diagrams1111NNLO}, and can be split into \textit{bulk} and \textit{boundary} diagrams.

It is important to understand that all the diagrams are evaluated in the conformal frame
\begin{equation}
    (\tau_1, \tau_2, \tau_3, \tau_4)
    =
    (0, x, 1, \infty)\,.
    \label{eq:ConformalFrame}
\end{equation}
In particular, the limit $\tau_4 \to \infty$ provides an important simplification for the calculations.
However, the diagrams are not conformal on their own; it should thus be understood that the results for each individual diagram is only correct for the conformal frame.
At the end of the calculation, we expect that summing up all the diagrams cancels the divergences and the spurious $\log \tau_4$ terms, and that the final result is conformal.

In all the expressions above, we ignore the trivial $R$-symmetry factors for readability.

\section{Bulk diagrams}
\label{subsec:BulkDiagrams}

We begin by calculating the bulk diagrams.
The results are given in terms of known integrals, for which the results can be found in Appendix \ref{app:Integrals}.

\subsection{Self-energy diagrams}
\label{subsubsec:SelfEnergyDiagrams}

The first line in Table \ref{table:Diagrams1111NNLO} consists of self-energy diagrams.
The manipulations in order to obtain the diagrams are elementary, and we obtain
\begin{align}
    &\DefectSSSSTwoLoopsSelfEnergyOne\
    +
    \DefectSSSSTwoLoopsSelfEnergyTwo\ \notag \\
    & \quad
    +
    \DefectSSSSTwoLoopsSelfEnergyThree\
    +
    \DefectSSSSTwoLoopsSelfEnergyFour\
    =
    - \frac{\lambda^4}{2} (I_{34} H_{23,11} + I_{34} H_{13,22} + I_{34} H_{12,33} + K_{44})\,.
\end{align}
The results for the $H$- and $K$-integrals are given in \eqref{eq:H2311}-\eqref{eq:H1233} and \eqref{eq:K44}, respectively.
Each of these terms is divergent.
Notice also that $K_{44}$ contains a term $\log \tau_4$, which ultimately must cancel with terms in other diagrams for the correlator to be conformal.

\subsection{XX diagrams}
\label{subsubsec:XXDiagrams}

The diagrams that contain two $X$-vertices are more intricate.
Taking into account the trace as well as the symmetry factors, their sum can be shown to be equal to
\begin{align}
    \DefectSSSSTwoLoopsXXOne\
    +
    \DefectSSSSTwoLoopsXXTwo
    &=
    - \frac{\lambda^4}{4} \int d^4 x_5\, ( I_{35} I_{45}\, X_{1255} + I_{15} I_{45}\, X_{2355} ) \notag \\
    &\sim
    - \frac{\lambda^4}{8}
    \biggl(
    I_{34} H_{23,11}
    +
    I_{34} H_{13,22}
    +
    I_{34} H_{12,33}
    +
    K_{44} \notag \\
    &\phantom{=\ }
    -
    \frac{1}{16 \pi^2} \log \tau_4^2\, I_{34} Y_{123}
    \biggr)\,.
    \label{eq:1111_XX}
\end{align}

\subsection{XH diagrams}
\label{subsubsec:XHDiagrams}

The diagrams with one $X$ and two $Y$ vertices, which we refer to as XH-diagrams, can be evaluated in the same way as the other diagrams.
Summing them up, we find
\begin{align}
    &\DefectSSSSTwoLoopsXHOne\
    +
    \DefectSSSSTwoLoopsXHTwo\ \notag \\
    &\qquad
    +
    \DefectSSSSTwoLoopsXHThree\
    +
    \DefectSSSSTwoLoopsXHFour
    =
    \frac{\lambda^4}{8}
    \biggl(
    I_{34} H_{23,11}
    +
    I_{34} H_{13,22}
    +
    I_{34} H_{12,33}
    +
    K_{44} \notag \\
    &\phantom{\qquad
    +
    \DefectSSSSTwoLoopsXHThree\
    +
    \DefectSSSSTwoLoopsXHFour =\ }
    -
    H_{12,13}
    -
    H_{13,23}
    -
    2 H_{12,23} \notag \\
    &\phantom{\qquad
    +
    \DefectSSSSTwoLoopsXHThree\
    +
    \DefectSSSSTwoLoopsXHFour =\ } 
    +
    \frac{1}{16 \pi^2} (4 - \log x (1-x) + \log \tau_4^2) I_{34} Y_{123}
    \biggr)\,.
    \label{eq:1111_XH}
\end{align}

\subsection{Spider diagram}
\label{subsubsec:SpiderDiagram}

We call spider diagram the diagram with a fermionic loop.
By using the fermionic star-triangle identity \eqref{eq:StarTriangle} and the trace identity
\begin{equation}
    \tr \slashed{x}_1 \slashed{x}_2 \slashed{x}_3 \slashed{x}_4
    =
    16
    [
    (x_1 \cdot x_2) (x_3 \cdot x_4)
    -
    (x_1 \cdot x_3) (x_2 \cdot x_4)
    +
    (x_1 \cdot x_4) (x_2 \cdot x_3)
    ]\,,
    \label{eq:TraceIdentity}
\end{equation}
the diagram can be shown to be equal to
\begin{align}
    \DefectSSSSTwoLoopsSpider
    &=
    \frac{\lambda^4}{4} \tr \int d^4 x_5 \int d^4 x_6\, I_{25} I_{46}\, \spd_6 \spd_5 Y_{156}\, \spd_5 \spd_6 Y_{356} \notag \\
    &=
    \frac{\lambda^4}{4}
    \biggl(
    I_{34} H_{23,11}
    +
    I_{34} H_{13,22}
    +
    I_{34} H_{12,33}
    +
    K_{44} \notag \\
    &\phantom{=\ }
    -
    2 H_{12,23}
    -
    \frac{1}{16 \pi^2} \log \tau_4^2\, I_{34} Y_{123}
    \biggr)\,,
    \label{eq:1111_Spider}
\end{align}
in the conformal frame \eqref{eq:ConformalFrame}.
In order to obtain this result, we also used the result for the kite integral given in \eqref{eq:KiteIntegral}.

\section{Boundary diagrams}
\label{subsec:BoundaryDiagrams1111}

We now treat the boundary diagrams of Table \ref{table:Diagrams1111NNLO}.
They all consist of one $X$- and one $Y$-vertex, with the insertion of a gluon field coming from the expansion of the Wilson line.
The results are expressed in terms of divergent integrals and Goncharov polylogarithms.

\subsection{First XY diagram}
\label{subsubsec:FirstXYDiagram}

The first such diagram is given by the two slices
\begin{align}
    \DefectSSSSTwoLoopsXYOne
    &\sim
    \frac{\lambda^4}{8} \tau_{13}^2 \tau_{24}^2 \int_{-\infty}^{\tau_2} d\tau_6\, \veps(1 3 6) \int d^4 x_5\, I_{25} I_{35} I_{45}\, \pd_{15} Y_{156} \notag \\
    &\sim
    \frac{\lambda^4}{4} I_{34} H_{23,11}
    -
    \frac{\lambda^4}{8} I_{34} H_{12,23} \notag \\
    &\phantom{=\ }
    +
    \frac{I_{34}}{2048 \pi^6}
    \biggl(
    (G(0,0,1,x)+(x - (1-x)) (G(0,1,0,x) + G(1,1,0,x)) \notag \\
    &\phantom{=\ }
    -
    2 x (G(0,1,1,x) + G(1,0,0,x)) + G(1,0,1,x)
    +
    \frac{\pi^2}{3} (1-x) G(1,x)
    \biggr)\,.
    \label{eq:1111_XY1}
\end{align}

\subsection{Second XY diagram}
\label{subsubsec:SecondXYDiagram}

For the second diagram, we obtain
\begin{align}
    \DefectSSSSTwoLoopsXYTwo
    &=
    \frac{\lambda^4}{8} \tau_{13}^2 \tau_{24}^2 \int_{\tau_1}^{\tau_3} d\tau_6\, \veps(2 4 6) \int d^4 x_5\, I_{15} I_{35} I_{45}\, \pd_{25} Y_{256} \notag \\
    &\sim
    \frac{\lambda^4}{4} I_{34} H_{12,33}
    -
    \frac{\lambda^4}{8} I_{34} H_{12,23} \notag \\
    &\phantom{=\ }
    +
    \frac{I_{34}}{2048 \pi^6}
    \biggl(
    x (2 (G(1,0,0,x)- G(0,1,1,x)) + G(1,0,1,x) + G(1,1,0,x)\notag \\
    &\phantom{=\ } + 3 \zeta_3)
    +
    (1-x) G(0,0,1,x) - (1+x) G(0,1,0,x)\notag \\
    &\phantom{=\ }
    +
    2 (G(0,1,1,x)-G(1,0,1,x))
    -
    \frac{\pi^2}{3} G(1,x)
    \biggr)\,.
    \label{eq:1111_XY2}
\end{align}

\subsection{Third XY diagram}
\label{subsubsec:ThirdXYDiagram}

The third XY-diagram yields
\begin{align}
    \DefectSSSSTwoLoopsXYThree
    &=
    \frac{\lambda^4}{8} \tau_{13}^2 \tau_{24}^2 \int_{\tau_2}^{\tau_4} d\tau_6\, \veps(1 3 6) \int d^4 x_5\, I_{15} I_{25} I_{45}\, \pd_{35} Y_{356} \notag \\
    &\sim
    \frac{\lambda^4}{4} I_{34} H_{13,22}
    -
    \frac{\lambda^4}{8} I_{34} (H_{12,23} + H_{13,23}) \notag \\
    &\phantom{=\ }
    -
    \frac{I_{34}}{2048 \pi^6}
    \biggl(
    2 x (G(1,0,1,x) + 3 \zeta_3) + 2 (1-x) (G(0,1,1,x) + G(1,0,0,x)) \notag \\
    &\phantom{=\ }
    -
    G(0,1,0,x)
    -
    G(1,0,1,x)
    -
    G(1,1,0,x)
    +
    (x - (1-x)) G(0,0,1,x) \notag \\
    &\phantom{=\ }
    +
    \frac{\pi^2}{3} x G(1,x)
    \biggr)\,.
    \label{eq:1111_XY3}
\end{align}

\subsection{Fourth XY diagram}
\label{subsubsec:FourthXYDiagram}

The last diagram is found to be
\begin{align}
    \DefectSSSSTwoLoopsXYFour
    &\sim
    \frac{\lambda^4}{8} \tau_{13}^2 \tau_{24}^2 \int_{-\infty}^{\tau_1} d\tau_6\, \veps(2 4 6) \int d^4 x_5\, I_{15} I_{25} I_{35}\, \pd_{45} Y_{456} \notag \\
    &\sim
    \frac{\lambda^4}{4} K_{44}
    -
    \frac{\lambda^4}{8} (K_{14} + K_{34}) \notag \\
    &\phantom{=\ }
    +
    \frac{I_{34}}{2048 \pi^6}
    \biggl(
    2 x G(0,0,x) + G(0,1,x) + G(1,0,x) + 2 (1-x) G(1,1,x) \notag \\
    &\phantom{=\ }
    +
    \biggl(\frac{2\pi^2}{3}-4\biggr) (x G(0,x) + (1-x) G(1,x))
    \biggr)\,.
    \label{eq:1111_XY4}
\end{align}

\chapter{Feynman diagrams for $\vev{1122}$}
\label{app:FeynmanDiagramsOf1122}

We now list the results for the diagrams of the channel $F_1$ of the correlator $\vev{1122}$ at next-to-next-to-leading order.
The relevant diagrams, after subtracting the factorized pieces, are summarized in Table \ref{table:Diagrams1122NNLO}, and can be classified into bulk and boundary diagrams, similarly to $\vev{1111}$.
As before, the final expressions are given in the conformal frame \eqref{eq:ConformalFrame}.

\section{Bulk diagrams}
\label{subsec:1122_BulkDiagrams}

We begin by calculating the bulk diagrams.

\subsection{XX-diagrams}
\label{subsubsec:1122_XXDiagrams}

The XX-diagrams can be calculated to give
\begin{align}
    \NNLOXXOne\
    +
    \NNLOXXTwo\
    &\sim
    \frac{\lambda^5}{4 \pi^2} (1 - \log \veps + \log \tau_4) I_{24} I_{34} Y_{123}\,,
\end{align}
where analytical expressions for all the integrals can be found in Appendix \ref{app:Integrals}.

\subsection{XH-diagrams}
\label{subsubsec:1122_XHDiagrams}

The XH-integrals can be evaluated as well, and they yield
\begin{align}
    \NNLOXHOne\
    +
    \NNLOXHOne\
    &\sim
    \frac{\lambda^5}{8 \pi^2} (3 - 2 \log \veps + 2 \log \tau_4) I_{24} I_{34} Y_{123}\,.
\end{align}

\section{Boundary diagrams}
\label{subsec:BoundaryDiagrams1122}

Boundary diagrams are more intricate, but they can also be evaluated analytically.
The first diagram gives
\begin{align}
    \NNLOXYSubtractOne\ \NNLOSubtractFactorOne
    &\sim
    \frac{\lambda^5}{2048 \pi^{6}}
    \log \veps\, I_{24} I_{34} Y_{123}
    +
    \frac{\lambda^5}{4096 \pi^{6}} \frac{I_{24} I_{34}}{x (1-x)}
    \biggl(
    \frac{x}{6} \bigl(3 (G(0,1,x)+G(1,0,x) \notag \\
    &\phantom{=\ }
    -
    2 G(1,1,x)+2 G(1,0,1,x)-2 G(1,1,0,x)+6 \zeta_3) \notag \\
    &\phantom{=\ }
    +
    2 \left(3-2 \pi^2\right) G(0,x)+2 \left(\pi^2-3\right)
    G(1,x)\bigr)+G(1,x)+G(1,1,x)
    \biggr)\,,
\end{align}
while the second evaluates to
\begin{align}
    \NNLOXYSubtractTwo\ \NNLOSubtractFactorOne
    &\sim
    - \frac{\lambda^5}{128 \pi^{2}}
    \biggl(2 + \frac{2\pi^2}{3} - \log \veps + 2 \log \tau_4 \biggr) I_{24} I_{34} Y_{123} \notag \\
    &\phantom{=\ }
    -
    \frac{\lambda^5}{8192 \pi^{6}} \frac{I_{24} I_{34}}{x (1-x)}
    \left(
    x (G(0,1,x)+G(1,0,x)) + 2 (1-x) G(1,1,x)
    \right)\,.
\end{align}
The last diagram is found to give
\begin{align}
    \NLOX\ \NNLOSubtractFactorTwo
    &\sim
    \frac{\lambda^5}{64 \pi^{2}}
    \biggl(
    \left(\frac{2 \pi ^2}{3}-1\right)
    +
    \log \veps - 2 \log \tau_4
    \biggr) I_{24} I_{34} Y_{123}\,.
\end{align}

\printbibheading
\addcontentsline{toc}{chapter}{Bibliography}

\printbibliography[heading=subbibliography, title={References for Preface}, section=2]
\printbibliography[heading=subbibliography, title={References for Chapter 1}, section=3]
\printbibliography[heading=subbibliography, title={References for Chapter 2}, section=4]
\printbibliography[heading=subbibliography, title={References for Chapter 3}, section=5]
\printbibliography[heading=subbibliography, title={References for Conclusions}, section=6]

\end{document}